\documentclass[rmp,aps,twocolumn,showkeys,floatfix]{revtex4}
\usepackage{float}
\usepackage{amssymb}
\usepackage{graphicx}
\usepackage{epsfig}
\usepackage{dcolumn}
\usepackage{bm}
\usepackage{amsmath}

\def\beq{\begin{equation}}
\def\eeq{\end{equation}}
\def\beqa{\begin{eqnarray}}
\def\eeqa{\end{eqnarray}}

\begin{document}
\title{Memory effects in complex materials and nanoscale systems}
\author{Yuriy V. Pershin$^{1}$}\email{pershin@physics.sc.edu}
\author{Massimiliano Di Ventra$^{2}$} \email{diventra@physics.ucsd.edu}
 \affiliation{$^1$Department of Physics and Astronomy and USC
Nanocenter, University of South Carolina, Columbia, SC, 29208 \\ $^2$Department of
Physics, University of California, San Diego, La Jolla, California
92093-0319}

\bigskip
{\bf This is a preprint of an article submitted for publication in Advances in Physics}
\bigskip

\begin{abstract}
Memory effects are ubiquitous in nature and are particularly
relevant at the nanoscale where the dynamical properties of
electrons and ions strongly depend on the history of the system,
at least within certain time scales. We review here the memory
properties of various materials and systems which appear most
strikingly in their non-trivial time-dependent resistive,
capacitative and inductive characteristics. We describe these
characteristics within the framework of memristors, memcapacitors
and meminductors, namely memory circuit elements whose properties
depend on the history and state of the system. We examine basic
issues related to such systems and critically report on both
theoretical and experimental progress in understanding their
functionalities. We also discuss possible applications of memory
effects in various areas of science and technology ranging from
digital to analog electronics, biologically-inspired
circuits, and learning. We finally discuss future research
opportunities in the field.\bigskip
\end{abstract}


\maketitle

\bigskip

\tableofcontents

\section{Introduction}
\label{sec:intro}

The concept of memory may assume various connotations in different
contexts. For instance, when referring to humans, we generally
mean the act of recalling past experiences. In the context of
computer architectures, we mean the ability to store digital
information for use in computation. In fact, if we look closer
to these two examples - or any other case in which the word
``memory'' is used - we realize that they all share a common
thread that allows us to define it as follows: {\it memory is the
ability to store the state of a system at a given time, and access
such information at a later time}. The physical process of storing
such information may be realized in a variety of ways. In the
brain, for instance, the information seems to be stored in the
synapses (or connections) that are established among different
neurons via the chemicals that are released when the synapses are
excited by ionic (action) potentials~\cite{neuralbook}. In
computer memories of present use a bit of information can be
either stored as a charge in a capacitor (or transistor gate)
\cite{Lai08a} or as spin polarization of certain magnetic materials
\cite{Bratkovsky08a}. Ultimately, all these examples show that a
memory state is related to some {\it dynamical} properties of the
constituents of condensed matter, namely electrons and ions.
Indeed, a closer look would show that history-dependent features are related to how
electrons and/or ions rearrange their state in a given material
under the effect of external perturbations. It is then not
surprising that understanding how memory arises in physical
systems requires us to analyze the properties of materials at the
nanoscale. In turn, as we will emphasize in this review, memory
will emerge quite naturally in systems of nanoscale dimensions:
the change of state of electrons and ions is not instantaneous, and it
generally depends on the past dynamics~\cite{Maxbook}. This means
that the resistive, capacitive and/or inductive properties of
these systems show interesting time-dependent (memory) features
when subject to time-dependent perturbations. In other words, the
systems we will be discussing in this review demonstrate properties of
{\it memristors}~\cite{chua71a},
{\it memcapacitors} or {\it meminductors}~\cite{diventra09a},
namely circuit elements whose resistance, capacitance and
inductance, respectively, depends on the past states through which
the system has evolved.

Our scope for this review is then two-fold. On the one hand, we will briefly examine the most likely microscopic
physical mechanisms that lead to storing of information in several complex
materials and nanoscale systems. Our choice of not delving too much into all possible explanations for each material and/or system is because, in many instances,
these mechanisms are still not completely clear or not agreed upon. Therefore, each case would require an extensive review by itself.
For instance, in some correlated oxides
it is not completely clear whether the dependence on the past
dynamics emerges only from the formation of structural
transitions, or an electronic correlated state forms as well, and
whether these two can, under certain conditions, be separated
\cite{marezio72a,Imada98a,Cavalleri04a}. In the same vein, the
spin response of certain systems shows a peculiar
history-dependence upon time-dependent bias \cite{pershin09a}.
However, the relation between spin and charge memory effects has
not been fully explored. Moreover, mechanisms of resistance
switching in some materials \cite{yang08a} are not yet fully
understood, and are sometimes explained based on totally different physical phenomena
\cite{yang08a,Wu09b}. We will briefly investigate these and other open
questions and critically analyze the results in the existing
literature. We will also point out possible future research
directions to explore them in more depth. Clearly, space limitations do not allow us to
cover the huge existing literature on materials/systems that exhibit memory features. We will then select specific
cases that are particularly instructive, and are representative of larger classes of memory systems.

Irrespective - and this is the main scope of this review - for almost all memory examples
considered we are in a position
to state the general framework
in which this memory fits. We will indeed show that essentially
all systems that exhibit memory fall in one or more of the above
categories of memory-circuit elements. This last aspect is not
just an academic exercise. Rather, the classification of physical
processes in the context of memory-circuit elements sheds more
light on the physical mechanisms at play, and suggests new ways
these systems could be combined to obtain new functionalities. For
instance, by realizing that the potassium and sodium channel
conductances in the classic nerve membrane model of neurons~\cite{hodgkin52a} can in
fact be both identified as memristive~\cite{chua71a}, has recently
inspired these authors to formulate a simple electronic memristive
circuit that accomplishes the tasks of learning and associative
memory of neurons and their networks~\cite{pershin09c}. Likewise,
the memcapacitive properties enabled by the metal-insulator
transition of certain oxides, like VO$_2$, have suggested ways to
tune the frequency response of metamaterials~\cite{driscoll09a}.

The review is then organized as follows: In section \ref{sec:1} we
introduce the notion of memory circuit elements, describe
their general properties and give one simple example for each memory circuit element.
This mathematical framework will be the
basis upon which all memory properties of the materials and
systems we introduce later in the review will be discussed. We
will then discuss actual systems that show memristive (section
\ref{sec:memristors}), memcapacitive (section
\ref{sec:memcapacitors}), and meminductive (section
\ref{sec:meminductors}) behavior based on different physical
properties that lead to memory: structure, charge and spin. In
section \ref{sec:structure}, we will consider memory systems whose
structure and/or dynamics require a more involved analysis with a
combination of the three memory device classes and other basic
circuit elements, and in section \ref{sec:applications} we discuss
applications ranging from information storage to
biologically-inspired circuits. Finally, we conclude in section
\ref{sec:conclusions} and present our outlook for the field.

\section{Definition and properties of memory circuit elements}
\label{sec:1}

We know from classical circuit
theory that there are three fundamental circuit elements
associated with four basic circuit variables, namely the charge,
voltage, current and flux (or time integral of the voltage). For
linear elements these relations take the following forms:
\begin{equation}
V=RI,\label{VIohm}
\end{equation}
for a resistor of resistance $R$, given the current $I$ and the
voltage response $V$.
\begin{equation}
q=C V_C,\label{qVohm}
\end{equation}
for a capacitor of capacitance $C$ that holds a charge $q$ and
sustains a voltage $V_C$ and
\begin{equation}
\phi=L I,\label{phiIohm}
\end{equation}
when the flux $\phi$ is generated by an inductor of inductance $L$
when a current $I$ flows across it. In the equations
(\ref{VIohm}-\ref{phiIohm}), the constants $R$, $C$ and $L$
describe the linear response of the resistor, capacitor and
inductor.

\subsection{General definition} \label{gen_defs}

From both formal and practical points of view the above relations
(\ref{VIohm}-\ref{phiIohm}) can be generalized to time-dependent
and non-linear responses. In addition, all responses may depend
not only on the circuit variables (such as current, charge,
voltage or flux) but also on other state variables. The latter
ones follow their own equations of motion and they provide memory to the system (we will
give explicit examples of these variables as we go along with the
review).

{\it Discrete memory elements -} If $u(t)$ and $y(t)$ are any two complementary constitutive circuit variables
(current, charge, voltage, or flux) denoting input and output of
the system, respectively, and $x$ is an $n$-dimensional vector of
internal state variables, we may then postulate the existence of
the following $n$th-order $u$-controlled memory element as that
defined by the equations~\cite{diventra09a}
\begin{eqnarray}
y(t)&=&g\left(x,u,t \right)u(t) \label{Geq1in}\\ \dot{x}&=&f\left(
x,u,t\right) \label{Geq2in}.
\end{eqnarray}
Here, $g$ is a generalized response, and $f$ is a continuous
$n$-dimensional vector function. Generally, the relation between
current and voltage defines a {\it memristive
system}~\cite{chua76a}, while the relation between charge and
voltage specifies a {\it memcapacitive system}~\cite{diventra09a},
and the flux-current relation gives rise to a {\it meminductive
system}~\cite{diventra09a}. Two other pairs (charge-current and
voltage-flux) are linked through equations of electrodynamics, and
therefore do not define any elements. Devices defined by the
relation of charge and flux (which is the time integral of the
voltage) are not considered as a separate group since such devices
can be redefined in the current-voltage basis~\cite{chua76a}. {\it
Memristors} (short for memory resistors)~\cite{chua71a}, {\it
memcapacitors} (for memory capacitors)~\cite{diventra09a} and {\it
meminductors} (memory inductors)~\cite{diventra09a} are special
{\it ideal} instances of memristive, memcapacitive and
meminductive systems whose definitions are given in what follows.

\begin{figure}[t]
 \begin{center}
\includegraphics[angle=0,width=7cm]{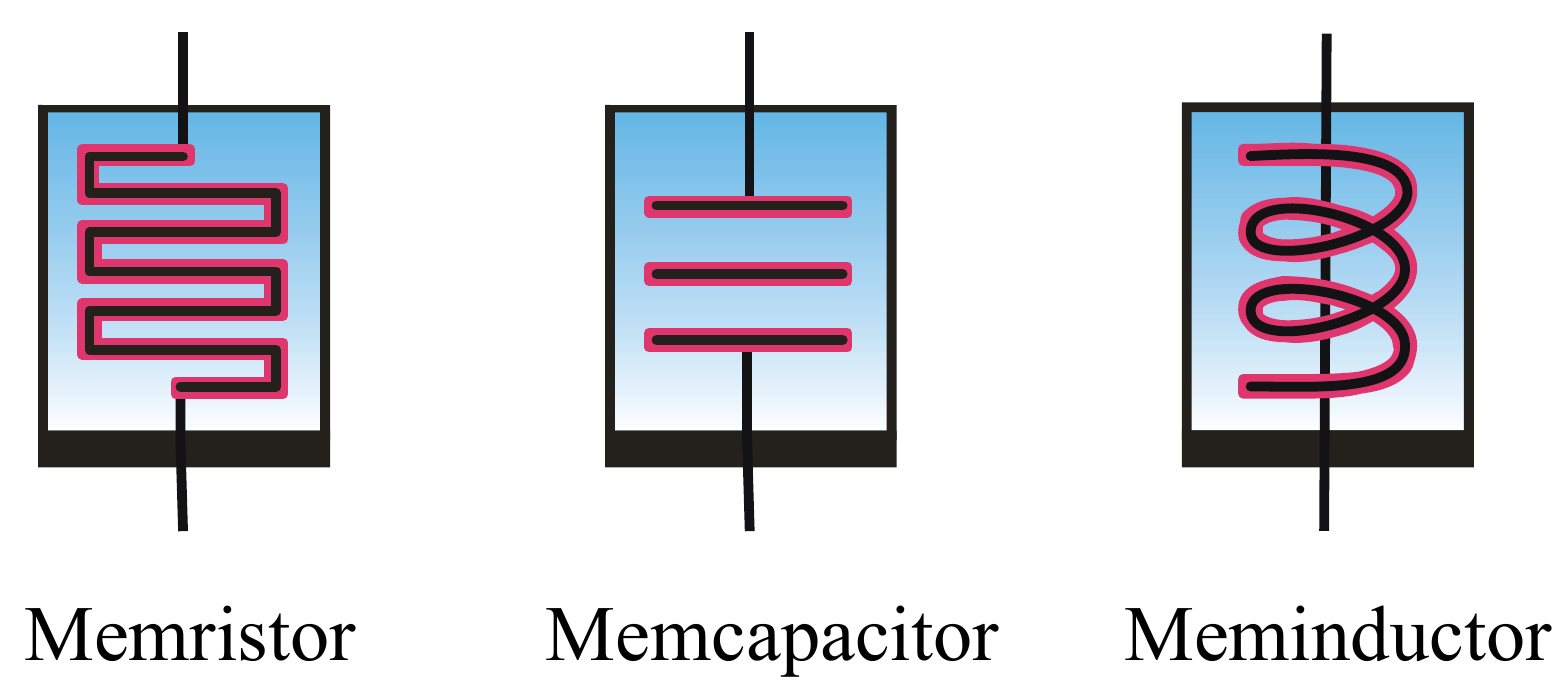}
\caption{\label{figsymbols} Symbols of the three
different devices defined in the text: memristor, memcapacitor,
and meminductor. The same symbols are used to represent the more general classes of
memristive, memcapacitive, and meminductive systems. These devices are generally asymmetric as
indicated by the black thick line. The following convention could
then be used whenever possible or appropriate: when a positive
voltage is applied to the second (upper) terminal with respect to
the terminal denoted by the black thick line, the memory device
goes into a state of high resistance, capacitance or inductance,
respectively. Correspondingly, the device goes into a state of low
resistance, capacitance or inductance when a negative voltage
(with respect to the lower terminal) is applied.}
 \end{center}
\end{figure}

{\it Continuous memory elements -} Furthermore, there are systems in which the internal state is described by a continuous function or functions
instead of discrete variables.
An example of such a situation is spintronics memristive systems \cite{pershin08a} whose spin polarization density (defining the internal state of the device) is a function of coordinates. Therefore, in order to cover such cases, the definition of memory elements given by equations (\ref{Geq1in}-\ref{Geq2in}) has to be generalized appropriately. This can be done in the following way
\begin{eqnarray}
y(t)&=&g\left(X(s,t),u,t \right)u(t) \label{Geq1inA}\\ \dot{X}(s,t)&=&f\left(
X(s,t),u,t\right) \label{Geq2inB}.
\end{eqnarray}
where, generally, $X(s,t)$ is a vector-function (such as a spin-polarization density), $s$ symbolically represents a set of
continuous variables (such as coordinates) and $g(...)$ and $f(...)$ are now functionals. We will call physical systems described by equations (\ref{Geq1inA}-\ref{Geq2inB}) {\it continuous} $u$-controlled memory elements.

{\it Stochastic memory elements -} Finally, the state variables - whether defining a continuous or a discrete set of states - may follow a {\it stochastic differential equation} rather than a deterministic one. In this case, assuming the input $u(t)$ to be deterministic, we may formally define $u$-controlled {\it stochastic} memory elements as
\begin{eqnarray}
\{y(t)\}_{\xi}&=&\{g\left(x,u,t \right)\}_{\xi}\,u(t) \label{Geq1ins}\\ \dot{x}&=&f\left(
x,u,t\right) +H(x)\xi(t)\label{Geq2ins},
\end{eqnarray}
for discrete states, and similarly for continuous states. Here, $H(x)$ is some $n\times n$ matrix function of the state variables allowing for coupling of the noise components, and $\xi(t)$ is an $n$-dimensional vector of noise terms defined by
\begin{equation}
\langle \xi_i(t) \rangle = 0,\;\;\;\;\;\langle \xi_i(t) \xi_j(t') \rangle = k_{ij}(t,t'),\;\;\;\;\;i,j=1,\cdots,n\,,
\end{equation}
where the symbol $\langle \cdots \rangle$ indicates ensemble average, and $k_{ij}(t,t')$ is the autocorrelation matrix. The symbol $\{\cdots \}_{\xi}$ has then the meaning of a realization of the stochastic
process $\xi(t)$. For white noise on each independent state variable we can choose
$k_{ij}(t,t')=\Gamma_i \delta_{ij}\delta (t-t')$ with $\Gamma_i$ some constants. For colored noise, one would have different autocorrelation functions. For instance, for uncoupled colored noises on each state variable we could
choose the noise to follow the stochastic differential equations ($i=1,\cdots,n$)
\begin{equation}
\dot{\xi}_i(t) = -\frac{1}{\tau_i} \xi_i(t) +l_i(t),
\end{equation}
with $l_i(t)$ white noise defined by $\langle l_i(t) \rangle = 0$ and $\langle l_i(t)\rangle = \Gamma_i\delta(t-t')$, and $1/\tau_i$ the frequencies (colors) of the noise. To the best of our knowledge, stochastic memory elements have not been thoroughly studied
in literature even though we anticipate many interesting and important effects due to noise in these systems.

Figure \ref{figsymbols}
shows the circuit symbols of memristor, memcapacitor, and meminductor. The same symbols are used for memristive, memcapacitive, and meminductive systems.
In the following sections we will discuss in detail these three
memory elements and their properties. Since the majority of examples we provide in this review regard {\it discrete} memory elements, we will drop the word ``discrete" in those cases, and explicitly specify
when the elements are continuous or stochastic.

\subsection{Hysteresis loops} \label{hystloops}

\begin{figure}[bt]
 \begin{center}
\includegraphics[angle=0,width=8cm]{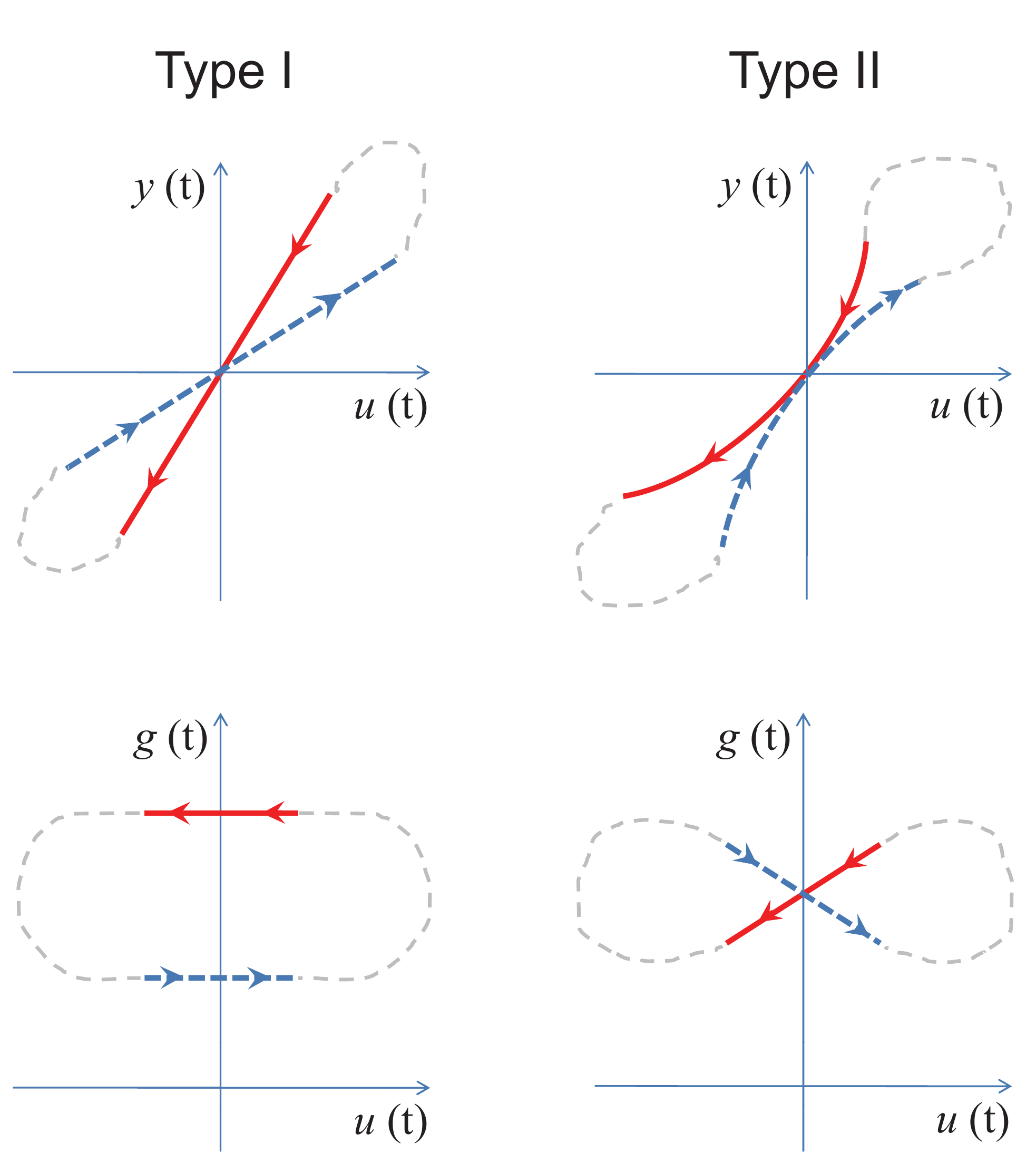}
\caption{The pinched hysteresis loop of memory
elements when subject to a periodic stimulus can be
``self-crossing'' (type I crossing behavior) or not (type II
crossing behavior). The latter property often (but not always) arises when the
state dynamics function $f$ and the response function $g$ are even
functions of the input variable $u$. \label{crossing} }
 \end{center}
\end{figure}

Here, we point out that a distinctive signature of memory devices
is a hysteresis loop. Such loops are very frequently reported in
experimental papers when the response function $g(t)$ or the
function $y(t)$ (or both) are plotted versus $u(t)$. The shape of a
loop is determined by both the device properties and the input $u(t)$
applied. In particular, it depends on both amplitude and frequency
of the input. Therefore, the input $u(t)$ should be fully
specified when measurement results are reported, which is
regrettably not always the case in the existing literature.

Let us now consider the situation in which hysteresis loops are well
defined in the sense that $y(t)$ and $g(t)$ are periodic with a
period $T$ of the applied ac voltage. This may not be always the
case, for example, if equation (\ref{Geq2in}) involves a
stochastic component, or when the switching gradually occurs only
in one direction. There are some common properties of these hysteresis
loops. First of all, we note that as it follows from equation
(\ref{Geq1in}) for well-defined generalized response functions $g$
(by ``well-defined'' we mean $g\neq 0$ and $g \neq \pm \infty$), the
function $y(t)$ hysteresis loop passes through the origin ($y$ is
zero whenever $u$ is zero and vice versa). This is called a
``pinched'' hysteresis loop. A pinched loop may be ``not self-crossing'' or
``self-crossing'' (see figure \ref{crossing}).

The symmetry of equations (\ref{Geq1in}-\ref{Geq2in}) does not
always define the type of crossing. However, as it follows
from examples considered in this review, ``not self-crossing'' loops
are very often observed when $g\left(x,u \right)$ and $f\left(x,u
\right)$ are even functions of $u$. We emphasize that this is not
a necessary condition for ``not self-crossing''. For instance, in
the case of unipolar resistance switching (section \ref{unipol}),
$g\left(x,u \right)$ and $f\left(x,u \right)$ are even functions
of $u$, but $I-V$ hysteresis loops show ``self-crossing''. In the
opposite case, when $f\left(x,u \right)$ is an odd function of
$u$, ``self-crossing'' behavior of $y-u$ curves is more common.

The feature of ``not self-crossing'' loops is observed, for
example, in the case of thermistors and elastic memcapacitive systems
(see, e.g., sections \ref{subsec:thermistors} and
\ref{sec:elasticmemc}). We define this hysteresis of {\it type} II
{\it crossing behavior} because it normally results in a double
loop in the response $g$ as a function of the input (see figure
\ref{crossing}, right panel). The other type of hysteresis - shown
in the left panel of figure \ref{crossing} - involves typically a
single loop in the response $g$ as a function of the input.
Therefore, we call such a hysteresis of {\it type} I {\it crossing
behavior}.

We also note that there are situations when the response function
$g$ becomes zero or infinite when $y=0$ or $u=0$ as in the example
of superlattice memcapacitive systems considered in section
\ref{superlatt_memcap}. In this situation, $y-u$ curves do not
pass through the origin. Moreover, in some systems, additional
crossings are possible at $u\neq 0$ as in the case of ionic
channels. We will discuss this example explicitly in section
\ref{ion_channels}.

\subsection{Some remarks on time scales} \label{timescale}

The ability to categorize experimental systems as memristive,
memcapacitive and meminductive in terms of general
equations (\ref{Geq1in})-(\ref{Geq2in}) (or their continuous and/or stochastic counterparts) provides the opportunity
to understand some of their general properties following
directly from the above mentioned equations. For clarity, let us consider discrete memory elements. The
steady-state values of $x$ can be found as a solution of the
algebraic equation
\begin{equation}
f\left( x,u_0,t\right)=0, \label{steady_st}
\end{equation}
which follows from Eq. (\ref{Geq2in}) assuming a constant value of
the external control parameter $u=u_0$ and the existence of a
steady-state. In particular, for non-volatile information storage,
Eq. (\ref{steady_st}) should provide at least two possible solutions
at $u_0=0$.

\begin{figure}[tb]
 \begin{center}
\includegraphics[angle=0,width=5.5cm]{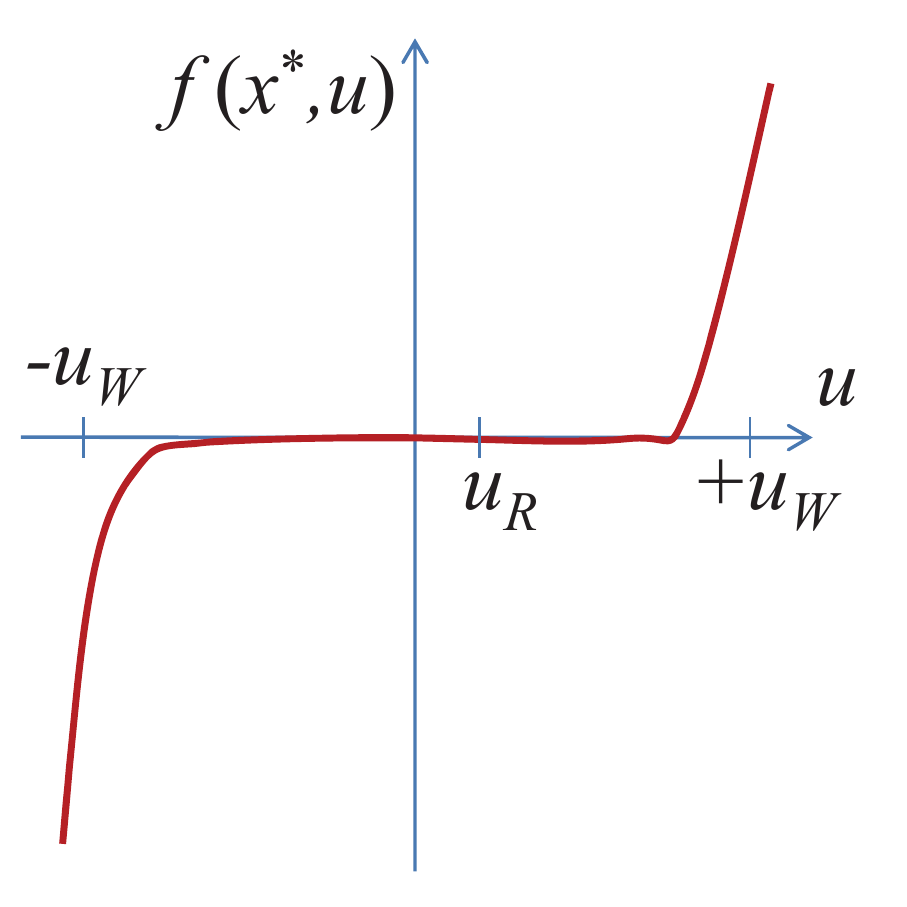}
\caption{ Schematics of the function $f(x^*,u)$ suitable for
non-volatile information storage. Here, $x^*$ is an intermediate
value of $x$, $u_W$ and $u_R$ stand for values of $u$ used to
write and read the information, respectively. For non-volatile memory
applications, it is desirable to have zero or very small $f(x,u_R)$ and finite
$f(x^*,\pm u_W)$. \label{time_scales} }
 \end{center}
\end{figure}

Moreover, the magnitude of $f\left( x,u,t\right)$ in Eq.
(\ref{Geq2in}) can be used to estimate the rates of change of the
internal state variables or, in other words, the relevant time
scales. This information is especially important for non-volatile
memory applications when fast writing and reading time scales, as
well as long-term data retention capability are required. Fig.
\ref{time_scales} shows a sketch of a one-dimensional $f(x,u)$
desirable for non-volatile information storage (for the sake of simplicity we assume that $f$
does not depend explicitly time $t$)
evaluated at a certain value of $x=x^*$ selected between
the minimum $x_{min}$ and maximum $x_{max}$ values of $x$
(assuming that such values do exist).

Generally, the function $f$ can be
asymmetric with respect to $u$, therefore, it can be reasonable to define
ON and OFF write times $t_{W,\pm}$ separately.
In particular, the writing time $t_{W,\pm}$ is of the order of
\begin{equation}
t_{W,\pm}\sim \frac{x_{max}-x_{min}}{f(x^*,\pm u_W)} \label{tW}
\end{equation}
while the change of $x$ during the reading is of the order of
\begin{equation}
\delta x\sim  f(x^*,u_R) t_{R},  \label{delta_x}
\end{equation}
where $t_R$ is the read time.  In addition, Eq. (\ref{delta_x}) can be used to
determine the amount of read cycles before the device state is
altered by consecutive read operation. Roughly, the allowed number
of read operations is of the order of
\begin{equation}
N_{R}\sim \frac{x_{max}-x_{min}}{\delta x} \label{number_reads}.
\end{equation}
We also note that sequential readings of information are not the
only source leading to changes in the internal state of the device. In some
cases, the state degradation can occur naturally, for example,
because of diffusion or certain stochastic processes. Both these effects, in
fact, can be taken into account explicitly using the formalism of stochastic memory circuit
elements given in section \ref{gen_defs}.

After these general considerations we are now ready to define the three different classes of memory elements.

\subsection{Memristors and Memristive Systems}\label{secmemristors}

\subsubsection{Definitions}

The relation between the charge and the flux has been recognized long ago as a missing connection between the four different circuit variables~\cite{chua71a}. For the sake of this completeness let us then postulate a relation of the type
\begin{equation}
\phi(t)=M(F(t)) q(t),\label{qphi}
\end{equation}
between the charge that flows in the system and the flux $\phi$ - which does not need to have a magnetic interpretation.
The proportionality function $M(F(t))$ is some response functional, with $F(t)$ a time-dependent function to be specified later.
Recalling that the flux is related to the voltage via
\begin{equation}
\phi(t)=\int_{-\infty}^t  dt' V(t'),
\end{equation} we can
re-write equation~(\ref{qphi}) as
\begin{equation}
\int_{-\infty}^t  dt' V(t')=M(F(t)) q(t).\label{qphi1}
\end{equation}
By differentiating the above relation with respect to time, and recalling that the current $I(t)=dq/dt$, we obtain
\begin{equation}
V(t)=\left.\frac{dM}{dt'}\right|_{t'=t}q(t) +M(F(t)) I(t).\label{VMI}
\end{equation}
Let us now define the quantity
\begin{equation}
R_M(t)\equiv I^{-1}(t)\left.\frac{dM}{dt'}\right|_{t'=t}q(t) +M(F(t)), \label{defM}
\end{equation}
which has the dimensions of a resistance, so that
equation~(\ref{VMI}) can be compactly written as
\begin{equation}
V(t)=R_M(t) I(t).\label{VMI0}
\end{equation}

If the response $M$ depends only on the charge ($F(t)=q(t)$), namely
$M=M(q(t))$,
 the simple chain rule
\begin{equation}
\frac{dM}{dt}=\frac{\delta M}{\delta q} \times \frac{dq}{dt}\,,
\end{equation}
with $\delta M/\delta q$ the functional derivative of $M$
with respect to the charge function $q(t)$, implies that equation~(\ref{defM}) simplifies as
\begin{equation}
R_M(q(t))= \frac{\delta M}{\delta q}q(t) +M(q(t)), \label{defMq}
\end{equation}
and equation~(\ref{defMq}) as
\begin{equation}
V(t)=R_M(q(t)) I(t)=R_M\left[ \int\limits_{-\infty}^t dt'
I(t')\right] I(t),\label{VMI1}
\end{equation}
where the last step is a simple substitution of $q(t)$ with the time integral of the current. The lower limit of integration can
also be chosen zero provided $\int_{-\infty}^0 I(t')dt'=0$.
Equation~(\ref{VMI1}) defines the relation between the current and
voltage but unlike its ohmic counterpart, equation~(\ref{VIohm}),
it is {\em non-linear} and the function $R_M$ depends not just on
the state of the system at a given time $t$, but {\em on the entire
history of states through which the system has evolved}. This is
in fact explicit in the functional derivative of the response $M$ with respect to the charge dynamics $q(t)$.

Relation~(\ref{VMI1}) is the definition of an {\it ideal
current-controlled} memristor~\cite{chua76a}. It has become
standard in circuit theory to represent it as in
figure~\ref{figsymbols}. It is clear that if the response function
$M$ does not dependent on time, we find from equation~(\ref{defM})
that also $R_M$ is independent of time, and equation~(\ref{VMI1})
reduces to the ohmic form~(\ref{VIohm}).

A memristor with $M$
depending only on the flux, that is $M=M(\phi (t))$, is instead called an
{\it ideal voltage-controlled} memristor. It is described by
\begin{equation}
V(t)=R_M\left[ \int\limits_{-\infty}^t dt' V(t')\right]
I(t),\label{VMI134}
\end{equation}
with the possibility of the lower limit of integration to be zero
provided $\int_{-\infty}^0 V(t')dt'=0$.

However, as we will discuss in the following sections, it may
occur - and this is probably the norm rather than the exception in
real systems - that the response function $M$ depends not just on
the charge that flows across the system but also on one or more
state variables that determine the state of the system at any
given time. This could be, e.g., the position of oxygen vacancies
in TiO$_2$ thin films~\cite{yang08a}, which determines the
resistance of the film, or the temperature of a
thermistor~\cite{chua76a}, or the degree of spin polarization in
certain structures~\cite{pershin08a,pershin09a}. Let us then group
in the symbol $x$ the set of $n$ possible state variables (related
to a particular device) of which we know their time evolution via
the equation
\begin{equation}
\frac{dx}{dt}=f(x,I,t) \label{dx}
\end{equation}
where $f$ is a continuous $n$-dimensional vector function.
The relation between voltage and current can then be written as
\begin{equation}
V(t)=R_M(x,I,t) I(t),\label{VMI2}
\end{equation}
and needs to be solved together with equation~(\ref{dx}) for the
state variables dynamics. These systems (with $f$ and/or $R_M$ depending on $I$) have been called
current-controlled {\em memristive systems}~\cite{chua76a}. The
voltage-controlled ones are those that satisfy the relations
\begin{eqnarray}
I(t)&=&G\left(x,V,t \right)V(t) \label{Condeq1}\\
\dot{x}&=&f\left( x,V,t\right), \label{Condeq2}
\end{eqnarray}
where $G$ is called the {\it memductance} (for memory conductance).

It might be well to point out that the classification of
memristive systems (as well as memcapacitive and meminductive
systems we discuss in the next sections) into current-controlled and voltage-controlled types is, in most cases,
rather a matter of mathematical (or experimental) convenience. In
particular, if we consider equation (\ref{VMI2}) describing a
current-controlled memristive system and algebraically solve it
with respect to the current $I$ (assuming a unique solution at any given time), then we actually get the
equation (\ref{Condeq1}) of the voltage-controlled memristive
system. The second equation (\ref{Condeq2}) can be obtained if we
substitute $I$ obtained from equation (\ref{VMI2}) into
equation (\ref{dx}). Thus, having equations of a current-controlled
memristive system, we can re-write them in the form of a
voltage-controlled one and vice-versa, provided that
a unique solution of equation (\ref{VMI2}) with respect to the current, or
a unique solution of equation (\ref{Condeq1}) with respect to the voltage
can be found.

As noted before, in real systems, it is very unlikely that the
memristor's state depends only on the charge that flows through the
system or on the integral over the applied voltage. In other words,
ideal memristors are probably rare - a notable exception is a small dissipative component in Josephson junctions, see
section~\ref{memJJ}. That is possibly the reason why many authors use these words
interchangeably so that the word ``memristor" indicates any resistive system with memory that
satisfies the coupled equations~(\ref{VMI2}) and~(\ref{dx})
or~(\ref{Condeq1}) and~(\ref{Condeq2}). In fact, even the authors of the
paper \cite{strukov08a} which has rejuvenated this field label their TiO$_2$ device as memristor while, in reality, it is a memristive system. Actually, an attempt to describe the response of such memristive systems using the ideal memristor's equation (\ref{VMI1}) or (\ref{VMI134}) can
result in inadequate  results, e.g., in switching at low applied biases or currents.
The experimental evidence of ideal memristor behavior should contain, for instance, a $\phi- q$ plot showing a
single (non-linear) curve when the device is driven by
a periodic input. In this review, in order to avoid additional confusion, we will
use the terms ``memristor'' and ``memristive system'' consistently with the
definitions given above. The same will be applied also to "memcapacitors" and "memcapacitive systems" as well as to "meminductors" and "meminductive systems".

\subsubsection{Properties of Memristors and Memristive Systems}
\begin{figure}[t]
 \begin{center}
\includegraphics[angle=0,width=7cm]{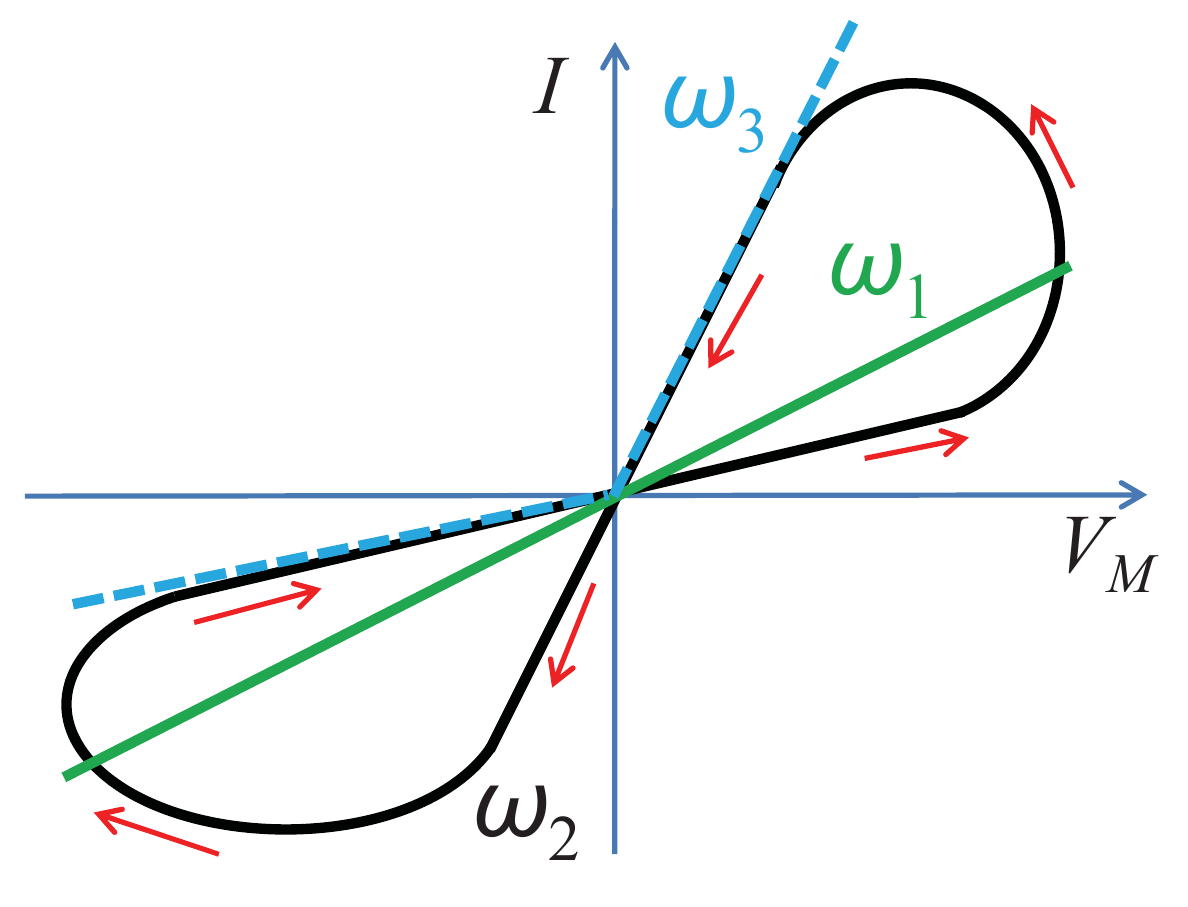}
\caption{Schematics of a pinched hysteresis loop of
a memristive system subject to a periodic stimulus. The size of
the hysteresis depends on the applied voltage frequency: at low
frequencies memristive systems typically behave as non-linear resistors,
at intermediate frequencies they exhibit pinched hysteresis loops,
and at high frequencies they typically operate as linear
resistors. On the plot, $\omega_1 \gg \omega_2 \gg \omega_3$. \label{memexample}
}
 \end{center}
\end{figure}

There are several properties that are relevant to identifying
materials and systems as memristors or memristive. As anticipated in section~\ref{hystloops}, arguably the most important
one is the appearance of a ``pinched hysteretic loop'' in the
current-voltage characteristics of these systems when subject to a
periodic input~\cite{chua76a}. This is obvious from
equations~(\ref{VMI2}) and~(\ref{dx}), because when $V=0$ then
$I=0$  (and vice versa), as it is illustrated in
figure~\ref{memexample}. In addition, if the
state equation~(\ref{dx}) has a unique solution at any given
time $t\geq t_0$, if the voltage (or current) is periodic,
then during each period the $I-V$ curve is a simple loop passing through the origin,
namely there may be at most two values of the current $I$ for a
given voltage $V$, if we consider a voltage-controlled device, or
two values of the voltage $V$ for a given current $I$, for a
current-controlled system.

Moreover, when subject to a periodic stimulus, a memristive system
typically behaves as a linear resistor in the limit of infinite
frequency, and as a non-linear resistor in the limit of zero
frequency (provided that $\dot{x}=f\left(x,I\right)=0$ in
equation~(\ref{dx}) has a steady-state solution) - see
figure~\ref{memexample}. These last two properties can be
understood quite easily. Irrespective of the physical mechanisms
that define the state of the system, at very low frequencies, the
system has enough time to adjust its value of resistance to the
instantaneous value of the voltage (or current), so that the
device behaves as a non-linear resistor. On the other hand, at
very high frequencies, there is not enough time for any kind of
resistance change during a period of oscillations of the control
parameter, so that the device operates as a usual (typically
linear) resistor~\cite{chua76a}.

If $R_M\left( x,I,t\right)\geq 0$, namely the resistance is positive at any given time, then
the quantity (which is the energy stored in the system)
\begin{equation}
U_M=\int\limits_{t_0}^{t}V(\tau)I(\tau)d\tau=\int\limits_{t_0}^{t}I^2(\tau)R_M\left( x,I,t\right)d\tau\geq 0 \label{energM}
\end{equation}
is always positive. In other words, the amount of energy removed
from a memristive system cannot exceed the amount of previously added
energy: memristors and memristive systems are {\it passive} devices. Typically, the electrical energy transforms into heat, which is
then dissipated due to a heat exchange with the environment. Another
important feature, that follows from the fact that the current is
zero when the voltage is zero (and vice versa), is the so-called
``no energy discharge property'' namely a memristive system can not store
energy, like a capacitor or an inductor. These are the main
properties that we will use to identify memristive systems when
discussing practical examples.

\begin{table*}
{ \renewcommand{\arraystretch}{1.6}
\begin{tabular}{| l | c | }
\hline
Physical system & Thermistor \\
\hline
Internal state variable(s) & Temperature, $x=T$ \\
\hline
Mathematical description & $V=R_0e^{\beta\left( \frac{1}{x}-\frac{1}{T_0}\right)}I$  \\
& $\frac{\textnormal{d}x}{\textnormal{d}t}=C_h^{-1}R_0e^{\beta\left(
\frac{1}{x}-\frac{1}{T_0}\right)}I^2+C_h^{-1}\delta \left(
T_{env}-x\right)$ \\
\hline
System type & First-order current-controlled
memristive system \\
\hline
\end{tabular}
}
\caption{Memristive model of thermistor.}
\label{table:term0}
\end{table*}

\subsubsection{Simple example: Thermistors} \label{subsec:thermistors1}

\begin{table*}
{ \renewcommand{\arraystretch}{1.6}
\begin{tabular}{| l | c | }
\hline
Physical system & Thermistor \\
\hline
Internal state variable(s) & Temperature, $x=T$ \\
\hline
Mathematical description & $I=\left[R_0e^{\beta\left( \frac{1}{x}-\frac{1}{T_0}\right)}\right]^{-1}V$  \\
& $\frac{\textnormal{d}x}{\textnormal{d}t}=C_h^{-1}\left[
R_0e^{\beta\left( \frac{1}{x}-\frac{1}{T_0}\right)}\right]^{-1}V^2+C_h^{-1}\delta \left(
T_{env}-x\right)$ \\
\hline
System type & First-order voltage-controlled
memristive system \\
\hline
\end{tabular}
}
\caption{Alternative memristive model of thermistor in terms of equations of voltage-controlled memristive system.}
\label{table:term1}
\end{table*}

\begin{figure*}
 \begin{center}
\includegraphics[angle=0,width=14cm]{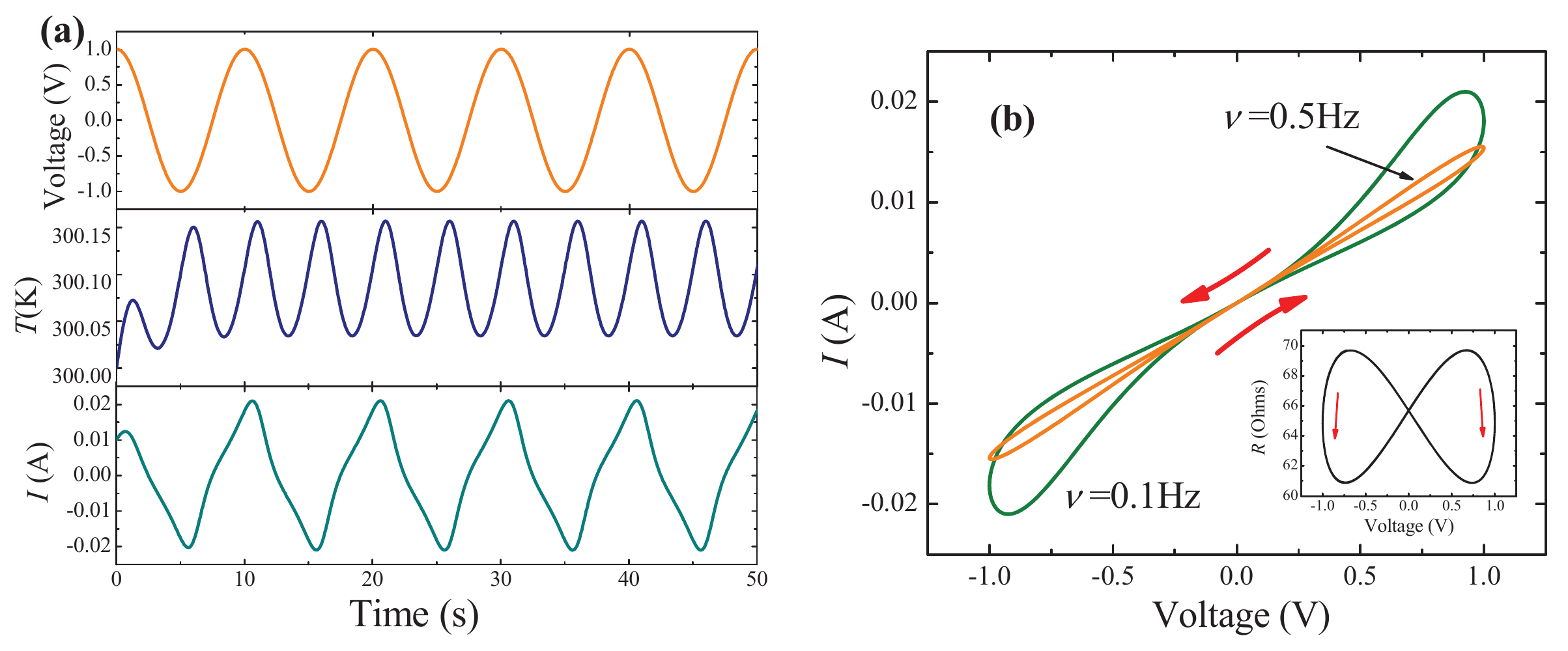}
\caption{Simulations of an ac-biased thermistor. It should be emphasized
that the $I-V$ lines in (b) do not self-cross at (0,0) (see also
figure \ref{crossing}). The plots were obtained using the
parameters values $T_0=T_{env}=300$K, $R_0=100\Omega$, $C_h=0.1$J/K,
$\delta=0.1$J/(s$\cdot$K), $\beta=5\cdot 10^5$K. The applied
voltage is $V(t)=V_0\cos \left(2\pi \nu t\right)$ with $V_0=1$V and
$\nu =0.1$Hz in (a). The inset in (b) shows the resistance versus voltage obtained
at $\nu=0.5$Hz.
 \label{fig:thermo}}
 \end{center}
\end{figure*}

Let us consider one of the first identified memristive systems:
the thermistor. The latter one is a temperature-sensitive
resistor. Although all resistances generally vary with
temperature, thermistors are built of semiconducting materials
that are especially sensitive to temperature, such as different oxides of
metals including manganese, iron, nickel, cobalt,
copper, and zinc. Memristive properties of thermistors are based on
self-heating and were noticed by Chua and Kang in 1976
\cite{chua76a}. To show that these are indeed memristive systems,
let us consider a negative temperature coefficient (NTC)
thermistor that is described by the current-voltage relation~\cite{sapoff63a}
\begin{equation}
V=R_0e^{\beta\left( \frac{1}{T}-\frac{1}{T_0}\right)}I \label{eq:thermistor}
\end{equation}
where the constant $R_0$ denotes the resistance at a certain
temperature $T_0$, $T$ is the absolute temperature of thermistor
(all temperatures are in Kelvin) and $\beta$ is a
material-specific constant. It follows from equation
(\ref{eq:thermistor}) that the resistance of NTC thermistors
decreases with temperature. The thermistor's temperature $T$
depends on the power dissipated in the thermistor and is described
by the heat transfer equation
\begin{equation}
C_h\frac{\textnormal{d}T}{\textnormal{d}t}=V(t)I(t)+\delta \left(
T_{env}-T\right) \label{eq:heatdissipation}
\end{equation}
where $C_h$ is the heat capacitance, $\delta$ is the dissipation
constant of the thermistor \cite{chua76a}, and $T_{env}$ is the
background (environment) temperature. In the memristive description of
thermistor, its temperature $T$ plays the role of the internal state variable.
In order to obtain a memristive model of thermistor, we substitute $V$ from
equation (\ref{eq:thermistor}) into equation
(\ref{eq:heatdissipation}). Table \ref{table:term0} summarizes the memristive model of thermistor.

As an example of the previously mentioned equivalence of current-controlled and voltage-controlled
memristive system's formulations, we would like to demonstrate that the thermistor can alternatively be described as a voltage-controlled
memristive system. For this purpose,
we can express the current $I$ from equation
(\ref{eq:thermistor}) and substitute it into equation
(\ref{eq:heatdissipation}). Table \ref{table:term1} provides a description of thermistor as a voltage-controlled memristive
system. Such a transformation between current-controlled and voltage-controlled description and vice-versa can be easily performed for many systems. Therefore, in the following,
we will avoid giving a dual description, since very often the transformation between different descriptions is trivial.

Figure \ref{fig:thermo} illustrates a solution of the equations
given in table \ref{table:term1}. It is worth
noting that, because of the symmetry of the equation
for the internal state variable with respect to the change of the
voltage sign, the $I-V$ lines in figure \ref{fig:thermo}(b) do not
self-intersect when passing through (0,0) and, therefore, are of type-II crossing
behavior (cf. figure
\ref{crossing}). This is an interesting distinctive feature of
thermistor's pinched hysteresis loops compared to others we will
discuss later. We also note that the $I-V$ curves for the thermistor demonstrate frequency-dependent hysteresis typical of memristive systems.

\subsection{Memcapacitors and Memcapacitive Systems}

\subsubsection{Definitions}

Let us now extend the above concept of
resistor with memory to the other two fundamental
circuit elements, namely capacitors and inductors, and define
corresponding memory circuit elements~\cite{diventra09a}. It is worth pointing out that
for these memory elements a microscopic
derivation based on response functions has not been developed yet.
We thus follow the axiomatic definitions of
reference~\cite{diventra09a}. We will then show in the following
sections that several systems do indeed fall within these
classifications.

From equations~(\ref{Geq1in}) and~(\ref{Geq2in}) we define a
{\it voltage-controlled memcapacitive system} by the equations
\begin{eqnarray}
q(t)&=&C\left(x,V_C,t \right)V_C(t) \label{Ceq1} \\
\dot{x}&=&f\left(x,V_C,t\right) \label{Ceq2}
\end{eqnarray}
where $q(t)$ is the charge on the capacitor at time $t$, $V_C(t)$
is the corresponding voltage, and $C$ is the {\it memcapacitance}
(for memory capacitance) which depends on the state and history of the system, as it is evident in its dependence on the state variables $x$.
The relation
\begin{eqnarray}
V_C(t)&=&C^{-1}\left(x,q,t \right)q(t) \label{CCeq1} \\
\dot{x}&=&f\left( x,q,t\right) \label{CCeq2}
\end{eqnarray}
defines a {\it charge-controlled
memcapacitive system}, where $C^{-1}$ is an inverse memcapacitance.

A subclass of the above systems corresponds to the case in which the capacitance depends only on
the full history of the voltage, namely
\begin{equation}
C(t)=C\left[\int\limits_{t_0}^tV_C\left( \tau \right)d\tau
\right]\,.
\end{equation}
This equation defines an ideal {\it voltage-controlled memcapacitor}. In this case, equations~(\ref{Ceq1}) and~(\ref{Ceq2}) reduce to

\begin{equation}
q(t)=C\left[\int\limits_{t_0}^tV_C\left( \tau \right)d\tau
\right]\, V_C(t) \label{VCMC}.
\end{equation}
Similarly, for a {\it charge-controlled memcapacitor}

\begin{equation}
V_C(t)=C^{-1}\left[\int\limits_{t_0}^tq\left( \tau \right)d\tau
\right]\, q(t), \label{CCMC}
\end{equation}
where in the above two equations, the lower integration limit
(initial moment of time) may be selected as $-\infty$, or 0 if
$\int_{-\infty}^0 V_C(\tau)d\tau=0$ (in equation~(\ref{VCMC})) and $\int_{-\infty}^0
q(\tau)d\tau=0$ (in equation~(\ref{CCMC})).

We note that in this review (similarly to the reference~\cite{diventra09a})
the name ``memcapacitor'' will be reserved for the special class defined
by equation~(\ref{VCMC}) or equation~(\ref{CCMC}) which represent
{\em ideal} memory capacitors and
the term ``memcapacitive systems'' will be used for the broader class of elements
defined by equations~(\ref{Ceq1}) and~(\ref{Ceq2}). Similarly to the case of memristive systems,
we expect memcapacitive systems to be the norm rather than the
exception. For the sake of clarity, we will not use the terms "memcapacitor" and "memcapacitive systems"
interchangeably.

\subsubsection{Properties of Memcapacitors and Memcapacitive Systems} \label{propmemcap}

\begin{figure}[b]
 \begin{center}
\includegraphics[angle=0,width=7cm]{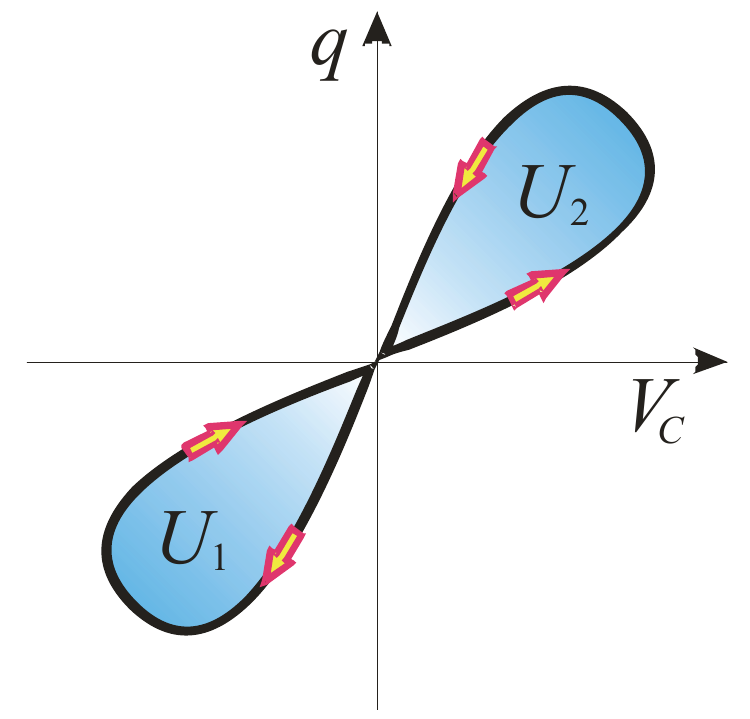}
\caption{Schematics of a pinched hysteresis loop of
a memcapacitive system subject to a periodic stimulus. The energy
added to/removed from the system is the area between the curve and
the $q$ axis (namely $\int V_C(q) dq$). The areas of shaded
regions $U_1$ ($U_2$) give the amount of added (removed) energy in
each half-period. The signs of $U_1$ and $U_2$ are determined by
the direction on the loop. For the direction shown here, $U_2$ is
positive and $U_1$ is negative. \label{figenergy} }
 \end{center}
\end{figure}

It follows from equation~(\ref{Ceq1}) or (\ref{CCeq1}) that, apart from specific
cases that we will discuss later (see, e.g., section~\ref{superlatt_memcap}), the charge is zero whenever the
voltage is zero. Note, however, that in this case, $q=0$ does not
imply $I=0$ (and vice versa), and thus this device can store
energy. Therefore, the latter can be both added to and removed
from a memcapacitive system. However, unlike memristive systems, it is
easy to see that in the present case the relation
\begin{equation}
U_C=\int\limits_{t_0}^{t}V(\tau)I(\tau)d\tau\geq 0 \label{energC1}
\end{equation}
does not always hold. This implies that equations~(\ref{Ceq1})
and~(\ref{Ceq2}) or equations~(\ref{CCeq1})
and~(\ref{CCeq2}) for the memcapacitive systems postulated above
may, in principle, describe both active and passive devices. To
see this point more clearly consider figure \ref{figenergy} where a
schematic memcapacitive system hysteresis loop passing through the
origin is shown when the system is subject to a periodic stimulus
of period $T$. The shaded areas give the energies
\begin{equation}
U_1=\int_0^{T/2} V_C(q) dq\,,
\end{equation}
and
\begin{equation}
U_2=\int_{T/2}^T V_C(q) dq\,,
\end{equation}
added to or removed from the system, respectively, when the integral runs over half of the period. The memcapacitive system can then be
\begin{equation}
U_1+U_2=0\,,\;\;\;\;\;\textrm{non-dissipative}\,,
\end{equation}
\begin{equation}
U_1+U_2>0\,,\;\;\;\;\;\textrm{dissipative}\,,
\end{equation}
or
\begin{equation}
U_1+U_2<0\,,\;\;\;\;\;\textrm{active}\,.
\end{equation}

The fact that a memcapacitive system may be dissipative can be
seen by looking at a practical realization of memory capacitance.
This can be accomplished in two different ways, or their
combination. The simplest realization is via a geometrical change
of the system (e.g., a variation in its structural shape like in
nano-electromechanical systems~\cite{evoy04a}). Alternatively, one
may rely on the quantum-mechanical properties of the free carriers
and bound charges of the materials composing the capacitor that
could give rise, for instance, to a history-dependent permittivity
$\varepsilon(t)$.
In either case, inelastic (dissipative) effects may be involved in
changing the capacitance of the system upon application of the
external control parameter, whether charge or voltage. These
dissipative processes release energy in the form of heating of the
materials composing the capacitor. We will discuss some of these
cases in section~\ref{sec:memcapacitors}.

On the other hand, an active device may be realized when, in order
to vary the capacitance, one needs energy from sources that
control the state variable dynamics, equation~(\ref{Ceq2}), other
than the energy from the control parameter. This energy could be,
e.g., in the form of elastic energy or provided by a power source
that controls, say, the permittivity of the system via a
polarization field. This energy can then be released in the
circuit thus amplifying the current.

Memcapacitive systems share with memristive systems the property that they typically
behave as linear elements in the limit of infinite
frequency, and as non-linear elements in the limit of zero
frequency, assuming that equations (\ref{Ceq2}) and~(\ref{CCeq2})
admit a steady-state solution. The origin of this behavior rests
again on the system's ability to adjust to a slow change in bias
(for low frequencies) and its inability to respond to extremely
high frequency oscillations.

In addition, if the state equation~(\ref{Ceq2}) has only a unique
solution at any given time $t\geq t_0$, then if $V_C(t)$ is
periodic, the $q-V_C$ curve is a simple loop during a period (as that shown in
Fig~\ref{figenergy}), namely there may be at most two values of the
charge $q$ for a given voltage $V_C$, for a voltage-controlled
device, or two values of the voltage $V_C$ for a given charge $q$,
for a charge-controlled system. This loop is also anti-symmetric
with respect to the origin if, for the case of
equations~(\ref{Ceq1}) and~(\ref{Ceq2}), $C\left(x,V_C,t
\right)=C\left(x,-V_C,t \right)$ and $f\left(x,V_C,t
\right)=f\left(x,-V_C,t \right)$.

Finally, we anticipate an important point we will come back to in
sections~\ref{fundamental} and~\ref{superlatt_memcap}. Equations~(\ref{Ceq1}) and~(\ref{Ceq2}) for
memcapacitive systems say nothing about whether these devices may
or may not be constructed as a combination of the basic circuit
elements (resistors, capacitors and inductors) and elements that
can, in fact, be derived from these.

Furthermore, some of these memcapacitive systems show interesting
features which are evident from the definition~(\ref{Ceq1})
and~(\ref{Ceq2}). In fact, there may be cases in which, at certain
instants of time, both the charge and the capacitance are zero,
and thus the voltage across the capacitor may be finite. Similarly, there may be times
when the capacitance {\it diverges} while the voltage
vanishes~\cite{martinez09a,krems2010a}. This gives rise to a
finite charge on the plate of the capacitor. Finally, nothing in
the definition~(\ref{Ceq1}) and~(\ref{Ceq2}) prohibits the voltage
to change sign while the charge does not. This situation
represents the case of {\it dynamical over-screening} and produces
a {\it negative} capacitance~\cite{martinez09a,krems2010a}, and this may occur
without the need of, e.g., ferroelectric materials.

\subsubsection{Simple example: elastic memcapacitive system} \label{sec:elasticmemc1}

\begin{figure}[t]
 \begin{center}
\includegraphics[angle=0,width=7cm]{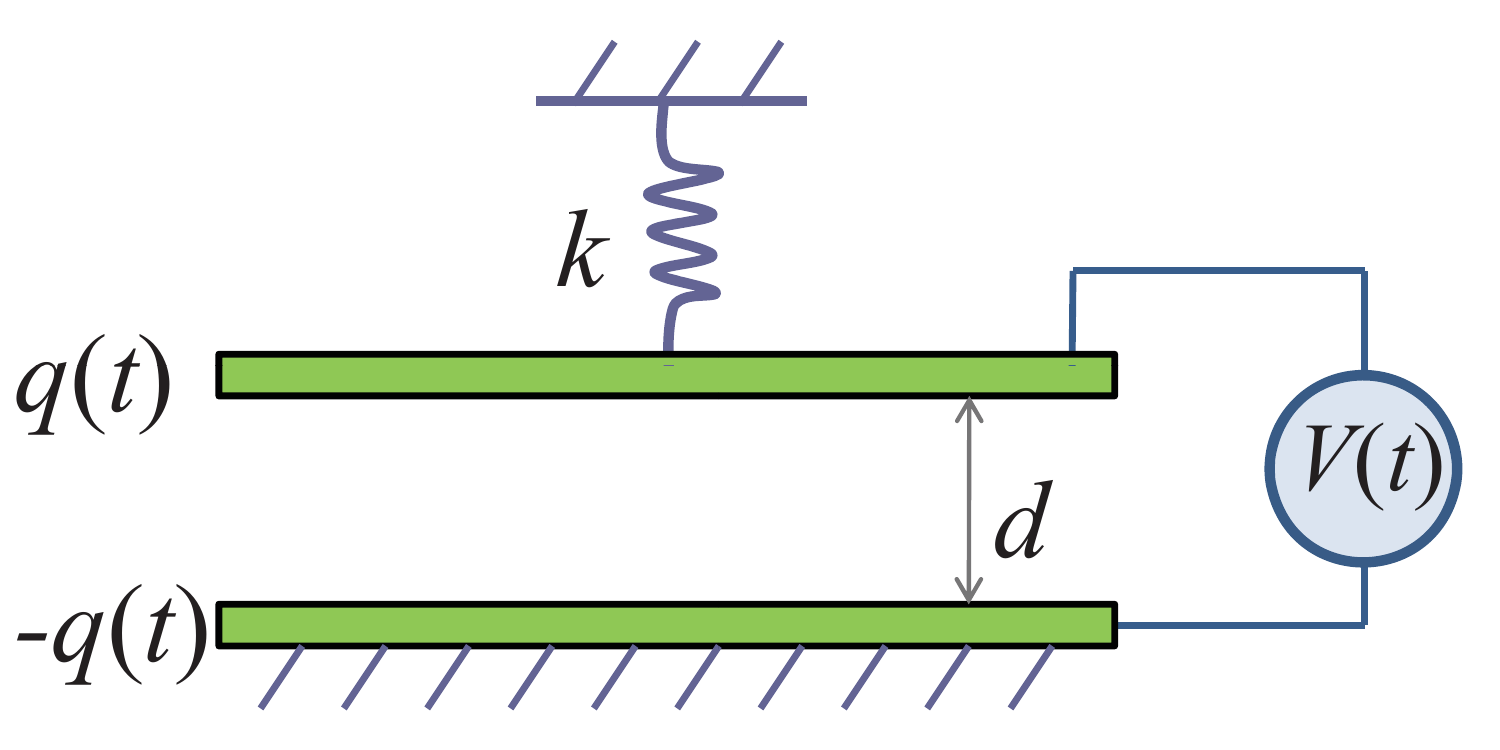}
\caption{Elastic memcapacitive system connected to a voltage source $V(t)$.} \label{elasticmemcap}
\end{center}
\end{figure}

Let us consider a model of a simple electro-mechanical device with
memory, we call {\it elastic memcapacitive system}. Some authors also call
such a system an elastic capacitor \cite{partensky2002-1}. An elastic
memcapacitive system is a parallel-plate capacitor with an elastically
suspended upper plate and a fixed lower plate as shown in figure
\ref{elasticmemcap}. When a charge is added to the plates, the
separation between plates changes as oppositely-charged plates
attract each other. The dynamics of the system depends on initial
conditions and time-dependent fields thus providing a memory
mechanism. Previously, the model of elastic memcapacitive system was used
in studies of lipid bilayers and to explain electrical breakdown
of biological membranes \cite{crowley73a,partensky2002-1}. From
the memory elements standpoint, the elastic memcapacitive system is an
important example of passive memcapacitive device.

The internal state variable $y$ of the elastic memcapacitive system is the
displacement of the upper plate from its equilibrium uncharged
position, $d_0$, under the action of a Coulomb interaction of oppositely charged plates. Mathematically,
the charge dynamics on the elastic memcapacitive system is described by a
parallel-plate capacitor model with variable separation between
the plates
\begin{equation}
q=\frac{C_0}{1+\frac{y}{d_0}}V_C, \label{eq:elastic}
\end{equation}
where $C_0$ is the equilibrium capacitance at $q=0$, and the dynamics of $y$ is given by the classical harmonic oscillator equation
including damping and driving terms:
\begin{equation}
\frac{\textnormal{d}^2y}{\textnormal{d}t^2}+\gamma \frac{\textnormal{d}y}{\textnormal{d}t} +\omega_0^2y+\frac{q^2}{2 \varepsilon_0 m S}=0.
\label{eq:elastic2}
\end{equation}
Here, $\gamma$ is a damping coefficient representing dissipation
of the elastic excitations, $\omega_0=\sqrt{k/m}$, $k$ is the
spring constant, $m$ is the upper plate's mass, $S$ is the plate's
area. It follows from equations
(\ref{eq:elastic}-\ref{eq:elastic2}) that the elastic memcapacitive system
is a second-order charge-controlled memcapacitive system. It is
dissipative when $\gamma>0$ and non-dissipative when $\gamma=0$. The model of
elastic memcapacitive system is summarized in the table \ref{table:elmemc}.

\begin{table*}
{ \renewcommand{\arraystretch}{1.6}
\begin{tabular}{| l | c | }
\hline
Physical system & Elastic memcapacitive system \\
\hline
Internal state variable(s) & Displacement and velocity of upper plate, \\
&$x_1=y$, $x_2=\frac{\textnormal{d}y}{\textnormal{d}t}$ \\
\hline
Mathematical description & $V_C=\left[\frac{C_0}{1+\frac{x_1}{d_0}}\right]^{-1}q$  \\
& $\frac{\textnormal{d}x_1}{\textnormal{d}t}=x_2$ \\
& $\frac{\textnormal{d}x_2}{\textnormal{d}t}=-\left( \gamma x_2 +\omega_0^2x_1+\frac{q^2}{2 \varepsilon_0 m S}\right)$ \\
\hline
System type & Second-order charge-controlled memcapacitive system \\
\hline
\end{tabular}
}
\caption{Model of elastic memcapacitive system.}
\label{table:elmemc}
\end{table*}

\begin{figure*}[tb]
 \begin{center}
\includegraphics[angle=0,width=12cm]{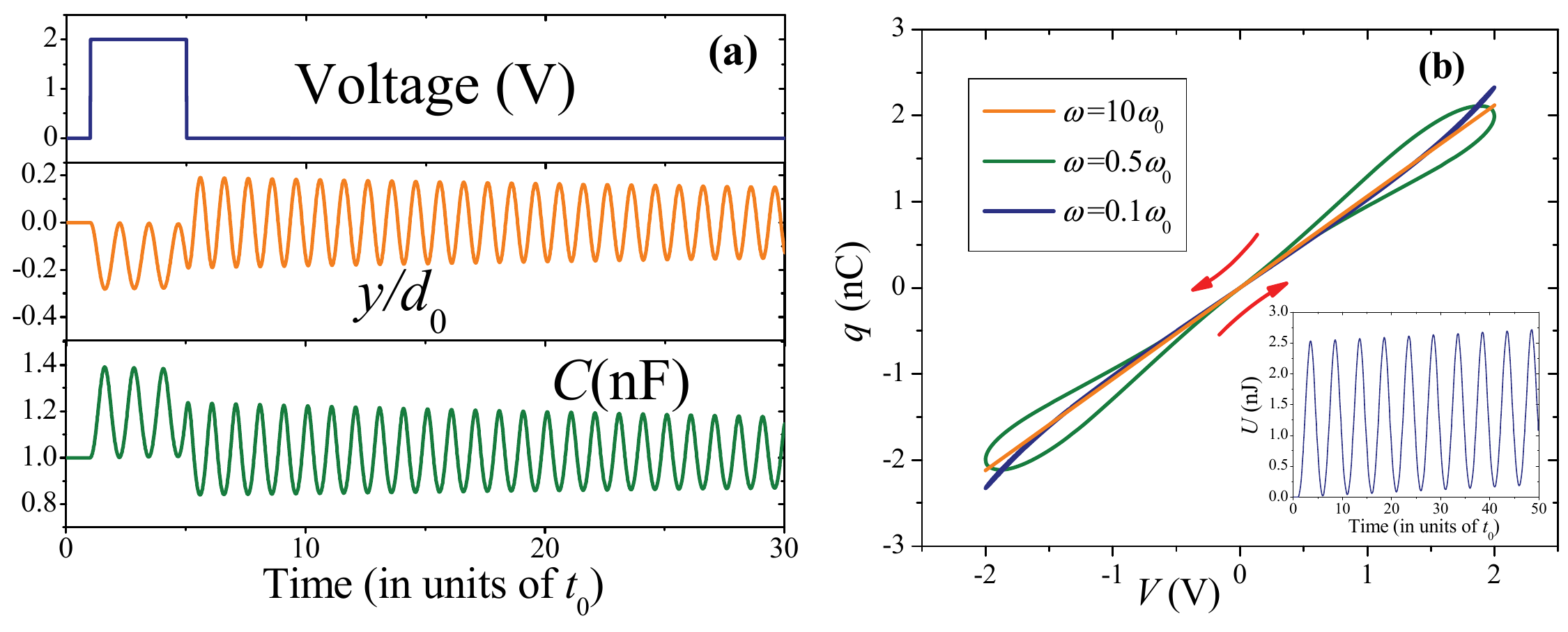}
\caption{Simulations of the elastic memcapacitive system excited by a single voltage pulse (a) and an ac-voltage (b).
We used $C_0=1$nF, $\gamma t_0=0.02$, $t_0=2\pi/\omega_0$, and the ac voltage
amplitude in (b) is 2V. It is interesting to note the absence of
self-crossing in (b) at (0,0) (cf. figure \ref{crossing}). The inset in (b) shows the capacitor energy,
equation~(\ref{energC1}), proving that this element is dissipative.}
\label{elasticmemcapsim}
\end{center}
\end{figure*}

Figure \ref{elasticmemcapsim} shows simulations of the elastic
memcapacitive system. When a single voltage pulse is applied  to the elastic memcapacitive system (figure
\ref{elasticmemcapsim}(a)), the upper plate begins oscillating.
These oscillations last for an extended period of time keeping the
memory about the pulse. In the equation of motion for the state
variable (equation (\ref{eq:elastic2})), the driving force is
proportional to $q^2$. As a consequence, the hysteresis $q-V$ curves of the elastic
memcapacitive system (Fig. \ref{elasticmemcapsim}(b)) do not self-intersect at (0,0), namely they are of
type II, similar to the $I-V$ curves of thermistor (section
\ref{subsec:thermistors1}, see also figure \ref{crossing}).

\subsection{Meminductors and Meminductive Systems}

\subsubsection{Definitions}
Let us now introduce the third class of memory devices, namely
that of memory inductors~\cite{diventra09a}. In order to do this,
let us first define the flux
\begin{equation}
\phi(t)=\int\limits_{-\infty}^t V_L(t')dt',
\end{equation}
where $V_L(t)$ is the induced voltage on the inductor (equal to
minus the electromotive force). In analogy with memristive and
memcapacitive systems we then define from equations~(\ref{Geq1in})
and~(\ref{Geq2in}) a {\it current-controlled meminductive system} one
that follows the relations
\begin{eqnarray}
\phi(t)&=&L\left(x,I,t \right)I(t) \label{Leq1} \\
\dot{x}&=&f\left(x,I,t\right) \label{Leq2}
\end{eqnarray}
where $L$ is the {\it meminductance}. Similarly, we define a
{\it flux-controlled meminductive system} the following

\begin{eqnarray}
I(t)&=&L^{-1}\left(x,\phi,t \right)\phi(t) \label{LLeq1} \\
\dot{x}&=&f\left( x,\phi,t\right) \label{LLeq2}
\end{eqnarray}
with $L^{-1}$ the inverse of the meminductance.

In this work, we will not use interchangeably the terms ``meminductive
system'' and ``meminductor'', reserving the latter for {\it ideal
meminductors}~\cite{diventra09a}. Those are a subclass of the above
systems, when equations~(\ref{Leq1}) and~(\ref{Leq2}) reduce to
\begin{equation}
\phi(t)=L\left[\int\limits_{t_0}^tI\left( \tau \right)d\tau
\right] \, I(t) \label{VCMC1}
\end{equation}
for a current-controlled system, or when equations~(\ref{LLeq1})
and~(\ref{LLeq2}) can be written as
\begin{equation}
I(t)=L^{-1}\left[\int\limits_{t_0}^t \phi\left( \tau \right)d\tau
\right]\, \phi(t),\label{CCMC1}
\end{equation}
for a flux-controlled meminductor.
In the above two equations, the lower integration limit may be
selected as $-\infty$, or 0 if $\int_{-\infty}^0 I(\tau)d\tau=0$
and $\int_{-\infty}^0 \phi(\tau)d\tau=0$, respectively.

\subsubsection{Properties of Meminductors and Meminductive Systems}

Meminductors and meminductive systems share essentially the same properties of
memcapacitive systems, such as an hysteretic loop that may or may
not be pinched - and situations in which the meminductance may
diverge and become negative, see, e.g., equations~(\ref{Leq1})
and~(\ref{Leq2}) - non-linearity at very low frequencies and
linearity at high frequencies, energy storage, and the possibility
to represent passive, dissipative or active elements.

The energy stored in the system can be evaluated as follows. For
the sake of definiteness, let us consider only the
current-controlled meminductive systems. Taking first the time
derivative of both sides of equation (\ref{Leq1}) yields
\begin{equation}
V_L=\frac{\textnormal{d}\phi}{\textnormal{d}t}=L\frac{\textnormal{d}I}{\textnormal{d}t}+
I\frac{\textnormal{d}L}{\textnormal{d}t},
\end{equation}
which shows that the second term on the right-hand side of the above equation is an
additional contribution to the induced voltage due to a
time-dependent inductance $L$. The energy stored in the current-controlled
meminductive system can then be calculated as

\begin{equation}
U_L(t)=\int\limits_{t_0}^t
V_L(\tau)I(\tau)d\tau=\int\limits_{t_0}^t \left[
L\frac{\textnormal{d}I}{\textnormal{d}t}+
I\frac{\textnormal{d}L}{\textnormal{d}t}\right]I(\tau)d\tau.
\label{Lenerg}
\end{equation}

When the inductance, $L$, is constant, we readily obtain the well-known expression
for the energy $U_L=LI^2/2$. This is interpreted as
the energy of the magnetic field generated by the current. However, in the meminductive case, the extra term
due to the time derivative of $L$ may be related to other processes that control the meminductance and which are
formally represented by the state variables $x$. These processes could be related, for instance, to the elastic energy required to
modify the shape of the inductor in time, or an external energy source that controls the
permittivity of the inductor as a function of time.

In general, we can formulate the passivity criterion of meminductive systems by stating that if
at $t=t_0$ they are in their minimal energy state
then $U_L(t)\geq 0$ at any subsequent moment of time, with the inequality sign indicating the presence of energy storage and dissipative processes.

\subsubsection{Simple example: elastic meminductive system} \label{elastic_meminductor}

\begin{figure*}[t]
 \begin{center}
 \centerline{
    \mbox{(a)}
    \mbox{\includegraphics[angle=0,width=8cm]{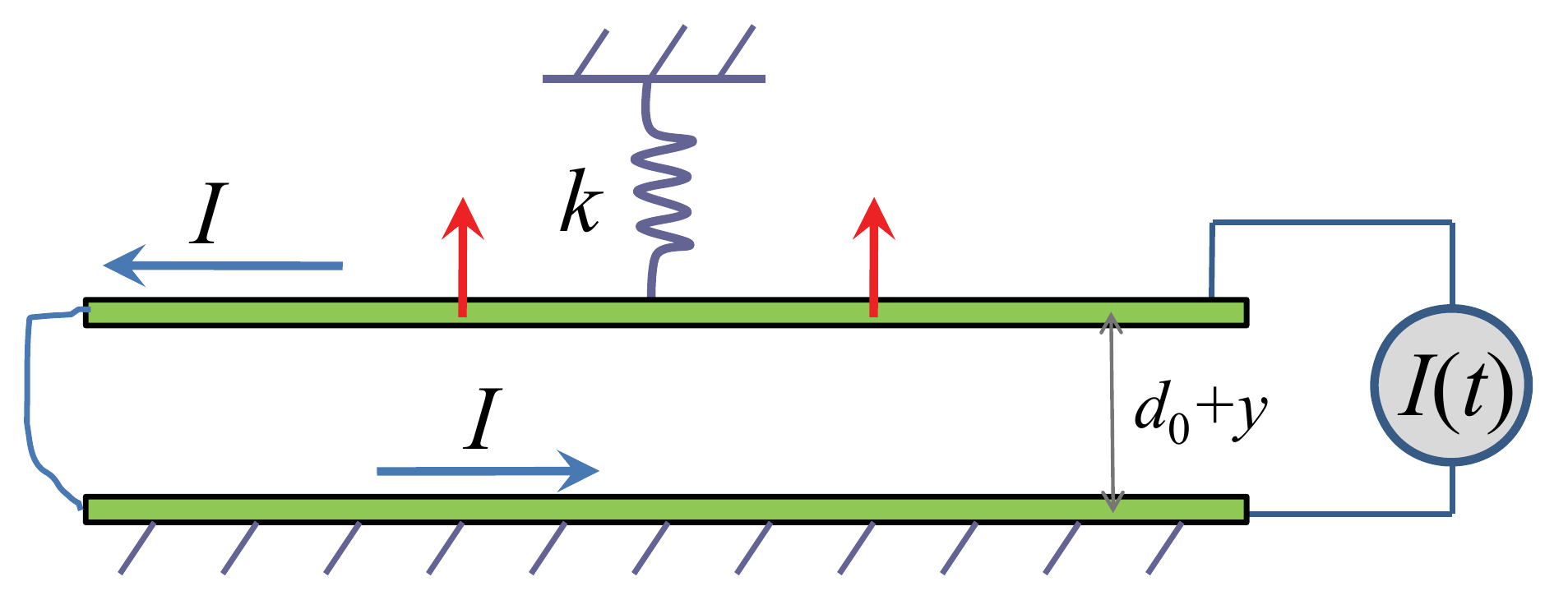}}
  }

  \centerline{
    \mbox{(b)}
    \mbox{\includegraphics[width=6.50cm]{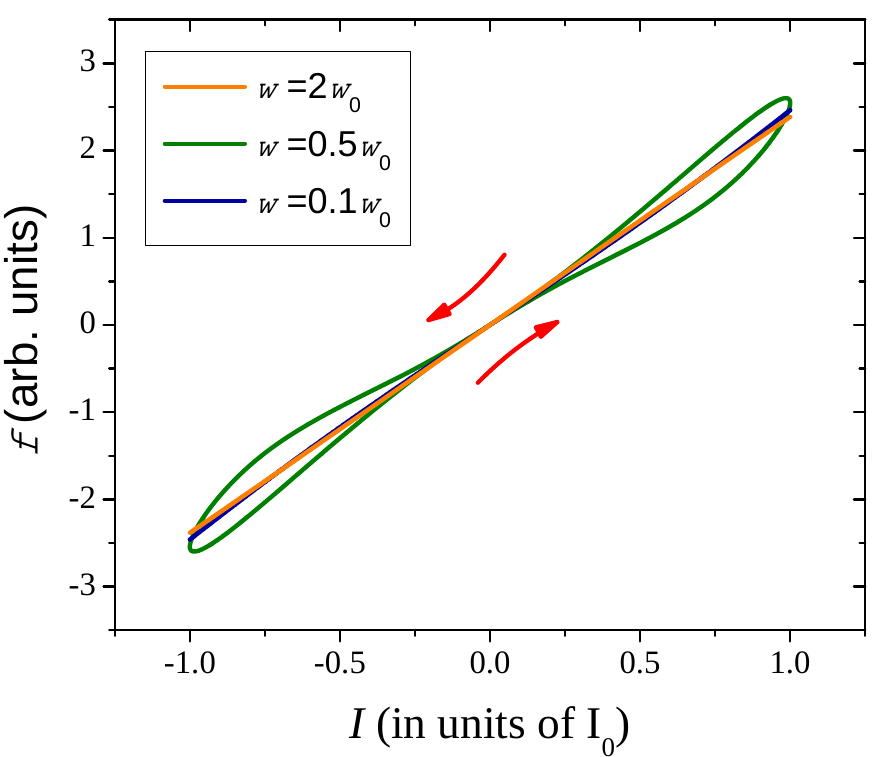}}
    \mbox{(c)}
    \mbox{\includegraphics[width=6.50cm]{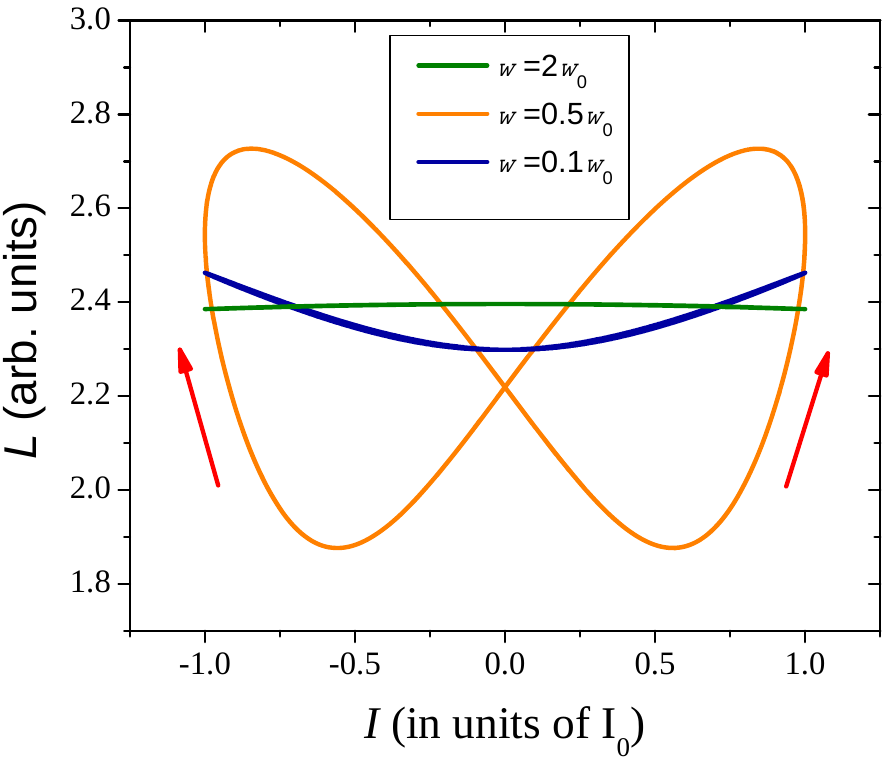}}
  }
\caption{ (a) Schematic of an elastic meminductive system whose operation
is based on repulsion (vertical red arrows represent a repulsive
force) of current-carrying wires of opposite current. The
inductance of the system increases with increasing wires
separation $d_0+y$ from the equilibrium distance $d_0$.  (b) and (c): $\phi-I$ and $L-I$ curves of
elastic meminductive system, respectively. The curves' shapes indicate
type-II hysteretic behavior. These plots were obtained using
parameter values $\gamma / \omega_0=0.2$, $r/d_0=0.1$, $\mu_0 l
I_0^2/(2 \pi m w_0^2 d_0^2)=0.2$. Here, $I_0$ is the ac-current
amplitude.} \label{memind}
\end{center}
\end{figure*}

We here introduce the model of {\it elastic meminductive system}, namely an
inductor which is formed by two parallel wires of length $l$
separated by a distance $d_0+y$ as we depict in figure
\ref{memind}(a). One of the wires is kept fixed while the other is
allowed to move under the effect of a spring, and we assume for
simplicity that both wires have the same radius $r$. The two wires are
connected together as illustrated in figure \ref{memind}(a) in
such a way that they carry currents of equal magnitude but opposite direction. This system
can in principle be realized using micro-electromechanical systems (MEMS) as we will
discuss in section~\ref{geomL}. As current flows through the wires, the
repulsive magnetic force pushes the top wire up compressing the
spring. In this way, the distance between the wires increases
resulting in a different value of inductance according to the
equation
\begin{equation}
L\left( y \right)=\frac{\mu_0
l}{\pi}\textnormal{ln}\frac{d_0+y}{r} \label{Lmemind},
\end{equation}
where $\mu_0$ is the vacuum permeability, $d_0$ is the equilibrium
distance between the wires when $I=0$, and $y$ is the displacement
of the top wire from its equilibrium position at $I=0$.

The classical equation of motion of the top wire, taking into
account damping and the currents interaction force, can be written as
\begin{equation}
\frac{\textnormal{d}^2y}{\textnormal{d}t^2}+\gamma
\frac{\textnormal{d}y}{\textnormal{d}t}+\omega_0^2y-\frac{\mu_0lI^2}{2 \pi m  \left( d_0+y\right)}=0
. \label{eq:elasticIn1}
\end{equation}
Here, $\gamma$ is a damping coefficient representing dissipation
of the elastic excitations, $\omega_0=\sqrt{k/m}$, $k$ is the
spring constant and $m$ is the upper wire's mass. It follows from
equation (\ref{eq:elasticIn1}) that the
elastic meminductive system is a second-order current-controlled
meminductive system. Its properties are presented in table \ref{table:elmemind}.

\begin{table*}
{ \renewcommand{\arraystretch}{1.6}
\begin{tabular}{| l | c | }
\hline
Physical system & Elastic meminductive system \\
\hline
Internal state variable(s) & Displacement and velocity of upper wire, \\
&$x_1=y$, $x_2=\frac{\textnormal{d}y}{\textnormal{d}t}$ \\
\hline
Mathematical description & $\phi=\frac{\mu_0
l}{\pi}\textnormal{ln}\frac{d_0+x_1}{r} I$  \\
& $\frac{\textnormal{d}x_1}{\textnormal{d}t}=x_2$ \\
& $\frac{\textnormal{d}x_2}{\textnormal{d}t}=-\left( \gamma
x_2+\omega_0^2x_1-\frac{\mu_0lI^2}{2 \pi m  \left( d_0+x_1\right)}
\right)$ \\
\hline
System type & Second-order current-controlled meminductive system \\
\hline
\end{tabular}
}
\caption{Model of elastic meminductive system.}
\label{table:elmemind}
\end{table*}

It is dissipative when $\gamma>0$, and
non-dissipative when $\gamma=0$. The overall form of the equation describing dynamics of elastic
meminductive system (Eq. (\ref{eq:elasticIn1}))
is similar to that of elastic memcapacitive system
(\ref{eq:elastic2}), with the difference that the
interaction in the case of the meminductive system is repulsive and depends on
the wires' separation (in the case of elastic memcapacitive system, the
interaction force is attractive and does not depend on the
separation between the plates).

Selected results of elastic meminductive system's simulations are
presented in figure \ref{memind}(b,c). These plots demonstrate
typical frequency dependencies of memory elements. It is
interesting to note the absence of self-crossing in the $\phi-I$
curves, and a self-crossing of $L-I$ curves. Therefore, following
our suggested classification scheme, such a system shows a type-II
hysteresis.

\subsection{Combination of memory features} \label{combined}
\begin{figure}[t]
 \begin{center}
\includegraphics[angle=0,width=6cm]{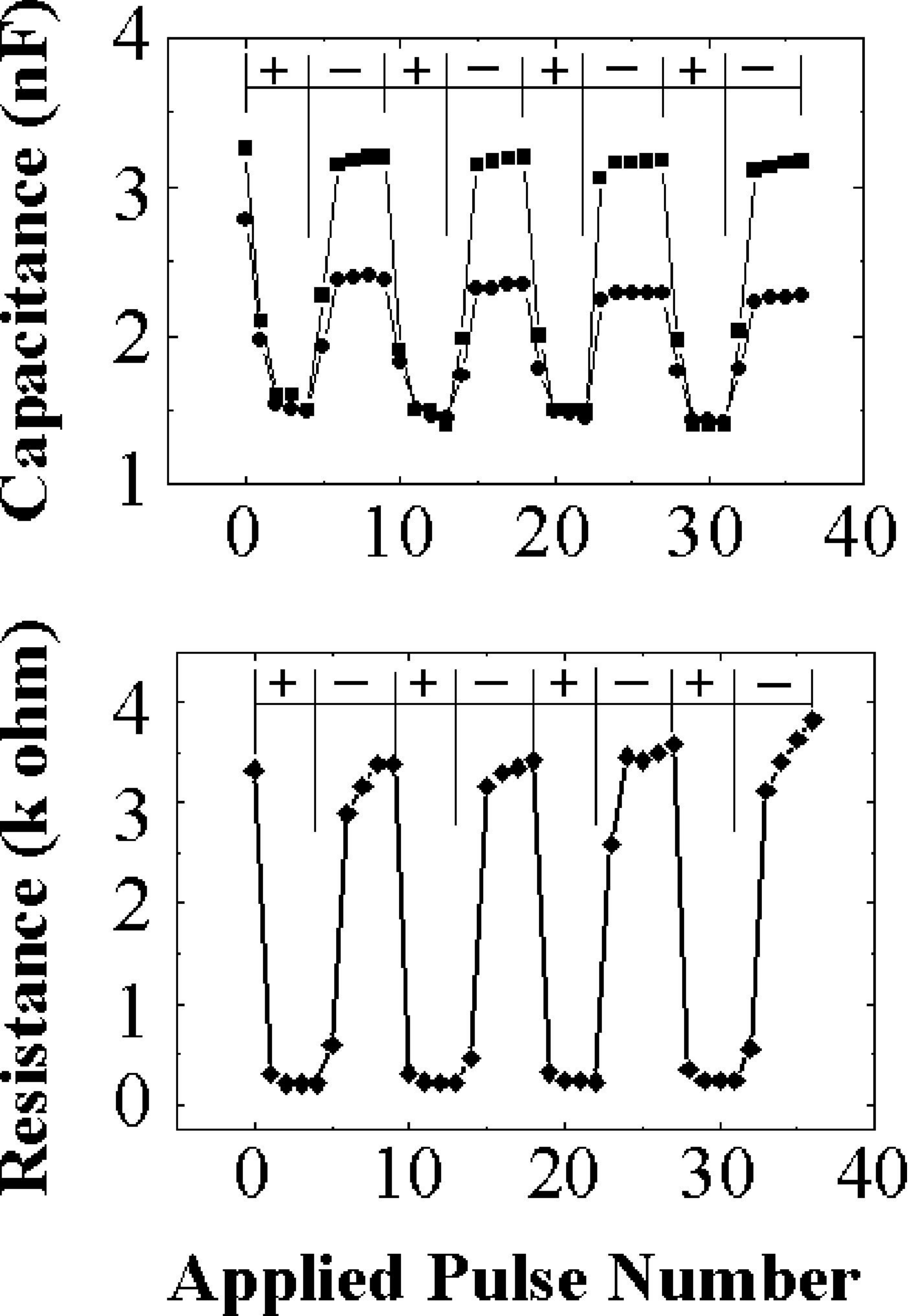}
\caption{ Nonvolatile capacitance and resistance changes versus number of applied pulses
for a Au/PCMO/YBCO/LAO sandwich structure at room
temperature. Here, PCMO is Pr$_{0.7}$Ca$_{0.3}$MnO$_3$, YBCO is
YBa$_2$Cu$)3$O$_{7-x}$ and LAO is LaAlO$_3$. The capacitance values shown by solid circles and square dots were measured at 200kHz and 20kHz, respectively.
Reprinted with permission from \cite{liu06a}. Copyright 2009, American Institute of Physics.\label{Ignat1RC} }
 \end{center}
\end{figure}

At this stage it is worth pointing out that in real systems, the
above three memory features may appear
simultaneously~\cite{diventra09a,martinez09a,driscoll09a}. This is
not surprising, especially at the nanoscale. To give just an
example we recall that when a current flows in a resistor, at the
microscopic level local resistivity dipoles form due to scattering
of carriers at interfaces~\cite{landauer57a}. Namely, charges
accumulate on one side of the resistor with consequent depletion
on the other side, with possible accumulation and depletion over
the whole spatial extent of the resistor~\cite{Maxbook}. It was
shown numerically that these dipoles develop {\em dynamically}
over length scales of the order of the screening
length~\cite{sai07a}. This means that the formation of these local
resistivity dipoles takes some time and is generally accompanied
by some energy storage related to the capacitance of the system.
It is also known that memristive and memcapacitive effects can be simultaneously observed in
the resistance switching memory cells \cite{liu06a}. The experimental data reproduced
in Figure \ref{Ignat1RC} from reference \cite{liu06a} clearly show that the changes in memristance and memcapacitance are correlated and, therefore, in this example they are most probably related to the same state variables.

We thus expect that in nanoscale systems memristive behavior is
always accompanied to some extent by a (possibly very small)
capacitative behavior. In this case the I-V hysteresis loop may
not exactly cross the origin at frequencies comparable to the
inverse characteristic time of charge equilibration processes. We
will give explicit examples of this effect in section
\ref{semic_spin_systems} when discussing the transverse voltage
memory response of the spin Hall effect in inhomogeneous
semiconductor systems~\cite{pershin09a}, and in section
~\ref{phase_change_memcap}, where the memristance and
memcapacitance induced by the metal-insulator transition of VO$_2$
is made clear in the optical response of
metamaterials~\cite{driscoll09a}.

\begin{figure}[t]
 \begin{center}
\includegraphics[angle=0,width=7cm]{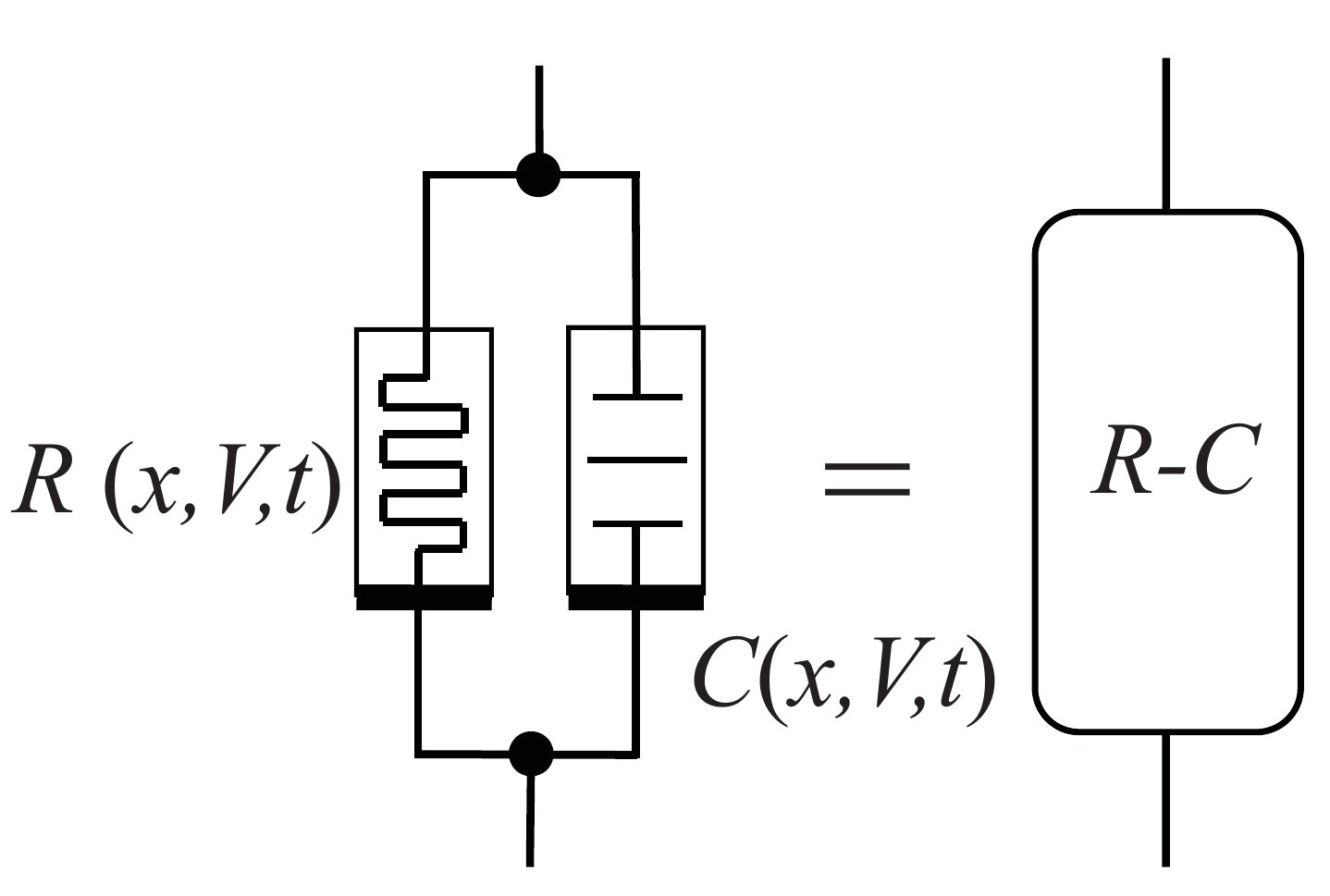}
\caption{The combination of two or more memory circuit elements in parallel or in series
may result in the compact description of a single circuit element with non-algebraic dependence
on the control parameters. In this particular case, a voltage-controlled memristive system and a
voltage-controlled memcapacitive system in parallel can be compactly described as a voltage-controlled
effective memristive system with non-algebraic dependence on bias.\label{MCparallel} }
 \end{center}
\end{figure}

The analysis of these systems may be done by combining a number of three
memory circuit elements in parallel and/or in series, possibly
together with standard resistors, capacitors and inductors. It is
worth mentioning that combination of these elements may result in
compact descriptions of memory systems with non-algebraic
dependence on the control parameters. As an example of this, let
us consider a voltage-controlled memristive system (described by
equations~(\ref{Condeq1}) and~(\ref{Condeq2})) and a
voltage-controlled memcapacitive system (equations~(\ref{Ceq1})
and~(\ref{Ceq2})) coupled in parallel (see figure~\ref{MCparallel}).

For simplicity we may
assume that the same state variables determine the memory of both
the resistor and the capacitor. This, however, may not always
hold. This latter case, however, can easily be accounted for by
expanding the range of state variables $x$ to include both sets.
Using Kirchhoff's law, the total current of the parallel circuit
is
\begin{equation}
I(t)=I_M +\frac{dq}{dt},
\end{equation}
where we indicate with $I_M$ the current that flows across the
memristive system. Introducing the definitions of the two memory elements
(and noticing that in a parallel circuit the voltage drop across
the resistor $V(t)$ is the same as that across the capacitor) we
then get
\begin{equation}
I(t)=G\left(x,V,t \right)V(t) + \frac{d}{dt}\left [ C\left(x,V,t \right)V(t)\right ].
\end{equation}
By defining the effective memductance
\begin{equation}
G_{eff}\left(x,V,t \right)=G\left(x,V,t \right)+\frac{1}{V(t)}\frac{d}{dt}\left [ C\left(x,V,t \right)V(t)\right ],
\end{equation}
we find that the whole circuit behaves like a voltage-controlled memristive system with current
\begin{equation}
I(t)=G_{eff}\left(x,V,t \right)V(t),
\end{equation}
and non-algebraic dependence on the bias $V$.

Along similar lines, Mouttet \cite{mouttet10a} has suggested a memadmittance (memory
admittance) model in order to describe simultaneous
memristive and memcapacitive properties of a system. This idea was
applied to thin-film memory materials. Based on this approach,
equations relating the cross-sectional area of conductive bridges
in resistive switching films to shifts in capacitance were derived
\cite{mouttet10a}. Moreover,  Riaza \cite{riaza10a} has proposed
a {\it hybrid memristor} that is described by a relation involving
all four circuit variables $q$, $\varphi$, $I$ and $V$. Such an element
may account for physical devices in which memory
effects of different nature coexist. Several examples of hybrid memristors as well as
corresponding extension of circuit theory is given in his work \cite{riaza10a}.

\subsection{Are memristors, memcapacitors and meminductors fundamental circuit elements?} \label{fundamental}

Finally, it is worth mentioning that the behavior of an {\it ideal} memristor,
namely that which is described by the relation~(\ref{VMI1}) (or equation (\ref{VMI134})) cannot
be simulated by any combination of ``standard" - namely two-terminal, time-independent, albeit possibly non-linear - resistors, capacitors
and inductors. In other words, there is no
possible finite combination of standard circuit elements that can
reproduce the dynamical properties of ideal memristors. This can be shown by recalling that a memristor does not store energy so that capacitors and inductors cannot possibly be used to simulate it, while a finite
number of non-linear resistors cannot reproduce, e.g., the current history as required by the definition~(\ref{VMI1}). It is for this reason that an ideal memristor is
sometimes called the ``fourth'' circuit element~\cite{chua71a}.

However, also an ideal
memcapacitor and an ideal meminductor (see, e.g., equations~(\ref{CCMC}) and~(\ref{VCMC1})) cannot be simulated by
combinations of standard resistors, capacitors and inductors, because in this case as well, the lack of time dependence does not allow to retain information on the full charge (memcapacitor) or current (meminductor) dynamics.
For instance, one could argue that since standard capacitors store information on the {\it current} history
(the integral of the current is the charge on the capacitor) they could be used to simulate the behavior of meminductors. However, if this were the case, the resulting circuit would have capacitative components and therefore
would not be an ideal meminductor. The same reasoning can be made for using inductors to simulate memcapacitors.

Therefore, it is in this sense that {\it ideal}
memristors, memcapacitors and meminductors can be
considered as ``fundamental'' circuit elements, and it would thus be tempting to call the last two
the ``fifth'' and ``sixth''
circuit elements. These authors, however, do not
share this view, and prefer to subscribe to the notion that there are
only three fundamental circuit elements: resistors, capacitors and
inductors, with or without memory.

We also note that the above considerations can not be extended to memristive, memcapacitive and meminductive
elements because in that case,
the internal
state variables could have a physical origin which could be simulated by standard (possibly non-linear) circuit
elements. For instance, as we will discuss in
section \ref{superlatt_memcap}, some memcapacitive systems may be
represented by a combination of basic circuit elements (capacitors and non-linear resistors; for examples of these, see, e.g.,
references~\cite{krems2010a,martinez09a}). Or one could envision a combination of non-linear resistors
with negative differential resistance \cite{sze_book1} to retain the history of the voltage or current and thus
simulate memristive systems. Irrespective, these types of
memory elements are still of great importance since they provide a
complex functionality within a single electronic
structure~\cite{martinez09a}.

After these general considerations we are now ready to discuss several systems that
exhibit memory features and the physical mechanisms that generate these phenomena.

\section{Memristive systems}
\label{sec:memristors}

\begin{figure}
 \begin{center}
\includegraphics[angle=0,width=6cm]{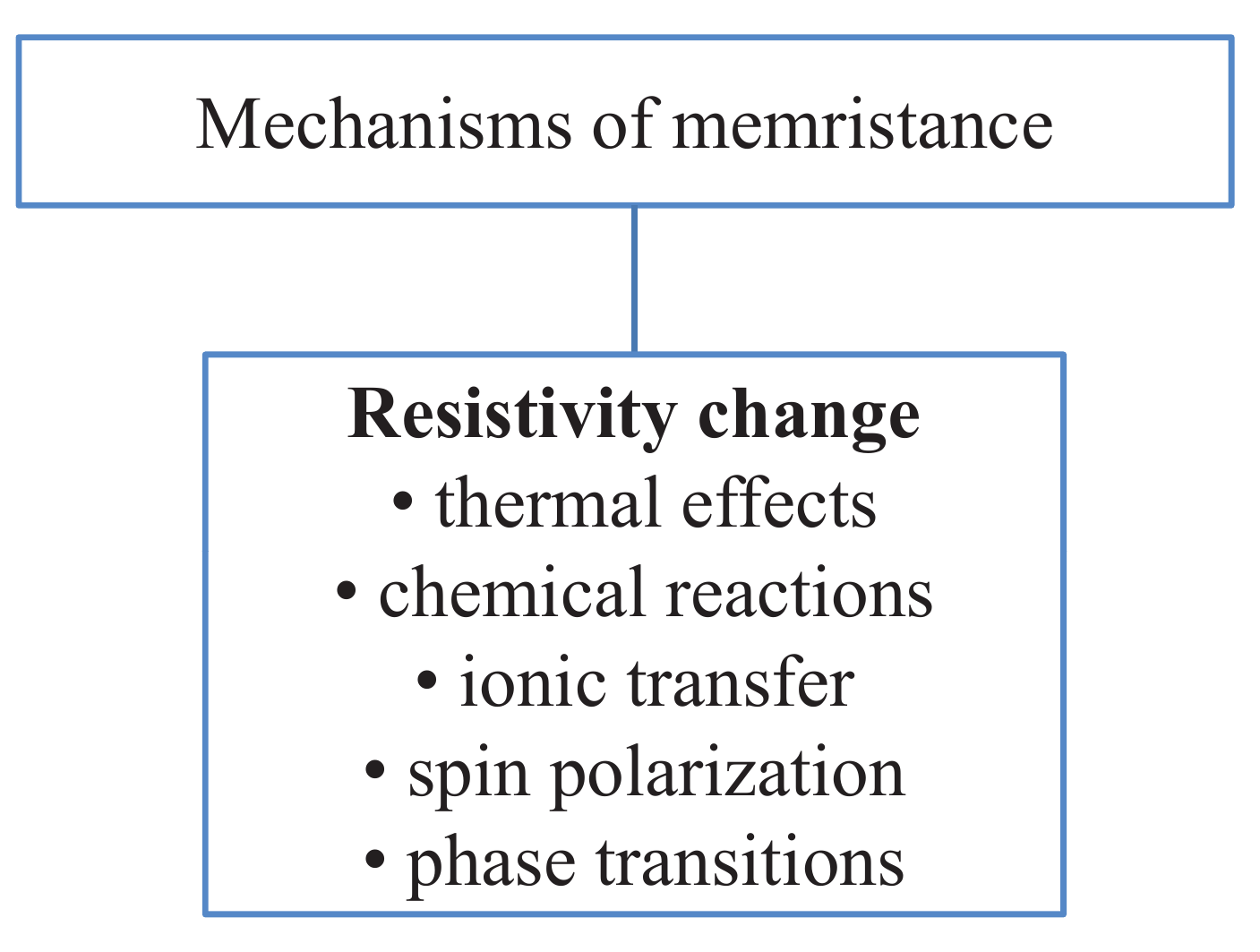}
\caption{Classification scheme of memristance mechanisms.}
\label{memrist}
\end{center}
\end{figure}

Many systems exhibit memristive behavior whose underlying physical
mechanisms vary considerably (see figure \ref{memrist} for a list
of common mechanisms). Examples that were identified early in
reference \cite{chua76a} include thermistors~\cite{sapoff63a} and
ionic systems, in particular membranes in neuron
cells~\cite{hodgkin52a}. For the reasons we have anticipated in
section \ref{sec:intro}, recent interest has been focused on the
dynamics of nanostructures. Some of them were intensively studied
in the context of the development of resistive-switching memory
during the last 10-15 years. To the best of our knowledge, this
area of research was initiated by Hickmott in 1962 by observing
hysteretic behavior in oxide insulators \cite{hickmott62a}. Later,
resistance switching was demonstrated in thin films of silicon
monoxide (SiO) between metal electrodes by Simmons and Verderber
\cite{Simmons67a} and in TiO$_2$ by Argall \cite{Argall68a}.
(These and other earlier works in the field were reviewed by
Dearnaley {\it et al.} \cite{Dearnaley70a}.) However, it was only
in 2008 that resistive switching devices were recognized as
memristive systems \cite{strukov08a}. Additional examples of
memristive systems include spintronic devices
\cite{pershin08a,wang09a}, phase-transition materials
\cite{driscoll09a,driscoll09b} and polaronic systems \cite{Alexandrov09a}. Below we discuss
memristive properties of a variety of physical systems in details.

\subsection{Thermistors} \label{subsec:thermistors}

Thermistors were considered in section~\ref{subsec:thermistors1}, and we refer the reader
to that section for details on a mathematical model to describe them.

\subsection{Resistance Switching Memory Cells } \label{RSoxides}

At the present time, it is well known that resistive switching
effects can be observed in such diverse classes of materials as
binary oxides (TiO$_2$, CuO, NiO, CoO, Fe$_2$O$_3$, MoO,
VO$2$)~\cite{yang08a,inoue08a,lee07a,seo04a,Shima07a,driscoll09a,driscoll09b},
nanogap systems on SiO$_2$ \cite{Yao09a}, metal nanogap junctions
\cite{Naitoh06a},
 perovskite-type oxides
(Pr$_{1-x}$Ca$_x$MnO$_3$,
SrTiO$_3$:Cr)~\cite{asamitsu97a,fors05a,kim06a,meijer05a,nian07a},
sulfides (Cu$_2$S,Ag$_2$S)~\cite{terabe05a,tamura06a,waser07a},
semiconductors (Si, GaAs, ZnSe-Ge)~\cite{Jo08a,Dong08a,Jo09a} and
organics~\cite{Stewart04a,lai05a,Ling08a}. The basic mechanisms of switching are not
yet well understood in all cases, however, quite generally, the related memory devices can be separated into the following most important categories: nanoionic, nanothermal (a sub-class of nanothermal memory devices, phase-change memory cells, are considered in section \ref{sec:PCM}), macromolecular memory and molecular effects memory devices \cite{ITRS09a}. Due to this wide range
of physical systems and memory mechanisms, we will provide specific examples with
particular focus on the state variables responsible for memory and
their theoretical description, if available. The reader interested
in some more comprehensive reviews on just resistive switching is
urged to look into the recently published ones
\cite{waser07a,Scott07a,Sawa08a,Karg08a,Waser09a}.

\begin{figure}
 \begin{center}
\includegraphics[angle=0,width=8cm]{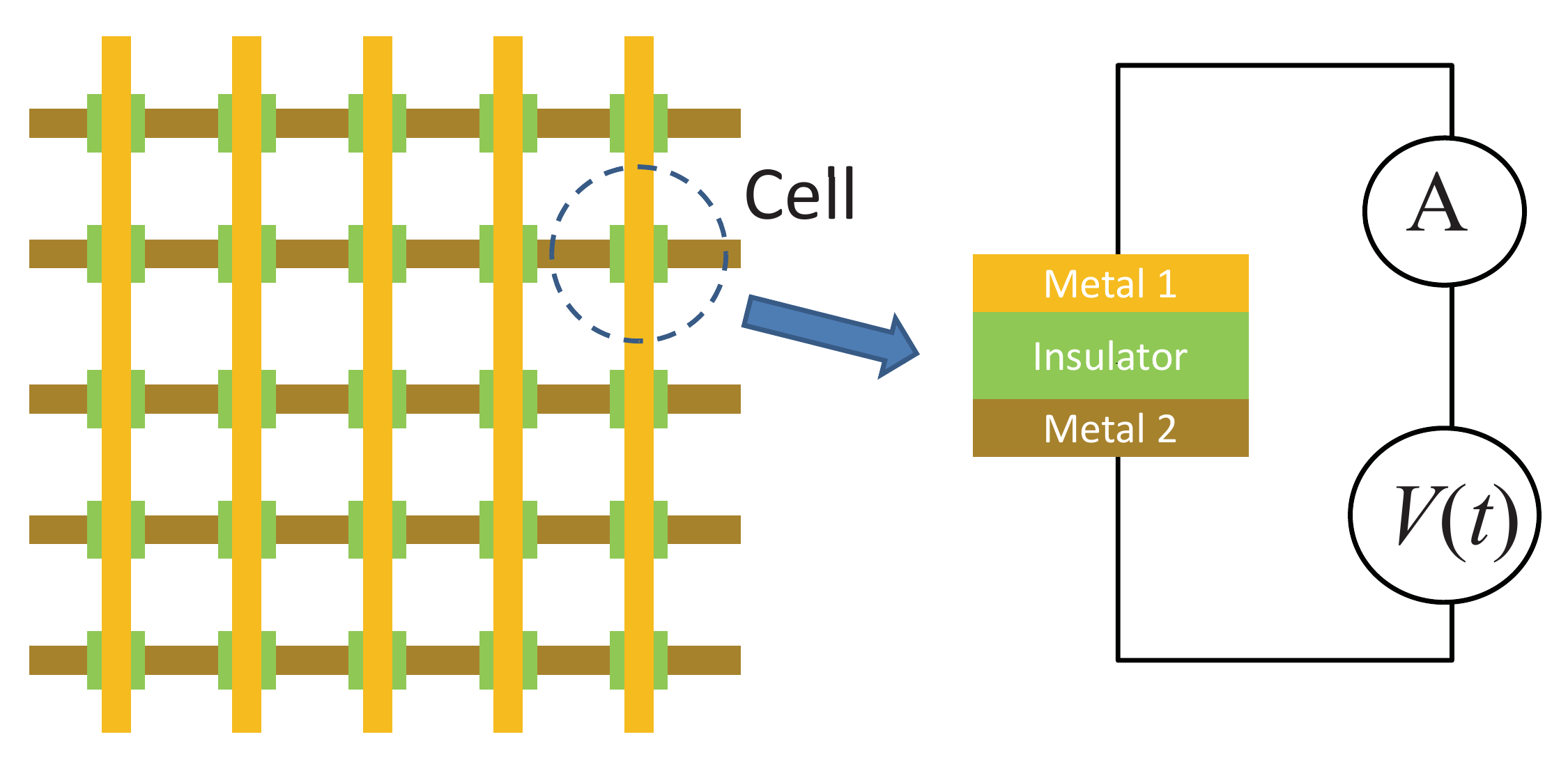}
\caption{ A single metal-insulator-metal memory cell (shown on the right) is often
fabricated as part of a crossbar array as shown on the left. Crossbar arrays
consist of two perpendicular sets of wires separated by an
insulating material. Each cell can be generally addressed independently, although the current, in a purely resistive structure, can flow through different cells as shown on the right. For example, in 2D crossbar arrays a voltage is applied to one of the lines and one to the perpendicular wires in the array. Here, ``A'' indicates an ammeter.
 \label{MIM}}
 \end{center}
\end{figure}

Very often, the experimental setup showing resistive switching involves an array (called a crossbar array)
of capacitor-like cells. Each capacitor-like cell contains a layer
of an insulating material sandwiched between two metal layers
(which may be made of the same or different materials) as depicted in figure
\ref{MIM}. In what follows, we will discuss behavior of separate cells. Although
each cell in the crossbar architecture can be addressed
independently by an appropriate selection of a word (say, horizontal) and
bit (vertical) lines, the current between such two lines can flow across many paths causing a potential
problem for the crossbar memory implementation. As a possible solution of this problem, additional individual access devices
(such as transistors or diodes) may be required.

Typically, an as-fabricated cell is in a highly-resistive state
and, in order to obtain resistance switching operation, an
electroforming step is needed. This is an important step of device fabrication
defining its future operation \cite{Jeong09b}. In this step, a high voltage
amplitude pulse is applied to the cell producing a nondestructive
(soft) breakdown, with the creation of single ``filaments'' inside
the main material matrix. This process of breakdown is usually
controlled by selecting a compliance current value (namely a
threshold current of the external input) that induces, at least in
some cases, the desired switching effect \cite{Jeong07a,Do09a}.
For example, the authors of reference \cite{Jeong07a} have
observed that the electroforming of a Pt/TiO$_2$/Ti structure with
a lower compliance current ($<$0.1 mA) results in a bipolar
hysteresis while a higher compliance current (1-10 mA)
leads to unipolar resistance switching (the meaning of these two
types of hysteresis will become clear in a moment). A detailed experimental study of the electroforming process in TiO$_2$ was
reported recently~\cite{Yang09c}. The dielectric breakdown in SiO$_2$ was investigated in
\cite{Li08b}. A deficiency of oxygen atoms along the breakdown path was observed, caused by large currents
through the percolation path. As a result, it is believed that the local energy gap could have collapsed
after the removal of oxygen atoms with consequent rearrangement of local atomic structure \cite{Li08b}.

\begin{figure}
 \begin{center}
\includegraphics[angle=0,width=8cm]{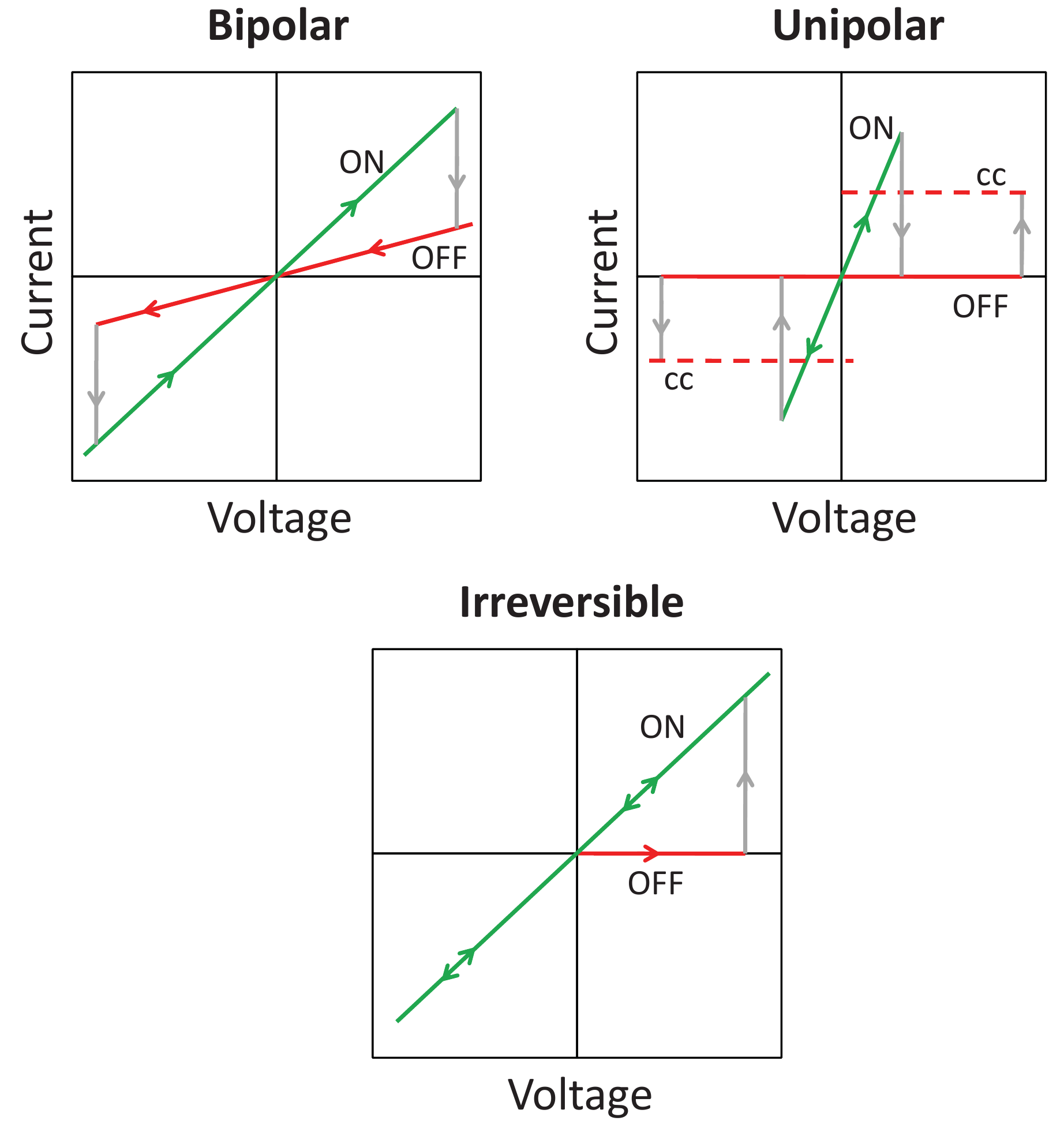}
\caption{ Three commonly observed types of $I-V$ curves in a
voltage sweeping experiment showing bipolar switching, unipolar
switching, and irreversible switching. The term ``cc'' means
``compliance current'' (see text for details).
 \label{looptypes}}
 \end{center}
\end{figure}

Let us now consider the different types of resistance switching reported
in the literature. Figure \ref{looptypes} illustrates the three most
general behaviors: bipolar, unipolar, and
irreversible. Such a classification of $I-V$ curves can be used in
addition to the categorization in terms of the crossing type at
the origin suggested in section \ref{hystloops}. Below we will
consider these three cases in more detail.

\subsubsection{Bipolar resistance switching} \label{bipolar}

In the case of {\it  bipolar switching} both voltage polarities
are required to switch  a device from a low resistance state (ON)
to a high resistance state (OFF) and back. A typical shape of
$I-V$ curve in the case of bipolar switching as shown in figure
\ref{looptypes} provides just a general topology of these curves.
Experimentally measured loops depend on both the system type
and measuring conditions, and can be thus quite different from this ideal topology. In the bipolar case
the hysteresis process often is of a threshold type: while a high applied
voltage is needed to change the device state, at low applied
voltages the resistance of the device remains unchanged. This is very often associated with ionic transport and
electrochemical reactions, even though the exact experimental cause is not always clear and may originate from a
combination of effects \cite{waser07a}. Moreover, we note that in most cases
the switching has a gradual character: the change of, say, resistance of a device proceeds continuously (not as an abrupt jump between two limiting values).
Such a property is promising for multi-state memory in which a single memory cell can store several bits of information.

\begin{figure*}
 \begin{center}
\includegraphics[angle=0,width=12cm]{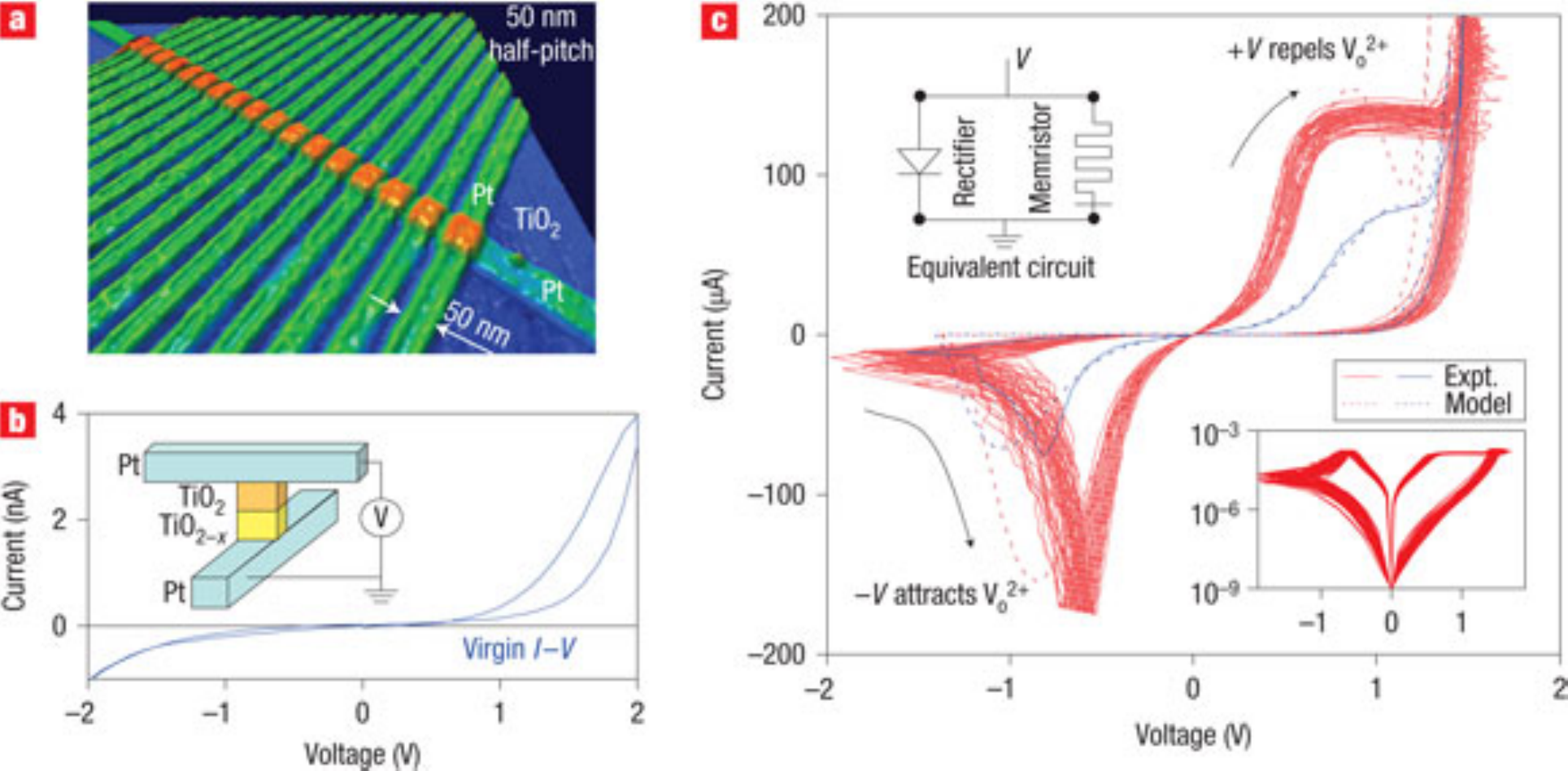}
\caption{ (a) An AFM image of 17 nano-crosspoint devices
containing a 50-nm-thick TiO$_2$ thin film sandwiched between Pt
electrodes. (b) The initial (before electroforming step) $I-V$
curve of the device and experimental scheme. (c) Experimental and
modeled $I-V$ curves. Reprinted with permission from Macmillan Publishers Ltd: Nature Nanotechnology \cite{yang08a}, copyright 2008.
 \label{tio2}}
 \end{center}
\end{figure*}

We now consider several examples of bipolar switching observed
in devices fabricated from different materials. Our first
example is a titanium dioxide (TiO$_2$) thin film sandwiched between metal electrodes (see figure \ref{tio2}).
This resistive feature has been known since 1968 \cite{Argall68a}. Recently, however,
the switching behavior of this binary oxide has been explained within the context of a
memristive model~\cite{strukov08a,yang08a}. In the case of
Pt/TiO$_2$/Pt, it was suggested that the switching involves
changes in the electronic barrier at the Pt/TiO$_2$ interface
induced by drift of oxygen vacancies under an applied electric
field. When vacancies drift towards the interface, they create
conducting channels that short-circuit the electronic barrier.
When vacancies drift away from the interface they eliminate these
channels, restoring the original electronic
barrier~\cite{yang08a}. This explanation, however, has been
recently challenged \cite{Wu09b}. In this alternative explanation,
the conductivity change of TiO$_2$ is associated with an
electrochemical reduction of Ti$^\textnormal{IV}$ oxide to the
much more conductive Ti$^\textnormal{III}$ oxide, analogous to a
solid-state redox reaction \cite{Wu09b}. A similar reaction-based mechanism
taking place at the interface between Pt and TiO$_2$ was considered in reference
\cite{Jeong09c}. The authors of this work state that the  electrochemical reactions modulating the Schottky barrier height at the interface play the dominant role in the device behavior.

Irrespective of the basic physical mechanism at play, this
material demonstrates promising characteristics for future
ultra-high density memory applications such as fast read/write
times ($\sim$10ns), high ON/OFF ratios ($\sim 10^3$), suitable
range of programming voltages, and possibility to fabricate small
size cells. The latter, in particular, is viewed as an advantage
in fabricating high-density neuromorphic circuits, namely circuits
that simulate neurological functions, because of the possibility
to reach structure densities comparable (or even larger) than the
density of neurons and synapses (connections among neurons) in the
human brain. We will come back to this type of applications in
section \ref{neuro_circuits}.

\begin{figure}
 \begin{center}
\includegraphics[angle=0,width=7cm]{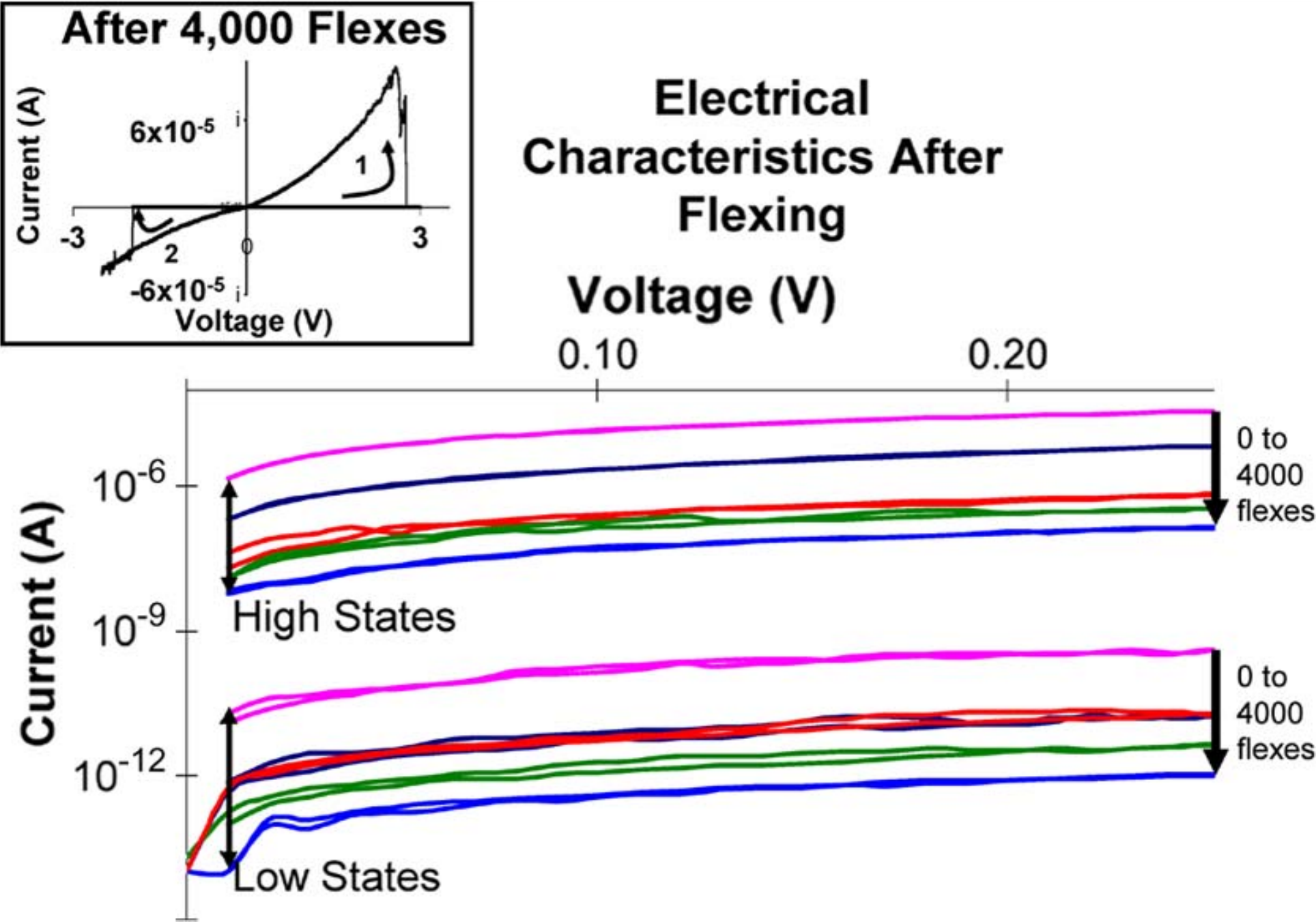}
\caption{ $I-V$ curves of flexible memristive system after 0, 100, 2000,
3000 and 4000 flexes. Inset: switching curve after 4000 flexes.
From \cite{Gergel09a} (\copyright 2009 IEEE).
 \label{flex}}
 \end{center}
\end{figure}

An interesting TiO$_2$-based device is a flexible memristive system
as presented in reference \cite{Gergel09a}. This device was
fabricated on an HP laser-jet transparency using an inexpensive
room temperature solution processing. The device demonstrates
ON/OFF ratios larger than $10^4$, about $10^6$s memory storage
potential, and the ability to operate after being physically
flexed 4000 times \cite{Gergel09a}. Multiple consecutive
mechanical deformations of the device lead to an increase of
resistance in both ON and OFF states, while keeping ON/OFF ratios almost
unchanged (see figure \ref{flex}). Such devices have the potential
for use in flexible lightweight portable electronics
\cite{Gergel09a}.

\begin{figure}
 \begin{center}
\includegraphics[angle=0,width=7cm]{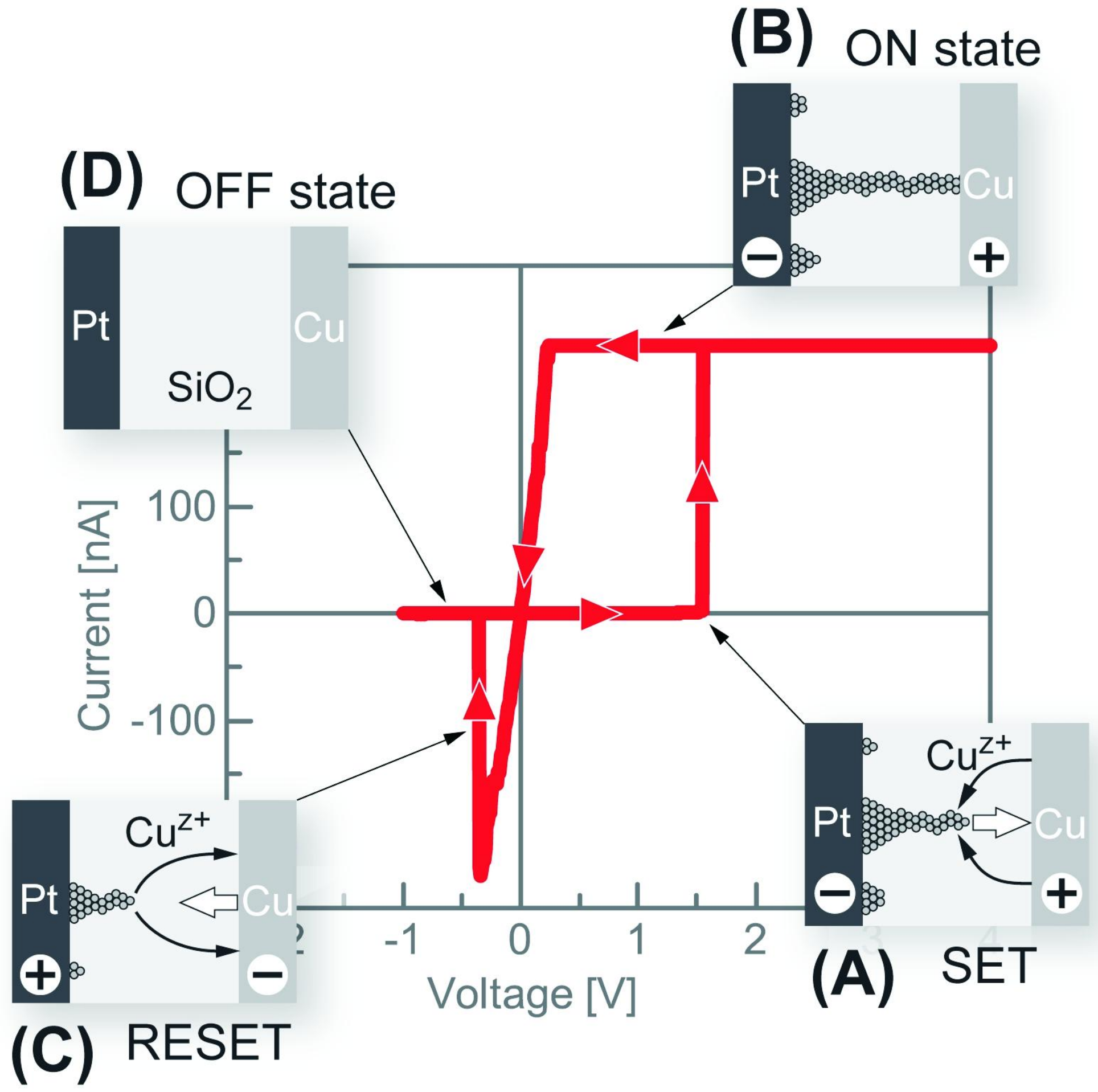}
\caption{ Current-voltage characteristic of a Cu/SiO2/Pt electrochemical metallization cell using a triangular voltage sweep.
 The positive bias current is limited by 250nA compliance current value. The insets show dynamics of metallic filament formation.
Reprinted with permission from \cite{Schindler09b}.
Copyright 2009, American Institute of Physics.
 \label{fig:EMC}}
 \end{center}
\end{figure}

In nanoionics-based memories \cite{waser07a}, a memory cell is composed by a couple of metal electrodes
separated by a thin film of insulating material playing the role of solid-state electrolyte, such as an oxide NiO \cite{seo04a}, SiO$_2$ \cite{Schindler07a,Schindler09b}, SrTiO$_3$ \cite{Szot06a}  or higher chalcogenides -
Ag$_2$S \cite{terabe05a}, Ge$_{x}$Se$1-x$ \cite{Kozicki05a,Schindler09a}. One of the electrodes is made of a
relatively inert metal (such as Pt) and the other electrode is electrochemically active (e.g., Ag or Cu).
A negative bias applied to the inert electrode forces a flow of metal ions in the electrolyte (originating from the now-positive active electrode)  toward the inert metal electrode. After a short period of time these ions compose a
filament that connects two metal electrodes. This filament dramatically reduces the resistance  of the cell.
Applied bias of the opposite polarity drifts the active electrode ions in the opposite direction destroying the filament.
The basic scheme of operation of such a cell is shown in figure \ref{fig:EMC}. Such memory technology is often called programmable metallization cell (PMC) \cite{Kozicki05a} or electrochemical metallization cell (ECM) or CBRAM (conductive-bridging random access memory) \cite{dietrich07a}. In figure \ref{fig:EMC} we show a pinched I-V hysteresis curve observed in Cu-SiO$_2$-based electrochemical metallization memory cells. This figure also shows dynamics of filament formation.

\begin{figure}
 \begin{center}
\includegraphics[angle=0,width=7cm]{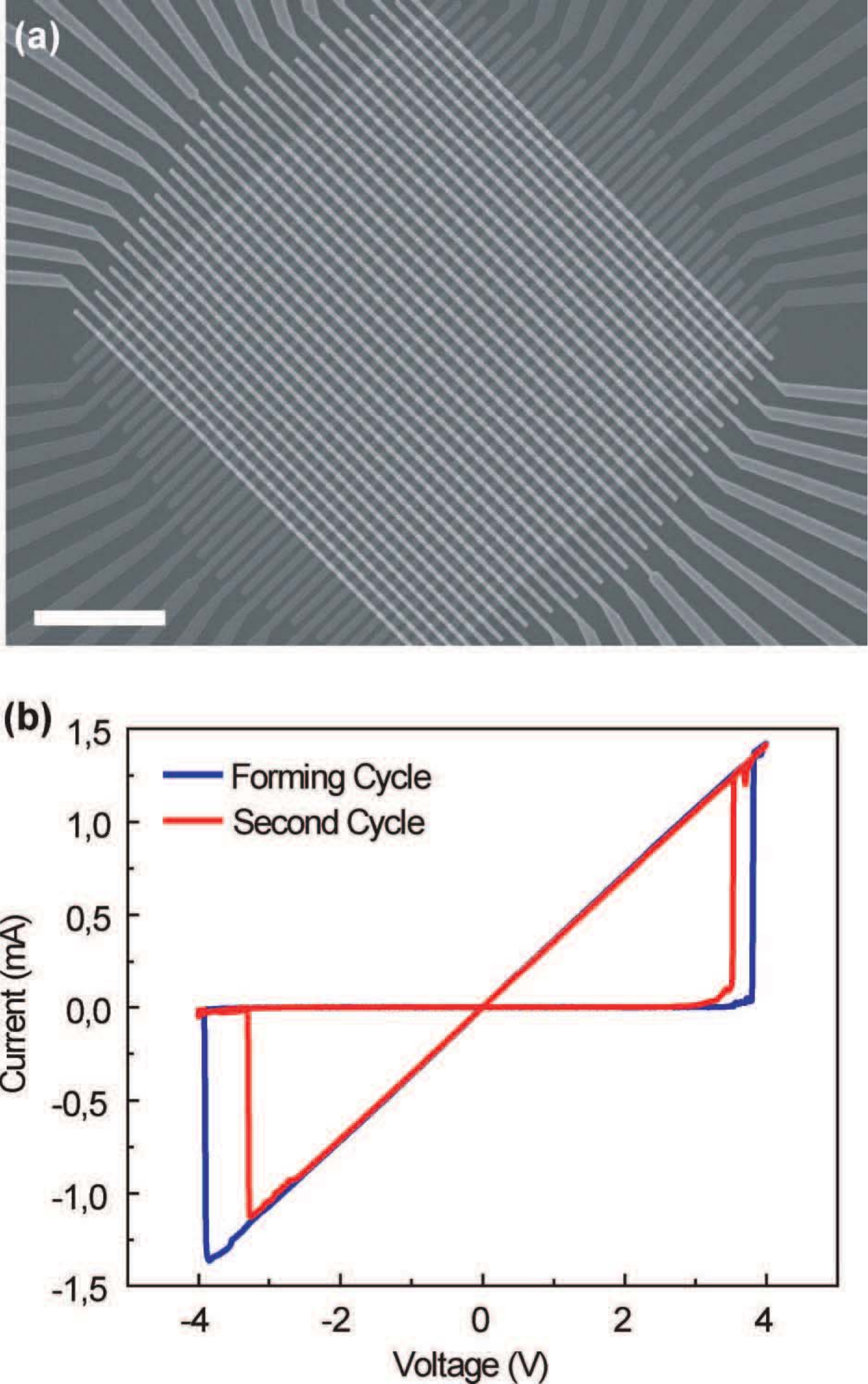}
\caption{(a) SEM image of Si-based memristive crossbar. (b) Switching
$I-V$ characteristics of an individual cell. Reprinted with permission from \cite{Jo09a}. Copyright 2009 American Chemical Society.
 \label{weilu}}
 \end{center}
\end{figure}

Recently, nanoionic resistive switching in silicon-based materials was
experimentally demonstrated
\cite{Jo08a,Dong08a,Jo09a}. In these experiments, amorphous-Si is
used as the switching medium, one of the electrodes is metallic
(typically Ag), and the other electrode is p-type Si. In this case the hysteresis is explained by the formation of conductive
filaments consisting of Ag particles inside the amorphous-Si. A
filament growth is a step-by-step process by which a new Ag
particle hops into a new trapping site. Such devices are CMOS (complementary
metal oxide semiconductor)
compatible, and offer promising characteristics such as fast
writing ($<$10ns), reasonable endurance ($>10^5$ cycles) and good
retention time ($\sim 7$ years). Moreover, a high-density crossbar
array based on such materials was fabricated \cite{Jo09a}. In
figure \ref{weilu} we show an image of this array as well as the
$I-V$ characteristics of a single cell. Such a structure exhibits
a symmetric bipolar switching together with a high ON/OFF
conductance ratio.

\begin{figure*}
 \begin{center}
\includegraphics[angle=0,width=11cm]{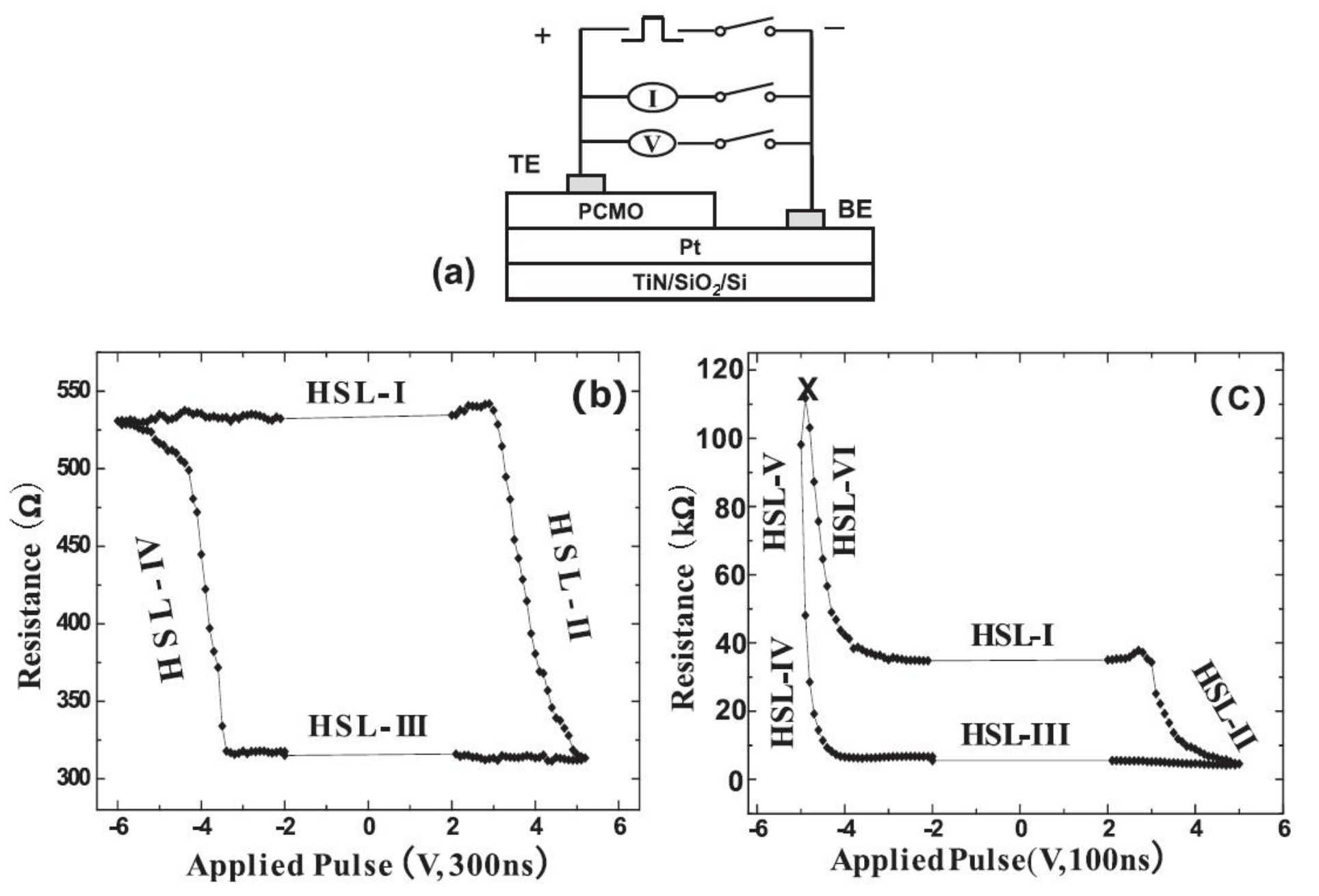}
\caption{ (a) Schematic of the experimental setup to observe resistive switching with perovskite oxide films PCMO. The top electrode
(TE) is made of Ag, while the bottom one (BE) is made of Pt films on TiN/SiO$_2$/Si substrate. (b,c) Resistance switching loops
(HSL) were measured by applying a sequence of pulses with a
constant voltage step. In (b), PCMO film grown in an oxygen
environment while in (c) the environment is oxygen-free. From
\cite{nian07a}.
 \label{ignat}}
 \end{center}
\end{figure*}

The mechanism of hysteresis in perovskite oxide films
Pr$_{0.7}$Ca$_{0.3}$MnO$_{3}$ (PCMO) was recently investigated
\cite{nian07a,Rozenberg10a}. These materials show a bipolar
switching as we demonstrate in figure \ref{ignat}. The authors of
Ref. \cite{nian07a} suggest an oxygen vacancies diffusion model to
explain the hysteresis phenomenon. Within this model, a
switching pulse results in a pileup of oxygen ions near metal
electrodes thus changing the resistance. Moreover, it is proposed
that this mechanism can explain the resistance switching in binary
transition-metal oxides and other complex oxides as well. The
authors of Ref. \cite{Rozenberg10a} suggest a discrete vacancies
diffusion model. Their study underlines the important role of
highly resistive dielectric-electrode interfaces and shows a
qualitative agreement with experimental data.

Bipolar resistance switching is also observed in structures
containing organic materials (see, e.g., references
\cite{Stewart04a, Cai05a,lai05a}), although its origin is not yet fully understood.
In particular, Stewart {\it et
al.} \cite{Stewart04a} have demonstrated that hysteresis
characteristics of different molecular materials sandwiched
between Pt and Ti electrodes are very similar suggesting a
molecular-independent hysteresis mechanism. Moreover, they have
shown that in a structure with similar (Pt) electrodes the
switching becomes irreversible. The authors suggest
\cite{Stewart04a} that a possible reason
could be related to formation of electromigration-induced filaments,
or incomplete electrochemical reaction of one electrode and the
molecular monolayer \cite{Stewart04a}.

\begin{figure*}
 \begin{center}
\includegraphics[angle=0,width=12cm]{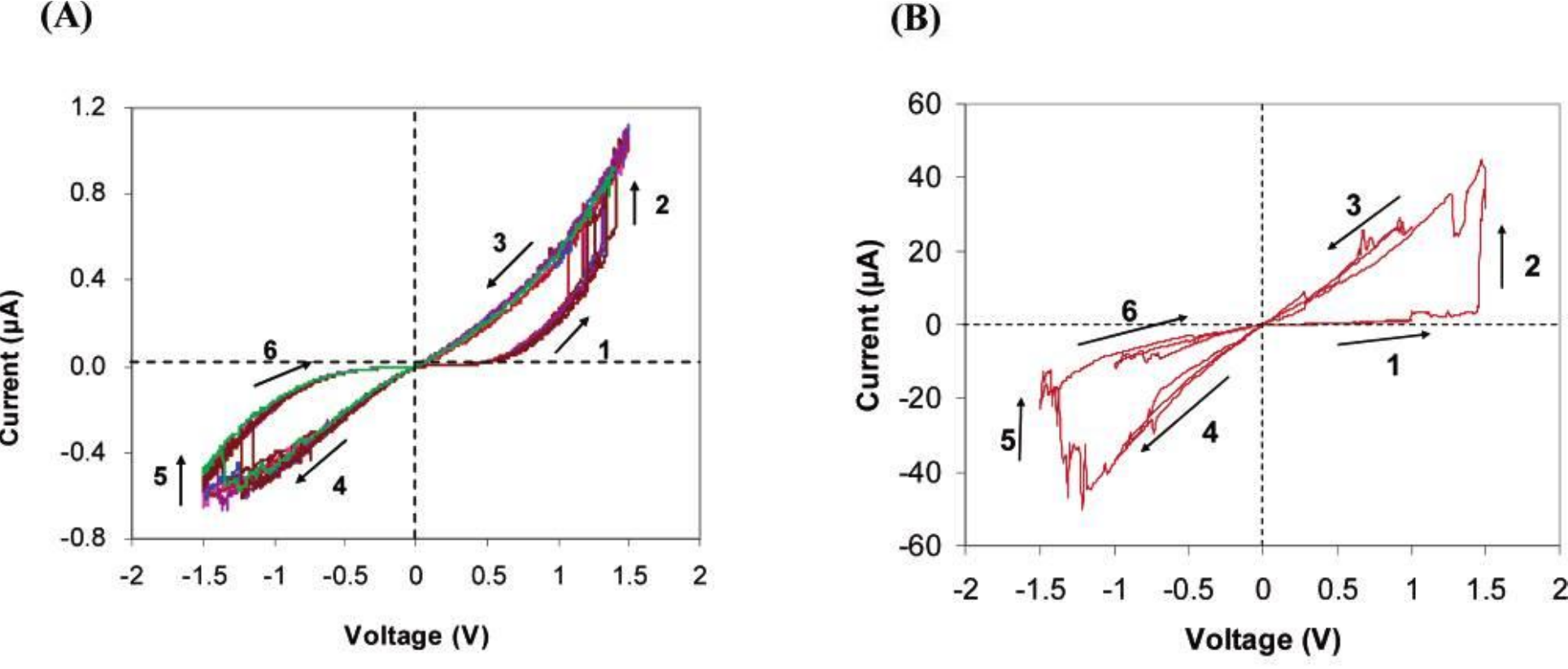}
\caption{(a) Reversible switching behavior in a Au-SAM-Pd in-wire
molecular junction, and (b) Pd-SAM-Pd molecular junction. Here, SAM stands for self-assembled monolayer of oligoaniline
dimers. Reprinted with permission from \cite{Cai05a}. Copyright 2005 American Chemical Society.
 \label{cai}}
 \end{center}
\end{figure*}

Cai {\it et al.} \cite{Cai05a} have reported a reversible
hysteresis in nanoscale thiol-substituted oligoaniline
molecular junctions. In their experiments, the switching
occurs at $\pm 1.5$V threshold voltage, with a high-to-low conductance
ratio of up to 50, and high-state storage times much longer than
22 hours. In figure \ref{cai} we show selected results of their
measurements.  These authors have concluded that reversible bistable switching in
their experimental system is an inherent molecular phenomenon,
most likely arising from a concerted shift in charge
delocalization and molecular conformation when the applied voltage
across the molecular junction exceeds a critical threshold value
\cite{Cai05a}.

\subsubsection{Unipolar resistance switching}\label{unipol}

The {\it unipolar switching} (see figure \ref{looptypes}) is
considered by some authors to be based on a thermal effect
\cite{waser07a}. The switching is defined by two voltages: the set (driving
OFF$\rightarrow$ON transition), and the reset (driving
ON$\rightarrow$OFF transition) voltages (see figure
\ref{looptypes}). The set voltage is always higher than the reset
one. Physically, in samples with this type of hysteresis, a weak
conducting filament with a controlled resistance is formed in the
electroforming process. During the reset operation, this filament
is partially destroyed because of the large amount of heat
released, similarly to a traditional household fuse
\cite{waser07a}. In the set operation, the filament is
reconstructed again. The nanoscale filament formation was recently
observed experimentally in such materials as TiO$_2$
\cite{Choi05a} and NiO \cite{Kim06c}.
\begin{figure}
 \begin{center}
\includegraphics[angle=0,width=7cm]{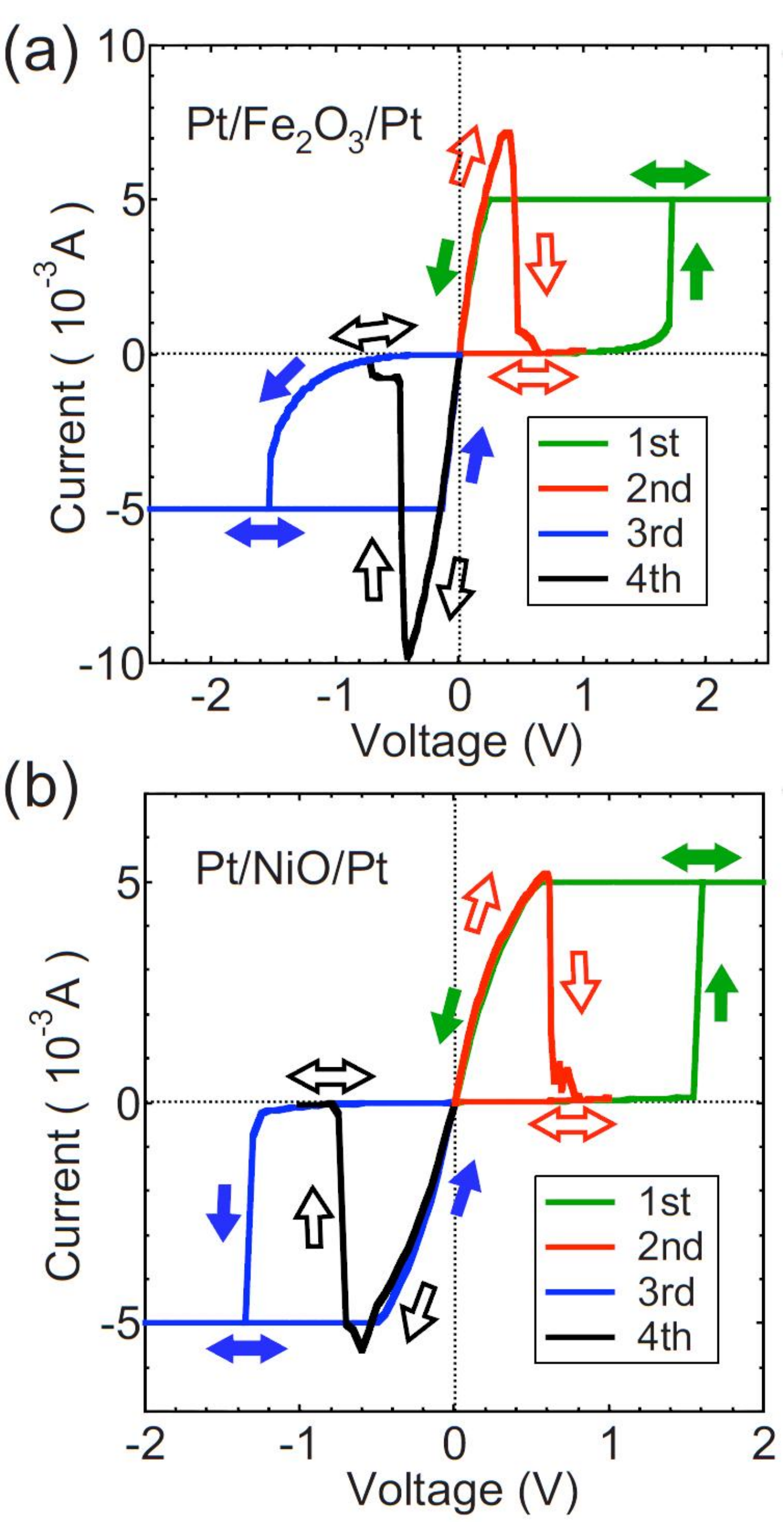}
\caption{$I-V$ characteristics of (a) Pt/Fe$_2$O$_3$/Pt and (b)
Pt/NiO/Pt devices showing unipolar switching. The oxide thickness is 100nm (Fe$_2$O$_3$)
and 60nm (NiO). From \cite{inoue08a}.
 \label{inoue}}
 \end{center}
\end{figure}

In figure \ref{inoue} we show examples of unipolar switching
behavior observed in metal/binary-transition-metal oxides/metal
structures based on Fe$_2$O$_3$ and NiO \cite{inoue08a}. In order to
clarify the hysteresis mechanism, the authors of
reference \cite{inoue08a} have studied several different
structures, in particular varying a top Pt electrode area. They
found that in the ON state the current does not depend on the area
of the top electrode, while in the OFF state the current is
proportional to the top electrode area. Based on these findings they
have suggested a ``faucet'' model of resistance switching in which
an ``electric faucet'' opens/closes in one or both interfaces
between metal electrodes and oxide when the device switches into the
ON/OFF state \cite{inoue08a}. However, a complete understanding of this effect is still lacking.

Our next example is the unipolar switching in a polymer material
poly(3,4-ethylene-dioxythiophene):polystyrenesulfonate, commonly
referred to PEDOT:PSS. The  hysteresis mechanisms in PEDOT:PSS
are still generally unknown, and those suggested in the literature are highly speculative \cite{Liu09a}.
This material, in fact, can exhibit all
types of switching behavior shown in figure \ref{looptypes}, depending on the type of material electrodes used.
Moller {\it et al.} \cite{Moller03a} have observed an irreversible
switching in a Au/PEDOT:PSS/Si p-i-n diode structure. In
experiments with ITO/PEDOT:PSS/Al devices~\cite{Ha08a}, bipolar
switching characteristics were observed. Finally, in recent papers
\cite{Liu09a,Ha09a} studying Au/PEDOT:PSS/Au and Al/PEDOT:PSS/Al
structures, respectively, the unipolar switching was reported.

\begin{figure*}
 \begin{center}
  \centerline{
    \mbox{\includegraphics[width=5.00cm]{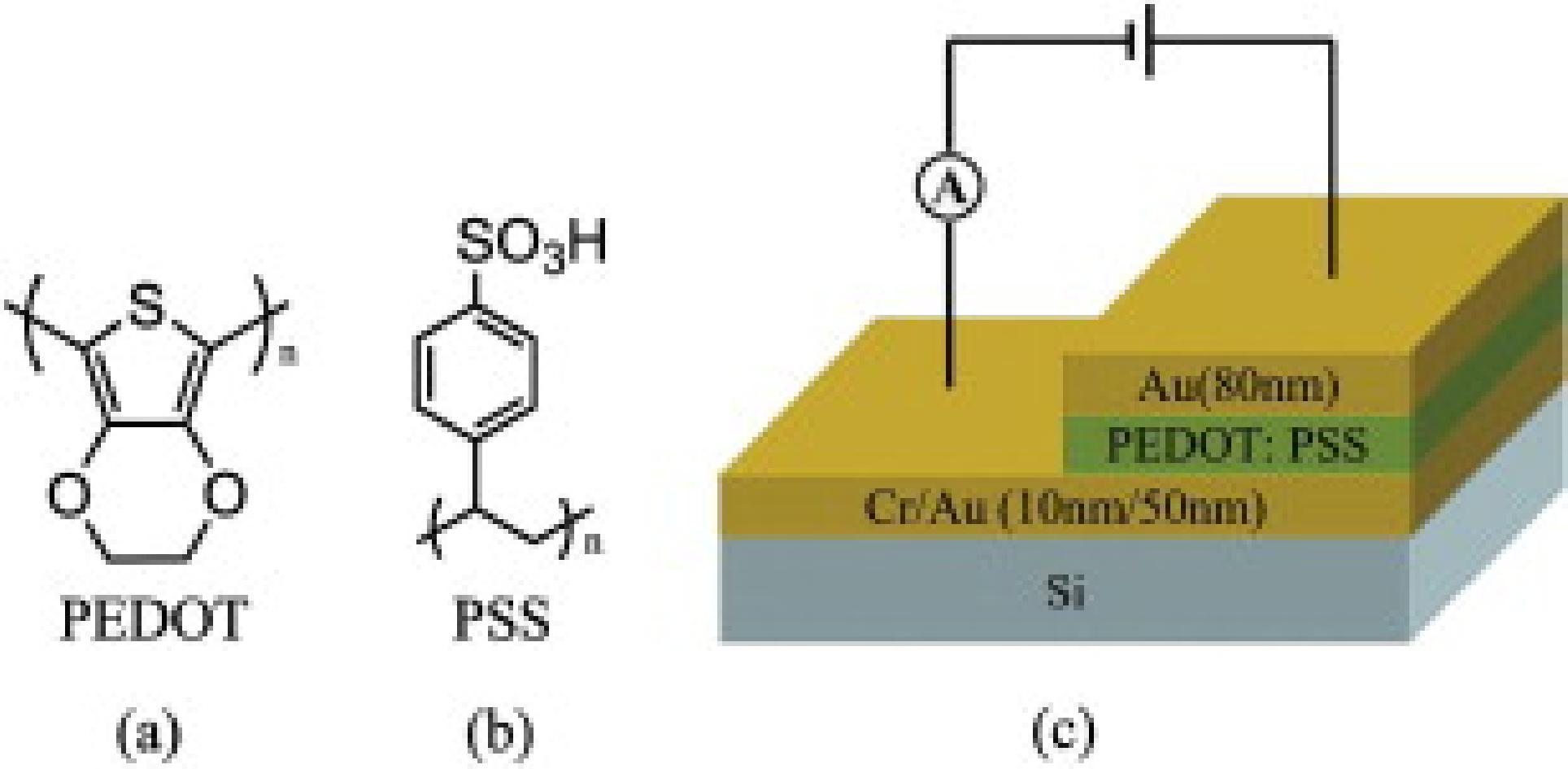}}
    \mbox{(d)}
    \mbox{\includegraphics[width=7.00cm]{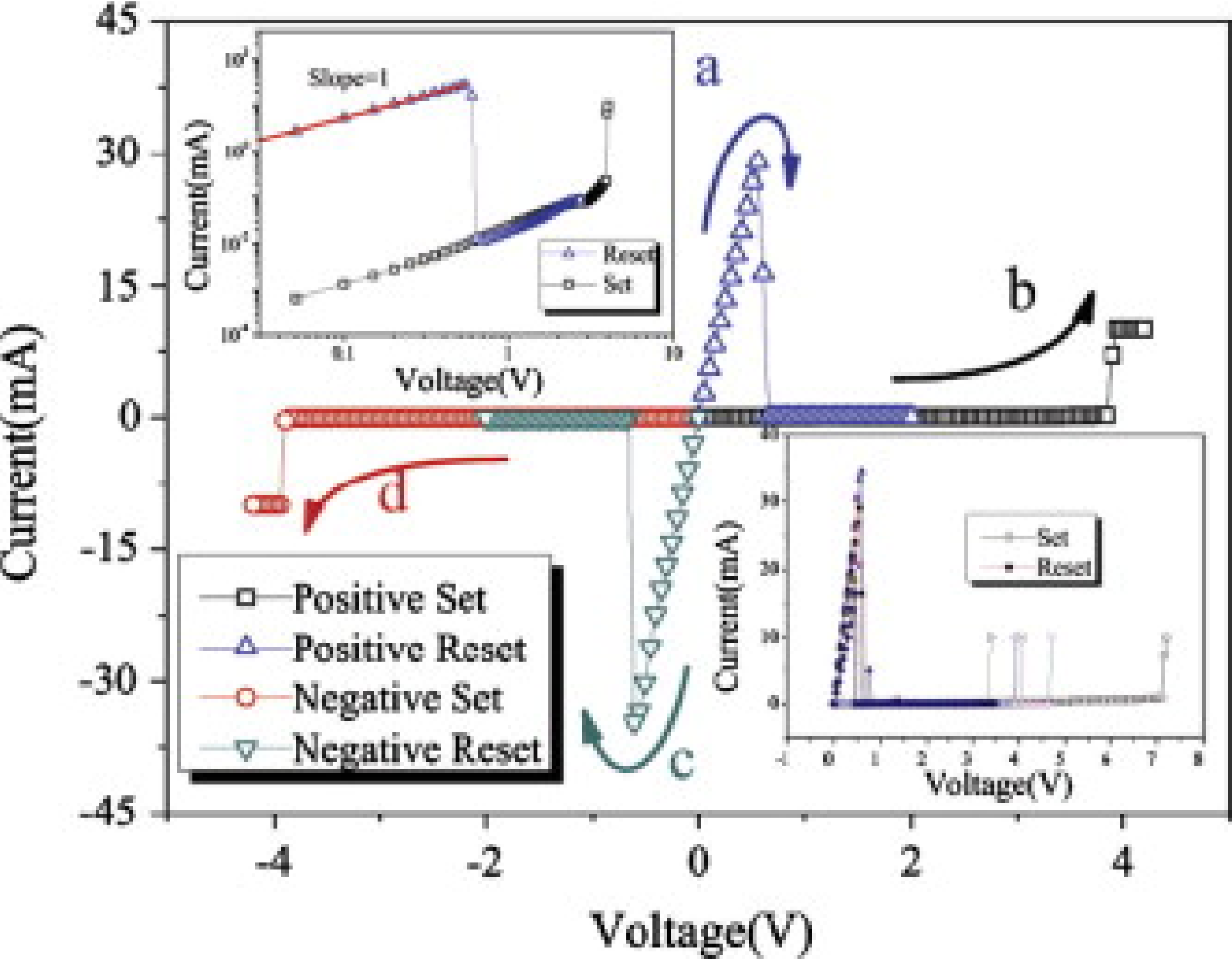}}
  }
\caption{Structure of (a) PEDOT and (b) PSS complexes. (c)
Schematic of experimental setup. (d) Unipolar switching
characteristics of Au/PEDOT:PSS/Au memory devices.
Reprinted from \cite{Liu09a}. Copyright 2009, with permission from Elsevier.} \label{PEDOT}
\end{center}
\end{figure*}

We show $I-V$ switching characteristics of a Au/PEDOT:PSS/Au
device \cite{Liu09a} in figure \ref{PEDOT}. In this plot, the
sweep sequence is indicated by the alphabet letters. The probable mechanism in this case is the forming and rupture of conductive paths, presumably related to the oxidation and
reduction of the PEDOT:PSS film \cite{Liu09a}. A resistive ratio of 10$^3$ and no significant
degradation over $10^4$s under continuous readout testing were
found.

\subsubsection{Irreversible resistance switching}

The third type of switching, the {\it irreversible switching}, is
of little interest from the point of view of memristive systems
since in this case the initially fabricated device can only be switched
toward a different state irreversibly (only in one direction). However,
it is evident that these devices are memristive since the resistance at any given moment of time is defined by
the history of the system (e.g., voltages applied to the device in the past). Of course, some typical features of memristive systems
such as repeatable hysteresis loops can not be observed with such devices.

\begin{figure*}
 \begin{center}
\includegraphics[angle=0,width=12cm]{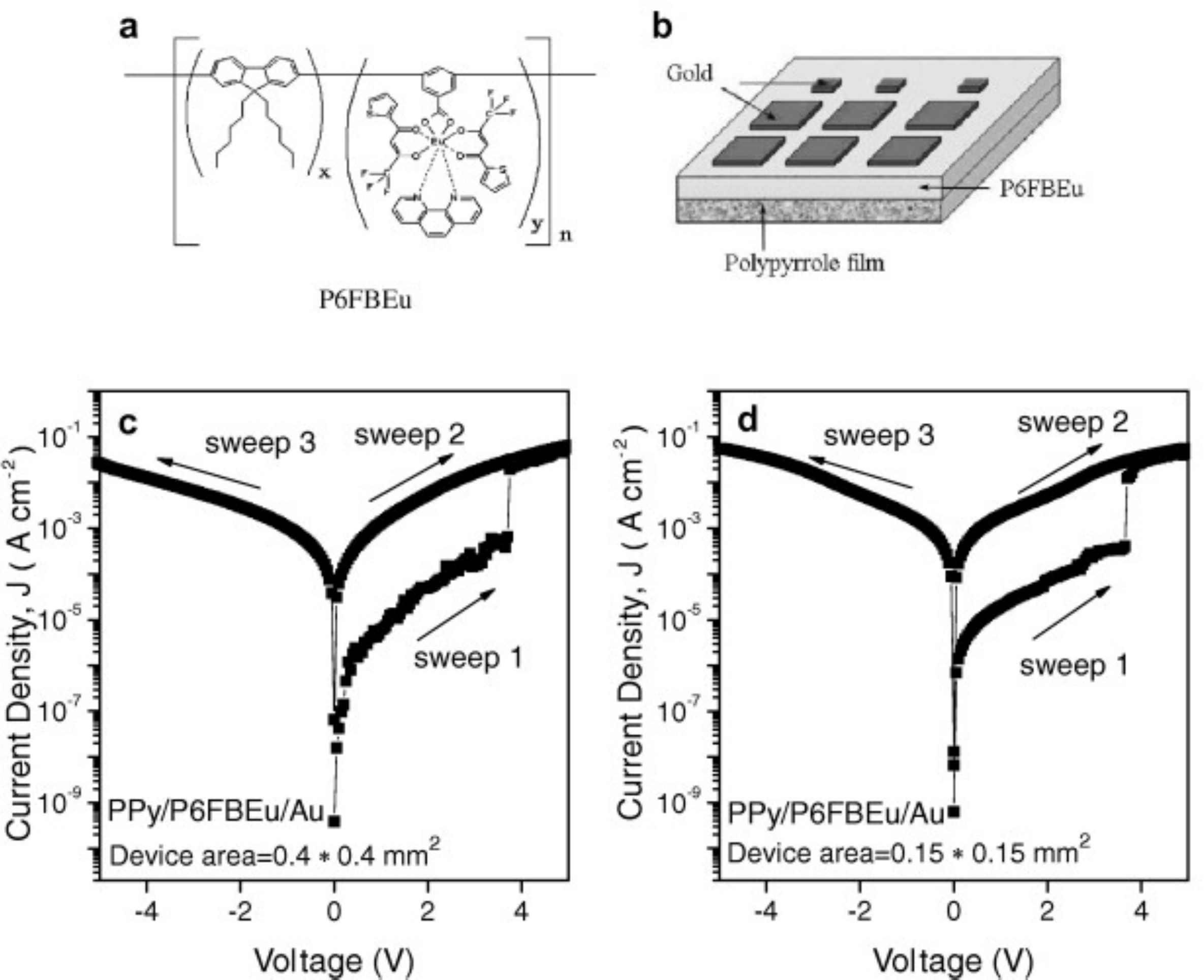}
\caption{Flexible WORM memory device demonstrating irreversible
OFF$\rightarrow$ON switching. (a) Molecular structure of P6FBEu
polymer. (b) Schematics of memory cells consisting of a thin film
($\sim$50nm) of P6FBEu located between a flexible PPy substrate
and gold top electrodes. (c,d) $I-V$ characteristics of cells with
different size of gold electrodes. Reprinted from \cite{Li2007401}, Copyright 2007, with permission from Elsevier.
 \label{WORM}}
 \end{center}
\end{figure*}

Systems with irreversible resistance switching represent a simple, non-volatile, WORM
(write-once-read-many-times) memory. Very often, this type of
memory is observed in organic devices
\cite{Moller03a,Stewart04a,Choi08b,Li2007401}. Figure
\ref{looptypes} shows a schematic of irreversible switching loop
when the device initially fabricated in the OFF state is switched into the ON
state. This type of behavior is indeed observed experimentally
\cite{Li2007401}. Figure \ref{WORM} shows an example of $I-V$
characteristics of a flexible polymer device based on the
conjugated copolymer 9,9-dihexylfluorene and benzoate with
chelated europium thenoyltrifluoroacetone ligand complex (P6FBEu).
This device shows an irreversible OFF$\rightarrow$ON switching at a bias of
$\sim 4$V, ON/OFF current ratio of 200, and stability
of the ON and OFF states up to $10^6$ reading cycles at 1V read voltage
\cite{Li2007401}. The authors of this work have suggested a charge-migration mechanism
to explain the hysteresis.

\begin{figure}
 \begin{center}
\includegraphics[angle=0,width=7cm]{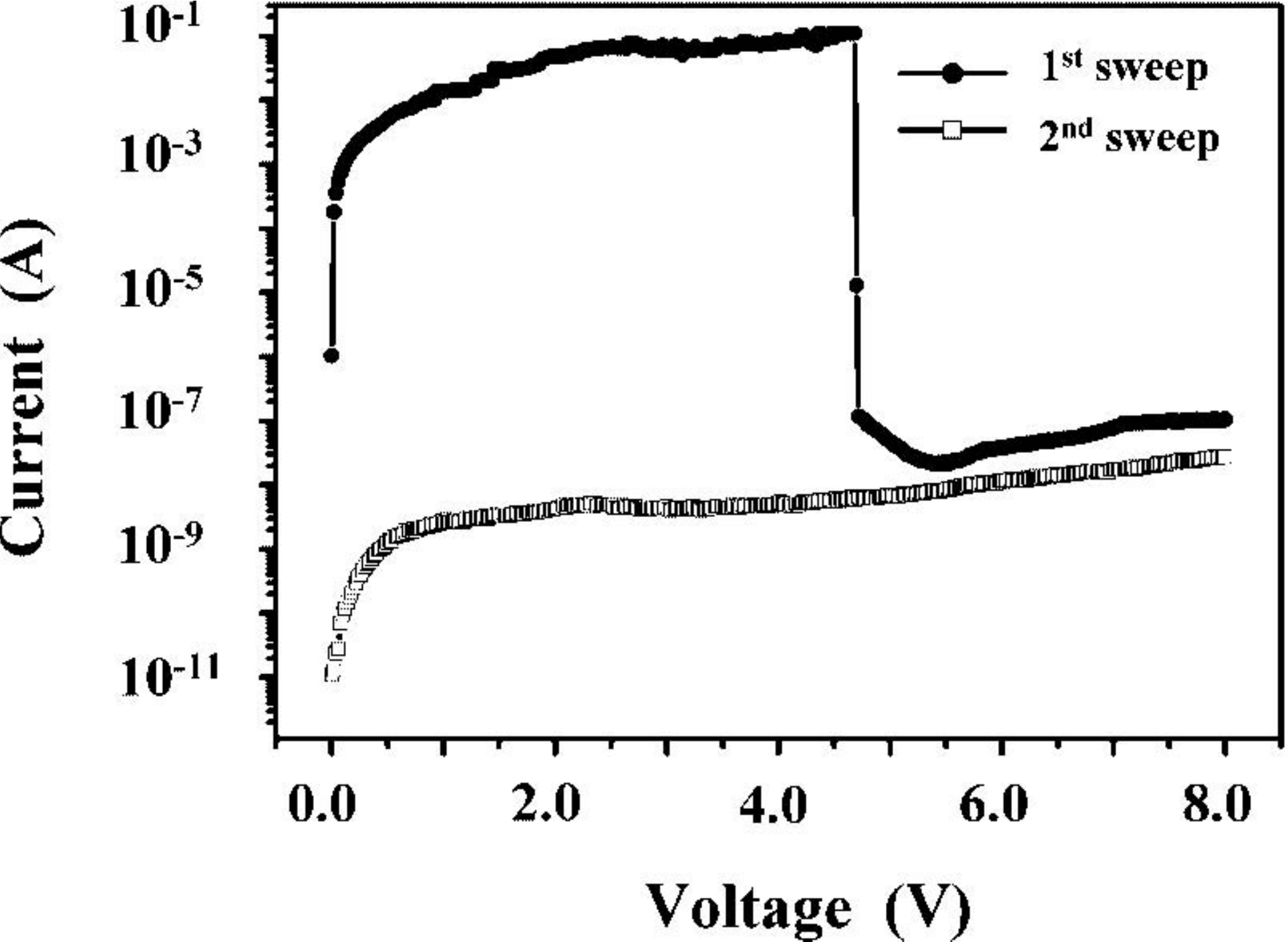}
\caption{Irreversible resistance switching in ITO/HCuPc/Ti
device. From \cite{Choi08b}, copyright Wiley-VCH Verlag GmbH $\&$ Co. KGaA.
Reproduced with permission.
 \label{WORM1}}
 \end{center}
\end{figure}

Moreover, the different direction of switching, when a device
fabricated in ON state is switched into an OFF state is also possible
\cite{Moller03a,Stewart04a,Choi08b}. For example, in figure
\ref{WORM1} we report $I-V$ curves of
indium-tin-oxide/hyperbranched copper phthalocyanine polymer/Ti
(ITO/HCuPc/Ti) device \cite{Choi08b}. Such a device shows an
irreversible ON$\rightarrow$OFF switching behavior at
approximately 4.75V with a titanium top electrode, and
at 2.5V with a gold electrode instead of Ti \cite{Choi08b}.
Moreover, the authors of this work report a stability of ON- and OFF-states
during $10^{12}$, 1V amplitude read pulses, and a stability of both
states after 1 year of storing. The switching into the OFF state is
explained by rupture of filaments which takes place when a high
voltage is applied \cite{Choi08b}. However, the origin of the
filaments in a virgin device is not elucidated and more work in this direction is thus
highly desirable.

\subsubsection{Models of resistance switching devices} \label{memristor_models}

Several models of resistance switching have been proposed
recently
\cite{Baikalov03a,rozenberg04a,Rozenberg05a,Rozenberg06a,nian07a,strukov08a,strukov09b,pershin09b,Tamura09a,joglekar09a,Jeong09a,shin10a}
including molecular dynamics simulations \cite{Savelev10a}.
Some of them were implemented in SPICE (Simulation Program with
Integrated Circuit Emphasis)
\cite{Biolek2009-1,Biolek2009-2,Benderli09a,rak10a,sharifi10a,Zhang10a}.
Here, we consider several models of memristive devices \cite{strukov08a,joglekar09a,pershin09b} that
elucidate the resistance switching from the point of view of
memristive theory \cite{chua76a}.

\begin{figure}
 \begin{center}
\includegraphics[angle=0,width=7cm]{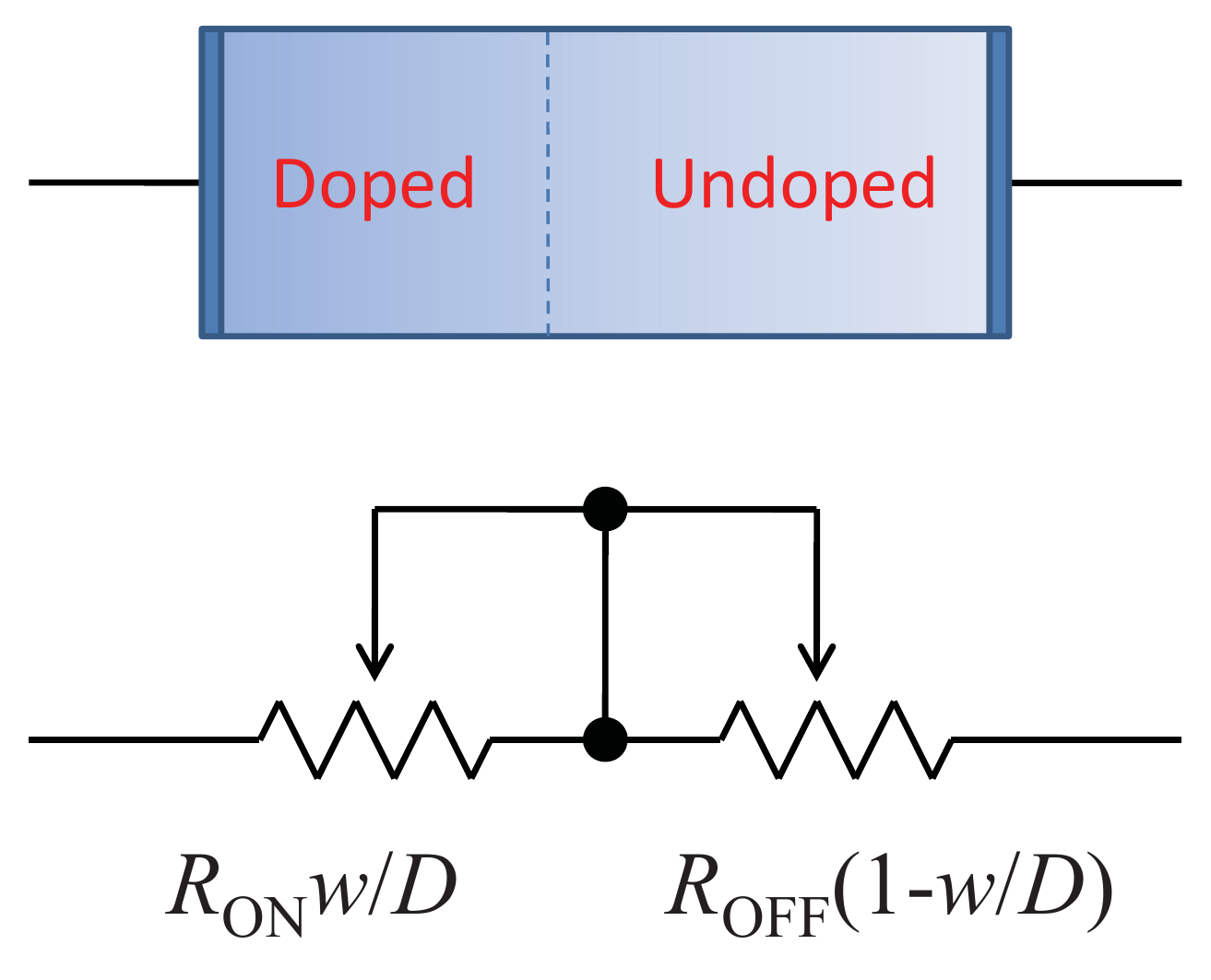}
\caption{Two variable resistors model of a resistance switching device:
$R_{\textnormal{ON}}$ and $R_{\textnormal{OFF}}$ are variable
resistors describing doped and undoped regions, respectively.
 \label{strukovSCH}}
 \end{center}
\end{figure}

A simple model of titanium dioxide memristive system was suggested by
Strukov and co-authors in their paper \cite{strukov08a}.
In this work, a semiconductor film of thickness $D$
sandwiched between two metal electrodes is modeled using two
variable resistors connected in series (see figure
\ref{strukovSCH} (a)). These resistors represent two spatial regions with a
high and low concentration of dopants. The application of the
external bias $V(t)$ causes charged dopants to drift thus moving
the boundary between the regions. This process is described by the
equations
\begin{eqnarray}
V(t)&=&\left(
R_{\textnormal{ON}}\frac{w(t)}{D}+R_{\textnormal{OFF}}\left[
1-\frac{w(t)}{D}\right]\right) I(t) \label{strukovcur}  \\
\frac{\textnormal{d}w(t)}{\textnormal{d}t}&=&\mu \frac{R_{\textnormal{ON}}}{D}I(t).
\label{strukovmob}
\end{eqnarray}
Here, $\mu$ denotes the ion mobility. Equation
(\ref{strukovmob}) gives
\begin{equation}
w(t)=\mu\frac{R_{\textnormal{ON}}}{D}q(t)
\end{equation}
which, inserted into equation (\ref{strukovcur}), in the limit of
$R_{ON} \ll R_{OFF}$ gives the memory resistance
\begin{equation}
R_M(q)=R_{\textnormal{OFF}}\left(1-\frac{\mu
R_{\textnormal{ON}}}{D^2}q(t) \right). \label{strukovmemr}
\end{equation}
The latter equation describes an ideal current-controlled
memristor (see section~\ref{secmemristors}). We note, however, that the derivation
of equation (\ref{strukovmemr}) is based on several assumptions such as current-controlled drift, constant
mobility, and does not account for the boundary conditions. Many of these assumptions may in fact not hold in the
actual experimental conditions.
Moreover, experimentally observed curves actually correspond to a
memristive system (in the sense of equations~(\ref{VMI2}) and~(\ref{dx})) rather than to an ideal memristor. The model of Strukov and co-authors
\cite{strukov08a} was later improved by Joglekar and Wolf
\cite{joglekar09a} who introduced a function that ensures no drift at the boundaries. Their model is summarized in the table \ref{table:Joglekar}.

\begin{table*}
{ \renewcommand{\arraystretch}{1.6}
\begin{tabular}{| l | c | }
\hline
Physical system & Solid state memristive device \\
\hline
Internal state variable(s) & Doped region size, $x=w$ \\
\hline
Mathematical description & $V=\left( R_{\textnormal{ON}}\frac{x}{D}+R_{\textnormal{OFF}}\left[
1-\frac{x}{D}\right]\right)I$  \\
& $\frac{\textnormal{d}x}{\textnormal{d}t}=\mu \frac{R_{\textnormal{ON}}}{D}I(t)F\left( \frac{x}{D}\right)$ \\
\hline
System type & First-order current-controlled
memristive system \\
\hline
\end{tabular}
}
\caption{Nonlinear drift model of a memristive device \cite{joglekar09a}. Here, $F(y)$ is the window function such that
$F(0)=F(1)=0$. This condition ensures no drift at the boundaries. The authors of Ref. \cite{joglekar09a} suggest a family of window functions $F_p(y)=1-(2y-1)^{2p}$, where $p$ is a positive integer number.}
\label{table:Joglekar}
\end{table*}

Several later papers by the same group
\cite{strukov09a,strukov09b,Pickett09a} report on further
development of resistance switching theory for TiO$_2$ devices. In
reference \cite{strukov09a} a model of threshold-type
switching is suggested. Within this model the ion drift in a
periodic potential is facilitated when the potential is
significantly tilted by the applied electric field. Pickett {\it et al.}
\cite{Pickett09a} have considered a model that involves an exponential
dependence of the switching time on the device current. A coupled
drift-diffusion equations approach to ionic and electronic
transport was instead considered in reference \cite{strukov09b}.

We have recently
introduced \cite{pershin09b} a different model of memristive system that
takes into account a threshold-type switching and boundary conditions (imposed on system's memristance). The former is frequently observed in the bipolar resistance switching devices considered
in section \ref{bipolar}. This model is based on the
assumption that the rate of the resistance change is small below a
threshold voltage $V_T$ and fast above $V_T$, and was inspired by the experimental work of reference \cite{strukov08a}. The mathematical formulation of this model is provided in the table \ref{table:threshold}.

\begin{table*}
{ \renewcommand{\arraystretch}{1.6}
\begin{tabular}{| l | c | }
\hline
Physical system & Solid state memristive device \\
\hline
Internal state variable(s) & Resistance, $x=R_M$ \\
\hline
Mathematical description & $I=x^{-1}V$  \\
& $\frac{\textnormal{d}x}{\textnormal{d}t}=\left(\beta V+0.5\left( \alpha-\beta\right)\left[
|V+V_T|-|V-V_T| \right]\right) \times$ \\
& $\theta\left( x-R_{min}\right) \theta\left( R_{max}-x\right) $ \\
\hline
System type & First-order voltage-controlled
memristive system \\
\hline
\end{tabular}
}
\caption{Threshold model of a memristive device \cite{pershin09b}. Here,  $x=R_M$ is the resistance of the memristive system and
$\theta(\cdot)$ is the step function indicating that
 the memristance acquires the limiting values $R_{min}$ and $R_{max}$.
 The parameters $\alpha$ and $\beta$ are constants defining the memristance rate
of change below and above the threshold voltage $V_T$; $V$ is
the voltage across the system.}
\label{table:threshold}
\end{table*}

\begin{figure}
 \begin{center}
\includegraphics[angle=0,width=7cm]{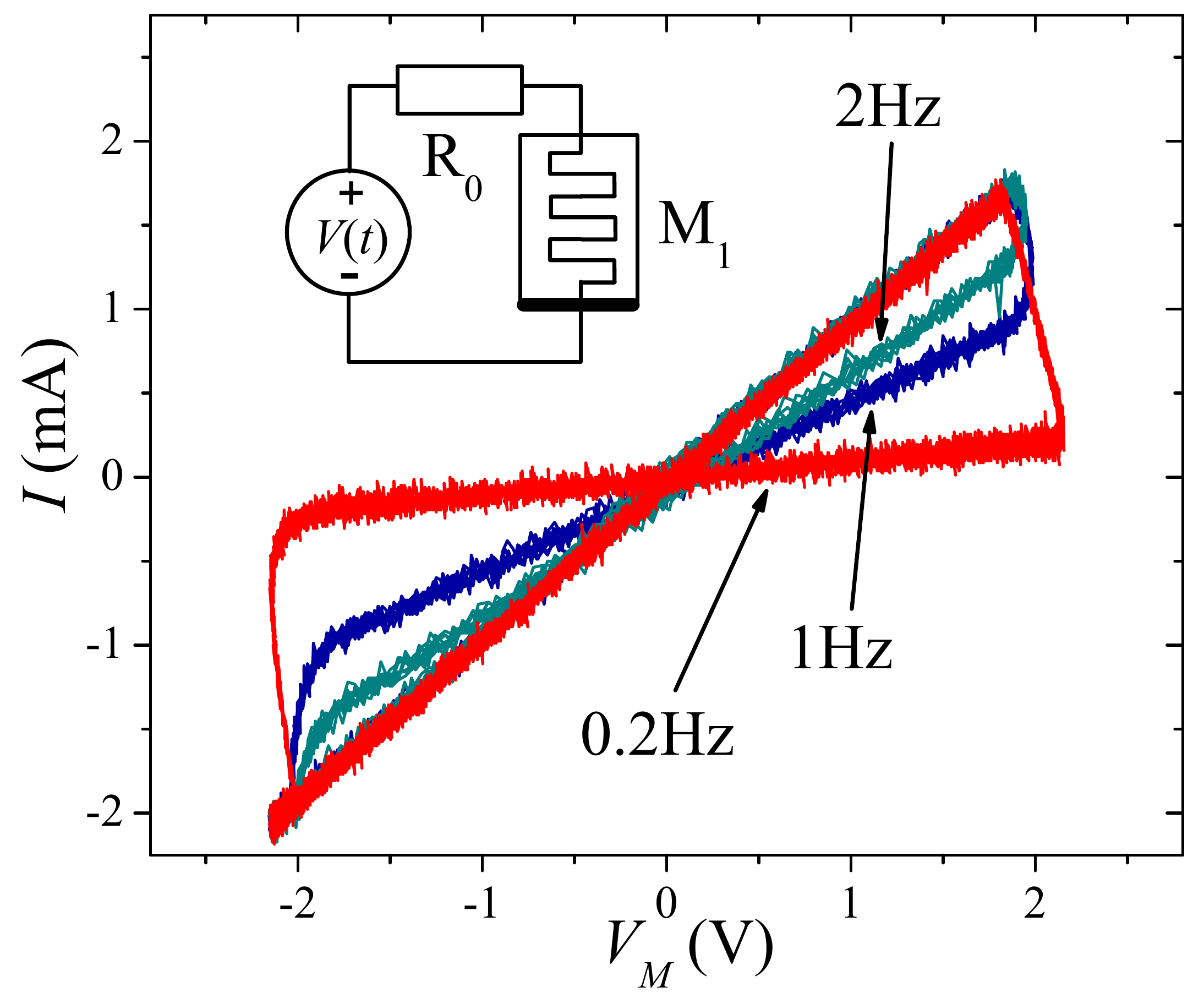}
\caption{Frequency-dependent $I-V$ curves obtained using
the model presented in the table \ref{table:threshold} preprogrammed into a memristor emulator (see section~\ref{emulators}).
These particular curves were obtained for $\alpha=0$. Such a choice
allows memristance change only when the applied voltage magnitude
is above the threshold voltage $V_T=1.75$V in the present case.  From
\cite{pershin09d} (\copyright 2010 IEEE).
 \label{analog}}
 \end{center}
\end{figure}

The model from table \ref{table:threshold} was used to
describe learning of simple biological organisms (see section
\ref{bio-insp}) as well as in several other publications
\cite{pershin09c,pershin09d,pershin10b}. In figure \ref{analog} we
demonstrate $I-V$ curves of a memristive system described by equations
shown in the table \ref{table:threshold} and obtained using a memristor emulator (see section~\ref{emulators}).
Note the similarity of these curves with
those, for example, in figure \ref{weilu}.

\subsection{Phase-change memory cells} \label{sec:PCM}

Phase-change memory cells utilize phase-change materials that can exist in at least
two different phases: amorphous and crystalline. These two different phases are characterized by
distinctive physical properties such as resistivity, optical reflectivity, etc. The idea of
phase-change memory (also known as PCM, PRAM, PCRAM, Ovonic Unified Memory, Chalcogenide RAM and C-RAM)
relies on abilities to electrically induce switching between amorphous and crystalline states by the Joule
heating due to the current flow and to probe the cell's state by measuring its resistance. Since both states are stable, no energy is required to store data.  Chalcogenides constitute most of the phase-change materials. They include Ge-Te \cite{Chen86a}, GeSeTe$_2$ \cite{Chung07a}, Ge$_2$Sb$_2$Te$_5$ \cite{Senkader04a}, AgSbSe$_2$ \cite{Wang05a}, Sb-Se \cite{Yoon06a}, Ag-In-Sb-Te \cite{Iwasaki93a}. Here, we would like to focus only on memristive aspects of a phase-change memory cell that is a unit element of phase-change memory. Additional information about properties of phase-change materials and phase-change memory
technology can be found in recent review papers \cite{Wuttig07a,Raoux08a,Raoux09a}.

\begin{figure*}
 \begin{center}
    \centerline{
    \mbox{(a)}
    \mbox{\includegraphics[width=6.00cm]{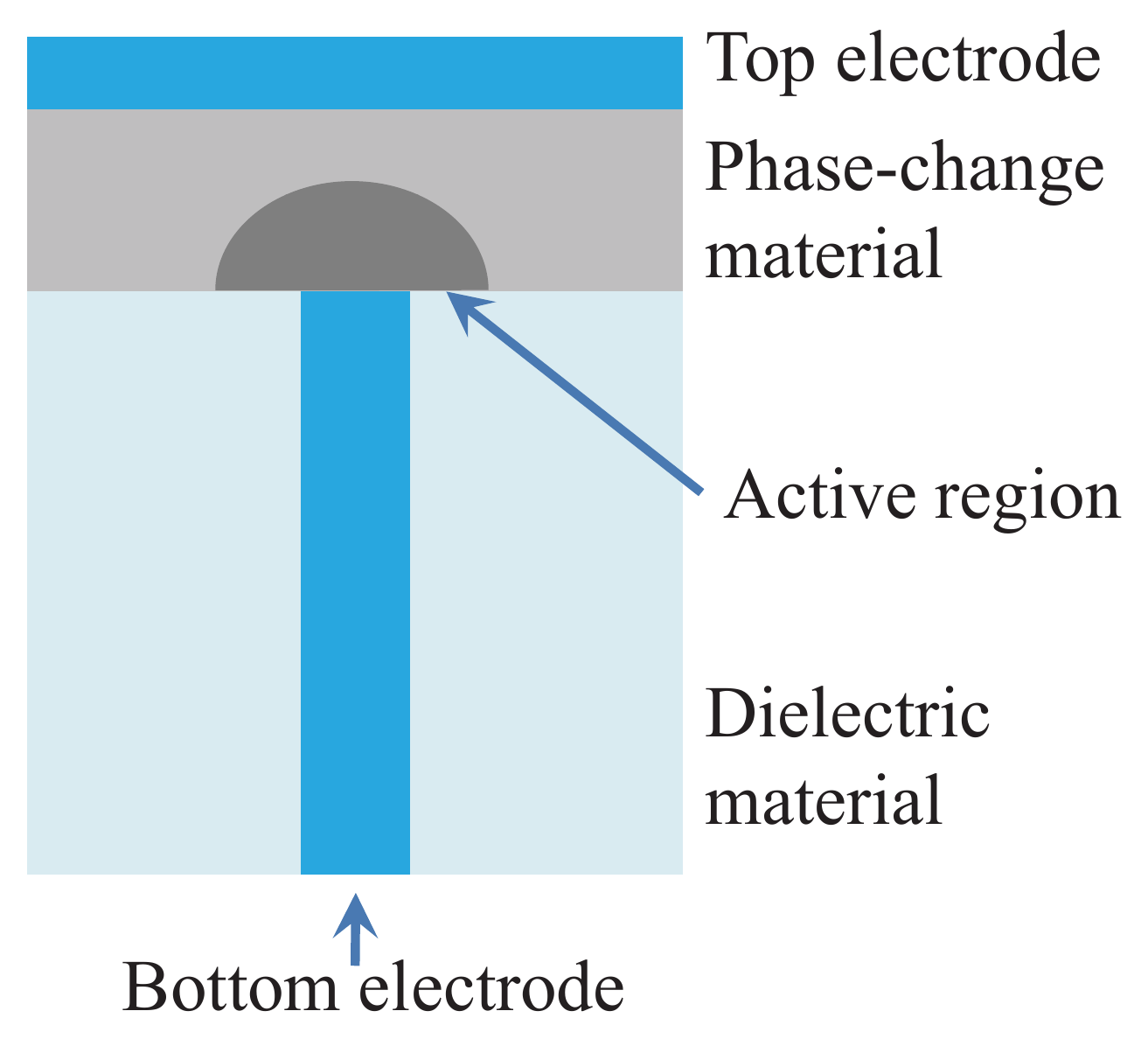}}
    \mbox{(b)}
    \mbox{\includegraphics[width=6.00cm]{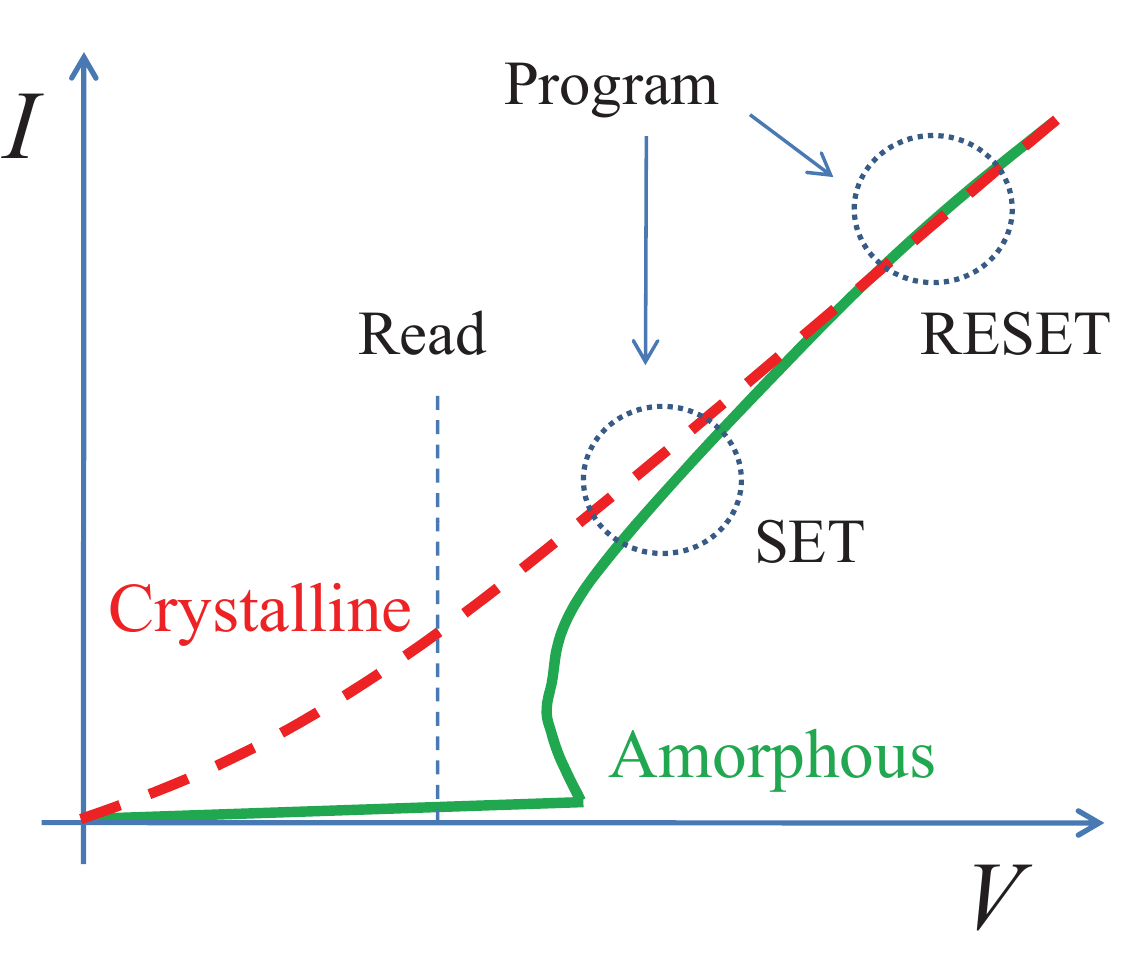}}
  }
\caption{ (a) Schematic cross section of a phase-change memory cell. (b) Shape of typical
$I-V$ curves for crystalline and amorphous phases of phase-change materials. \label{fig:PCM}}
\end{center}
\end{figure*}

Fig. \ref{fig:PCM}(a) shows schematically a so-called contact-minimized ("mushroom") phase-change memory cell \cite{Raoux08a}. It consists of
a phase-change material sandwiched between large area (top) and small area (bottom) electrodes. The active (switchable) region
of the phase-change material is located right above the bottom electrode where the current density is high. The cell design and
operation regime normally do not allow the active region to contact the top electrode. Typical experimentally-measured $I-V$ curves \cite{Pirovano04a} of phase-change materials are shown schematically in Fig. \ref{fig:PCM}(b). The switching from crystalline (low-resistive) to amorphous (high-resistive)) state (RESET operation) is performed by melting and quenching the material quickly enough. In this case, the material solidifies in the amorphous state. The switching from amorphous to crystalline state (SET operation) occurs when the material is heated above its crystallization temperature (which is below its melting temperature) for sufficiently long time. SET and RESET operations are achieved by {\it unipolar} pulses of appropriate amplitude and duration. Moreover, we note that the switching from amorphous (highly-resistive) to crystalline state would not be possible without a threshold-type increase of conductivity above a certain strength of the applied electric field (see Fig. \ref{fig:PCM}(b)). Such an effect is not yet well understood and is currently related to the interplay between impact ionization and carrier recombination \cite{Raoux08a,Pirovano04a}.

 Theoretical and numerical modeling \cite{Pirovano04a,Kang03a,Ielmini07a,Kim07b,Sonoda08a} of phase-change memory cells was previously reported focusing on either the total device operation or just some particular processes within a device (such as crystallization dynamics). In table \ref{table:PCM} we summarize a memristive model of phase-change material cell based on the idea suggested by Sonoda and co-authors \cite{Sonoda08a}. This model employs  three rate equations to describe the following factors: temperature of active region, crystallization dynamics, and threshold behavior of conductivity of amorphous state. From the electrical point of view, the cell can be represented as three resistors $R_{bottom}$, $R_{gst}$ and $R_{top}$ connected in series \cite{Sonoda08a}. Here, $R_{gst}$ describes variable resistance of active region, while $R_{bottom}$ and $R_{top}$ the resistances below and above the
 active region, respectively (see figure~\ref{fig:PCM}(a)). In the table \ref{table:PCM}, we provide the memristive model with respect to the voltage drop on $R_{gst}$ as in Ref. \cite{Sonoda08a}. As the total cell consists of only resistive elements, it is memristive \cite{chua76a}. We also note that phase-change memory cells have potential for multi-level applications \cite{Ovshinsky04a,Ventrice07a,Yin08a,Lee08a}.

\begin{table*}
{ \renewcommand{\arraystretch}{1.6}
\begin{tabular}{| l | c | }
\hline
Physical system & Phase-change memory cell \\
\hline
Internal state variable(s) & Temperature, $x_1=T_b$ \\
& Amorphous ratio, $x_2=C_a$\\
& Switching variable, $x_3=F$\\
\hline
Mathematical description &  $I=G(x_2,x_3,V_{gst})V_{gst}$  \\
&  $\frac{\textnormal{d}x_1}{\textnormal{d}t}=C_t^{-1}\left( P_t(x_2,x_3,V_{gst})-\frac{x_1-T_0}{R_t} \right)$ \\
&  $\frac{\textnormal{d}x_2}{\textnormal{d}t}=f_2(x_1,x_2)$ \\
&  $\frac{\textnormal{d}x_3}{\textnormal{d}t}=-\frac{x_3-\theta (V_{gst}-V_t)}{\tau_f}$ \\
\hline
System type & Third-order voltage-controlled memristive system \\
\hline
\end{tabular}
}
\caption{Memristive model of phase-change memory cells. Here, $T_b$ is the temperature of active layer,
$C_a$ is the amorphous ratio that changes between 0 and 1. The complete (cumbersome) expression for $G(x_2,x_3,V_{gst})$ and $f_2(x_1,x_2)$ can be read from Ref. \cite{Sonoda08a}. $P_t$ is the dissipated electric power \cite{Sonoda08a}, $T_0$ is the temperature of environment, $R_t$ is the total thermal resistance, $C_t$ is the thermal capacitance, $V_t$ is the threshold voltage (we control $F$ by voltage instead of current as in Ref. \cite{Sonoda08a}) and $\tau_f$ is the switching time.
 }
\label{table:PCM}
\end{table*}

\subsection{Metal-insulator phase transition memristive systems}\label{VO2res}

\begin{figure*}
\centering
\includegraphics[width=14cm]{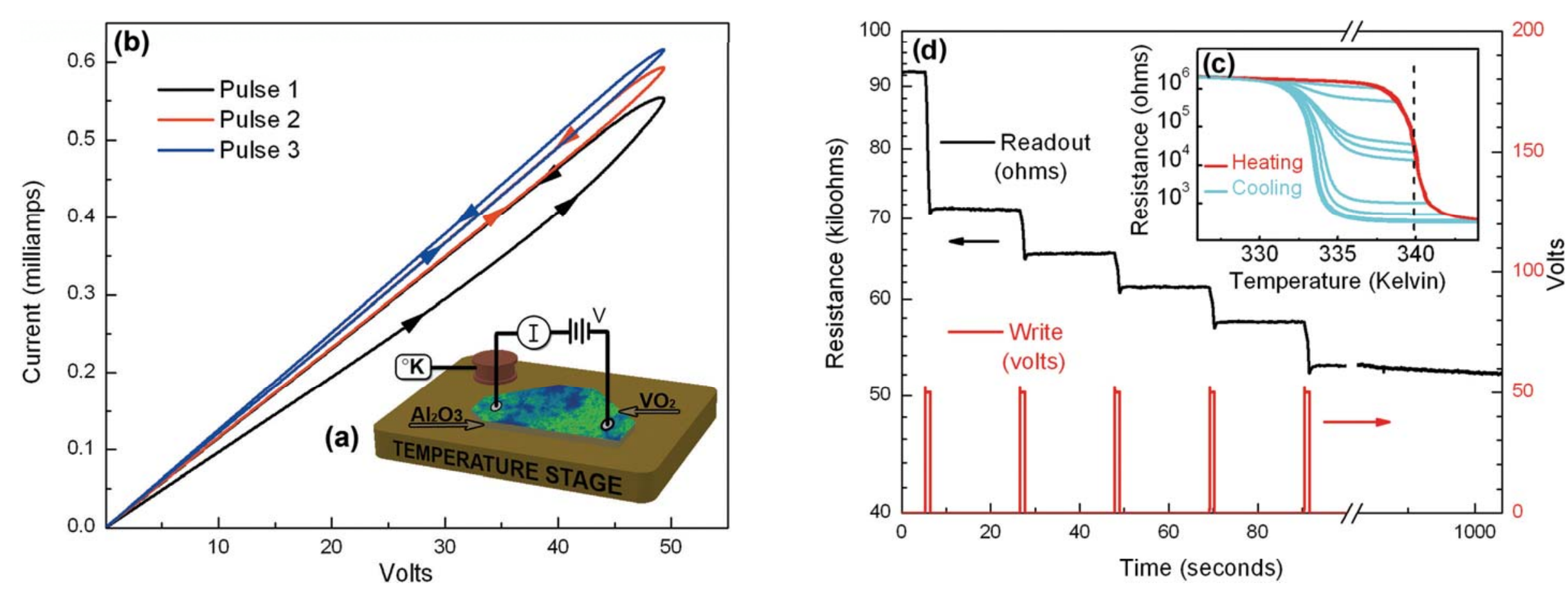}
\caption{(a) Phase-transition driven memristive
device. (b) Hysteretic $I-V$ curves under application of three
ramped voltage pulses. The ramp used for each pulse is 50V in 5s. (c)
The insulator-metal transition in VO$_2$ device. (d) Analog
information storage in memristive VO$_2$ film: resistance changes
in steps with each 50V pulse. Reprinted with permission from \cite{driscoll09b}. Copyright 2009, American Institute of Physics.
\label{VO2R}}
\end{figure*}

In strongly interacting electron materials, a complex interplay between charge, lattice and spin degrees of freedom
can result in the resistance switching effect \cite{Quintero07a,kim06a}.
Driscoll and co-workers have recently used a vanadium dioxide
based device to demonstrate memristive \cite{driscoll09b} and
memcapacitive behavior \cite{driscoll09a} of systems comprising
this oxide. In both types of experiments, the behavior is driven
by the metal-insulator phase transition (MIT) of this material
\cite{morin59a,kim06b,kim07a,krubler07a,arcangeletti07a,Qazilbash2006a,Qazilbash2008a,claassen10a}.
Here, we focus on the memristive behavior. The memcapacitive
behavior of VO$_2$ is discussed in section
\ref{phase_change_memcap}.

\begin{figure}
\centering
\includegraphics[width=8cm]{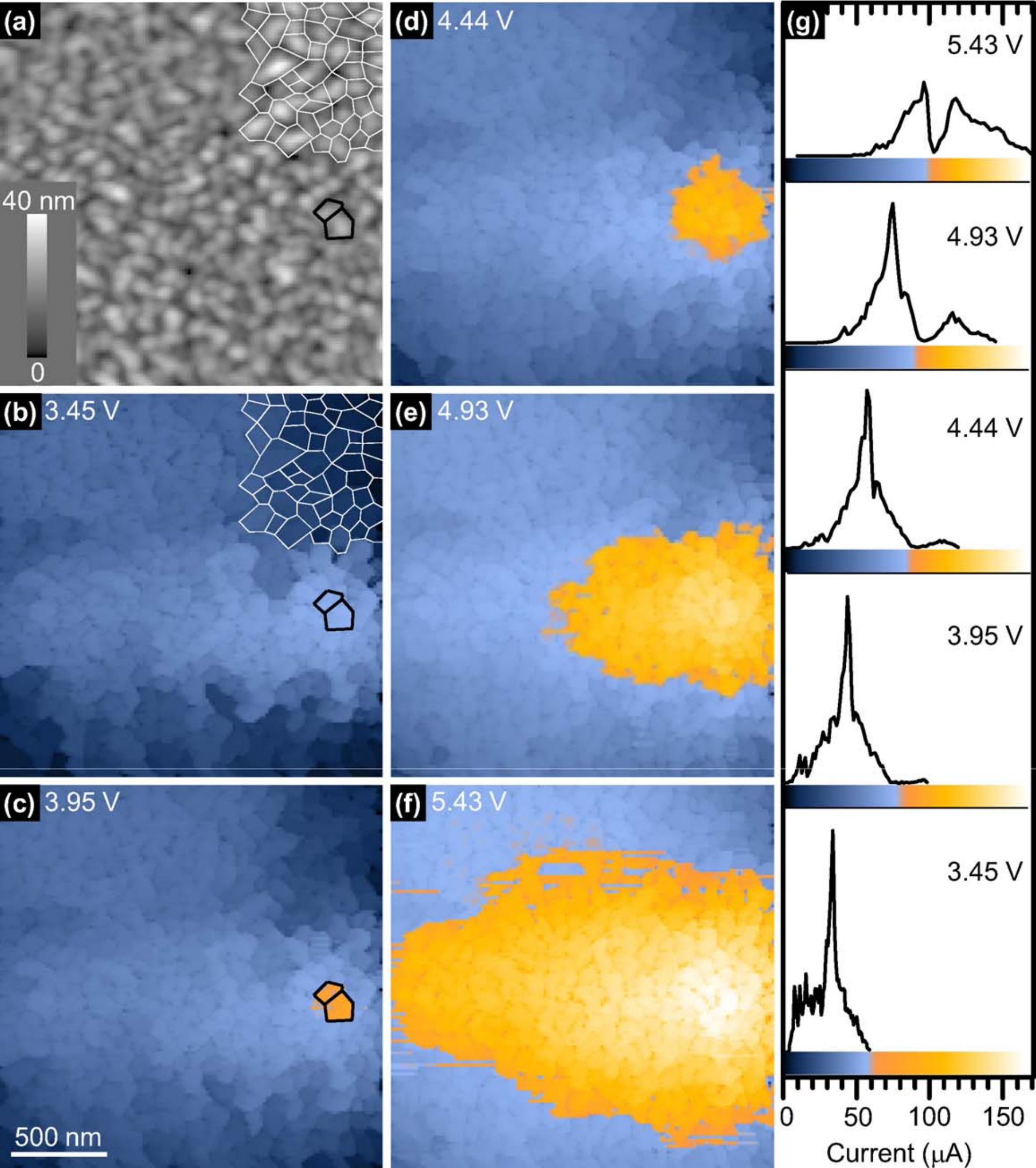}
\caption{(a) Conducting AFM topography of a VO$_2$ surface. (b-f)
Current maps at different biases showing a metallic paddle (yellow
color) increasing in size with bias. (g) Current distributions at
different biases showing the transition between insulating and
metallic state. Reprinted with permission from \cite{Kim10a}. Copyright 2010, American Institute of Physics. \label{VO2microimage}}
\end{figure}

Using the experimental setup schematically shown in figure
\ref{VO2R} where a film of VO$_2$, deposited on a sapphire substrate, is connected to two electrodes, the memristive behavior was demonstrated as follows.
The operation temperature was selected near the onset of the phase
transition (340K) where VO$_2$ properties are very
sensitive to temperature changes. High amplitude voltage pulses
(50V) were used to increase VO$_2$ temperature for short periods
of time thus promoting the MIT.
Triggered by each pulse, nanoscale metallic regions develop within
the insulating host, increasing in number and size to form a
percolative transition \cite{sharoni08a} (see figure~\ref{VO2microimage} for an AFM image of these metallic regions).

Figure \ref{VO2R}(b,d)
shows experimental results when ramping and step pulses are
applied.
The experimentally observed memristive properties of vanadium
dioxide are related to power dissipated in VO$_2$
\cite{driscoll09b}. Therefore, in such a device $f(x,I,t) \propto
I^2$ and it can be classified as an even-function
current-controlled memristive system. Consequently, the resistance
rate of change does not depend on the applied signal polarity and
can be switched only in one direction. Nonetheless, memory storage duration of
more than several hours and the possibility to store up to $2^{10}$
resistance values in a single device have been reported
\cite{driscoll09b}, thus making this system a good model exhibiting multibit information storage.

We now introduce a mathematically-transparent model of vanadium dioxide memristive system. It
can be developed assuming that the hysteresis in $R-T$ curves is
well defined \cite{almeida02a,ramirez09a,lopez02a}. This is a
reasonable approximation taking into account that the resistance
drift at a fixed temperature in the transition region is not very
strong \cite{claassen10a}. The local temperature $T$ of the
vanadium dioxide material can then be described by a heat transfer
equation (similar to the heat equation of thermistor
(\ref{eq:heatdissipation}))
\begin{equation}
C_h\frac{\textnormal{d}T}{\textnormal{d}t}=R_MI^2+\delta \left(
T_{stage}-T\right) \label{eq:VO2R1},
\end{equation}
while the equation for the resistance dynamics can be written as
\begin{eqnarray}
\frac{\textnormal{d}R_M}{\textnormal{d}
t}=\theta\left[R_M-R_1(T)\right]\frac{R_1(T)-R_M}{\tau}+ \nonumber \\
\theta\left[R_2(T)-R_M\right]\frac{R_2(T)-R_M}{\tau}
\label{eq:VO2R2},
\end{eqnarray}
where $C_h$ is the heat capacitance, $\delta$ is the dissipation
constant, $T_{stage}$ is the temperature of the thermal stage,
$\theta [...]$ is the step function, $R_1(T)$ and $R_2(T)$ are
ramp up and down $R(T)$ curves,  and $\tau$ is the resistance
relaxation time that can be considered to be much shorter than
characteristic thermal times. The first term on the right-hand
side of equation (\ref{eq:VO2R2}) describes a resistance decrease
when temperature increases, and the second term is responsible for
the opposite process. As the resistance relaxation time $\tau$ is
short, equation (\ref{eq:VO2R2}) represents very fast transition
$R_M\rightarrow R_1(T)$ if $R_M>R_1(T)$ and $R\rightarrow R_2(T)$ if
$R_M<R_2(T)$. It is clear that, according to equation
(\ref{eq:VO2R2}), $R$ does not change if its value falls inside of the
hysteresis loop.

Thus, our description of MIT in VO$_2$ involves two state
variables, the temperature and resistance. The time dependence of
state variables is given by equations
(\ref{eq:VO2R1},\ref{eq:VO2R2}). Since the current enters into the
right-hand-side of equation (\ref{eq:VO2R1}), our model describes
a second-order current controlled memristive system.

\begin{figure*}
\centering
\includegraphics[width=14cm]{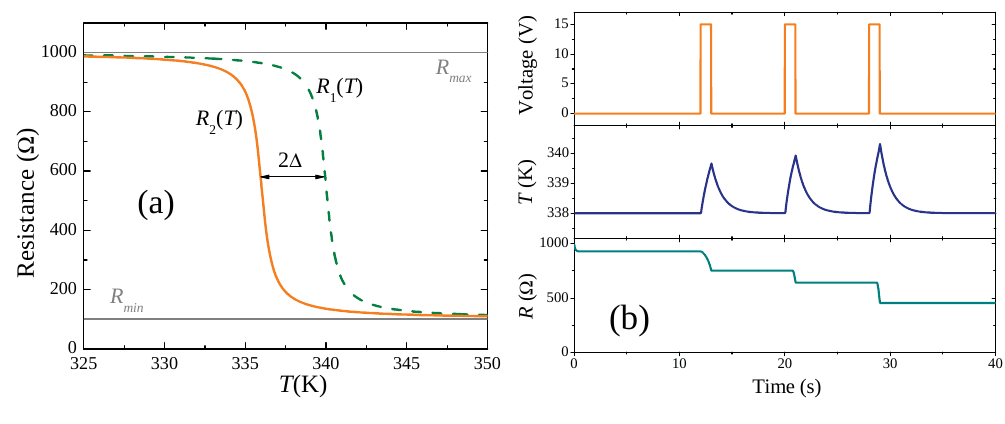}
\caption{(a) $R-T$ hysteresis loop obtained using
equations (\ref{eq:VO2sim1}-\ref{eq:VO2sim3}). (b) Simulations of
vanadium dioxide memristive system response to three 15V pulses. These
plots were obtained using parameter values $R_{max}=1000\Omega$,
$R_{min}=100\Omega$, $T_0=338$K, $\eta=0.5$K, $\Delta=4$K,
$R(t=0)=1000\Omega$, $T(t=0)=338K$, $\tau=0.1$s, $C_h=0.1$J/K,
$\delta=0.1$J/(s$\cdot$K). No attempt has been made here to
perfectly match the
experimental curves. \label{VO2Rsim}}
\end{figure*}

The experimentally measured profile of the hysteresis
\cite{driscoll09a,driscoll09b} suggests that the functions
$R_1(T)$ and $R_2(T)$ can be selected of the same shape displaced
by $2\Delta$, where $2\Delta$ is the hysteresis width. For example,
in our numerical simulations of memristive properties of vanadium
dioxide shown in figure \ref{VO2Rsim} we have employed the following profile for
$R_1(T)$ and $R_2(T)$
\begin{eqnarray}
R_1(T)&=&F(T-\Delta) \label{eq:VO2sim1} \\ R_2(T)&=&F(T+\Delta) \label{eq:VO2sim11} \\
F(x)&=&\left(\frac{1}{\pi}\textnormal{arctan} \left[
\frac{T_0-x}{\eta}\right]+0.5\right)\left[R_{max}-R_{min}
\right]+ \nonumber \\
R_{min} \label{eq:VO2sim3}
\end{eqnarray}
where $R_{max}$ and $R_{min}$ are asymptotic values of resistance
below and above the transition temperature, respectively, $T_0$ is
the average transition temperature and $\eta$ is the parameter
defining the sharpness of the resistance step at the
metal-to-insulator transition. An example of a hysteresis loop
based on equations (\ref{eq:VO2sim1}-\ref{eq:VO2sim3}) is
demonstrated in Figure \ref{VO2Rsim}(a). We emphasize that the
choice of $R_1(T)$ and $R_2(T)$ given by equations
(\ref{eq:VO2sim1}-\ref{eq:VO2sim3}) is not unique and possibly
better fits to experimental curves can be found.

\begin{table*}
{ \renewcommand{\arraystretch}{1.6}
\begin{tabular}{| l | c | }
\hline
Physical system & Vanadium dioxide memristive device \\
\hline
Internal state variable(s) & Temperature and resistance, $x_1=T$, $x_2=R_M$ \\
\hline
Mathematical description & $V=x_2I$  \\
& $\frac{\textnormal{d}x_1}{\textnormal{d}t}=C_h^{-1}\left[ x_2I^2+\delta \left(
T_{stage}-x_1\right) \right]$ \\
& $\frac{\textnormal{d}x_2}{\textnormal{d}t}=\theta\left[x_2-R_1(x_1)\right]\frac{R_1(x_1)-x_2}{\tau}+$ \\
& $\theta\left[R_2(x_1)-x_2\right]\frac{R_2(x_1)-x_2}{\tau}$ \\
\hline
System type & Second-order current-controlled
memristive system \\
\hline
\end{tabular}
}
\caption{Memristive model of vanadium dioxide device. The notations are given in the text, and the functions $R_1(y)$ and $R_2(y)$ are defined by Eqs. (\ref{eq:VO2sim1}-\ref{eq:VO2sim11}).}
\label{table:vanadium}
\end{table*}

The application of voltage pulses to vanadium dioxide memristive system
results in a local heating as demonstrated in figure
\ref{VO2Rsim}(b). Since the resistance of the system decreases with
each pulse, the amount of heat released due to subsequent pulses
increases. This can be clearly seen in the increasing pulses'
magnitude in the $T(t)$ curve. This mechanism permits the progress of
the insulator-to-metal transition using pulses of fixed amplitude
and width. The resulting $R_M(T)$ steps shown in figure
\ref{VO2Rsim}(b) are similar to those observed experimentally
\cite{driscoll09b} (compare with figure~\ref{VO2R}).

\begin{figure}
\centering
\includegraphics[width=8cm]{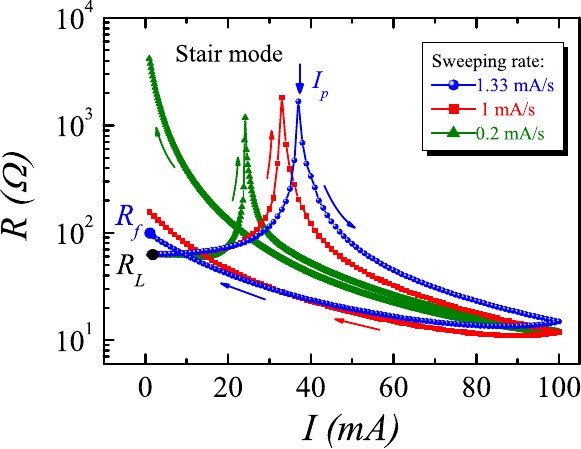}
\caption{ Resistance-current characteristics of a La$_{0.225}$Pr$_{0.4}$Ca$_0.375$MnO$_3$ device measured under the current sweeping in the stair mode with different rates as indicated. All curves start with the same initial resistance. The strongest resistance change is observed at the slowest sweeping rate. Reprinted with permission from \cite{Yan09a}. Copyright 2009, American Institute of Physics.
\label{yan}}
\end{figure}

The reversible resistance switching was recently observed in a manganite La$_{0.225}$Pr$_{0.4}$Ca$_{0.375}$MnO$_3$
at low temperatures \cite{Yan09a}. In this material, the resistance switching effect is explained by a Joule
heat-induced transition between charge-ordered insulator and ferromagnetic metal phases \cite{Yan09a}. The authors show that such a transition
is bidirectional and can be uniquely controlled by current pulse pairs \cite{Yan09a}. Below 30K, the system
enters into a frozen state in which the phase separation is blocked. Under this condition, low and high resistance states
become nonvolatile. Figure \ref{yan} demonstrates that the resistance switching in La$_{0.225}$Pr$_{0.4}$Ca$_{0.375}$MnO$_3$ has a pronounced
frequency dependence. This feature is typical for memristive systems and  can be potentially used for multi-state memory. However, we emphasize that the hysteresis in this particular material is not suitable for room-temperature applications.

\subsection{Spintronic systems}

Memory is not necessarily confined to structural or charge
properties but may also arise from the spin degree of freedom.
This is a particularly important result in view of the possible
development of devices that operate with low dissipation and show
high reliability under ac-bias conditions. In the following, we
then discuss memory effects in the two major areas of spintronic
research: semiconductor \cite{zutic04a} and metal
\cite{Bratkovsky08a,Brataas06a} spintronics.

\subsubsection{Semiconductor spintronic systems} \label{semic_spin_systems}

It was shown by Pershin and Di Ventra~\cite{pershin08a} that
memory effects are an intrinsic feature of many semiconductor spintronic
systems. These effects have a spin-related origin and,
consequently, can not be observed in spin unpolarized systems. The general idea behind these memory phenomena is as follows. If we
consider a structure (such as, for example, a semiconductor/ferromagnet junction)
driven by a time-dependent external control
parameter (such as applied voltage or current) then, when the
external control parameter changes, it takes some time for the electron
spin polarization to adjust to a new control parameter value - typically, the equilibration is governed by electron-spin
diffusion and relaxation processes \cite{pershin08a}. Within the time scale of this
``spin-polarization adjustment'', the system keeps its memory on the past dynamics and, when, in addition,
the level of electron spin polarization influences the system's resistance,
it exhibits memristive behavior.

\begin{figure*}
 \begin{center}
\includegraphics[angle=0,width=6cm]{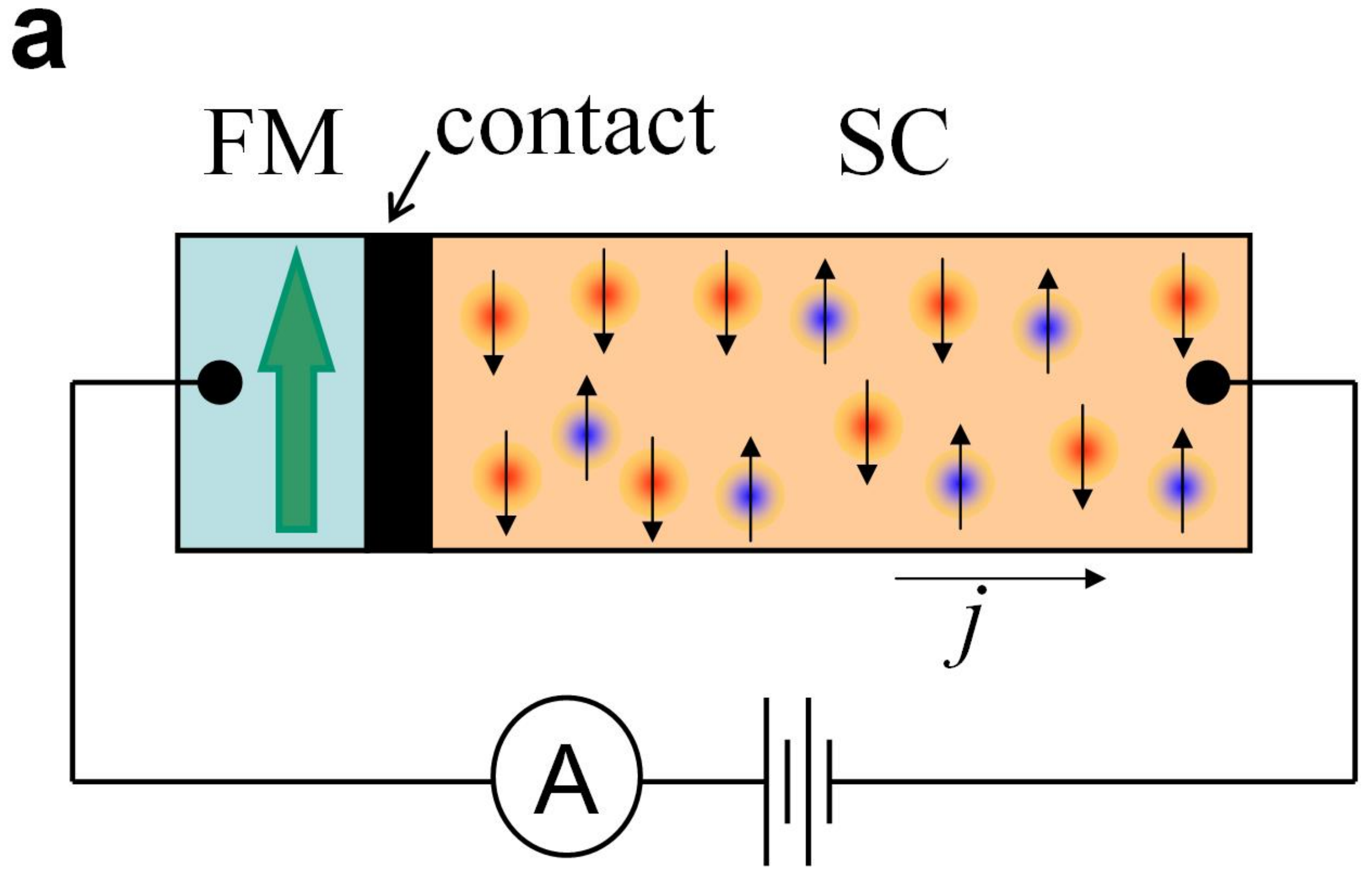}
\includegraphics[angle=0,width=12cm]{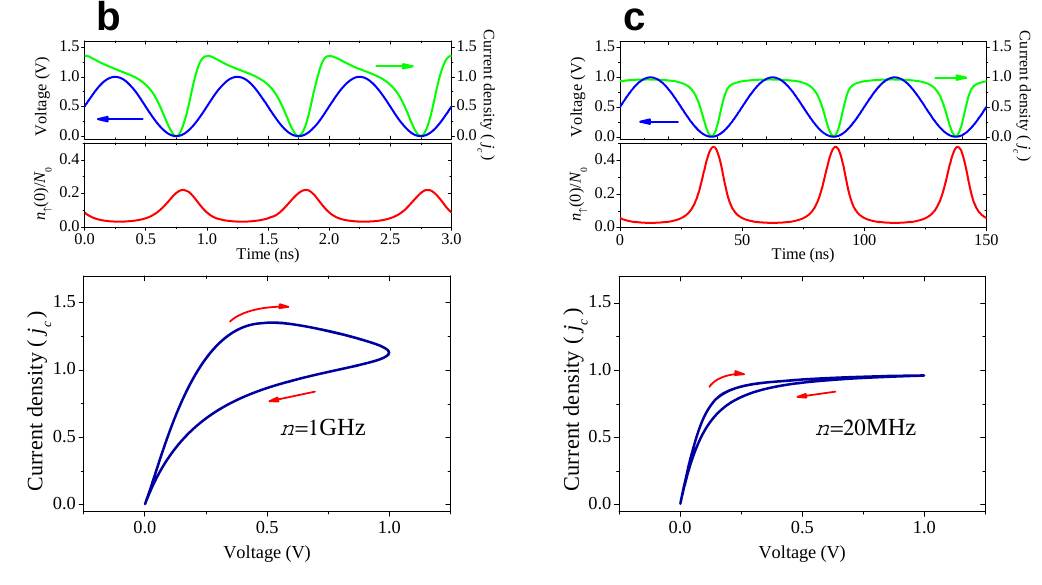}
\caption{Memristive effects in
semiconductor/ferromagnet junctions: (a) schematic representation
of the circuit containing an interface between a semiconductor
material and a half-metal; (b) and (c) - simulations of ac
response of the system. The applied voltages (blue lines) are
$V=V_1+V_2 \textnormal{sin}\left( 2 \pi \nu t \right)$ with
$V_1=V_2=0.5$V, $\nu=10^9$Hz in (b) and $\nu=2\cdot 10^7$Hz in
(c). In both cases the current densities (in units
of the critical current density, equation~(\ref{criticalcurr})) and spin-up electron densities
near the contact (in units of the electron density in the semiconductor) show saturation typical for spin
blockade \cite{pershin08b}. It is clearly seen that the $I-V$
hysterisis is significantly reduced in the low-frequency case. For
more details see reference \cite{pershin08a}. \label{spin_mem}}
 \end{center}
\end{figure*}

As an example, let us consider current flowing through a
semiconductor-half-metal (namely, a perfect ferromagnet \cite{Attema06a})
junction (see figure \ref{spin_mem}(a)) in such a way that the
electron flow is directed from the semiconductor into the
ferromagnet (this process is called spin extraction). It is known
that spin extraction is accompanied by the phenomenon of spin
blockade~\cite{pershin07a} characterized by a saturated $I-V$
curve~\cite{pershin08b}. Physically, the outflow of majority-spin
electrons from the semiconductor leaves a cloud of minority-spin
electrons near the junction. This minority-spin cloud can limit
the majority-spin current creating a pronounced spin-blockade at a
critical current density~\cite{pershin07a}
\begin{equation}
j_c=eN_0\sqrt{\frac{D}{2\tau_{sf}}},\label{criticalcurr}
\end{equation}
where $N_0$ is the electron density in the semiconductor, $D$ is
the diffusion coefficient, and $\tau_{sf}$ is the spin-relaxation
time.

The time evolution of the electron spin polarization in a non-degenerate semiconductor can
be described by the two-component drift-diffusion model~\cite{yu02a}
\begin{equation}
e\frac{\partial n_{\uparrow (\downarrow)}}{\partial
t}=\frac{\partial j_{\uparrow (\downarrow)}}{\partial y}
+\frac{e}{2\tau_{sf}}\left(n_{\downarrow (\uparrow)}-n_{\uparrow
(\downarrow)} \right), \label{contEq}
\end{equation}
\begin{equation}
j_{\uparrow (\downarrow)}=\sigma_{\uparrow
(\downarrow)} E+eD\nabla  n_{\uparrow
(\downarrow)} , \label{currentEq}
\end{equation}
which are accompanied by an equation for the total voltage
drop~\cite{pershin07a}
\begin{equation}
V=\left[ \rho_s L+ \rho_c^0\frac{N_0}{2n_\uparrow (0)}\right]j
\label{Vtotal}
\end{equation}
where $-e$ is the electron charge, $n_{\uparrow (\downarrow)}$ is
the density of spin-up (spin-down) electrons, $j_{y,\uparrow
(\downarrow)}$ is the corresponding current density, $\sigma_{\uparrow
(\downarrow)}=en_{\uparrow
(\downarrow)}\mu$ is the
conductivity, $\mu$ is the mobility, $L$ is the length of the semiconductor region, $E$
is the electric field, $\rho_s$ is the semiconductor resistivity,
and $\rho_c^0$ is the contact resistivity at $V \rightarrow 0$.

\begin{figure*}
 \begin{center}
\includegraphics[angle=0,width=14cm]{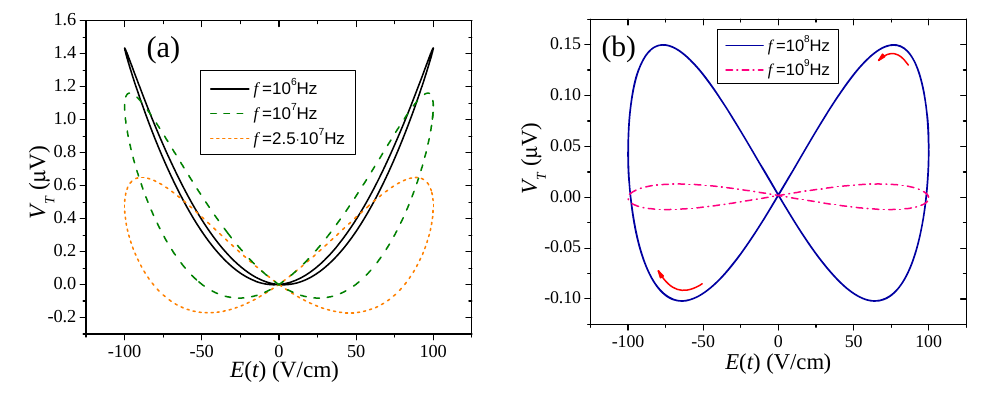}
\caption{Spin memristive effects in a semiconducting system with
inhomogeneous electron density in the direction perpendicular to
main current flow~\cite{pershin09a}. Here, we show the transverse
voltage as a function of applied electric field at different
applied field frequencies as indicated. From \cite{pershin09a}.}
\label{fig:spintransverse}
\end{center}
\end{figure*}

Clearly, equations (\ref{contEq},\ref{Vtotal}) define a
continuous current-controlled memristive system (note that by inverting equation~(\ref{Vtotal}) it can also be redefined as a voltage-controlled
memristive system). In such a system, the majority-spin density at
the boundary $n_\uparrow (0)$ defines (through equation
(\ref{Vtotal})) the resistance of the device. The details of the memristive model of semiconductor/half metal junctions (in the assumption of constant electron density) is summarized in the table \ref{table:semspin}. The differential equation for the continuous state variable $X=n_\uparrow (y,t)$ ($y$ is the direction perpendicular to
the interface) should be solved with the boundary conditions specified in the caption of table \ref{table:semspin}. In figure \ref{spin_mem}(b,c)
we show a numerical solution of the above equations with
appropriate boundary conditions~\cite{pershin08a}. The pinched
hysteresis loops demonstrate typical frequency behavior for
memristive systems.

\begin{table*}
{ \renewcommand{\arraystretch}{1.6}
\begin{tabular}{| l | c | }
\hline
Physical system & Semiconductor/half-metal junction \\
\hline
Internal state variable(s) & Density of spin-up electrons, $X(y,t)=n_\uparrow (y,t)$ \\
\hline
Mathematical description & $V=\left[ \rho_s L+ \rho_c^0\frac{N_0}{2X (0,t)}\right]j$  \\
& $\frac{\partial X}{\partial t}=\frac{j}{eN_0} \frac{\partial X}{\partial y}+D\frac{\partial^2 X}{\partial y^2}+
\frac{N_0-2X}{2\tau_{sf}}$ \\
\hline
System type & Continuous current-controlled
memristive system \\
\hline
\end{tabular}
}
\caption{ Memristive model of transport through a semiconductor/half-metal junction.
It is assumed that the total electron density in the
semiconductor is constant, that is $n_\uparrow+n_\downarrow=N_0$. Correspondingly, the electric field in the semiconductor region is homogeneous. The boundary conditions for $X$ follow from the equations $j_{\uparrow}(0)=j$ and $j_{\uparrow}(\infty)=j/2$.}
\label{table:semspin}
\end{table*}

In addition to the above results, in certain semiconductor
spintronic systems, spin memory effects can be observed directly
in the transverse voltage. For example, this occurs in spin Hall
effect systems \cite{Dyakonov71a,Wunderlich05a} with an
inhomogeneous electron density in the direction perpendicular to
the direction of main current flow~\cite{pershin09a}. Figure
\ref{fig:spintransverse} shows transverse voltage hysteresis loops
that demonstrate typical memristive behavior: non-linear
dependence at low frequencies, pronounced hysteresis at higher
frequencies, and hysteresis collapse at very high
frequencies~\cite{pershin09a}. The physical origin of this effect
is similar to what we have discussed above: the time it takes for
the electron spin polarization to relax to its instantaneous
equilibrium configuration is finite, thus leading to a
history-dependent observable.

\subsubsection{Metal spintronic systems}

\begin{figure*}
 \begin{center}
\includegraphics[angle=0,width=12cm]{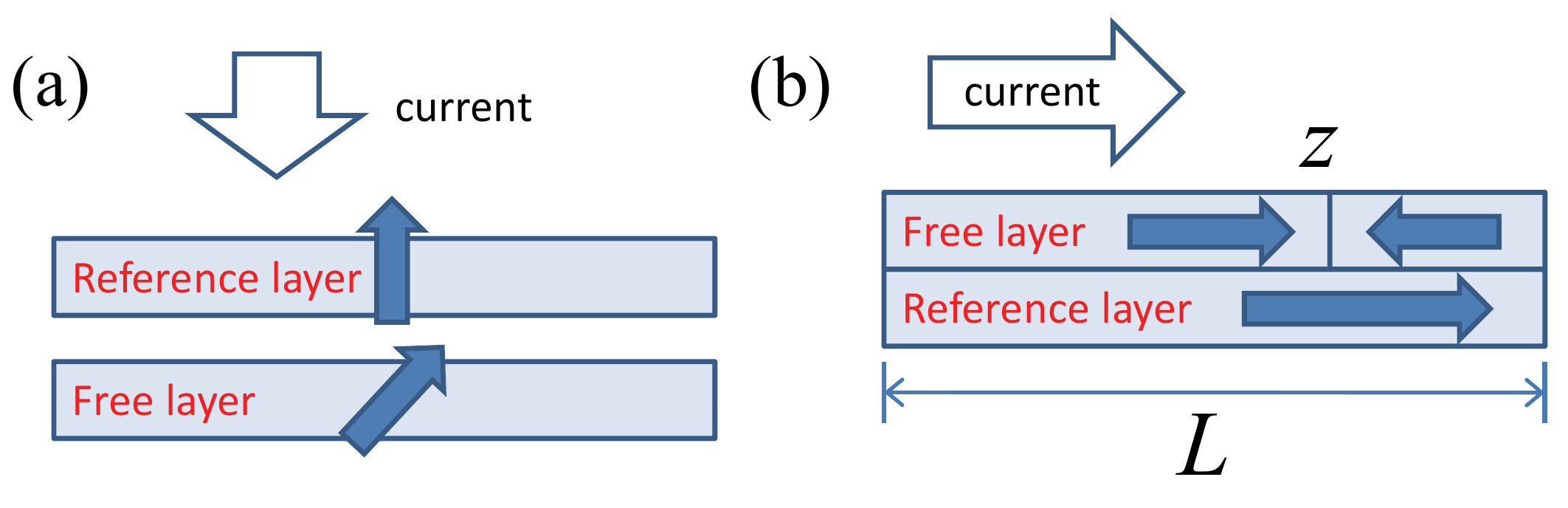}
\caption{All-metal spintronic memristive systems (a) magnetic tunnel junction with spin-torque-induced magnetization switching, and (b)
spin valve of width $L$ with spin-torque-induced domain-wall motion \cite{wang09a}. The arrows indicate the direction of spin polarization and $z$ the position of the domain wall.} \label{metal_memr}
\end{center}
\end{figure*}

A different class of spin-based memristive systems takes advantage of all-metal
spintronic devices \cite{Bratkovsky08a,Brataas06a}. The operation
of such devices can employ spin-torque-induced magnetization
switching or magnetic-domain-wall motion \cite{wang09a}. Figure
\ref{metal_memr} depicts both schemes.

{\it Spin-torque transfer systems -} One realization (figure
\ref{metal_memr}(a)) is based on spin-torque
transfer~\cite{berger78a,slonczewski96a,tsoi98a,myers99a}. In
spin-torque transfer systems, the resistance is determined by the
relative magnetization between opposite sides of a magnetic tunnel
junction. Current flowing through the junction induces spin
torque, in turn changing the relative magnetization.

Quite generally, considering injection of spin-polarized electrons
from a reference layer (with a fixed direction of magnetization)
into a free layer (whose magnetization direction can change), the
magnetization dynamics of the free layer can be described by the
Landau-Lifshitz-Gilbert equation
\cite{landau35a,gilbert55a,slonczewski96a}. Parameterizing the
magnetization direction of the free layer by two angles $\theta$
and $\phi$ (according to \cite{sun00a}, $\theta$ is the angle
between the magnetization direction and a uniaxial anisotropy axis
assumed to be in the interface plane, and $\varphi$ is the angle in
the plane perpendicular to the uniaxial anisotropy axis) the
Landau-Lifshitz-Gilbert equation can be written as \cite{sun00a}
\begin{equation}
\frac{\textnormal{d}}{\textnormal{d}t}\left( \begin{matrix} \theta
\\ \varphi \end{matrix} \right)= \left[ \begin{matrix}
f_1(\theta,\varphi) \\ f_2(\theta,\varphi) \end{matrix} \right],
\label{eq:spin_general}
\end{equation}
where the functions $f_1(\theta,\phi)$ and $f_2(\theta,\phi)$ on
the right hand side take into account different factors
influencing magnetization dynamics such as uniaxial anisotropy,
easy-plane anisotropy, magnetic field, spin-torque and dissipation
\cite{sun00a}. The resistance of the magnetic tunnel junction is
determined by the angle between magnetization directions of the
free and reference layer that can be expressed using $\theta$ and
$\phi$. Consequently, from the point of view of memory systems,
the magnetic tunnel junctions can be categorized as second-order
current-controlled memristive systems.

To illustrate the origin of memory behind these concepts, let us
consider a simple system of a free layer with uniaxial anisotropy
only. In this case, the Landau-Lifshitz-Gilbert equation involves
only $\theta$ and the order of the memristive system is reduced.
For this specific situation, equations (\ref{eq:spin_general}) are
written as \cite{wang09a,sun00a}
\begin{equation}
\frac{\textnormal{d}\theta}{\textnormal{d}t}=\alpha \gamma H_k \left(-\sin \theta\cos\theta+
p\sin\theta \right) \label{eq:theta}
\end{equation}
where $\theta$ is the angle between the free-layer magnetization direction and the
pinned-layer (reference layer) magnetization direction, $\gamma$ is the gyromagnetic ratio,
$\alpha$ is the damping parameter, $H_k$ is the perpendicular anisotropy of the
free layer, and $p=(\eta\hbar I)/(2\alpha eM_sH_kV)$ describes the effect of
the polarized-current ($I$) spin-torque. Here, $\eta$ is the polarization efficiency,
$V$ is the film-element volume and $M_s$ is the magnetization saturation.

Resistance of
the corresponding magnetic tunnel junction (MTJ) is given by
\begin{equation}
R_M(\theta)=\frac{1}{G_0\left( 1+\frac{TMR}{TMR+2}\cos\theta\right)}  \label{eq:R}
\end{equation}
where $G_0$ is the MTJ conductance when $\theta=\pm \pi/2$, and
the tunneling magnetoresistance, $TMR$, is the ratio of the
difference between high and low conductance to low conductance
(high conductance occurs when both layers magnetizations are
parallel to each other and low conductance when the
magnetizations are antiparallel). The model of spin-torque transfer device discussed above
is condensed in the table \ref{table:STT}. Its memristive origin is evident. In fact, an interesting feature of this system is the periodicity of the angle when it is changed by $2\pi$. The state variable $\theta$, actually, can not record  such a change. This feature should be taken into account in studies of such structures.

\begin{table*}
{ \renewcommand{\arraystretch}{1.6}
\begin{tabular}{| l | c | }
\hline
Physical system & Spin torque transfer device \\
\hline
Internal state variable(s) & Magnetization direction angle, $x=\theta$ \\
\hline
Mathematical description & $V=\frac{1}{G_0\left( 1+\frac{TMR}{TMR+2}\cos x\right)}I$  \\
& $\frac{\textnormal{d}x}{\textnormal{d}t}=\alpha \gamma H_k \left(-\sin x \cos x+
\frac{\eta\hbar I}{2\alpha eM_sH_kV}\sin x \right)$ \\
\hline
System type & First-order current-controlled memristive system \\
\hline
\end{tabular}
}
\caption{ Memristive model of a spin torque transfer device \cite{wang09a}.}
\label{table:STT}
\end{table*}

{\it Spin-torque-induced domain-wall motion -} In the second
realization of metallic spintronics (figure \ref{metal_memr}(b)) a
long spin-valve structure is realized with domain-wall motion in
the free layer induced by the current \cite{wang09a}. In this
geometry, the current flows in both free and reference layers. The
resistance of such a structure depends on the domain-wall position
$z$ (along the direction of current flow) as
\begin{equation}
R_M(z)=\frac{R_lz}{L}+\frac{R_h(L-z)}{L}. \label{eq:resist_spin}
\end{equation}
Here, $L$ is the free-layer length, $R_l$ is the low resistance when magnetizations of free and reference layers are parallel and $R_h$ is the high resistance when magnetizations of both layers are antiparallel. In a linear approximation, the domain wall velocity is proportional to the current strength
\begin{equation}
\frac{\textnormal{d}z}{\textnormal{d}t}=\Gamma I, \label{eq:domain_vel}
\end{equation}
where $\Gamma$ is a proportionality coefficient. Equation
(\ref{eq:domain_vel}) should be solved with the boundary
conditions $0 \le z \le L$. It follows from equations
(\ref{eq:resist_spin},\ref{eq:domain_vel}) that the device is
described by general equations
\begin{equation}
V=R_M\left(z\right)I, \quad \frac{\textnormal{d}z}{\textnormal{d}t}=f\left( I\right).
\end{equation}
Therefore, this spin-valve structure with domain-wall motion is
also a first-order current-controlled memristive system. When the position
of the domain wall $z$ is confined inside the free layer
($0<z(t)<L$), we can integrate equation (\ref{eq:domain_vel})
obtaining $z(t)=\Gamma q(t)$. In this regime, the device behaves
as an ideal memristor whose memristance is given by
\begin{equation}
R_M= R_h -\frac{(R_h-R_l)\Gamma \int\limits_{-\infty}^t
I(t')dt'}{L} . \label{eq:memr_domain}
\end{equation}
In reality, the domain wall motion is a complex process.
Consequently, equation (\ref{eq:memr_domain}) provides only a
first approximation \cite{wang09a} to the device response.

\begin{table*}
{ \renewcommand{\arraystretch}{1.6}
\begin{tabular}{| l | c | }
\hline
Physical system & Domain wall \\
\hline
Internal state variable(s) & Position, $x=z$ \\
\hline
Mathematical description & $V=\left[ R_h -\frac{(R_h-R_l)\Gamma \int\limits_{-\infty}^t
I(t')dt'}{L} \right]I$  \\
\hline
System type & Current-controlled memristor \\
\hline
\end{tabular}
}
\caption{Memristive model of spin-torque induced domain wall motion.}
\label{table:domwallmemr}
\end{table*}

Wang {\it et al.} have suggested to use the temperature dependence
of the domain wall mobility - which is observed within a certain
range of parameters - to sense an external temperature
\cite{wang10a}. Generally, the domain wall velocity as a function
of the driving current has a threshold-type dependence
\cite{Zapperi98a}. Around the critical current value, the thermal
fluctuations play an important role in the domain wall de-pinning
from crystallographic defects thus providing a basis for thermal
sensitivity \cite{wang10a}. The resulting {\it spintronic
memristor temperature sensor} has the same device structure as
shown in figure \ref{metal_memr}(b). For temperature sensing, it
is suggested to apply a voltage pulse of a constant magnitude to
the device. The resistance difference before and after the
voltage pulse is measured and calibrated to sense the temperature.
Theoretically, the domain wall motion in the critical current
region is described by stochastic differential equations
\cite{duine07a}. Therefore, this system is an example of a {\it stochastic memory circuit element} introduced in  section \ref{gen_defs}.

\begin{figure}
 \begin{center}
\includegraphics[angle=0,width=7cm]{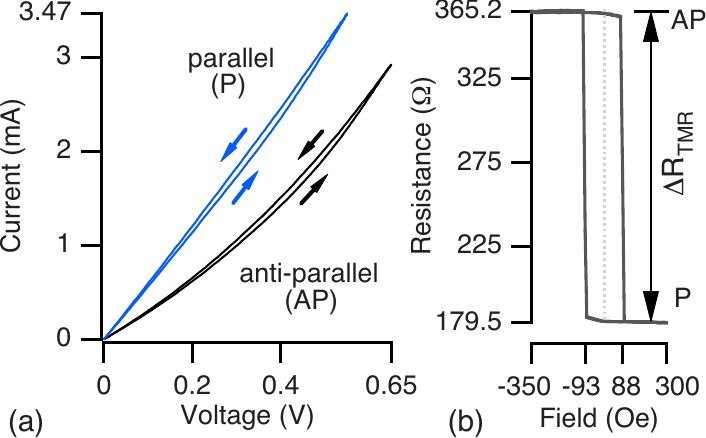}
\caption{Simultaneous manifestation of resistive and
magnetoresistive switching. (a) Due to resistive switching,
splitted $I-V$ curves are observed for both magnetic
configurations (parallel and anti-parallel) of TMJ device. (b)
Magnetic hysteresis loop showing magnetoresistive switching of the
device. Reprinted with permission from \cite{krzysteczko09a}. Copyright 2009, American Institute of Physics.} \label{MTJ}
\end{center}
\end{figure}

A different approach to memristive switching of magnetic junctions
was followed by Krzysteczko {\it et al.} \cite{krzysteczko09a}. These
authors have fabricated a MgO modified magnetic tunnel junction
with a resistive switching material inside. As shown in figure
\ref{MTJ}, resistive switching results in a splitting of
magnetoresistance curves. The authors observed a tunnel
magnetoresistance of about 100\% ratio, and bipolar resistive
switching of about 6\%. Five resistance states were demonstrated.
This approach is thus a promising alternative to create multi-bit
states for storage and logic.

\subsection{Ionic channels} \label{ion_channels}
Another important memristive system pertains to biological neural
networks, and in particular to the functioning of membranes in
axon cells. In fact, in 1952 Hodgkin and Huxley suggested a model
of action potentials \cite{hodgkin52a} in neurons that employs
history-dependent channel conductances that are essentially
memristive. This model is one of the most significant conceptual
achievements in neuroscience \cite{hausser00a}. In its original
formulation, the nerve membrane (specifically, the membrane of the
squid giant axon) was described by three types of ion channels:
leakage channels (primarily carrying chloride ions), Na channels
and K channels. Leakage channels are mainly responsible for the
resting membrane potential and have a relatively low constant
conductance. The conductance of the other two channels changes as
a function of time and voltage, and here lies the memristive
behavior we are interested in. Hodgkin and Huxley have
demonstrated that step depolarizations of the squid axon trigger a
rapid inward current across the membrane carried by Na$^+$ ions,
followed by an outward current due K$^+$ ions.

\begin{figure}
 \begin{center}
\includegraphics[angle=0,width=7cm]{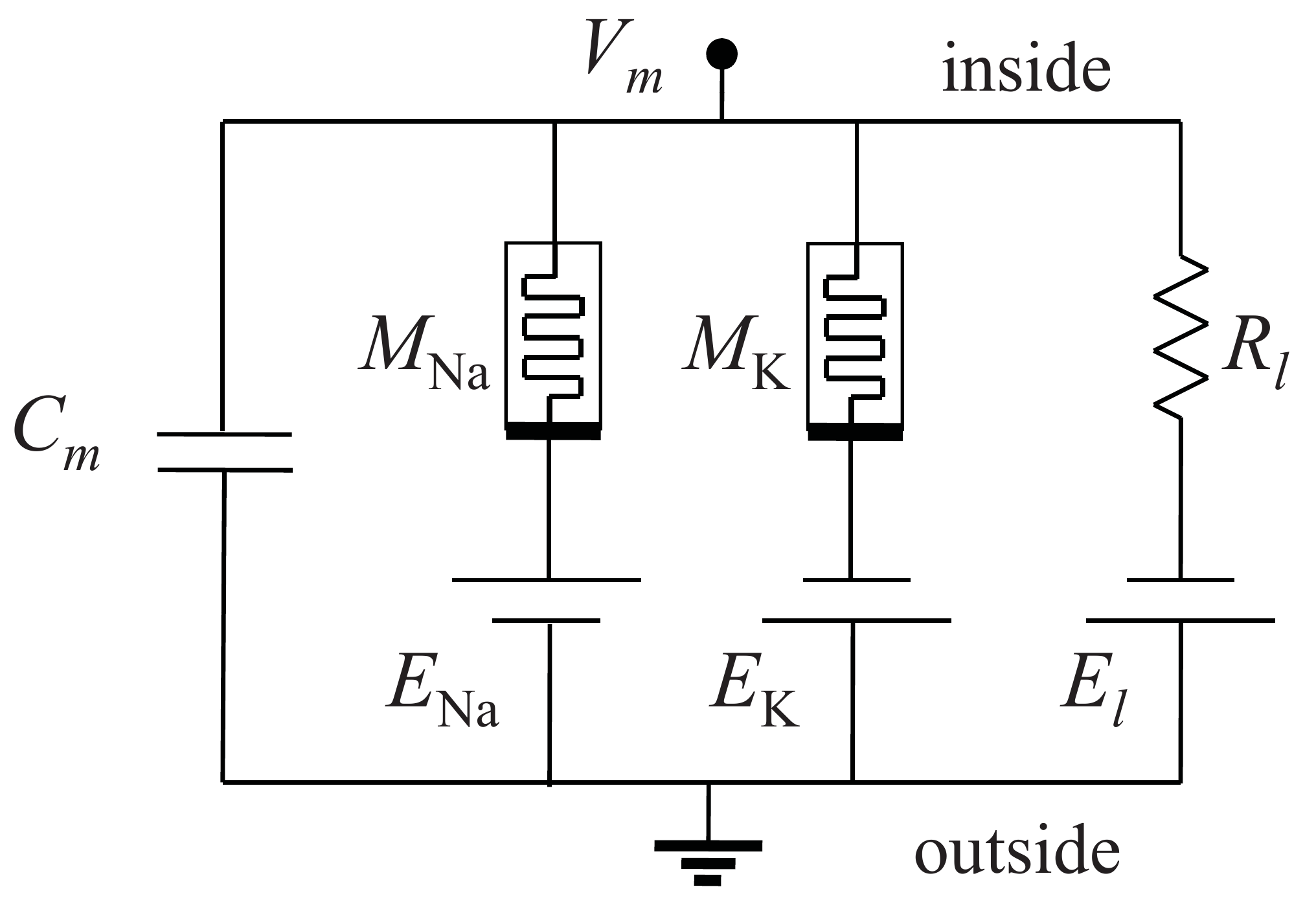}
\caption{Equivalent electrical circuit for a short
segment of squid axon membrane (modified from reference
\cite{hodgkin52a}). Here, $C_m$ is the membrane capacitance, $V_m$
is the membrane potential, $M_{\textnormal{Na}}$ and
$M_{\textnormal{K}}$ are memristive systems describing conductivity of Na
and K channels, respectively, $R_l$ is the leakage resistance,
$E_{\textnormal{Na}}$, $E_{\textnormal{K}}$ and $E_l$ are reverse
ion channel potentials, and by ``inside'' and ``outside'' we mean
the interior of the axon and its exterior with the membrane
delimiting the two sides. \label{fig:HHmodel}}
 \end{center}
\end{figure}

In order to quantitatively understand the experimental data, they
have then suggested an equivalent circuit model of the membrane as
shown in figure \ref{fig:HHmodel}. In this plot, we use symbols of
memristors in order to denote variable conductance channels.
Mathematically, the membrane current is written as
\cite{hodgkin52a}
\begin{eqnarray}
I=C_m
\frac{\textnormal{d}V_m}{\textnormal{d}t}+M_{\textnormal{Na}}^{-1}\left(
V_M-E_\textnormal{Na}\right)+M_{\textnormal{K}}^{-1}\left(
V_M-E_\textnormal{K}\right)+ \nonumber \\
R_l^{-1}\left(V_m-E_l \right)
\label{eq:Icell}
\end{eqnarray}
where the memory conductances
$M_{\textnormal{Na}}^{-1}=\overline{g}_\textnormal{Na}m^3h$,
$M_{\textnormal{K}}^{-1}=\overline{g}_\textnormal{K}n^4$, with
$\overline{g}_\textnormal{Na}$, $\overline{g}_\textnormal{K}$,
constants, $R_l$ describes the leakage resistance, and all other
circuit quantities can be read from the circuit
diagram~\ref{fig:HHmodel}. The time-dependencies of
voltage-dependent gating variables $n$, $m$ and $h$ are given by
the equations
\begin{eqnarray}
\frac{\textnormal{d}n}{\textnormal{d}t}=\alpha_n(1-n)-\beta_n n,
\label{eq:Icellc1}
\\ \frac{\textnormal{d}m}{\textnormal{d}t}=\alpha_m(1-m)-\beta_m
m, \\
\frac{\textnormal{d}h}{\textnormal{d}t}=\alpha_h(1-h)-\beta_h h,
\label{eq:Icellc3}
\end{eqnarray}
where $\alpha_{n(m,h)}$ and $\beta_{n(m,h)}$ are voltage-dependent
constants \cite{{hodgkin52a}} defined as
\begin{eqnarray}
\alpha_n=0.01\frac{\left( V_m+10\right)}{e^\frac{V_m+10}{10}-1},
\\ \beta_n=0.125e^\frac{V_m}{80}, \\ \alpha_m=0.1\frac{\left(
V_m+25\right)}{e^\frac{V_m+25}{10}-1}, \\
\beta_m=4e^\frac{V_m}{18},
\\ \alpha_h=0.07e^\frac{V_m}{20}, \\
\beta_h= \frac{1}{e^\frac{V_m+30}{10}+1}.
\end{eqnarray}
In the above equations, the voltage is in mV and time is in ms.
The quantities $n$, $m$ and $h$ take values between 0 and 1 thus
representing variation of channels' conductances in time.

\begin{figure*}
 \begin{center}
\includegraphics[angle=0,width=14cm]{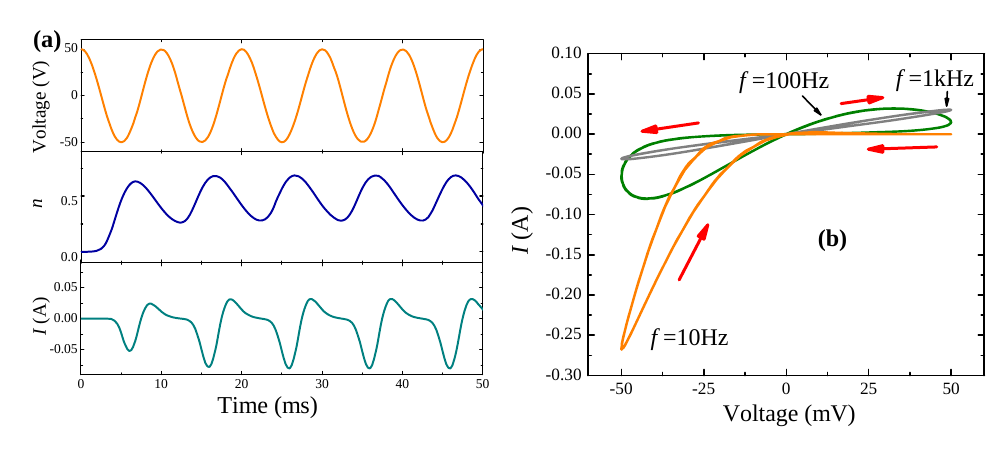}
\caption{Simulations of the potassium channel
memristive system ($M_\textnormal{K}$) response. We have used
$\overline{g}_\textnormal{K}=10$mS. The applied voltage is
$V(t)=V_0\cos \left(2\pi f t\right)$ with $V_0=50$mV and
$f=0.1$kHz in (a). \label{fig:neuro1}}
 \end{center}
\end{figure*}

\begin{figure*}
 \begin{center}
\includegraphics[angle=0,width=14cm]{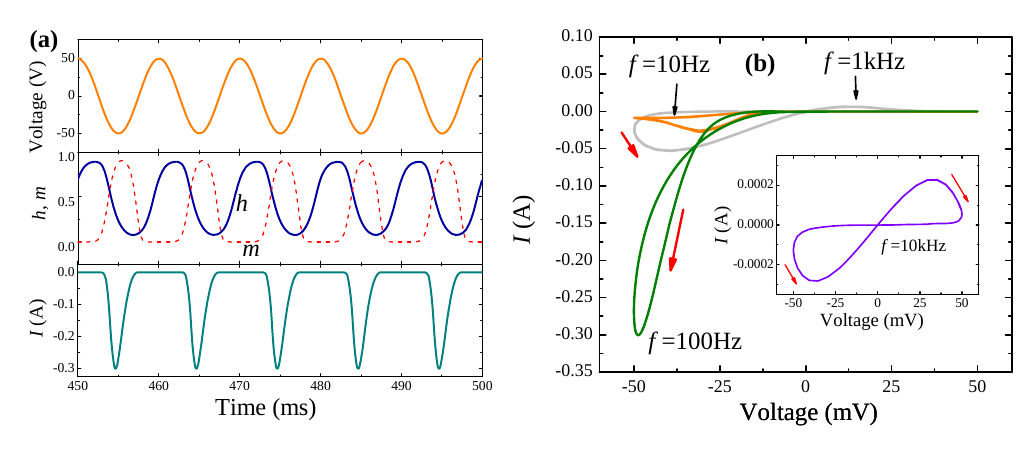}
\caption{Simulations of the sodium channel
memristive system ($M_\textnormal{Na}$) response. We have used
$\overline{g}_\textnormal{Na}=33$mS. The applied voltage is
$V(t)=V_0\cos \left(2\pi f t\right)$ with $V_0=50$mV and
$f=0.1$kHz in (a). \label{fig:neuro2}}
 \end{center}
\end{figure*}

It follows from the expressions for $M_{\textnormal{Na}}$ and
$M_{\textnormal{K}}$ (given below equation (\ref{eq:Icell})) and
equations (\ref{eq:Icellc1}-\ref{eq:Icellc3}) that the potassium
channels can be classified as first-order voltage-controlled
memristive systems and the sodium channels as second-order
voltage-controlled memristive systems~\cite{chua76a}. In figures
\ref{fig:neuro1}, \ref{fig:neuro2} we plot simulations of
potassium and sodium channel memristive systems biased by ac-voltage.
These plots demonstrate frequency-dependent $I-V$ curves typical
of memristive systems. There is, however, an interesting feature that can
be seen in figure \ref{fig:neuro2}(b) for the $f=100$Hz curve: at
negative voltages, the curve has a self-crossing. To the best of
our knowledge we are not aware of experimental results for these systems that employ ac biases,
and thus this self-crossing feature remains to be verified.

Modern models of neuronal dynamics are based on similar equations,
but often involve many more ion channel types, with the ion
channels possibly located on different parts of a spatially
extended neuron~\cite{neuralbook1}. Therefore, a single neuron
description may involve a large number of memristive systems.

\section{Memcapacitive systems} \label{sec:memcapacitors}

\begin{figure}[t]
 \begin{center}
\includegraphics[angle=0,width=8cm]{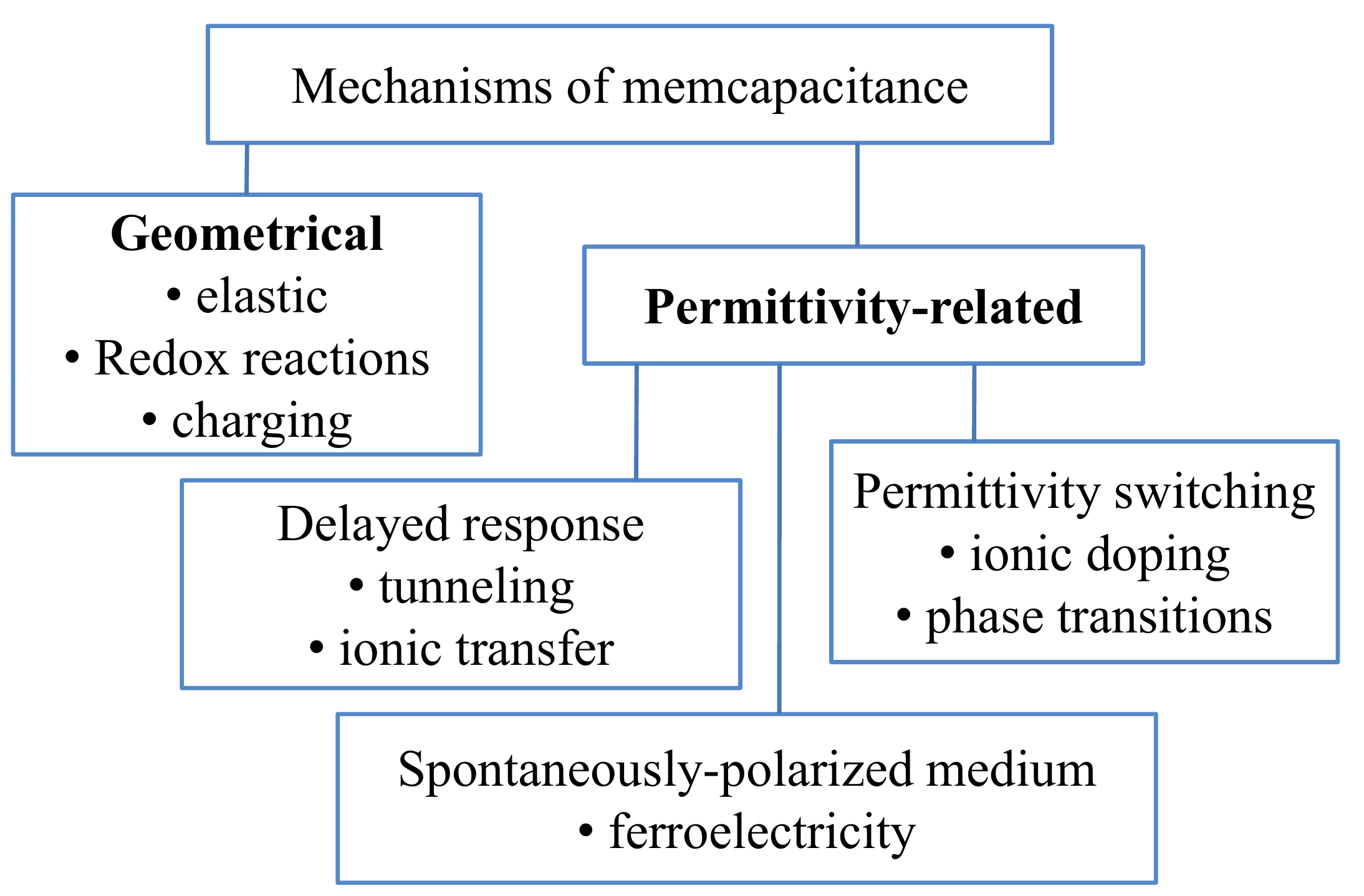}
\caption{Classification scheme of memcapacitance mechanisms.}
\label{memcap}
\end{center}
\end{figure}

Having discussed experimental realizations of memristive systems, in this section we consider systems showing memcapacitive
behavior. Under the term capacitor, we understand a general
electronic device capable of storing charge and energy. Such a
device normally includes a couple of external metal plates having
negligible resistance and a dielectric medium between the plates.
In capacitors, memory effects can originate from changes in the
geometry and/or permittivity (figure \ref{memcap}).

Under {\it geometrical mechanisms} of memcapacitance we understand
situations when geometrical morphology of the plates changes in
time (e.g., their relative distance and/or shape). In
permittivity-related mechanisms, dielectric properties of the
material between the plates provide the memory. We can identify
three most probable permittivity-related mechanisms (figure
\ref{memcap}): {\it delayed-response mechanism}, when dielectric
permittivity dynamics involves a time delay, {\it
permittivity-switching mechanism}, when the dielectric constant
changes its value under the external input (but the response is fast), and {\it
spontaneously-polarized medium mechanism}, in which a
spontaneously-polarized material (ferroelectric) is used in the
capacitor structure.

We discuss below physical systems demonstrating these different mechanisms of
memcapacitance. Known mathematical models of some of these systems are also presented.
We also note that a methodology for SPICE modeling of memcapacitive systems has been developed recently~\cite{biolek10a}.

\subsection{Geometrical memcapacitive systems}

\subsubsection{Micro- and nano-electro-mechanical systems}

Micro-electro-mechanical system (MEMS) and nano-electro-mechanical
system (NEMS) capacitors
\cite{rfmemsbook,rfmemsbook1,evoy04a,luo06a,jang08a,herfst08a,Nieminen2002-1}
are variable capacitors whose operation is based on an interplay
of mechanical and electrical properties of micro- and nano-size
systems. Such elements are key components in many radio-frequency (RF)
applications such as tunable filters, impedance matching circuits
and voltage-controlled oscillators \cite{rfmemsbook,rfmemsbook1}.
In addition, these structures, on the nanoscale, are considered
for memory applications \cite{jang08a} and sensitive measurements
\cite{LaHaye09a}. Normally, capacitors built from these systems
utilize a diaphragm-based, a microbridge-based or a
cantilever-based structure fabricated via micromachining
\cite{rfmemsbook,rfmemsbook1}.

\begin{figure}[b]
\centering
\includegraphics[width=7cm]{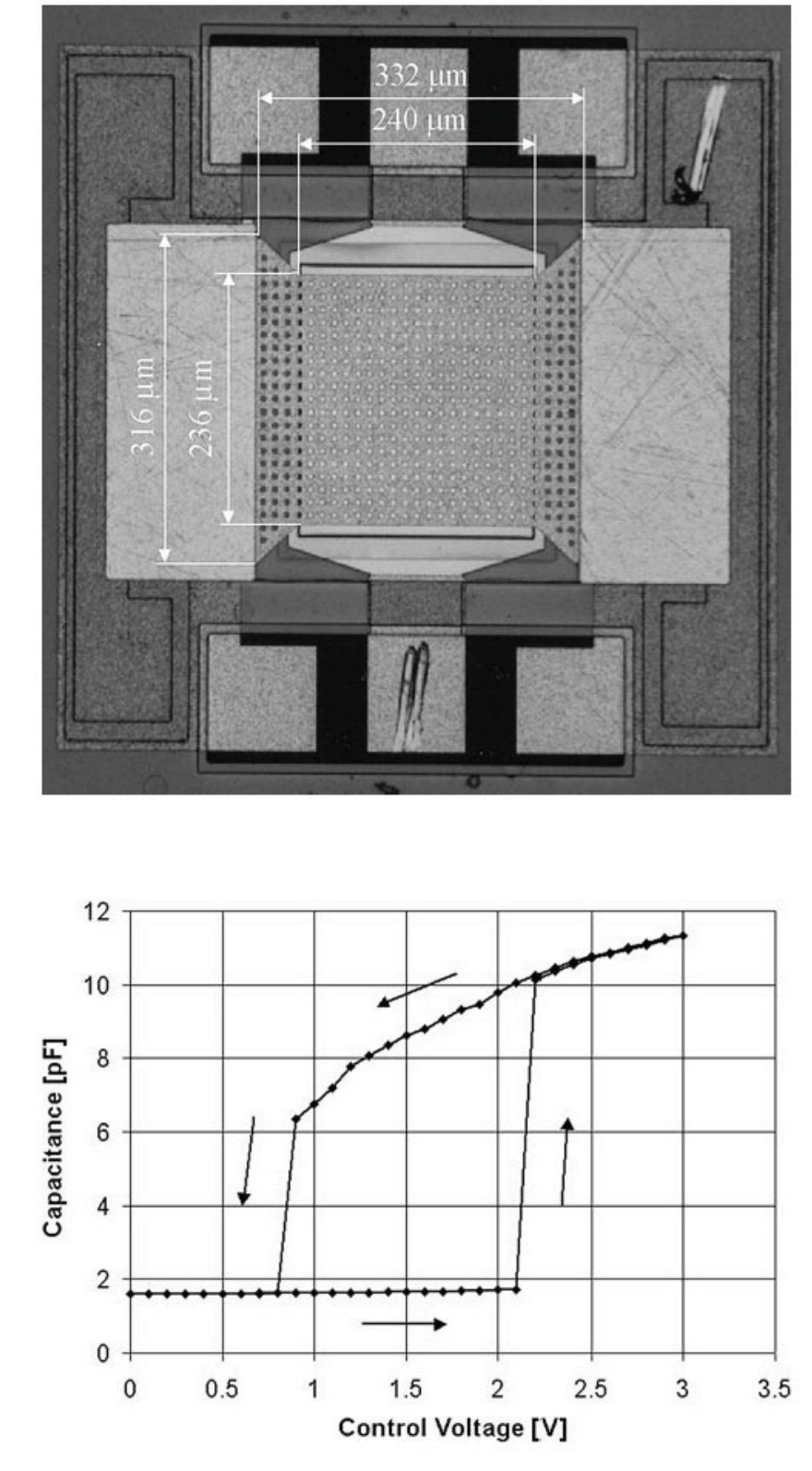}
\caption{Image of a two-state memcapacitive system constructed from MEMS, and its capacitance as
a function of applied voltage. From \cite{Nieminen2002-1}.
\label{nieminen}}
\end{figure}

Figure \ref{nieminen} shows an experimental image of a
memcapacitive system constructed from MEMS and measured capacitance as a function of voltage. The
capacitance curve shows a well defined hysteresis which is a
manifestation of memory effects in the structure. Memory effects
are related to displacement of a top capacitor plate that can be
considered to be suspended by a spring \cite{Nieminen2002-1}.
Below, we consider a model of {\it elastic memcapacitive system} that
(with appropriate boundary conditions) can be used to described
the capacitor features of the MEM system shown in figure
\ref{nieminen}. For more details on MEMS and NEMS we refer
to \cite{rfmemsbook,rfmemsbook1,memsmodelingbook}.

\subsubsection{Elastic memcapacitive systems}\label{sec:elasticmemc}

\begin{figure}[t]
 \begin{center}
\includegraphics[angle=0,width=7cm]{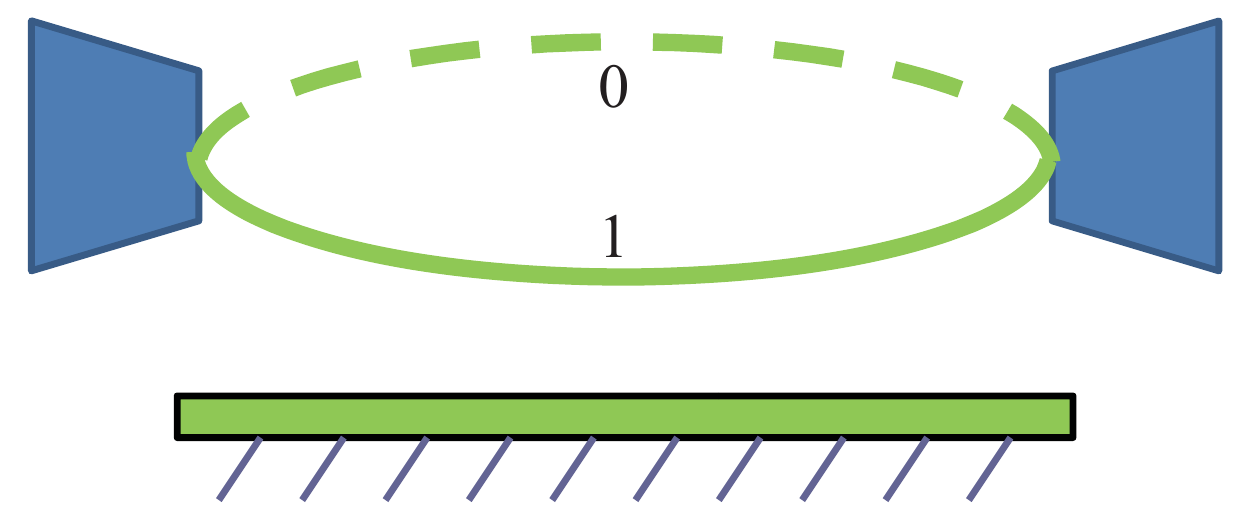}
\caption{Strained membrane memcapacitive system. The top
plate of a regular parallel-plate capacitor is replaced by a
flexible strained membrane. Because of two equilibrium positions
of the membrane (whose potential energy can be modeled by a double-well potential), stable high and
low capacitance configurations are possible in such a system. Both configurations are perfectly stable
providing a non-volatile information storage capability. However, since the system
is bistable, an analog value can not be stored in such a device.} \label{strained_membrane}
\end{center}
\end{figure}

A model of a simple elastic memcapacitive system is discussed in Sec. \ref{sec:elasticmemc1}
of this review. Here, we would like to mention about a different elasticity-based memcapacitive system \cite{pershin10a} whose
schematics is shown in Fig. \ref{strained_membrane}.
In this realization, the capacitor
is formed by a strained membrane (upper plate) and a flat fixed
lower plate. In this model two equilibrium states (up-bent and
down-bent membrane) coexist that can be used for a non-volatile
storage of a bit of information. The possibility of switching
between the two states as well as chaotic behavior of such a
system in a certain range of parameters has also been demonstrated
\cite{pershin10a}. For reference, we provide the model of strained membrane memcapacitive system in the table \ref{table:membr}.

\begin{table*}
{ \renewcommand{\arraystretch}{1.6}
\begin{tabular}{| l | c | }
\hline
Physical system & Strained membrane memcapacitive system \\
\hline
Internal state variable(s) & Displacement and velocity of upper plate, \\
&$x_1=z$, $x_2=\frac{\textnormal{d}z}{\textnormal{d}t}$ \\
\hline
Mathematical description & $V_C=\left[\frac{C_0}{1+\frac{x_1}{d}}\right]^{-1}q$  \\
& $\frac{\textnormal{d}x_1}{\textnormal{d}t}=x_2$ \\
& $\frac{\textnormal{d}x_2}{\textnormal{d}t}=-\omega_0^2x_1\left[ \left( \frac{x_1}{z_0}\right)^2-1\right]- \gamma x_2 -\frac{q^2}{2 \varepsilon_0 m S}$ \\
\hline
System type & Second-order charge-controlled memcapacitive system \\
\hline
\end{tabular}
}
\caption{Model of strained membrane memcapacitive system. Here, $C_0=\varepsilon_0 S/d$, $S$ is the plate's
area, $d$ is the separation between the bottom plate and middle
position of the membrane, $\omega_0$ is the natural angular frequency of the system,
  $\gamma$ is a damping coefficient representing dissipation
of the elastic excitations, and $m$ is the upper plate's mass \cite{pershin10a}.}
\label{table:membr}
\end{table*}

\subsubsection{Other geometrical memcapacitive systems}

There are other possible realizations of geometrical
memcapacitive systems. In many memristive systems, for example, the
morphology of conducting regions changes in time \cite{waser07a}.
Therefore, in addition to a resistance change, the capacitance of
such systems changes as well. In particular, co-existence of
memristive and memcapacitive behavior has also been observed in
perovskite thin films \cite{liu06a} (see also figure \ref{Ignat1RC}).

\subsection{Delayed-response memcapacitive systems}

\subsubsection{Superlattice memcapacitive systems} \label{superlatt_memcap}

\begin{figure}[t]
\centering
\includegraphics[width=8cm]{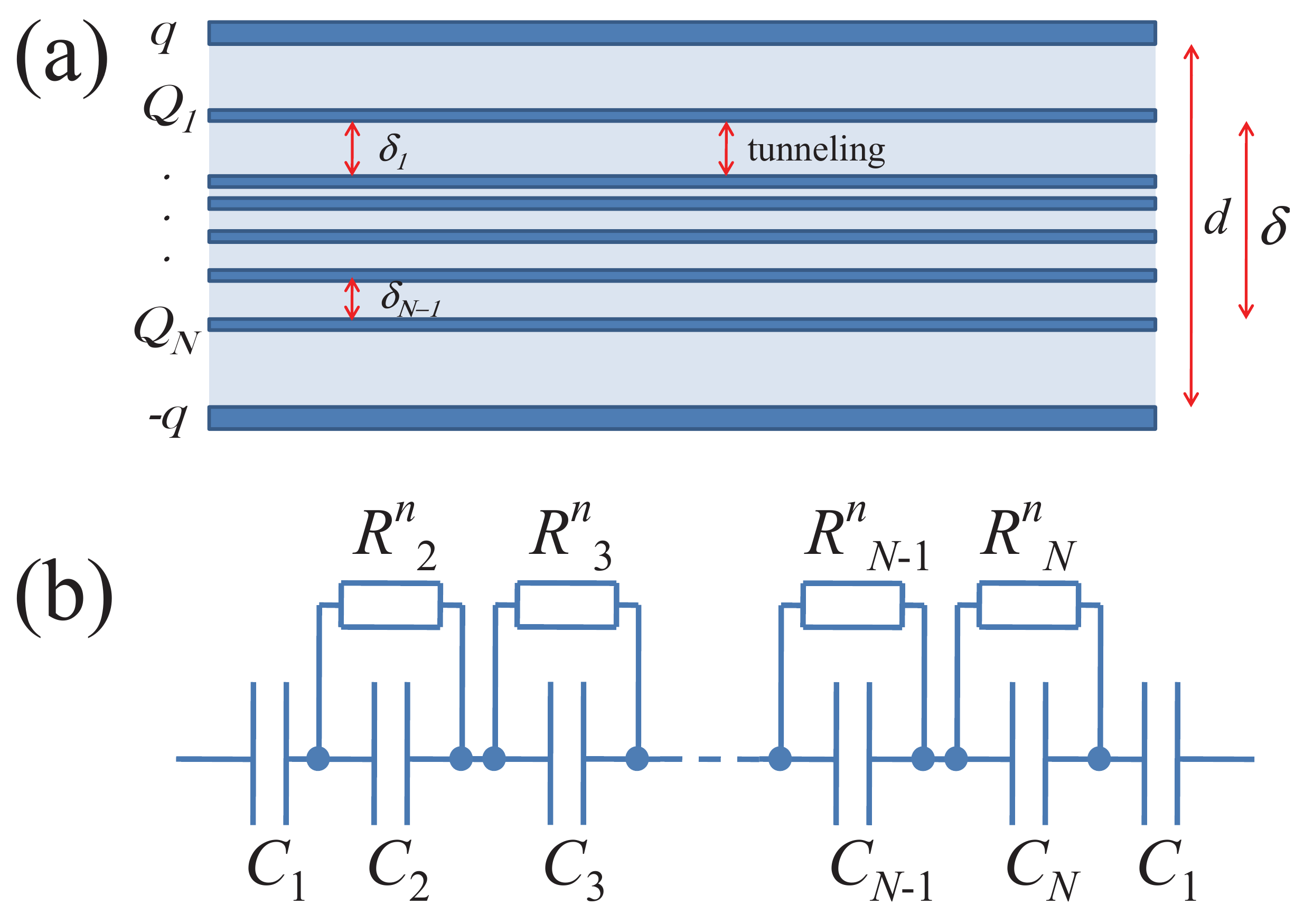}
\caption{(a) Schematics of a superlattice
memcapacitive system. A superlattice medium consisting of $N$ metal layers
embedded into an insulator is inserted between the plates of a
parallel-plate capacitor. $\pm q$ and $Q_k$ are charges on
external plates and on internal metal layers, respectively. (b)
Equivalent circuit model of $N$-layer memcapacitive system. From
 \cite{martinez09a}. \label{memcapJulian}}
\end{figure}

A solid-state memcapacitive system based on the slow polarization
rate of a medium between plates of a regular parallel-plate
capacitor has been recently proposed \cite{martinez09a}. The key idea is to use non-linear electron transport
(tunneling) for fast writing and long storage capabilities. Figure
\ref{memcapJulian}(a) shows a particular example of memcapacitive system
in which $N$ metal layers are embedded into an insulating material
between external capacitor plates. The structure is designed in
such a way that the electron transport between external plates and
internal layers is not possible. Therefore, the internal charges
$Q_k$ can only be redistributed between the internal layers
creating a medium polarization. Correspondingly, there is a
constrain imposed on the total internal charge:
\begin{equation}
\sum\limits_{i=1}^NQ_i=0 \label{Qconstr}.
\end{equation}

The capacitance of the total structure is given by
\cite{martinez09a}
\begin{equation}
C=\frac{q}{V_C}=\frac{2\,C_0}{2+\sum_{i=1}^{N}\left[\Delta-2\Delta_{i-1}\right]\frac{Q_i}{q}}.
\label{eqC1}
\end{equation}
where $\Delta$, $\Delta_i$ are geometry-related parameters. The
dynamics of the charge at a metal layer $k$ is determined by the
currents flowing to and from that layer:
\begin{equation}
\frac{dQ_k}{dt}=I_{k-1,k}-I_{k,k+1}, \label{eq:charge_dynamics}
\end{equation}
where $I_{k,k+1}$ is the tunneling electron current flowing from
layer $k$ to layer $k+1$ (for more details see reference
\cite{martinez09a}). It follows from equations
(\ref{Qconstr},\ref{eqC1},\ref{eq:charge_dynamics}) that such a superlattice
memcapacitive system is an $N-1$ order charge-controlled memcapacitive
system. Table \ref{table:superl} presents summary of the superlattice memcapacitive system.

\begin{table*}
{ \renewcommand{\arraystretch}{1.6}
\begin{tabular}{| l | c | }
\hline
Physical system & Superlattice memcapacitive system \\
\hline
Internal state variable(s) & Charges on internal metal plates, \\
&$x_i=Q_i$ ($i=1,..., N-1$) \\
\hline
Mathematical description & $V_C=\frac{1+\sum_{i=1}^{N-1}\left[\Delta_{N-1}-\Delta_{i-1}\right]\frac{x_i}{q}}{C_0}q$  \\
& $\frac{dx_k}{dt}=I_{k-1,k}-I_{k,k+1}$ \\
\hline
System type & ($N-1$)-order charge-controlled memcapacitive system \\
\hline
\end{tabular}
}
\caption{Superlattice memcapacitive system. Here, $N$ is the number of the embedded internal metal layers.}
\label{table:superl}
\end{table*}

\begin{figure*}[b]
\centering \includegraphics[width=14cm]{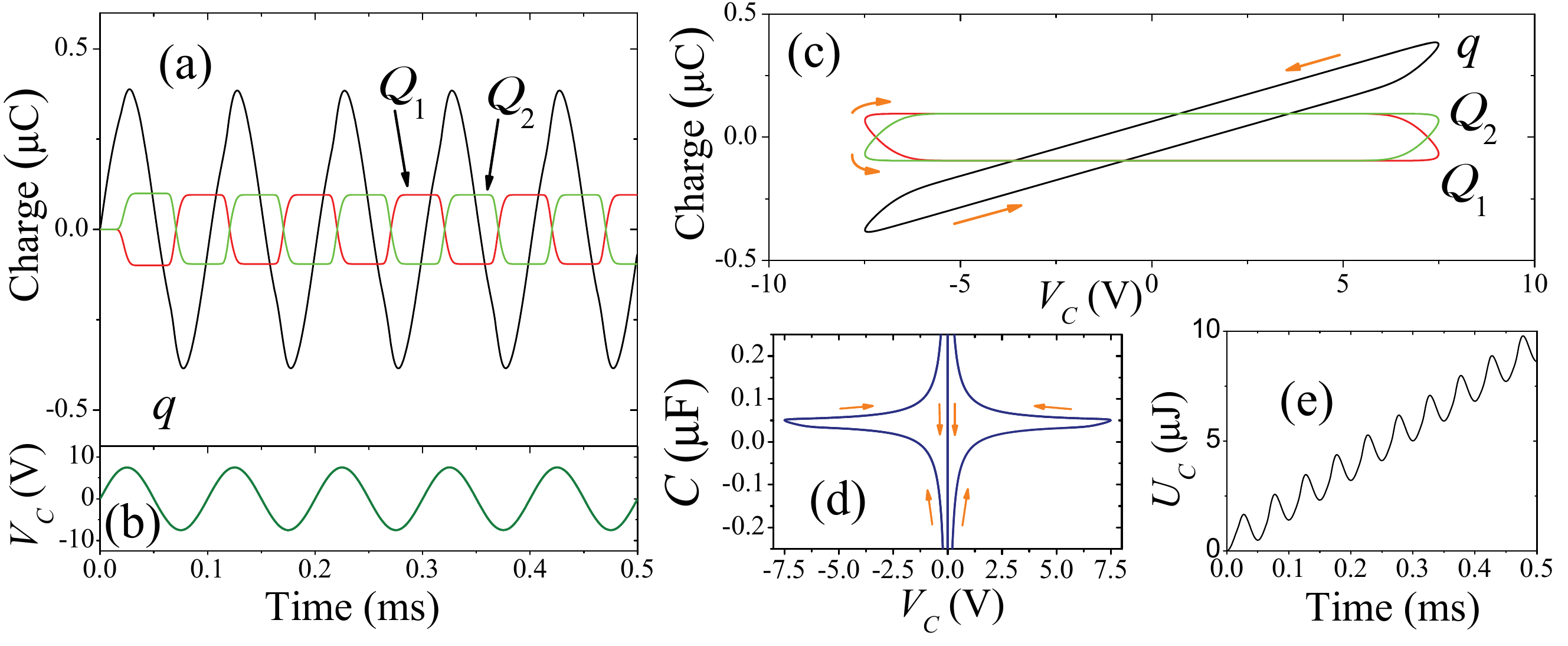}
\caption{(a) The charge on internal metallic layers
and superlattice memcapacitive system plates as a function of time $t$. (b) Voltage on
memcapacitive system, $V_C$, as a function of time $t$. Charge-voltage (c)
and capacitance-voltage (d) plots. (e) Added/removed energy as a
function of time $t$. These plots were obtained using the
parameter values $V_0=7.5$V, $f=10$kHz, $d=100nm$,
$\delta=66.6$nm, $S=10^{-4}$m$^2$, $\varepsilon_r=5$, $U=0.33$eV,
$R=1\Omega$. From \cite{martinez09a}.} \label{memcapJulian1}
\end{figure*}

Simulations of a two-layer memcapacitive system with symmetrically positioned
internal layers is depicted in figure \ref{memcapJulian1}.
Interesting features of this memcapacitive system include non-pinched
$q-V_C$ hysteresis loop (figure \ref{memcapJulian1}(c)), and both
negative and diverging capacitance (figure
\ref{memcapJulian1}(d)). Moreover, since electron tunneling between
the layers is accompanied by energy loss, the memcapacitive system is
dissipative (figure \ref{memcapJulian1}(e)). Using a set of
non-linear resistors and usual capacitors, an equivalent model of
the given memcapacitive system can be formulated (see figure
\ref{memcapJulian}(b)). This is an example of the situation - anticipated in section~\ref{fundamental} - in which a
memcapacitive system can be constructed from classical circuit elements. We re-iterate here that
this is not a trivial point because one can use a unified set of equations, (\ref{Geq1in}) and~(\ref{Geq2in}), describing a single
memory system to represent a
complex functionality as that reported here.

\subsubsection{Ionic memcapacitive systems}

Memcapacitive behavior can be observed also in certain ionic
systems because of their relatively slow dielectric response. A
typical nanopore sequencing setup \cite{branton08,zwolak08} is an
example of such a system. Let us then consider two chambers with
ionic solution separated by a membrane with a nanopore
\cite{krems2010a}. When a varying voltage is applied to electrodes
located in different chambers, ions redistribute with a time lag
affecting the total system capacitance. In particular, it has been
shown \cite{krems2010a} that the ac response of such a system
demonstrates non-pinched $q-V_C$ hysteresis loops, and both
negative and divergent capacitance. Moreover, the equivalent
scheme of this setup can also be modeled using a set of classical
circuit elements \cite{krems2010a}. Since nanopores are ubiquitous on
membranes of biological cells (e.g., the nerve cells) we expect these phenomena to be
observable (at appropriate frequencies) even in those cases.
We are not aware, though, of either theoretical or experimental work along these lines.

\subsection{Permittivity-switching memcapacitive systems}

\subsubsection{Polymer-based memcapacitive systems}

\begin{figure*}[t]
\centering
\includegraphics[width=14cm]{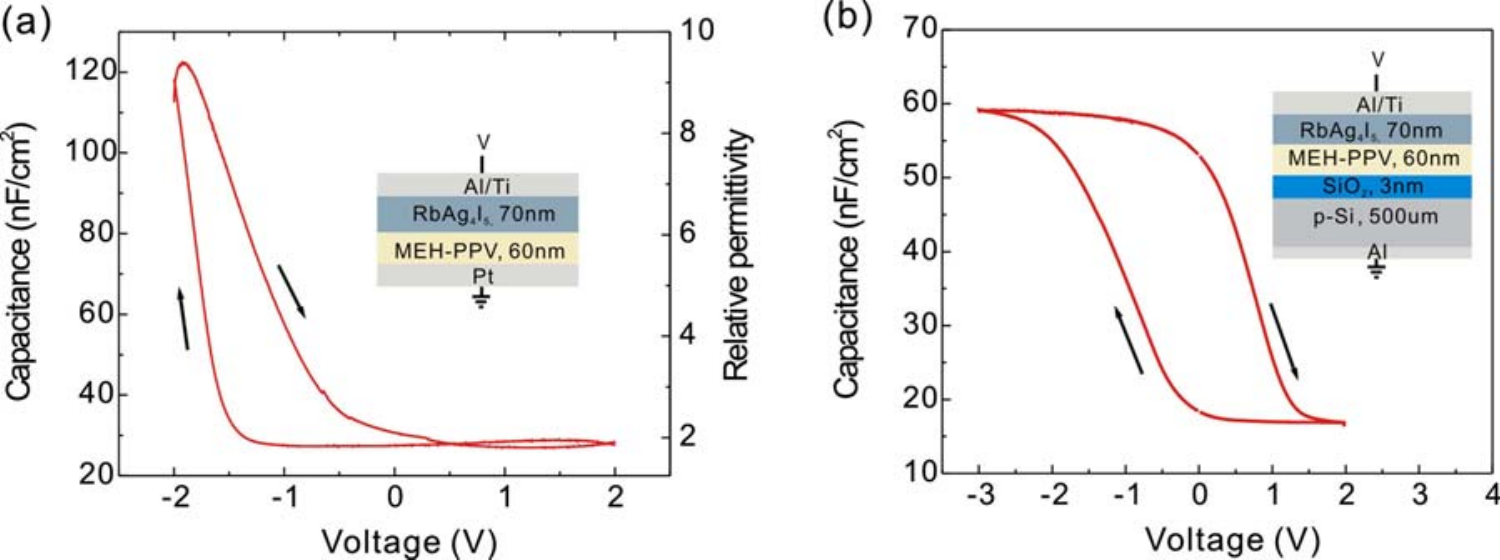}
\caption{Programmable device capacitances as a
function of the applied voltage for the device structures shown in
the insets. A better device performance is achieved when a
3-nm-thick SiO$_2$ insulating layer blocking the leakage current
is inserted as in (b). Reprinted with permission from \cite{lai09a}. Copyright 2009, American Institute of Physics. \label{analog_cap1}}
\end{figure*}

An analog memory capacitor has been reported in reference
\cite{lai09a}. In this work the programmable capacitance was
achieved in a field-configurable doped polymer, in which the
modification of ionic concentrations induces a nonvolatile change
in the polymer dielectric properties. Several device structures
were investigated and two of them are shown in figure
\ref{analog_cap1}. In both structures, the RbAg$_4$I$_5$ ionic
conductor functions as an ionic source.  It contains Ag cations
having a higher mobility and iodine anions whose mobility is lower
and, in a polymer, is of a threshold type. The latter is because
the iodine anions, having a much larger radius, can form a large
ionic complex I$^-_3$, and/or chemically bond with the MEH-PPV
polymer used in the memory capacitor.

Negative voltage pulses above a threshold were used to inject iodine
anions into the polymer, while positive voltage pulses (above the
threshold) were used to extract iodine anions from the polymer.  Some
of Ag cations follow the iodine anions because of electrostatic
interactions. When in the polymer, the anions and the cations form ionic
dipoles increasing the polymer permittivity and device capacitance
\cite{lai09a}. Since a voltage above a threshold is needed to
overcome the ionic bonding with the polymer, the devices provide
reasonably nonvolatile memory characteristics \cite{lai09a}. In
particular, it has been reported that after the analog capacitance was configured to a certain
value, it changed by less then 10$\%$ under continuous reading for
5 days.

\subsubsection{Phase-transition memcapacitive systems} \label{phase_change_memcap}

During the process of a metal-to-insulator transition (MIT) both resistance (as we discussed in section
\ref{VO2res}) and dielectric properties
\cite{qazilbash07a,driscoll08a} are affected. This latter property was
used to fabricate a memory metamaterial in which a persistent
electrical tuning of a resonant frequency was demonstrated
\cite{driscoll09a}. In this particular case, vanadium dioxide
has been used as the memory material undergoing the MIT,
which we have discussed in section \ref{VO2res} in the context of
memory resistance. In the experimental setup, a split-ring
resonator array was patterned on a 90-nm thin film of vanadium
dioxide connected by two electrodes to a voltage source. The
device was mounted on a temperature-controlled stage and a
temperature of 338.6 K was selected. At this temperature, the
slope of the resistance as the function of temperature is the
steepest because of the proximity to the MIT. Therefore, even
small amplitude voltage pulses have a notable effect on material
characteristics \cite{driscoll09a}. Such electrical pulses
directly applied to vanadium dioxide were used to modify its
dielectric properties. The latter ones were registered by measuring the
modification of the resonant frequency of the split-ring resonator
array.

\begin{figure*}[b]
\centering
\includegraphics[width=14cm]{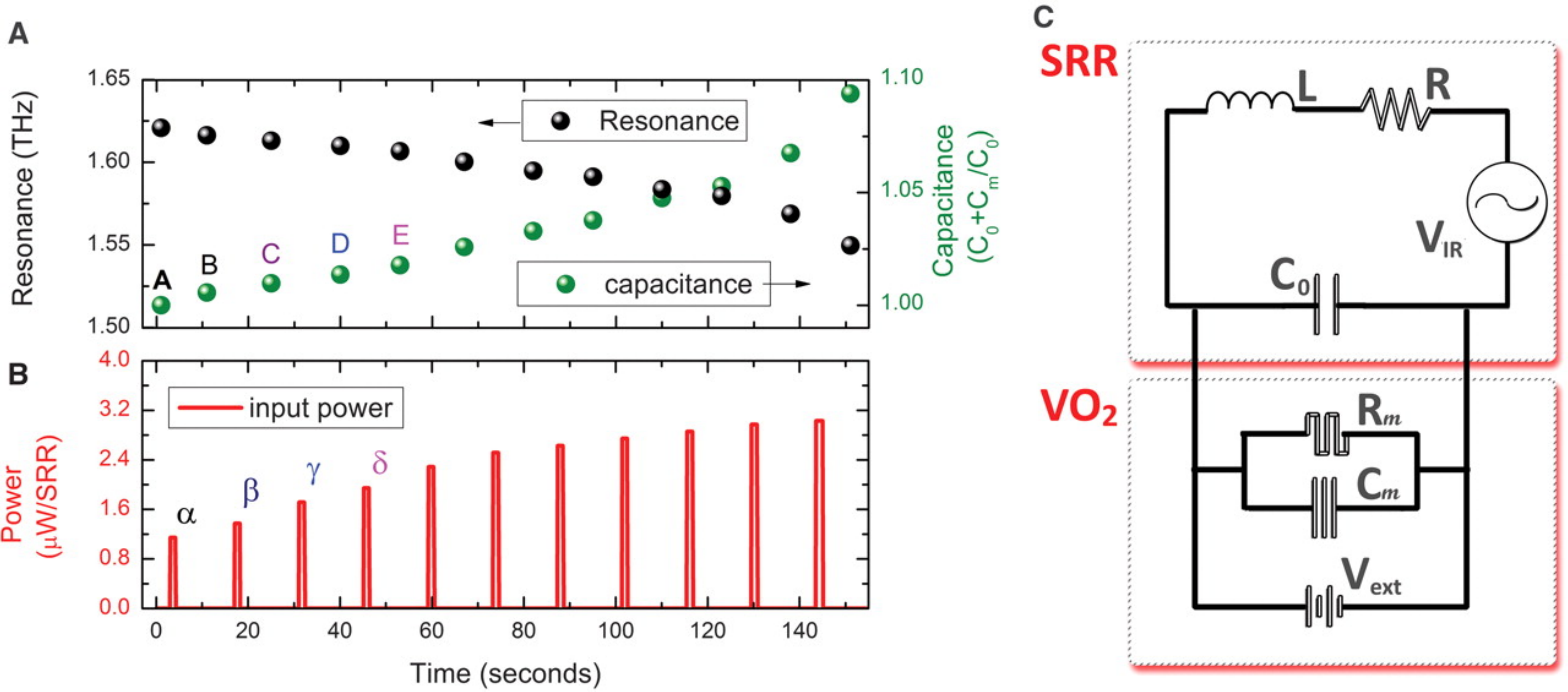}
\caption{Electrical tuning of a metamaterial: (a)
Modification of resonance frequency and capacitance by electrical
pulses of increasing power (b). (c) Equivalent circuit model in
which split-ring resonator (SRR) is described by traditional
circuit elements $L$, $R$, $C_0$ and vanadium dioxide material is
represented by symbols of memristor $R_m$ and memcapacitor $C_m$. From \cite{driscoll09a}. Reprinted with permission from AAAS. \label{vo2C}}
\end{figure*}

Figure \ref{vo2C}(a) demonstrates the experimentally measured
increase in the metamaterial capacitance with each applied voltage
pulse. As the input power per pulse increases with pulse number
(see figure \ref{vo2C}(b)), the MIT in vanadium dioxide progresses
causing the permittivity increase. Overall, the resonant frequency
can be red-shifted by as much as 20\% from its spectral maximum at
$\omega_0=1.65$THz \cite{driscoll09a}. It is also worth noting
that the effective circuit model of split-ring resonator deposited
on vanadium dioxide involves both memristive $R_m$ and memcapacitive $C_m$ elements (figure
\ref{vo2C}). As already anticipated, memristive and memcapacitive
properties co-exist very often, and in the metamaterial
configuration \cite{driscoll09a} they play together an important
role in the device operation.

\subsection{Spontaneously-polarized medium memcapacitive systems}

\subsubsection{Ferroelectric memcapacitive systems}

Another interesting concept is the use of ferroelectric materials
\cite{valasek20a,dawber05a,ferroelectricitybook} as the dielectric
medium of a memory capacitor. Ferroelectric materials are composed
of domains with a non-zero average electrical polarization. The
polarization of ferroelectric materials shows hysteresis as a
function of electric field, revealing two well-defined
polarization states. These states are used in the ferroelectric
random-access memory (RAM) technology \cite{scott89a} having functionality
similar to Flash memory.

\begin{figure}[b]
\centering
\includegraphics[width=7cm]{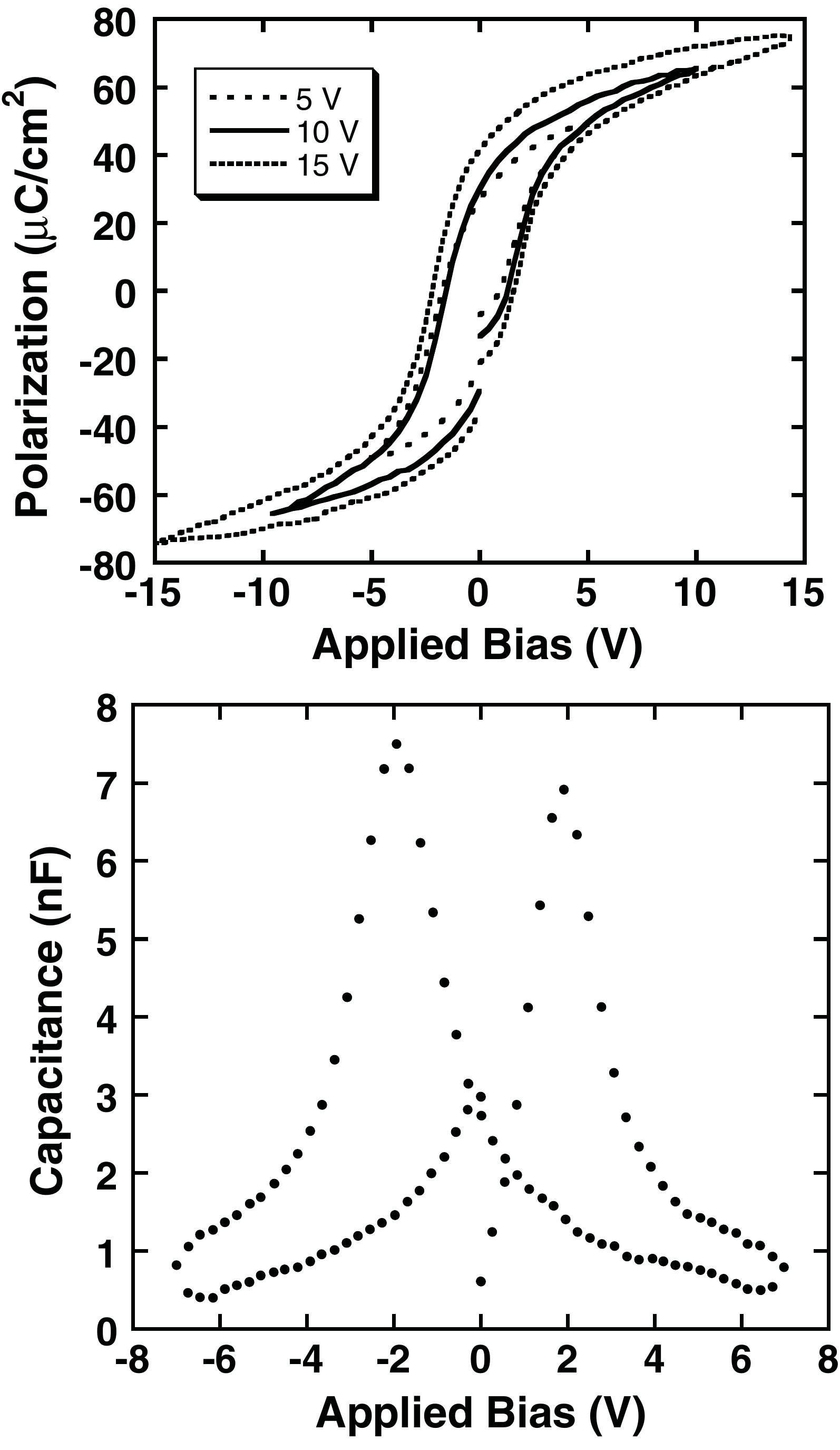}
\caption{Ferroelectric behaviour of Pt/PZT/Pt thin film
capacitors. Polarization-electric field and capacitance-voltage
measurements are plotted. From \cite{cagin07a}. \label{cagin}}
\end{figure}

Experimentally, when ferroelectric materials are inserted into
capacitor structures, a hysteretic $C-V$ behavior is observed
(see, for example, references
\cite{kholkin02a,kim07c,cagin07a,ducharne07a}). In particular, in
figure \ref{cagin} we plot the polarization-voltage and
capacitance-voltage characteristics of metal-ferroelectric-metal
capacitor Pt/PZT/Pt where PZT is (Pb,Zr)TiO$_3$ \cite{cagin07a}.
In this plot, the $C-V$ curve demonstrates the characteristic
``butterfly'' shape related to ferroelectric material and can be thus
characterized as a type-II memcapacitive system (cf. figure~\ref{crossing}). In this experiment, the
dielectric constant of the PZT film ranged from
$\varepsilon_r=83\varepsilon_0$ to $330\varepsilon_0$, where
$\varepsilon_0$ is the vacuum permittivity. The peak capacitance
occurring at $\pm 2.5$V corresponds to the coercive field of the
ferroelectric material.

\subsection{Other memcapacitive systems}

\subsubsection{MOS capacitors with nanocrystals}

\begin{figure}[t]
\centering
\includegraphics[width=8cm]{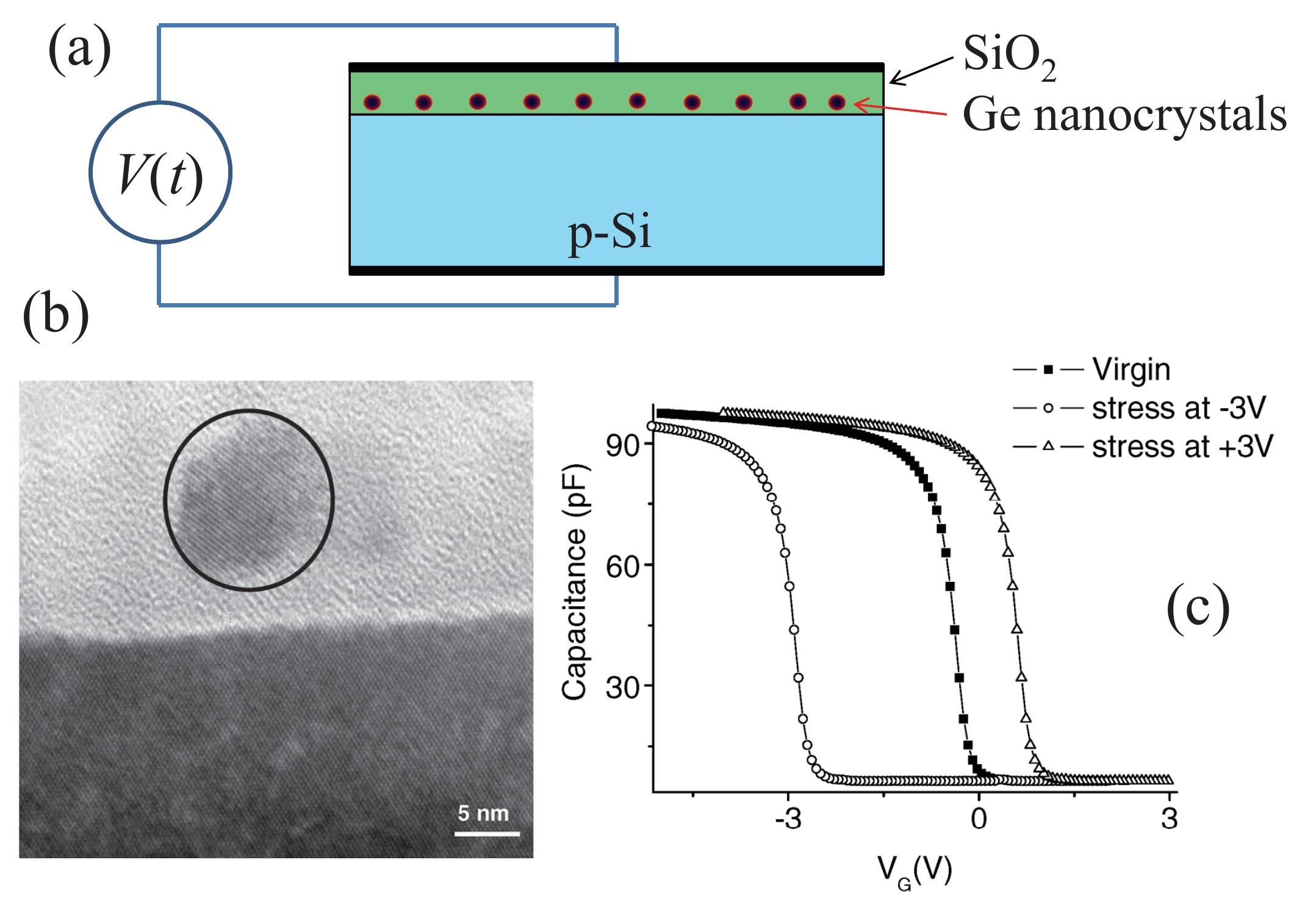}
\caption{(a) Schematic of MOS capacitor with
embedded germanium nanocrystals. (b) High-resolution transmission
electron microscopy image of an isolated Ge nanocrystal with a
mean size of 8.5nm. (c) $C-V$ curves measured after applying a
stress bias. (b) and (c) are
reprinted from \cite{kanoun06a}, Copyright 2006, with permission from Elsevier.
\label{kanoun}}
\end{figure}

Metal-Oxide-Semiconductor (MOS) structures with embedded
nanocrystals (figure \ref{kanoun}(a)) have been much investigated
recently
\cite{kobayashi97a,shi98a,dimitrakis00a,kim01a,king01a,kim02a,wang04a,wang04b,su05a,kanoun06a,lee06a,zhu07a,choi08a,li10a}.
These devices are promising candidates to replace floating-gate
Flash memory. The latter, in fact, has long programming times and poor endurance.
Many different materials such as Si
\cite{tiwari96a,shi98a,dimitrakis00a,kim01a,king01a,su05a}, Ge
\cite{kobayashi97a,kim01a,kim02a,wang04a,kanoun06a,lee06a}, SiGe
\cite{zhu07a,li10a}, Au \cite{paul03a} and Ag \cite{wang04b} have
been considered as candidates for the nanocrystals that store
charge. Currently, Ge nanoscrystals seem to be the most promising
ones because of a better data retention due to the smaller
band-gap compared to Si \cite{lee06a}.

It is known that $C-V$ curves of usual MOS capacitors demonstrate
a non-linear behavior \cite{mosbook} (see the "Virgin" line in
figure \ref{kanoun}(c)). Nanocrystals added to a MOS structure
(an actual Ge nanocrystal image is shown in figure
\ref{kanoun}(b)), provide a mechanism controlling the displacement
of the $C-V$ curve. When charge is transferred to nanocrystals,
the $C-V$ curve shifts by the amount $\Delta V_{FB}$ determined
according to the equation \cite{kanoun06a,tiwari96a}
\begin{equation}
\Delta V_{FB}=\frac{-qnD}{\varepsilon_{ox}}\left(
t_{CO}+\frac{1}{2}\frac{\varepsilon_{ox}}{\varepsilon_{Ge}t_{dot}}\right)
\label{eq:kakoun},
\end{equation}
where $q$ is the elementary charge, $n$ is the number of charges
per nanocrystal, $D$ is the density of Ge  nanocrystals, $t_{CO}$
is the control oxide thickness, $t_{dot}$ is the mean diameter of
Ge nanocrystals, $\varepsilon_{Ge}$ and $\varepsilon_{ox}$ are
dielectric permittivities of Ge nanocrystals and oxide,
respectively. Figure \ref{kanoun}(c) demonstrates displacement
of $C-V$ curves induced by positive and negative stress biases.

From the point of view of memory elements, the amount of
transferred charge $n$ plays the role of the state variable
defining the capacitance $C\left(V,n\right)$. The dynamics of $n$
can be described by a rate equation. However, the total equivalent
scheme of such device should include a resistor in series with a
capacitor as charge transfer to nanocrystals involves also
dissipation processes. Indeed, such an equivalent
resistor-capacitor circuit of MOS capacitors with nanocrystals was
recently discussed \cite{kwok08a}. Therefore, MOS capacitors with
nanocrystals do not manifest a purely memcapacitive behavior,
although the memcapacitive component in these devices seems to be
the dominant one. This is thus an example of a combined system as those discussed in section \ref{combined}.

\section{Meminductive systems}
\label{sec:meminductors}

\begin{figure}
 \begin{center}
\includegraphics[angle=0,width=8cm]{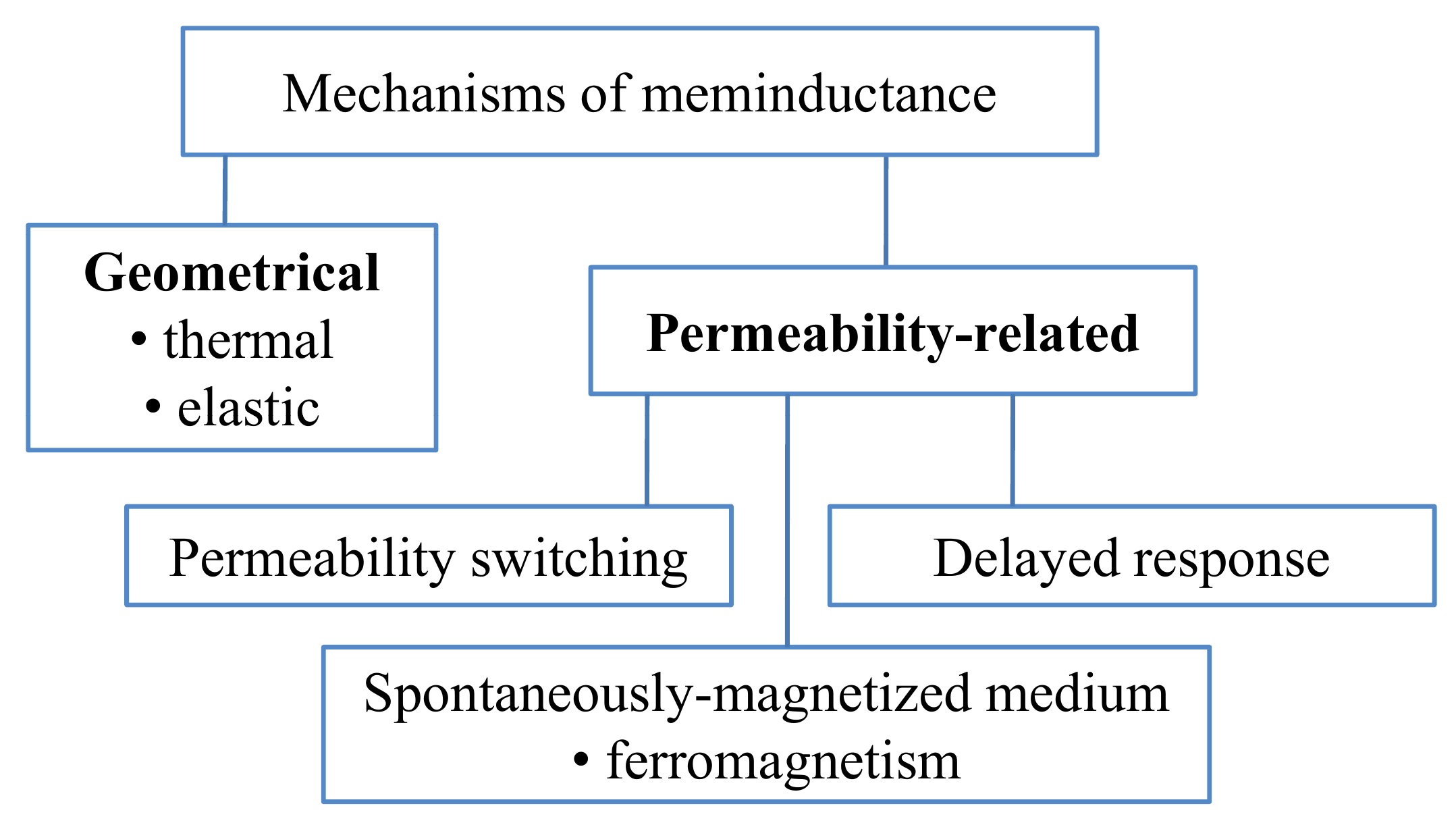}
\caption{Classification scheme of meminductance mechanisms.}
\label{meminduct}
\end{center}
\end{figure}
We are now left to consider systems that show
meminductive behavior. These are systems whose inductance depends on the past
dynamics and which can store energy. In this case, the energy stored may be a combination
of magnetic energy and energy due to other degrees of freedom (e.g., elastic kinetic energy).
Similar to the case of memcapacitive systems, memory effects in inductors can originate from a
geometrical variation of their structure, or be related to their
permeability (see figure \ref{meminduct}). Although at the present time
meminductors and meminductive systems are the least studied members of the family of circuit elements with memory, there are nonetheless several interesting examples of such
devices, and owing to their practical importance, we expect more will be discovered/designed in the future. Moreover, several SPICE models of meminductive systems have been suggested \cite{biolek10c}. Below, we overview experimental realizations of
meminductive systems based on the bimorph effect, introduce a model of an
elastic meminductive system, and discuss some other possible realizations
of these elements.

\subsection{Geometrical meminductive systems}\label{geomL}

\subsubsection{Bimorph meminductive systems}

Like for memcapacitive systems' realizations, the micro-electro-mechanical system (MEMS) technology provides an
opportunity to fabricate devices whose operation is based on an
interplay of mechanical, electrical, magnetic and thermal
properties. Several designs of inductors based on tunable MEMS that employ
the bimorph effect \cite{chu93a} have been reported
\cite{lubecke01a,zine04a,chang06a}. The bimorph effect refers to composite materials that show electrothermal actuation, namely they show reversible
mechanical deformation under electrothermal effects. The tunability of the
inductors based on this effect is then achieved by the difference in thermal expansion
coefficients of the two materials forming the inductor. When a voltage is
applied to such a structure, its temperature changes due to
heating and the structure deforms in a controllable manner. As a
finite time is required for heating and cooling, the inductance is
history-dependent. Therefore, such a device behaves as a {\it meminductive system}.

\begin{figure*}
 \begin{center}
  \centerline{
    \mbox{(a)}
    \mbox{\includegraphics[width=7.00cm]{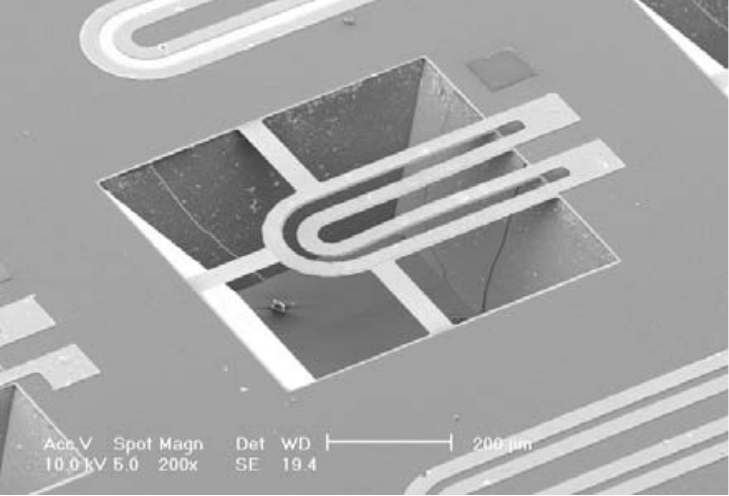}}
    \mbox{(b)}
    \mbox{\includegraphics[width=7.00cm]{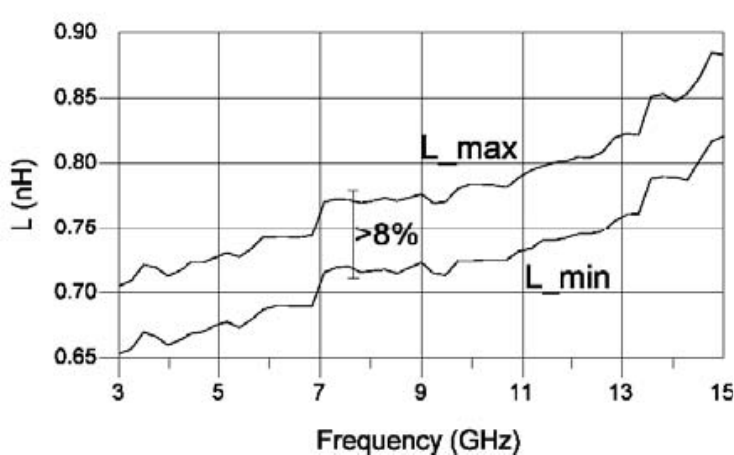}}
  }
\caption{ (a) An SEM image of a two-layer gold/silicon nitride
meminductive system. The inner inductor deflects down as current heats the
structure. (b) Inductance versus frequency curves showing a tuning
range of 8\%. From \cite{zine04a} (\copyright 2004 IEEE).  }\label{varind}
\end{center}
\end{figure*}

\begin{figure*}
 \begin{center}
    \centerline{
    \mbox{(a)}
    \mbox{\includegraphics[width=7.00cm]{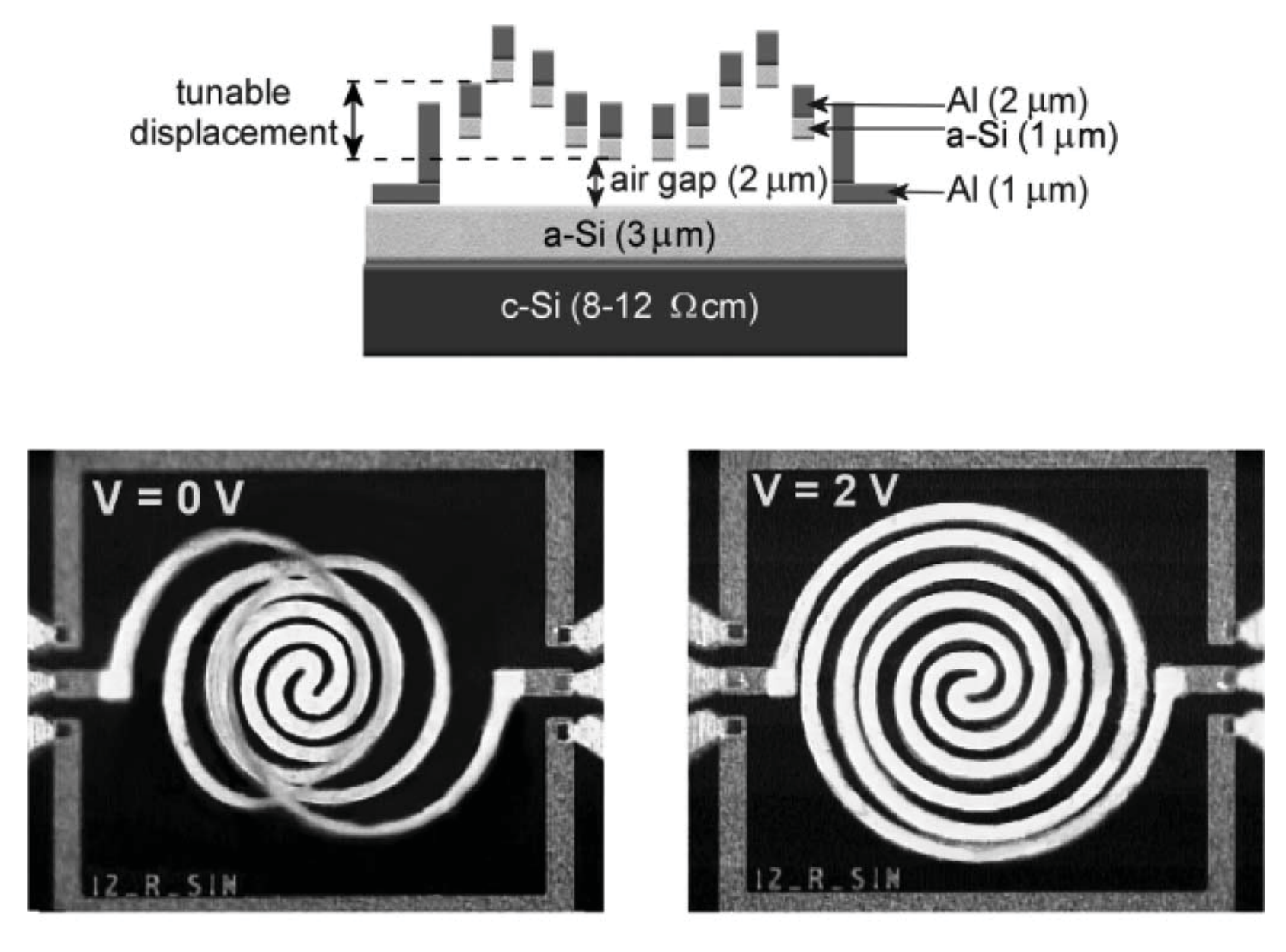}}
    \mbox{(b)}
    \mbox{\includegraphics[width=7.00cm]{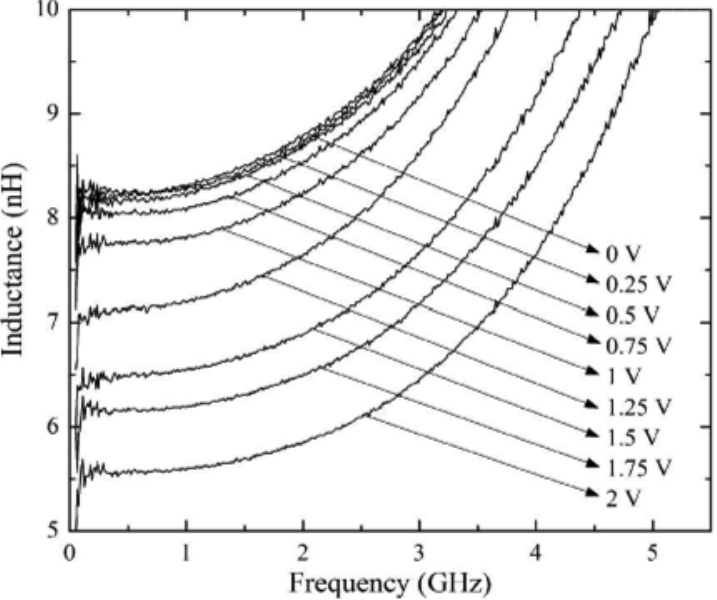}}
  }
\caption{ (a) Schematic cross section and micrograph image of an
a-Si/Al meminductive system and (b) its $L-f$ curves. In (a) one sees that
the meminductive system is 3D at 0V applied bias and completely flat at
2V. An inductance tuning of 32\% is reported for this structure. From \cite{chang06a} (\copyright 2006 IEEE).} \label{varind1}
\end{center}
\end{figure*}

Two experimental designs of bimorph-effect based meminductive system are shown
in figure \ref{varind}(a) and figure \ref{varind1}(a). The first
structure (figure \ref{varind}(a)) consists of two inductors
(fabricated of gold/silicon nitride bi-layer) connected in
parallel. The outer loop inductor is fixed by nitride beams while
the inner inductor is free to move. The heating of the structure
by current causes the inner inductor to deflect downwards changing
the structure's inductance. A variation of the inductance as much
as 8\% was reported \cite{zine04a}. A larger tuning range of 32\%
is achieved in the structure shown in Figure \ref{varind1}(a).
Here, the meminductive system is made of an amorphous-silicon/aluminum
(a-Si/Al) coil. Figure \ref{varind1}(a) shows that an applied dc
voltage flattens the initial 3D structure.

The modeling of such meminductive systems should be based on the heat
transfer equation, similar to the case of thermistors (see section
\ref{subsec:thermistors}). In the simplest case, we can assume
that the device state is defined by a single temperature $T$, that
is,
\begin{equation}
\phi \left( t \right)=L\left( T \right) I\left( t\right) ,
\label{meminductbimorph}
\end{equation}
where $\phi \left( t \right)$ is the magnetic flux piercing the inductor, $I(t)$ is the current that
flows in it, and $T$ satisfies
\begin{equation}
C_h\frac{\textnormal{d}T}{\textnormal{d}t}=R I^2+\delta
\left( T_{env}-T\right) \label{eq:heatdissipationInd}
\end{equation}
where $C_h$ is the heat capacitance, $\delta$ is the dissipation
constant of the device, and $T_{env}$ is the background
(environment) temperature. It follows from equations
(\ref{meminductbimorph},\ref{eq:heatdissipationInd}) that
meminductive systems based on the bimorph effect are first-order
current-controlled meminductive systems. Moreover, since the state
function (equation (\ref{eq:heatdissipationInd})) and the response
function (equation (\ref{meminductbimorph})) are even functions of
the current, these systems (under these simplified assumptions)
should exhibit type-II hysteresis loops under a periodic stimulus
(see figure \ref{crossing}). Table \ref{table:bimorph} summarizes properties of
bimorph meminductive systems.

\begin{table*}
{ \renewcommand{\arraystretch}{1.6}
\begin{tabular}{| l | c | }
\hline
Physical system & Bimorph meminductive system \\
\hline
Internal state variable(s) & Temperature, $x=T$ \\
\hline
Mathematical description & $\phi=L(x)I$  \\
& $\frac{\textnormal{d}x}{\textnormal{d}t}=C_h^{-1}RI^2+C_h^{-1}\delta \left(
T_{env}-x\right)$ \\
\hline
System type & First-order current-controlled
meminductive system \\
\hline
\end{tabular}
}
\caption{Bimorph meminductive system.}
\label{table:bimorph}
\end{table*}

\subsubsection{Elastic meminductive system}

A model of meminductive system whose operation is based on an interplay of elastic forces with current-current interaction can be found in Sec. \ref{elastic_meminductor} of this review.

\subsection{Other meminductive systems}

Having discussed geometry-based meminductive systems, we now turn to other
possible realizations that, as it was mentioned above, can be
based on peculiarities in the permeability response. In
particular, a meminductive behavior can be realized using the core
material whose response to the applied magnetic field depends on
its history. As an example, we can think about ferromagnetic
materials exhibiting a magnetic hysteresis such as iron or
iron-based alloys. Such inductors can be easily realized in
practice, and the desired orientation of the ferromagnetic core easy
axis with respect to the coil direction can be selected
\cite{vroubel04a}. Moreover, a mathematical framework to model
iron-core inductors is reported in \cite{matsuo99a}. As of now,
however, a clear connection of such inductors with the theory of
memory elements has not been developed. In addition, memory
effects in inductance (such as lagging) were revealed in the
context of superconducting circuits \cite{shevchenko08a}. An
interesting - yet unexplored - direction is to employ field-induced
ion motion in solid state electrolytes for non-volatile
realizations of meminductive systems.

\section{Other systems with memory}
\label{sec:structure}

In this section we consider several systems with memory that can
not be categorized as pure memristive, memcapacitive or
meminductive systems. Structural peculiarities of such devices (such as,
for example, the presence of a third terminal or complex behavior)
require a more involved description containing, in some cases,
several basic circuit elements. For example, the equivalent
structure of Josephson junctions considered below involves four
different elements: resistor, capacitor, non-linear inductor and
memristor. However, our goal here is not to provide a complete
list of complex device structures with memory. Rather, we want to
discuss some important examples that are presently investigated
for practical applications, and show the wide variety of systems
where memory may occur.

\subsection{Three-terminal devices}

\subsubsection{Electrochemical cell ``memistor''}

A transistor-like three-terminal structure, that was named {\it memistor} (not to be confused
with the term ``memristor'' we have employed so far), was
designed by Widrow and co-workers \cite{widrow60a,widrow62a} in
the early sixties for demonstration in neural networks. In Widrow's
memistor, the resistance between two of the terminals was controlled
by the time integral of the current in the third terminal
\cite{widrow60a}. The particular realization of the memistor
employed the phenomenon of ``electroplating'' whereby the amount of metal
deposited on a resistive substrate is determined by the control
current. In the first successful attempt, a memistor was realized
using electroplating of copper from a copper sulfate-sulfuric acid
bath upon an ordinary pencil lead \cite{widrow60a}.

\begin{figure*}
 \begin{center}
 \centerline{
    \mbox{(a)}
    \mbox{\includegraphics[angle=0,width=13cm]{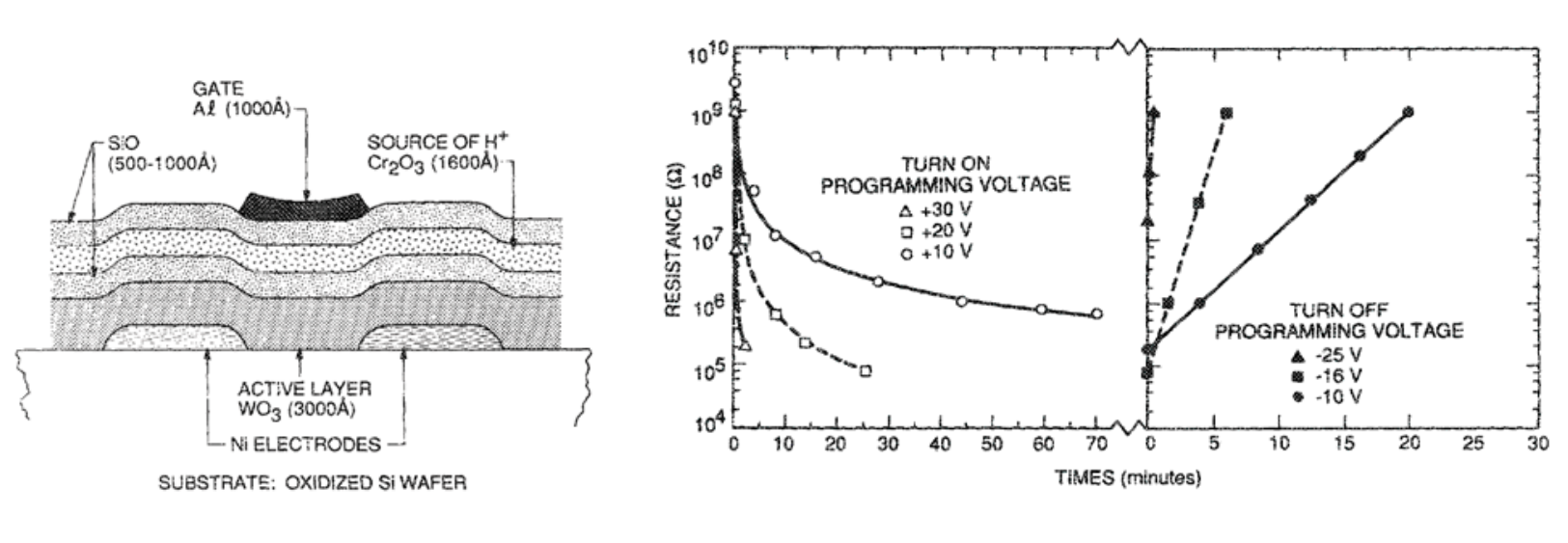}}
  }

  \centerline{
    \mbox{(b)}
    \mbox{\includegraphics[width=5.00cm]{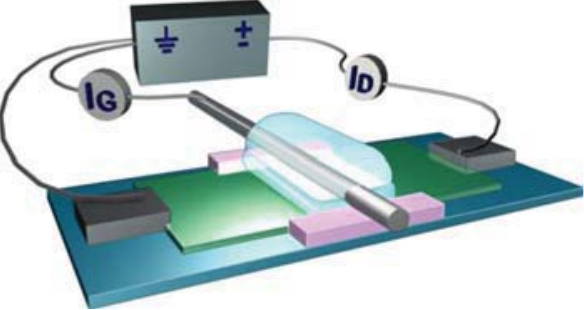}}
    \mbox{(c)}
    \mbox{\includegraphics[width=7.00cm]{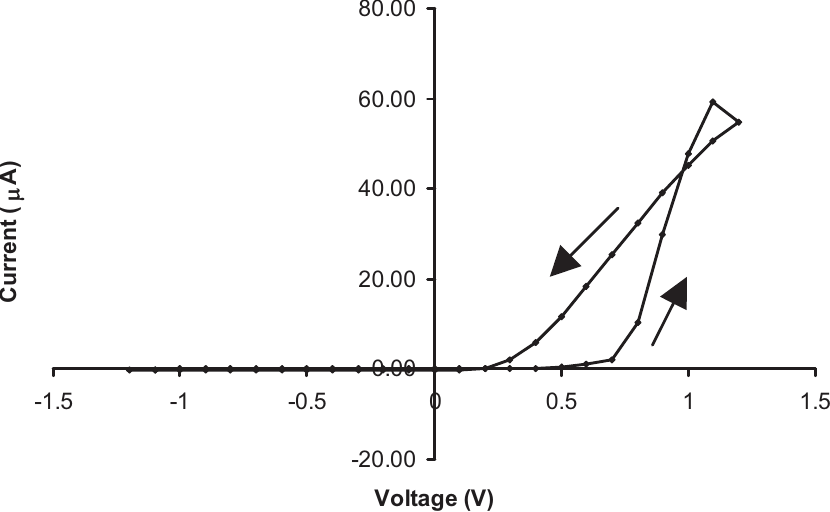}}
  }
\caption{ (a) Cross-section of a three terminal WO$_3$ ``memistor''
device and its programming characteristics (Reprinted with permission from \cite{thakoor90a}.
Copyright 1990, American Institute of Physics). (b) Schematic representation of polymeric
transistor with memory connected to a battery and (c) its $I-V$
characteristics (Reprinted with permission from \cite{berzina09a}. Copyright 2009, American Institute of Physics).} \label{memistor}
\end{center}
\end{figure*}

\subsubsection{Solid-state ``memistor''}

A solid-state realization of the memistor was reported by Thakoor
{\it et al.} \cite{thakoor90a} in 1990. In the solid-state
memistor device schematically shown in figure \ref{memistor}(a),
the programming voltage is applied to the gate electrode with
respect to the two read electrodes positioned at the bottom of the
structure. The read electrodes are separated by a layer of
tungsten trioxide, WO$_3$. The resistance of tungsten trioxide was
varied due to a transfer of hydrogen ions H$^+$ from a thin film
of hydroscopic chromium oxide Cr$_2$O$_3$ into the WO$_3$ active
layer. It was suggested that a redox reaction (with $x$ describing
the amount of H$^+$ ions transferred into WO$_3$)
\begin{equation*}
\textnormal{WO}_3+x\textnormal{H}^++ex\rightarrow\textnormal{H}_x\textnormal{WO}_3
\end{equation*}
converting the WO$_3$ film into a conducting bronze during
``turn-on'' and the reverse reaction during ``turn-off'' is
responsible for the variation of resistance \cite{thakoor90a}. The
possibility of continuous analog programming of resistance in a
wide range of values from $\sim 10^5$ to $\sim 10^9\Omega$ was
demonstrated. Experimentally measured $R(t)$ curves under
programming conditions (see figure \ref{memistor}(a)) show a
power-law relationship for the ``turn-on'' process, and an
exponential relationship for ``turn-off'' process
\cite{thakoor90a}. Such a different form of ``turn-on'' and
``turn-off'' curves was related to a generated emf that is in the
same direction with the applied voltage in the ``turn-off''
process, and in the opposite direction during the ``turn-on'' one,
thus facilitating and opposing ion transfer, respectively
\cite{thakoor90a}.

\subsubsection{Polymeric transistor}

In several recent papers
\cite{erokhin05a,smerieri08a,berzina09a,berzina10a} a polymeric
transistor with memory was investigated. Although the demonstrated
transistor is a three-terminal device,  it was called a memristor
in some of these publications. Here, in order to avoid confusion,
we prefer to use the name ``transistor with memory'' because of
the three terminals configuration.

The operation of polymeric transistors with memory is based on the
possibility to reversibly change the conductivity of the
conductive polymer emeraldin base polyaniline (PANI) between
oxidized and reduced states using ion drift in a solid electrolyte
(LiCl was used in initial studies \cite{erokhin05a}). The observed
variation of conductivity was about two orders of magnitude
\cite{erokhin05a} under bi-polar operational conditions. In figure
\ref{memistor}(c) we show an example of $I-V$ curve for a polymeric transistor with
memory.

It is interesting that in such devices the $I-V$ shape is
sensitive to electrolyte composition \cite{berzina10a}. On this
ground, an optimal electrolyte composition was determined
\cite{berzina10a}. Moreover, possible applications of polymeric
transistor  with memory in adaptive networks mimicking to some
extent the learning behavior of biological systems were suggested
\cite{erokhin07a} (see also section \ref{neuro_circuits} where we
discuss applications of three-terminal devices in neuromorphic
circuits).

\subsection{Memristive component in Josephson junctions}\label{memJJ}

\begin{figure}
 \begin{center}
 \includegraphics[angle=0,width=6cm]{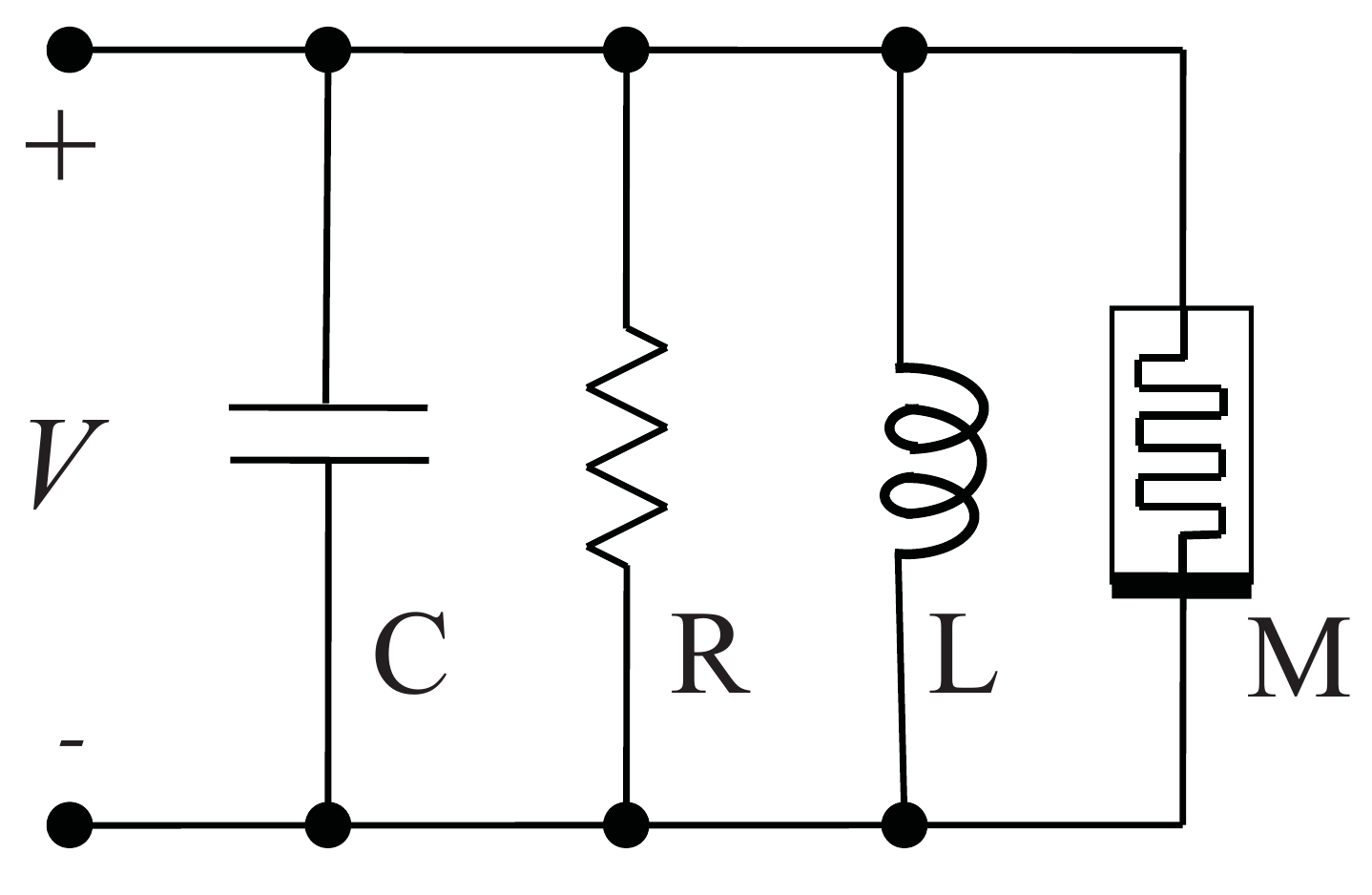}
\caption{ Equivalent scheme of Josephson junction involving resistor, capacitor, non-linear inductor and memristor \cite{chua03a}.}
\label{josephson}
\end{center}
\end{figure}

In a realistic model of a Josephson junction \cite{chua03a}, the
latter is approximated by an equivalent scheme shown in figure
\ref{josephson} involving a linear resistor $R$, linear capacitor
$C$, non-linear inductor $L$ and memristor $M$ connected in
parallel (this model is also discussed in reference
\cite{jeltsema10a}). In this scheme, the memristor takes into
account a small current component due to interference among
quasi-particle pairs \cite{chua03a}. The current through the
memristor (we do not consider here the rest of the circuit since
its modeling is standard, see, e.g., \cite{likharev79a}) is given
by \cite{overhauser00a,josephson62a}
\begin{equation}
I_M=G_0\cos\left(k_0\varphi \right)V \label{eq:joseph}
\end{equation}
where $G_0$ and $k_0$ are device-related constants \cite{chua03a},
and $\varphi$ is the superconducting phase difference across the junction.

Since in the ac Josephson effect the phase difference evolves according to ($e$ is the electron charge)
\begin{equation}
\frac{\textnormal{d}\varphi}{\textnormal{d}t}=\frac{2eV}{\hbar},
\end{equation}
the equation (\ref{eq:joseph}) results in
\begin{equation}
I_M=G_0\cos\left(\frac{2 e k_0}{\hbar}\int\limits_{-\infty}^t
V(t')dt' \right)V \label{eq:joseph1},
\end{equation}
which describes an ideal voltage-controlled memristor. We thus see
that Josephson junctions contain a memristive component. This
component is however generally small, and in most cases of
practical interest it can be neglected. Nevertheless, this example is important since
the memristive component in Josephson junctions is a rare example of {\it ideal} memristor.

\begin{table}[H]
{ \renewcommand{\arraystretch}{1.6}
\begin{tabular}{| l | c | }
\hline
Physical system & Josephson junction \\
\hline
Internal state variable(s) & Phase difference, $x=\varphi$ \\
\hline
Mathematical description & $I=G_0\cos\left(\frac{2 q k_0}{\hbar}\int\limits_{-\infty}^t
V(t')dt' \right)V$  \\
\hline
System type & Voltage-controlled memristor \\
\hline
\end{tabular}
}
\caption{Memristive component in Josephson junctions.}
\label{table:JJ123}
\end{table}

\section{Application of memory elements}
\label{sec:applications}

There are several already well established applications of memory
elements such as the use of thermistors or tunable capacitors
employing MEMS. Here, we would like to focus instead on future
potential applications. Such future applications of memory circuit
elements are envisioned in both digital and analog domains. These
mostly concern memristive systems, while those involving memcapacitive and
meminductive systems are still at an early stage, even though certain
applications discussed below (e.g., biologically-inspired circuits or logic)
could be equally realized with these classes of systems.

In the digital domain, applications of memristive systems involve
non-volatile solid-state memory, signal processing and
programmable logic. Analog applications of memristive systems are based on
the possibility to continuously vary their physical response and,
therefore, on their ability to store more (theoretically infinite) information than in the
digital regime. In this respect, these are related
to analog signal processing, learning circuits, programmable
analog circuits and neuromorphic circuits. Below we discuss such
cases.

\subsection{Digital applications}

\subsubsection{Digital memory}

The digital binary non-volatile memory is the most straightforward
and developed application of memristive systems. The current research
efforts are primarily focused on this technological direction
\cite{Karg08a,Csaba09a,Cagli09a,Vontobel09a,Strukov09c,lian10a,zhuge10a,Wu05a,Green07a}.
It is expected that the resistive random access memory (ReRAM) may
replace Flash memory \cite{Lai08a} in some years. Physically, a
bit of information can be easily encoded in the memristive system's state
assigning, for example, the low resistance state to 1 and the high
resistance state to 0. Moreover, the possibility of continuous
variation of device's resistance offers an opportunity for a
multistate memory cell \cite{nagashima10a,Moreno10a}.

In table \ref{memrFlashcomp} we compare some basic technical
characteristics of NAND (not AND gate-type) Flash memory and nanoionic
ReRAM \cite{ITRS09a}. Most of the demonstrated nanoionic ReRAM parameters
are extracted from Ref. \cite{dietrich07a} that reports fabrication of 2-Mbit
CBRAM (Conductive Bridging Random Access Memory) utilizing 90nm technology. In this architecture,
each memory cell contains one transistor and one memristive junction based on formation/disruption of a
conductive bridge formed by Ag atoms in a germanium selenide chalcogenide material.
The main advantage of ReRAM  over Flash memory is in the significantly shorter
read/write times.
In addition, ReRAM technology is suitable for
higher integration density circuit architecture by using a stack
of crossbar arrays. The drawbacks of ReRAM include a low read voltage and
relatively high write energy.

\begin{figure*}
 \begin{center}
  \centerline{
    \mbox{(a)}
    \mbox{\includegraphics[width=10.00cm]{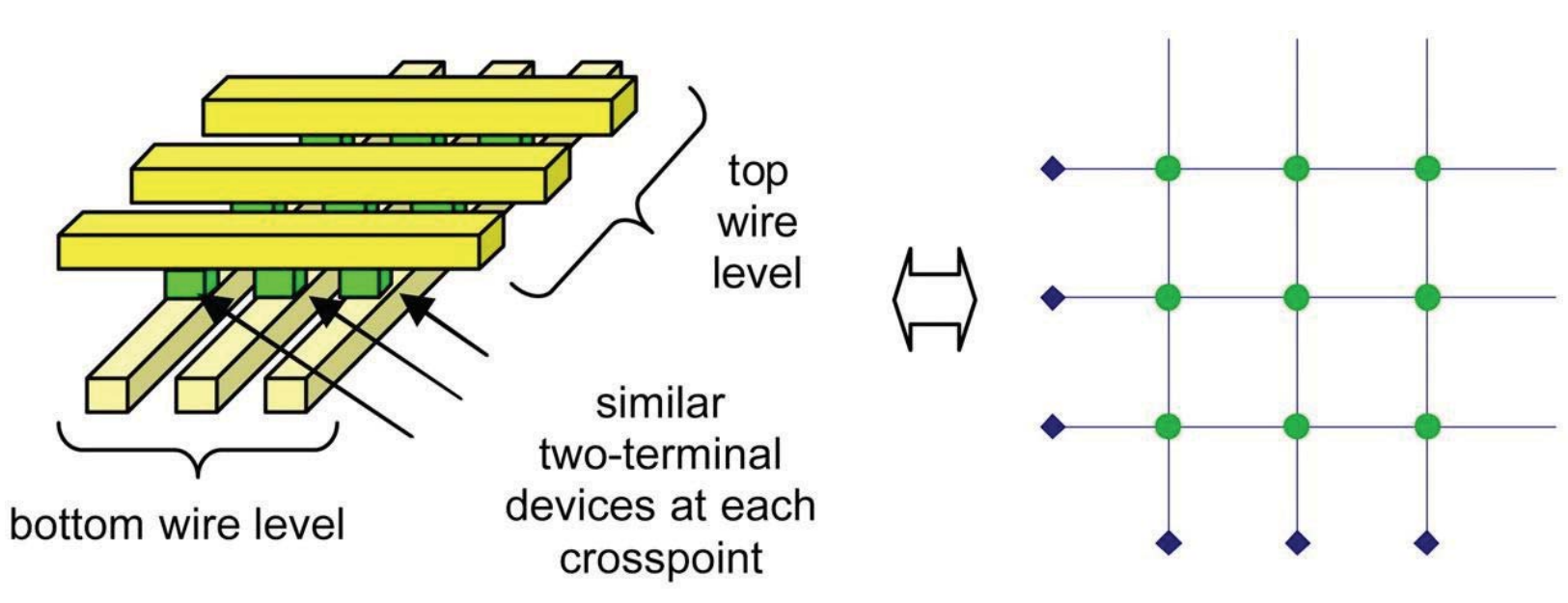}}
   }
   \centerline{
    \mbox{(b)}
    \mbox{\includegraphics[width=13.00cm]{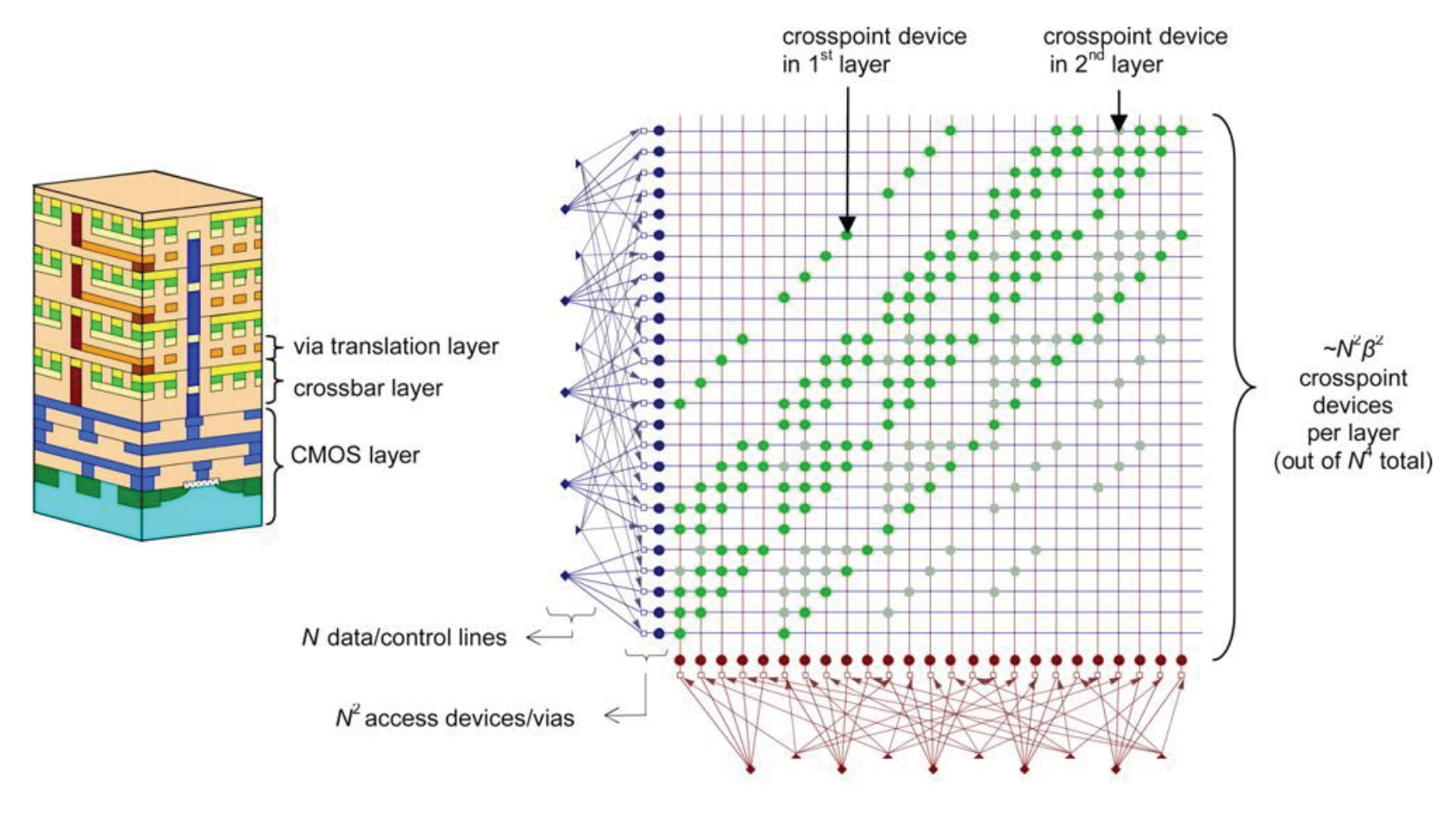}}
    }
\caption{ (a) Single crossbar array and its equivalent circuit
representation. (b) Three-dimensional hybrid CMOS/crossbar circuit
and its equivalent circuit diagram. We note that in this case each
crossbar layer has a connectivity pattern different from that in
(a). Reprinted with permission from \cite{Strukov09c}.} \label{stackedcrossbars}
\end{center}
\end{figure*}

A single crossbar array ReRAM architecture is
called CMOL \cite{Heath98a,Likharev05a} (Cmos+MOLecular-scale
devices) combining a single crossbar layer (see figure
\ref{stackedcrossbars}(a)) with a conventional CMOS (complementary
metal oxide semiconductor) layer. In the recently suggested
three-dimensional extension of CMOL technology \cite{Strukov09c},
a single CMOS layer is located underneath multiple crossbar
layers separated by translation layers (figure
\ref{stackedcrossbars}(b)). In such an architecture, each
memristive element can be accessed via a unique four-dimensional address
(for a single crossbar layer, the address is two-dimensional)
\cite{Strukov09c}.

\begin{table*}[t]
\begin{center}
\begin{tabular}{  l | c c | c c }
  \hline\hline
Property & \multicolumn{2}{c|} {NAND Flash} & \multicolumn{2}{c} {Nanoionic ReRAM} \\
     \hline
Year/status & 2009 & 2024 & Demonstrated & Best projected \\
Feature size F (nm) & 90 & 18 & 90 & 5-10 \\
Cell area & 5F$^2$ & 5F$^2$ & 8F$^2$ & 8/5F$^2$ \\
Read time & 50 ns& 8ns& $<$50ns & $<$10ns \\
Write/Erase time & 1/0.1ms&1/0.1ms & 5ns$/$5ns & $<$20ns \\
Retention time & $>$10y & $>$10y & $>$10y & $>$10y \\
Write cycles & $>$1E5 & $>$1E5 & $>$1E9 & $>$1E16 \\
Write operating voltage (V) & 15 & 15 & 0.6/-0.2 & $<$0.5 \\
Read operating voltage (V) & 2 & 1 & 0.15 & $<$0.2 \\
Write emergy (J$/$bit) & $>$1E-14 & $>$1E-15 & 5E-14 & 1E-15 \\
  \hline\hline
\end{tabular}
\caption{ Current and projected characteristics of NAND Flash
memory and nanoionic ReRAM from the ITRS 2009 edition \cite{ITRS09a}. Here, $F$ is the smallest lithographic
dimension.} \label{memrFlashcomp}
\end{center}
\end{table*}

As we have noted in section \ref{RSoxides}, the implementation of crossbar array memory
may be complicated as a crossbar array represents a resistive network so that the current
between any two selected word and bit lines can flow through many memristive devices. One of the approaches
to overcome this problem is to use individual access devices such as diodes or transistors. Another approach
was recently suggested by Wang and co-authors \cite{Wang10b}. In order to write information into a
specific cell, they propose to apply $V/2$ and $-V/2$ voltages to the corresponding word and bit lines and zero voltage to all other lines.
In this way, only the selected cell will be subjected to $V$ voltage amplitude and all others cell will experience $\pm V/2$ or $0$ bias. Assuming that the switching occurs when the applied voltage amplitude is above $V/2$, the change of only selected cell becomes possible. A negative
side of this approach is that the current still will flow through all memristive elements connected to two selected lines dramatically increasing
the power dissipation when utilizing many-elements crossbar arrays.

A detailed overview of emergent storage memory technologies can be
found in the review paper \cite{burr08a}.

\subsubsection{Logic}

Another important application of memristive systems is in the field of
digital logic circuits. On the one hand, memristive systems can serve as
configuration bits and switches in a data routing network, and on
the other hand they can be used to perform logic operations. In
particular, Strukov and Likharev \cite{strukov05a,strukov07a} have
shown the potential of using CMOL circuits in the areas of
field-programmable gate arrays (FPGAs) \cite{strukov05a} and image
processing \cite{strukov07a}. The efficiency analysis shows that
memristive FPGAs are much faster and more energy efficient in
comparison with similar traditional devices based on CMOS
technology \cite{Cabe09a}. Hybrid reconfigurable logic circuits
\cite{Xia09a}, and logic circuits with a ``self-programming''
capability \cite{Borghetti09a} (namely, with a capability of a
circuit to reconfigure itself) were demonstrated. In both works,
titanium-dioxide thin-film memristive systems were used as basic device
structures.

\begin{figure}
 \begin{center}
    \includegraphics[width=8.00cm]{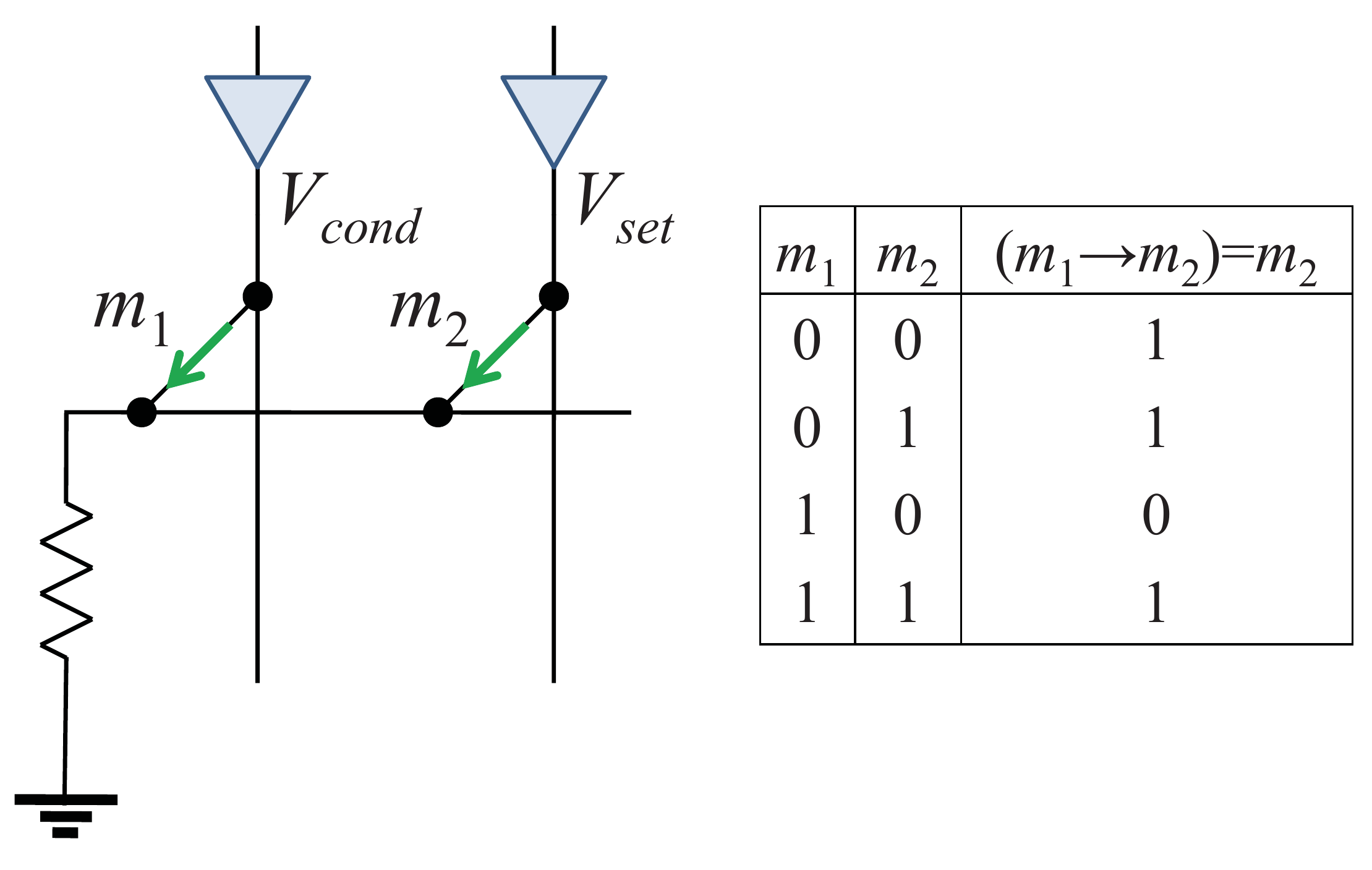}
\caption{ Material implication with memristive systems. State 0 (``false'') corresponds to the high-resistance
state of memristive system, and state 1 (``true'') is the low-resistance
state. The triangle symbols indicate the external driver circuits
that set the voltages $V_{cond}$ and $V_{set}$. The memristive systems are connected to a load resistor. Memristance $m_1$
value does not change from its initial state during operation,
while memristance $m_2$ changes according to the value of $m_1$.
The final result is stored in $m_2$. All this is indicated by the
symbol $(m_1 \rightarrow m_2)=m_2$, namely the material
implication logic operation $(m_1 \rightarrow m_2)$ is stored
($=$) into $m_2$. Adapted from \cite{Lehtonen09a} (\copyright 2009 IEEE).} \label{logic}
\end{center}
\end{figure}

In stateful logic architecture, the ``state'' of the memristive system acts
as both a logic gate and a latch (bi-stable circuit element
capable to hold one bit of information). The concept of crossbar
latch was discussed in reference \cite{Kuekes05a} where a device
storing a logic value, enabling logic value restoration and
inversion was demonstrated. It was concluded that such
functionality, in combination with resistor/diode logic gates,
enables universal computing \cite{Kuekes05a}, namely given a description
of any other computer or program and some data, it can perfectly
emulate this second computer or program. Recently, Borghetti {\it
et al.} have experimentally demonstrated a realization of material
implication and NAND operation with memristive elements
\cite{borghetti10a}. In addition, Lehtonen and Laiho
\cite{Lehtonen09a,Lehtonen10a} have analyzed how many additional
memristive elements, in addition to memristive systems holding initial values, are
needed to compute any Boolean function. Their conclusion is that
two memristive systems indeed suffice to compute all Boolean functions
\cite{Lehtonen10a}.

To understand logic operations with memristive systems, let us consider
a realization of material implication shown in figure \ref{logic}.
In this circuit, voltages $|V_{cond}|<|V_{set}|$ are used in
circuit operation,
state 0 corresponds to the high-resistance state of memristive systems, and
state 1 is the low-resistance state, and the memristive systems are connected to a load resistor.  The circuit operation is based on the
threshold-type behavior of experimentally realizable memristive systems (see, e.g., the equation for $\dot x$ in table \ref{table:threshold}). The calculation result is stored in $m_2$.

First of all, we note that the amplitudes of $V_{cond}$ and $V_{set}$ are selected in
such a way that the voltage on $m_1$ never exceeds the threshold. Therefore, the state of $m_1$
remains unchanged. When $m_1$ is in the high resistance state ($m_1$=0) and $V_{cond}$ and $V_{set}$ are applied,
the voltage on $m_2$ exceeds the threshold leading to the transition of $m_2$ from 0 to 1.
However, under the same conditions but with $m_1$=1,  the voltage drop on $m_2$ is reduced and
the transition of $m_2$ from 0 to 1 is prohibited. Symbolically, material implication operation can be
expressed as
\begin{equation}
(m_1\rightarrow m_2)=m_2,
\end{equation}
namely the logic operation $m_1\rightarrow m_2$ is stored in
$m_2$. Material implication - together with the possibility of
reset (by changing polarity of $V_{set}$) - are functionally
complete, thus providing an opportunity to compute any Boolean
function \cite{Lehtonen09a}.

However, to the best of our knowledge, the actual sequences of
operations to realize other logic gates were not reported anywhere
in the literature. Therefore, below we provide the realization of
the basic logic functions NOT, OR, and AND. In these operations,
an additional work memristive element $m_3$ is used, which is also used to
store the results.
\begin{align}
\textnormal{NOT}(m_1) = m_2:\; m_2&=0;\;(m_1\rightarrow m_2)= m_2, \\
\textnormal{OR}(m_1,m_2) = m_2 :\; m_3&=0;\; (m_1\rightarrow m_3)=m_3;\notag \\ (m_3\rightarrow m_2)=m_2, \\
\textnormal{AND}(m_1,m_2)=m_2 : \; m_3&=0;\; (m_2\rightarrow m_3)=m_3;\notag \\ (m_1\rightarrow m_3)=m_3;
  m_2&=0;\; (m_3\rightarrow m_2)=m_2.
\end{align}

Based on material implication, an algorithm for adding two numbers
was suggested \cite{Lehtonen09a}. However, this operation requires
a large number of steps. In order to overcome this deficiency, the present authors
have suggested a modification of the circuit shown in Fig. \ref{logic} by adding
a memcapacitive element \cite{pershin10c}. In such a configuration, "input" memristive
systems can be distinguished from "output" ones. The operation of the modified circuit is based on charging
the memcapacitive system through the input memristive system and discharging it through output ones. This protocol
requires smaller number of steps to perform both main logic and arithmetic operations. In fact,
the basic logic operations (NOT, AND, OR) as well as addition of two one-bit numbers were experimentally
demonstrated using memristor emulators \cite{pershin10c} (see also section~\ref{emulators}).

We would like to point out that several hybrid circuit layouts
 were patented that could perform these types of operations (see, e.g., references \cite{snider2004a,mouttet2006a}).
We also note that, since memory circuit elements are
intrinsically analog devices (they acquire a continuous set of
values within a certain range), they could also be used to
generate ``fuzzy logic'' \cite{fuzzybook}, namely non-Boolean
logic operations, such as statements of the type (IF variable IS
property THEN action), $m_1>m_2$, etc. However, we are not aware
of any experimental (or theoretical) work in this direction.

\subsection{Analog applications}

\subsubsection{Neuromorphic circuits} \label{neuro_circuits}

Another potential exciting application of memristive systems - and, possibly, the
most important among analog applications - is in neuromorphic circuits.
Neuromorphic circuits are circuits whose operation is meant to
mimic that of the (human or animal) brain. In these circuits,
memristive systems (and possibly also memcapacitive systems) can be used as
synapses whose role is to provide connections between neurons and
store information. As anticipated in section~\ref{bipolar}, the small size of solid-state memristive systems is
highly beneficial for this application since the density of memristive systems in a chip
can be of the same order of magnitude as the density of synapses
in human brains ($\sim 10^{10}$ synapses/cm$^2$)~\cite{snider08a}. Therefore, using memristive systems, the
fabrication of an artificial neural network of a size comparable to that of a
biological brain becomes possible.

An important feature of biological synapses is the spike-timing-dependent plasticity \cite{Levy83a,Markram97a,Bi98a,Froemke02a}.
In fact, when a post-synaptic signal reaches the synapse {\it before} the action potential of the pre-synaptic neuron, the synapse shows long-term depression (LTD), namely its strength decreases (smaller connection between the neurons) depending on the time difference between the post-synaptic and the pre-synaptic signals. Conversely, when the post-synaptic action potential reaches the synapse {\it after} the pre-synaptic action potential, the synapse undergoes a long-time potentiation (LTP), namely the signal transmission between the two neurons increases in proportion to the time difference between the pre-synaptic and the post-synaptic signals. These general features of biological synapses can be implemented using different types of memristive systems \cite{pershin10c}. We can distinguish three general approaches to STDP realization in artificial neural networks: using an overlap of asymmetric pulses \cite{snider08a,parkin10a,pershin10c}, employing additional CMOS circuitry to track pulse timing \cite{Jo10a}, and utilizing higher-order memristive systems with intrinsic pulse-timing tracking capability \cite{pershin10c}.

A model of such a higher-order memristive system that is capable to track time separation between
post- and pre-synaptic action potentials was suggested by the present authors in reference \cite{pershin10c}.
We provide the model's details in table \ref{table:STDP}. In there,
$\gamma$ is a constant, $V_t$ is a threshold voltage, $y_t$ is a threshold value of the internal variable $y$, $\tau$ is a constant defining the time window of STDP. It is assumed that short (e.g., $\sim 1$ms width) pre-synaptic and post-synaptic square pulses of the same polarity are applied to the second-order memristive system. According to the equation for $R_M=x_1$, the memristance can change when $|y|\geq y_t$. The change of $y=x_2$ is described by an equation whose right-hand side contains excitation terms involving $\theta$-functions and a relaxation term $-y/\tau$. Therefore, after being excited, the decay of the variable $y$ occurs with a decay constant $\tau$. The particular combination of $\theta$-functions in this equation defines the excitation rules: i) the excitation is possible only when $|V|>V_t$ and ii) the variable $y$ excited by a certain polarity of the voltage applied to the memristive system ($V$ is given by a difference of pre-synaptic and post-synaptic potentials) can not be re-excited by a pulse of opposite polarity if $|y|> y_t$. We also note that the change in memristance described by the equations in table \ref{table:STDP} is constrained between $R_{min}$ and $R_{max}$. These constraints are similar to those previously reported in table \ref{table:threshold}.

\begin{table*}
{ \renewcommand{\arraystretch}{1.6}
\begin{tabular}{| l | c | }
\hline
Physical system & Artificial memristive synapse \\
\hline
Internal state variable(s) & Resistance and timing variable, $x_1=R_M$, $x_2=y$ \\
\hline
Mathematical description & $I=x_1^{-1}V$  \\
& $\frac{\textnormal{d}x_1}{\textnormal{d}t}=\gamma \left[ \theta (V-V_t) \theta (x_2-y_t) +  \theta (-V-V_t) \times \right.$ \\
& $ \left. \theta (-x_2-y_t) \right] x_2  \theta\left( x_1-R_{min}\right) \theta\left( R_{max}-x_1\right) $ \\
& $\frac{\textnormal{d}x_2}{\textnormal{d}t}=\frac{1}{\tau} \left[ -V \theta (V-V_t) \theta (y_t-x_2) - \right.$\\
 & $\left. V\theta (-V-V_t) \theta (x_2+y_t) -x_1 \right]$ \\
\hline
System type & Second-order voltage-controlled memristive system \\
\hline
\end{tabular}
}
\caption{Memristive model of an artificial memristive synapse with timing-tracking capability \cite{pershin10c}.
All parameters are explained in the text.}
\label{table:STDP}
\end{table*}

\begin{figure}
 \begin{center}
    \includegraphics[width=7.00cm]{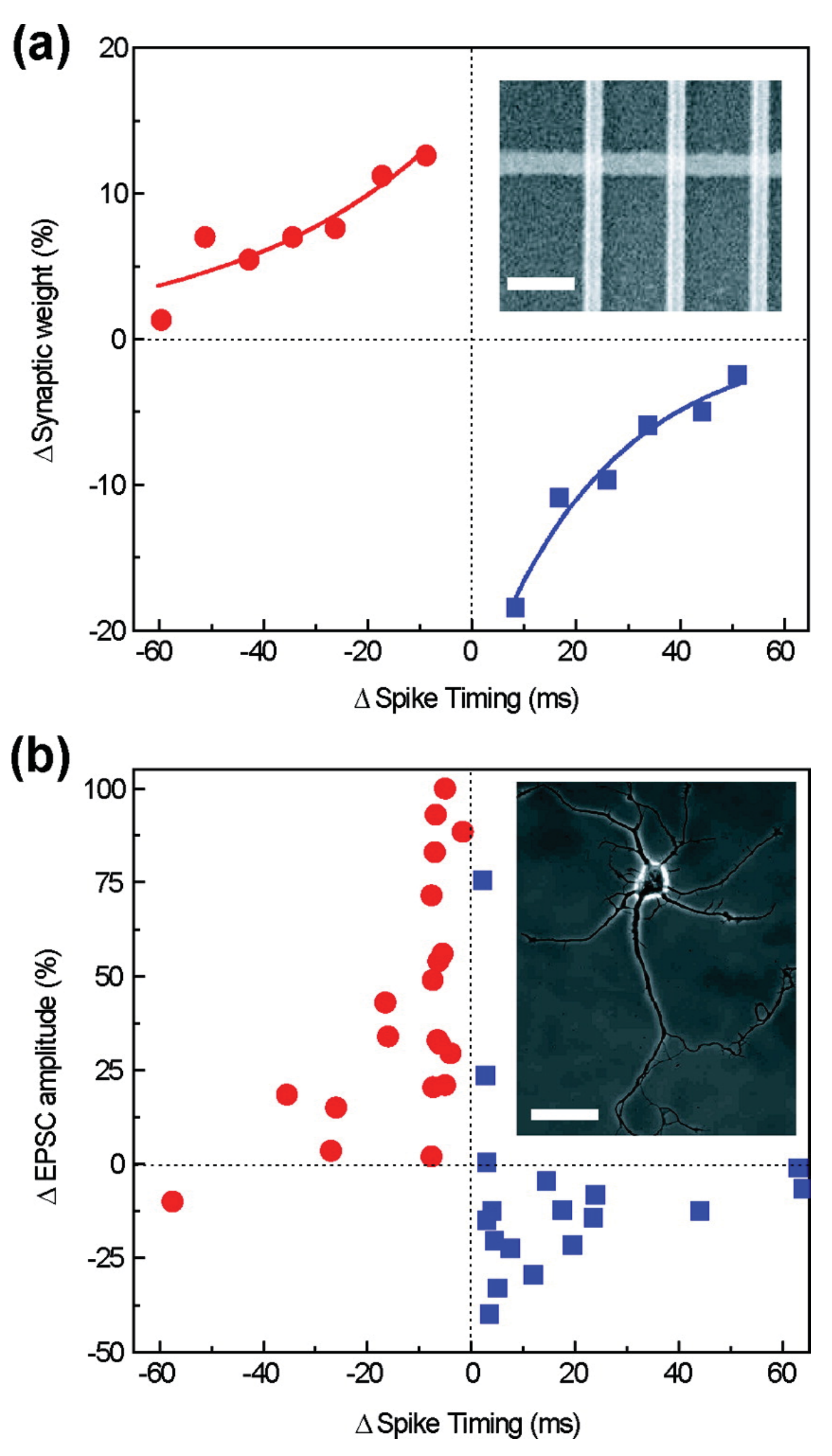}
\caption{ (a) Measured change in the synaptic weight versus spike separation. Inset: SEM image of
the memristive crossbar array, scale bar is 300 nm. (b) Measured change in excitatory postsynaptic current of rat
neurons after repetitive correlated spiking versus relative spiking timing. Inset: image of a hippocampal neuron (the image was adapted
with permission from reference \cite{Bi98a}). Scale bar is 50 $\mu$m. Reprinted with permission from \cite{Jo10a}. Copyright 2010 American Chemical Society.} \label{STDP}
\end{center}
\end{figure}

Experimentally, several circuits showing neuromorphic behavior of
different types and complexities were demonstrated
\cite{pershin09c,Jo10a,Choi09a,Alibart10a,Lai10a}. These circuits
were based on two-terminal \cite{pershin09c,Jo10a,Choi09a} and
three-terminal \cite{Alibart10a,Lai10a,Zhao10a} memristive
devices. Jo {\it et al.} \cite{Jo10a} have demonstrated a
spike-timing-dependent plasticity (STDP) using a crossbar array of
Ag-based memristive synapses. In their architecture, CMOS elements were used to analyze timing of pre- and post-synaptic
pulses in order to generate positive or negative memristive device
programming pulses. In figure \ref{STDP} we show a comparison of
STDP of artificial synapses with changes in excitatory
post-synaptic current of rat hippocampal neurons \cite{Jo10a}.
 Choi and co-authors \cite{Choi09a} have built a crossbar array of
GdO$_x$/Cu-doped MoO$_x$ memory and demonstrated a weight sum
operation important for neural networks.

\begin{figure*}
 \begin{center}
  \centerline{
      \mbox{(a)}
    \mbox{\includegraphics[width=6.00cm]{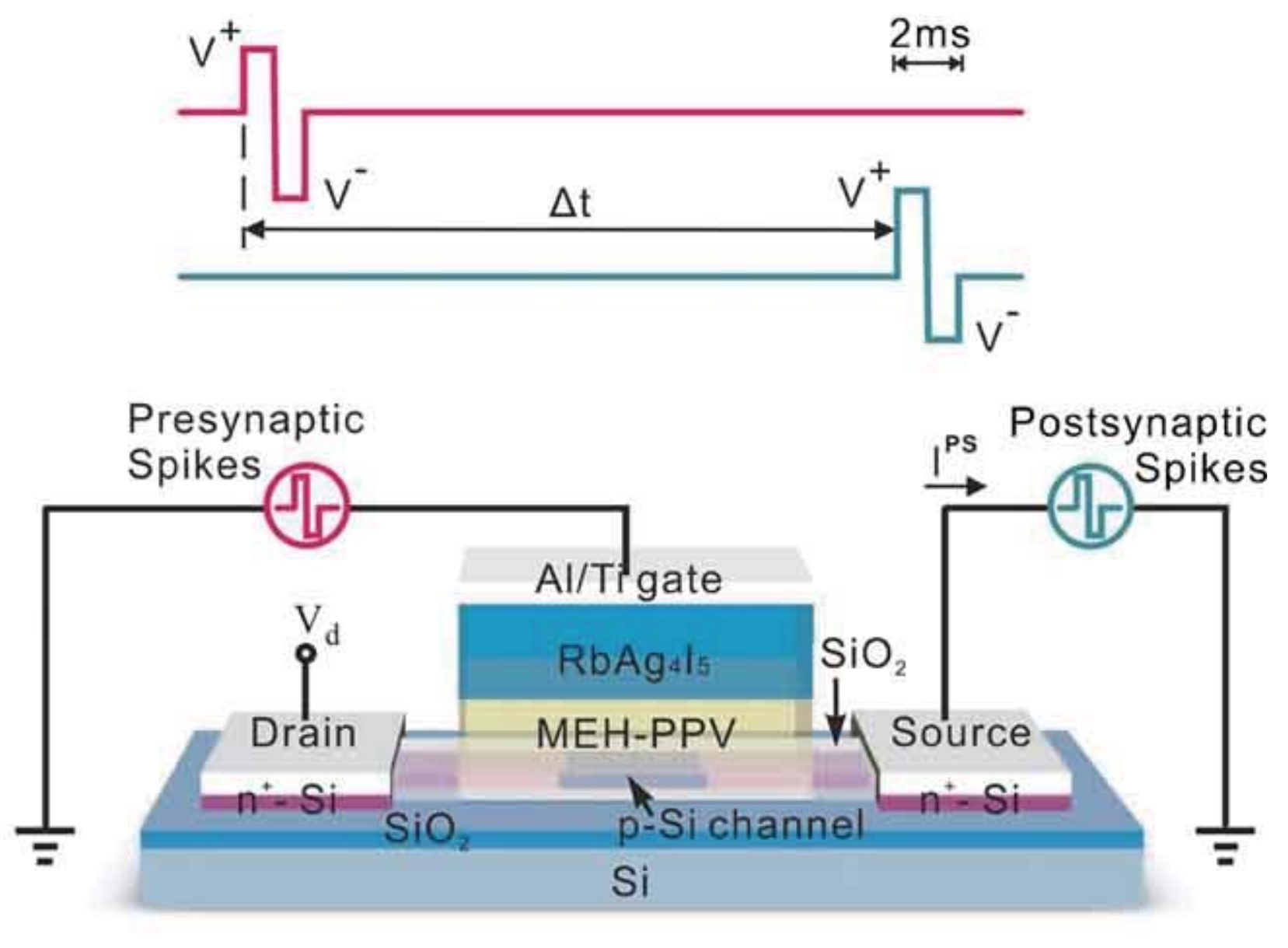}}
        \mbox{(b)}
    \mbox{\includegraphics[width=8.00cm]{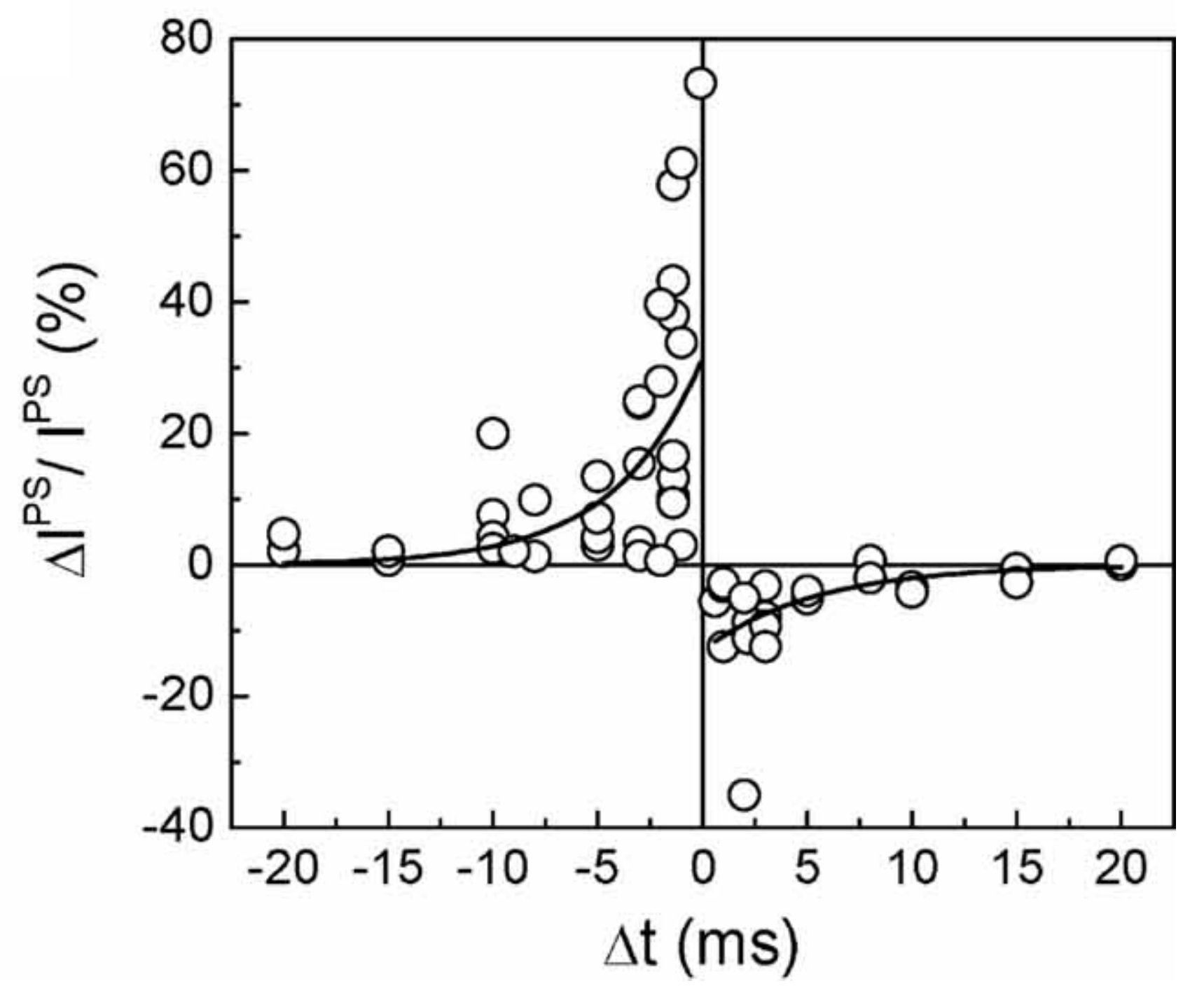}}
   }
\caption{ (a) Synaptic transistor structure. (b) The relative
changes of the postsynaptic currents measured after application of
120 pairs of temporally correlated pre- and post-synaptic spikes. From \cite{Lai10a}, copyright Wiley-VCH Verlag GmbH $\&$ Co. KGaA.
Reproduced with permission.} \label{Lai}
\end{center}
\end{figure*}

Three-terminal memristive devices employed in recent experiments \cite{Alibart10a,Lai10a}
were of a transistor structure in which the gate electrode was typically used to induce
resistance changes probed by a current through two other electrodes.
Lai and co-workers \cite{Lai10a} have demonstrated a synaptic transistor showing STDP.
They have used a conventional MOS transistor structure with a conjugated polymer
(MEH-PPV) layer and a layer of ionic conductor RbAg$_4$I$_5$ placed below the gate electrode. Peculiar features of their experiments are paired
spikes composed of 1ms positive and 1ms negative voltage pulses. In figure \ref{Lai} an experimentally measured STDP of such synaptic devices is demonstrated.
Alibart {\it et al.} \cite{Alibart10a} have fabricated a nanoparticle organic memory field-effect transistor (NOMFET)
whose memory response is due to charging gold nanoparticles embedded into a layer of
pentacene thin film. Using this NOMFET they have observed facilitating and depressing synaptic behavior. Possible drawbacks of
this approach include charge leaking limiting retention time in the range of a few seconds to a few thousand of seconds
and high amplitude voltage ($\sim 50$V) needed for circuit operation \cite{Alibart10a}.
In a recent paper \cite{Zhao10a}, a crossbar array architecture integrating optically-gated carbon nanotubes was also suggested
for neuromorphic applications.

\begin{figure}
 \begin{center}
    \includegraphics[width=7.00cm]{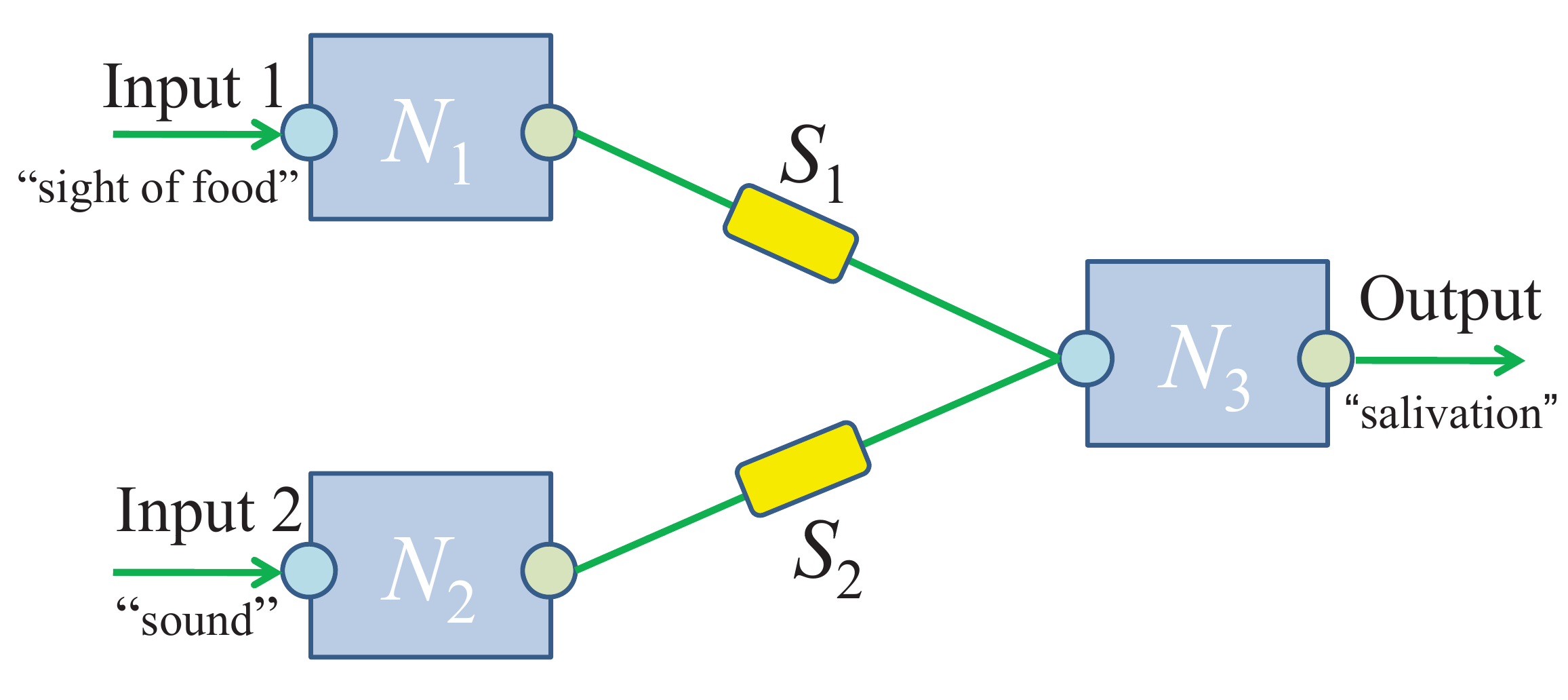}
\caption{ Artificial neural network for associative memory composed of three neurons ($N_1$, $N_2$ and $N_3$) coupled by two
memristive synapses ($S_1$ and $S_2$). The output signal is
determined by input signals and strengths of synaptic connections
which can be modified when learning takes place. Reprinted from \cite{pershin09c}. Copyright 2009, with permission from Elsevier.} \label{ournetwork}
\end{center}
\end{figure}

Pershin and Di Ventra \cite{pershin09c} have reported the experimental
realization of a neural
network based on memristor emulators (see section~\ref{emulators}
for their electronic scheme) demonstrating associative learning.
The neural network used in that experimental study contains two
memristive synapses and three neurons as shown in figure
\ref{ournetwork}. Associative memory is a fundamental property of
the animal brain closely related to the so-called Pavlovian
training~\cite{pavlov27a} in which a particular response to a
given stimulus develops. The most notable experiment of
associative memory is that of a dog to which food is shown and, at
the same time, the tone of a bell is rang so that, with time, the
dog salivates at the ring of the bell only. In the electronic
three-neuron network described in reference \cite{pershin09c}, two
input neurons were responsible for the ``sight of food'' and
``sound'' events, while the output neuron generated a
``salivation'' command.

\begin{figure*}
 \begin{center}
  \centerline{
    \mbox{\includegraphics[width=7.00cm]{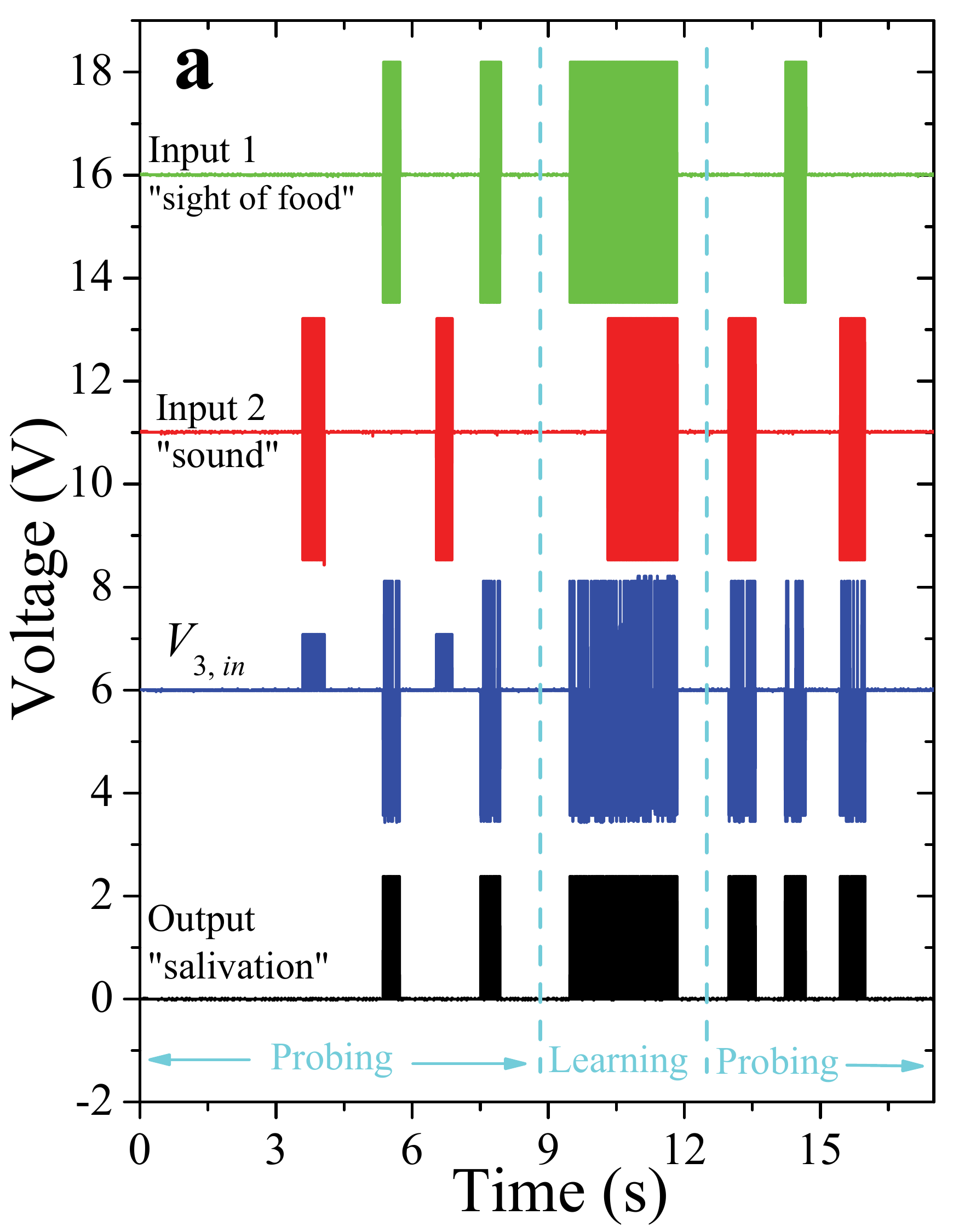}}
    \mbox{\includegraphics[width=7.00cm]{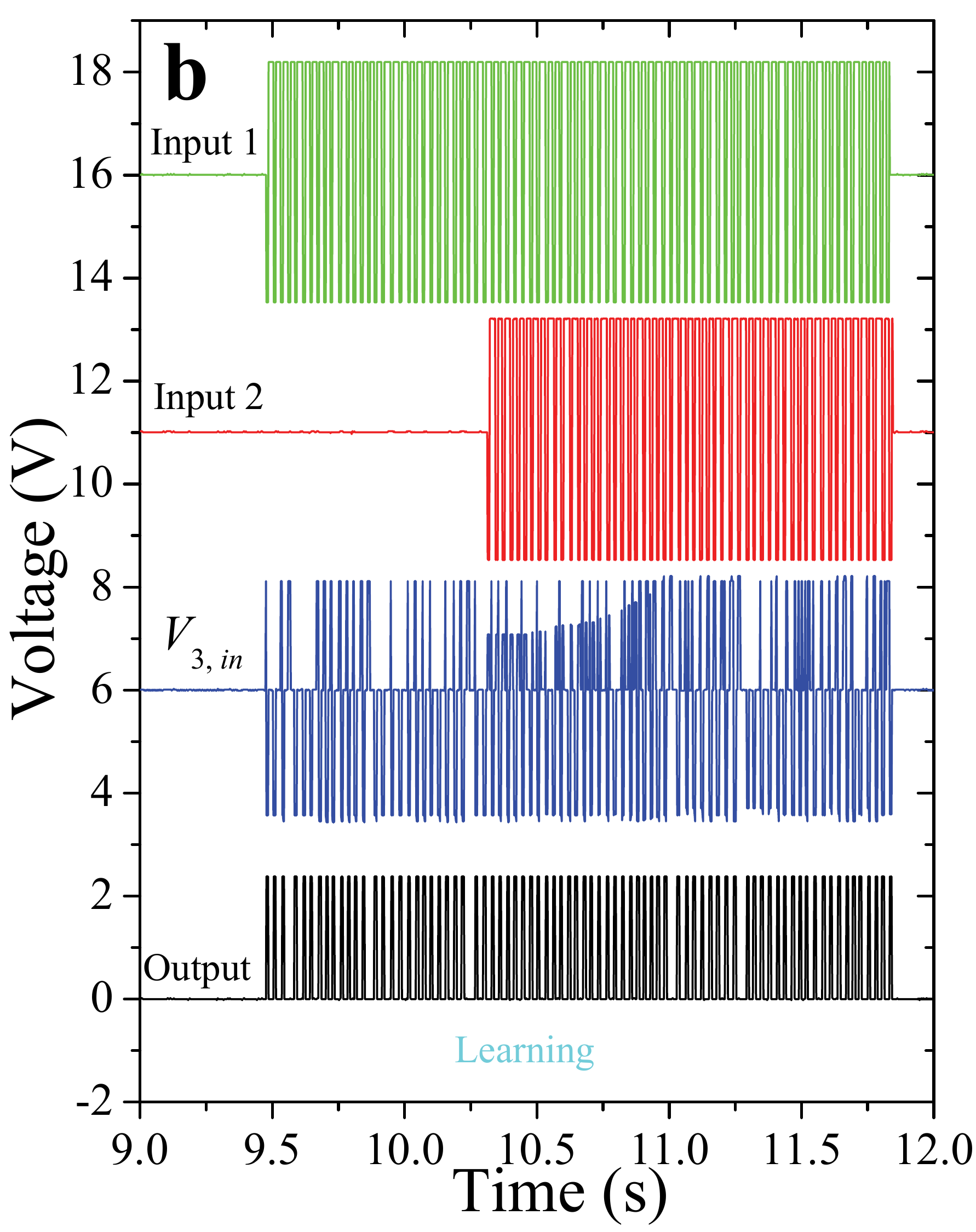}}
   }
\caption{Development of associative memory. {\bf a},
Signal patterns in an electronic circuit corresponding to the neural network shown
in figure \ref{ournetwork}. Such a circuit employs electronic neurons
and electronic synapses (memristor emulators) (see reference \cite{pershin09c} for more details).
At the initial moment of time, $S_1$ synapse is in a low-resistance state
while $S_2$ synapse is in a high-resistance state.
 Correspondingly, in the first probing phase, when
Input 1 and Input 2 signals are applied without overlapping, the
output signal develops only when Input 1 signal is applied. In the
learning phase (see also {\bf b} for more detailed picture), Input
1 and Input 2 signals are applied simultaneously. According to
Hebbian rule, simultaneous firing of input neurons leads to
development of a connection, which, in the present case, is a transition
of the second synapse $S_2$ from high- to low-resistance state.
This transition is clearly seen as a growth of certain pulses in
the signal $V_{3,in}$ (voltage at the input of third (output)
neuron) in the time interval from 10.25s to 11s in {\bf b}. In the
subsequent probing phase, we observe that ``firing'' of any input
neuron results in the ``firing'' of output neuron, and thus an
associative memory realization has been achieved. The curves in
{\bf a} and {\bf b} were displaced for clarity. Reprinted from \cite{pershin09c}. Copyright 2009, with permission from Elsevier.} \label{ournetworkres}
\end{center}
\end{figure*}

Figure \ref{ournetworkres} demonstrates experimental results of
the neural network operation. The functioning of the electronic
circuits is essentially based on pulse overlapping. When Input 1
is excited, the first neuron $N_1$ sends forward positive-polarity
pulses that excite the third neuron $N_3$ which starts sending
positive pulses in the forward direction and negative pulses
backward. If, at the same time, the second neuron $N_2$ is also
excited, the overlap of positive pulses from this second neuron
and negative pulses from the third neuron over $S_2$ memristive
synapse results in a voltage on $S_2$ exceeding its threshold. In
this way, the $S_2$ resistance decreases, or, in other words, an
association develops.

If we return to figure \ref{ournetworkres} again, it clearly shows
that, if one starts with an untrained state of the synapse
connecting the input ``sound'' neuron and output neuron and
exposes the neural network to ``sight of food'' and ``sound''
signals simultaneously, then an association between ``food'' and
``sound'' develops and an output signal is generated when only the
``sound'' input signal is applied. This is exactly the associative
memory behavior discussed above.

\subsubsection{Quantum computing with memory circuit elements}

Memcapacitive and meminductive systems may find useful applications
in electronic circuits with superconducting qubits \cite{Clarke08a,zagoskin07a}.
Since typical superconducting qubit circuits involve usual capacitors and
inductors \cite{Clarke08a,zagoskin07a}, certain memcapacitive and meminductive elements are ideal
for such circuits. What are the advantages of using memory elements in superconducting qubit circuits?
One of many possible applications is {\it field-programmable quantum computing} that was recently discussed by the  present authors \cite{pershin10c}. The idea is to replace capacitive (inductive) elements that provide coupling between different qubits by non-dissipative memcapacitive (meminductive) elements, and introduce additional voltage sources to control the state of these memory elements. In this way, the coupling strength between qubits can be selected. If we consider $N$ simultaneously interacting qubits then a variation of coupling between two of them will result in absolutely different interaction Hamiltonians thus leading to a different system evolution. Quantum computation algorithms will thus benefit from such novel quantum hardware functionality because of the many (practically infinite) interaction schemes that can be implemented within a single circuit architecture \cite{pershin10c}. We would like to
note, however, that for such application we envision memcapacitive and meminductive elements that are non-dissipative (at least within certain time scales),
so that they do not introduce additional qubit relaxation/decoherence.

It might be well to point out that Josephson junctions coupled with
nanomechanical resonators were studied in the past \cite{Cleland04a,Buks06a,Zhou06a,Zhang09a,Pugnetti10a}.
For example, it was shown in reference \cite{Cleland04a} that a nanomechanical resonator consisting of a
piezoelectric crystal sandwiched between split metal
electrodes can be used to provide coupling between phase qubits. Such a resonator has signatures of
a memcapacitive system in which elastic vibrations plays an important role. However, the coupling provided by the resonator is of a constant
strength \cite{Cleland04a} and, therefore, it can not be directly used for the field-programmable quantum computing
idea discussed above.

\subsubsection{Learning circuits}\label{bio-insp}

\begin{figure*}
 \begin{center}
  \centerline{
      \mbox{(a)}
    \mbox{\includegraphics[width=5.50cm]{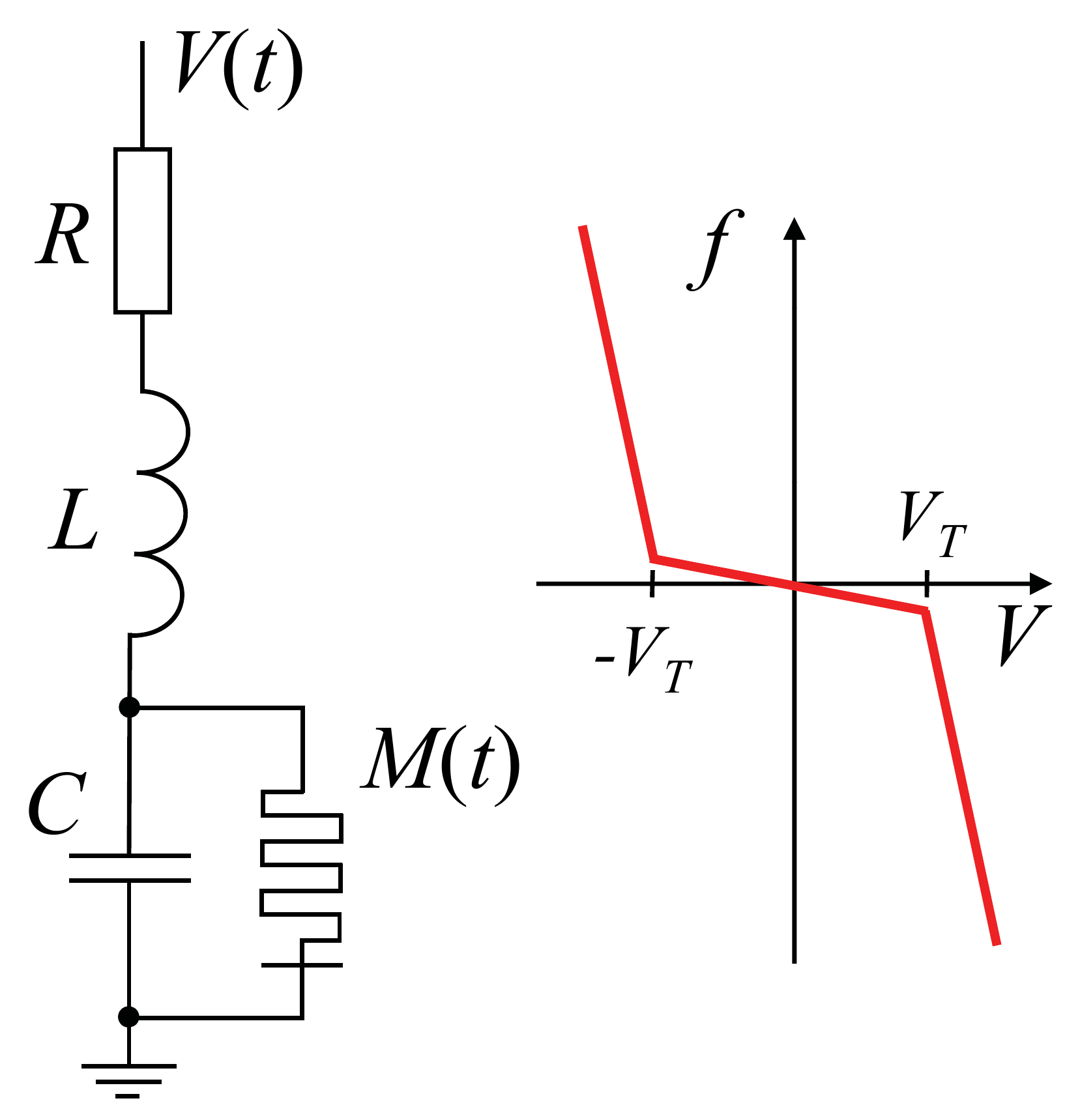}}
        \mbox{(b)}
    \mbox{\includegraphics[width=8.00cm]{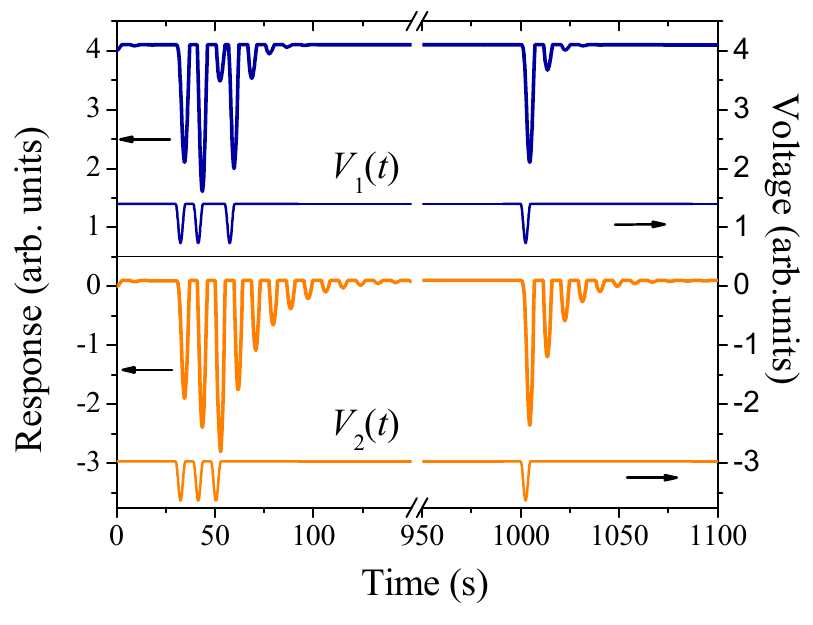}}
   }
\caption{(a) Schematic representation of the learning circuit and
sketch of the selected rate function $f$. The learning
circuit contains four two-terminal circuit elements: resistor $R$,
inductor $L$, capacitor $C$ and memristive system $M$. The function $f$ responsible
for memristive system dynamics depends on the voltage applied to the
system accordingly to $f(V)=-\beta V+0.5 (\beta-\alpha )\left(
|V+V_T|-|V-V_T| \right)$, where $\alpha$ and $\beta$ are positive
constants and $V_T$ is a threshold voltage. (b)
 Modeling of the circuit response. This plot demonstrates that stronger
and long-lasting responses for both spontaneous in-phase slowdown
and spontaneous in-phase slowdown after one disappearance of the
stimulus \cite{saigusa08a} are observed only when the circuit was
previously ``trained'' by a periodic sequence of three equally
spaced pulses as present in $V_2(t)$. The applied voltage $V_1(t)$
is irregular and thus the three first pulses do not ``train'' the
circuit (see reference \cite{pershin09b} for more details).  The
lines were displaced vertically for clarity. From
\cite{pershin09b}.} \label{amoeba}
\end{center}
\end{figure*}

Quite generally, a ``learning circuit'' is an electronic circuit
whose response at a given time adapts according to signals applied
to the circuit at previous moments of time \cite{pershin09b}. All
three memory circuit elements are ideal components for such a
circuit since they provide non-volatile information storage and
compatibility (as time-dependent devices) with other circuit
elements. The present authors have recently suggested
\cite{pershin09b} a learning circuit which mimics adaptive
behavior of a slime mold {\it Physarum polycephalum} from the
group of amoebozoa. This work was inspired by recent experimental
observations on this unicellular organism~\cite{saigusa08a}. These
experiments have shown that when {\it Physarum polycephalum} is
exposed to a pattern of periodic environment changes (of
temperature and humidity), it ``learns'' and adapts its behavior
in anticipation of the next stimulus to come. We have
shown~\cite{pershin09b} that such behavior can be described by the
response of a simple electronic circuit as that shown in figure
\ref{amoeba}(a). The circuit is composed of an LC contour and a
memristive system in parallel with the capacitor. The memristive system was
described by the equations presented in table \ref{table:threshold} that was introduced in section
\ref{memristor_models}. This model takes into account the
non-linear memristance rate of change defined by two parameters
$\alpha$ and $\beta$ (see also figure \ref{amoeba}(a)).

Figure \ref{amoeba}(b) represents simulation results of the response of the circuit.
When a periodic signal
is applied to the learning circuit, the voltage across the
capacitor significantly changes and can exceed the threshold
voltage of the memristive system. This leads to an increase in the
resistance of the memristive system and, consequently, in a smaller damping of
the LC contour. Therefore, the LC contour oscillations are
maintained for a longer period of time in analogy with the same
type of behavior observed when the amoeba is subject to periodic
environment changes.

Very recently, an electronic circuit \cite{driscoll10a} having a close similarity to the learning
circuit in Fig. \ref{amoeba}(a) was experimentally implemented. In this circuit, a vanadium dioxide memristive system
was placed into an $LC$ contour. It was demonstrated that the application of specific frequency signals sharpens the quality factor of its resonant response, and thus the circuit "learns" according to the input waveform.

\subsubsection{Programmable analog circuits}

In programmable analog circuits, memristive systems that operate under
threshold conditions (as those described in the previous section)
can be used as digital potentiometers \cite{pershin09d}. The main
idea is to apply small amplitude voltages to memristive systems when they
are used as analog circuit elements, and high amplitude voltage
pulses for the purpose of resistance programming.
Since the state of memristive system appreciably changes only when the
voltage applied to it exceeds a certain threshold
\cite{strukov09a}, its resistance is constant in the
analog mode of operation, and changes by discrete values with each
voltage pulse. Using this idea, several programmable analog
circuits demonstrating memristive system-based programming of threshold,
gain and frequency were demonstrated by these authors
\cite{pershin09d}.

\subsubsection{Emulators of memristive, memcapacitive and meminductive systems}\label{emulators}

\begin{figure}
 \begin{center}
\includegraphics[angle=0,width=7.0cm]{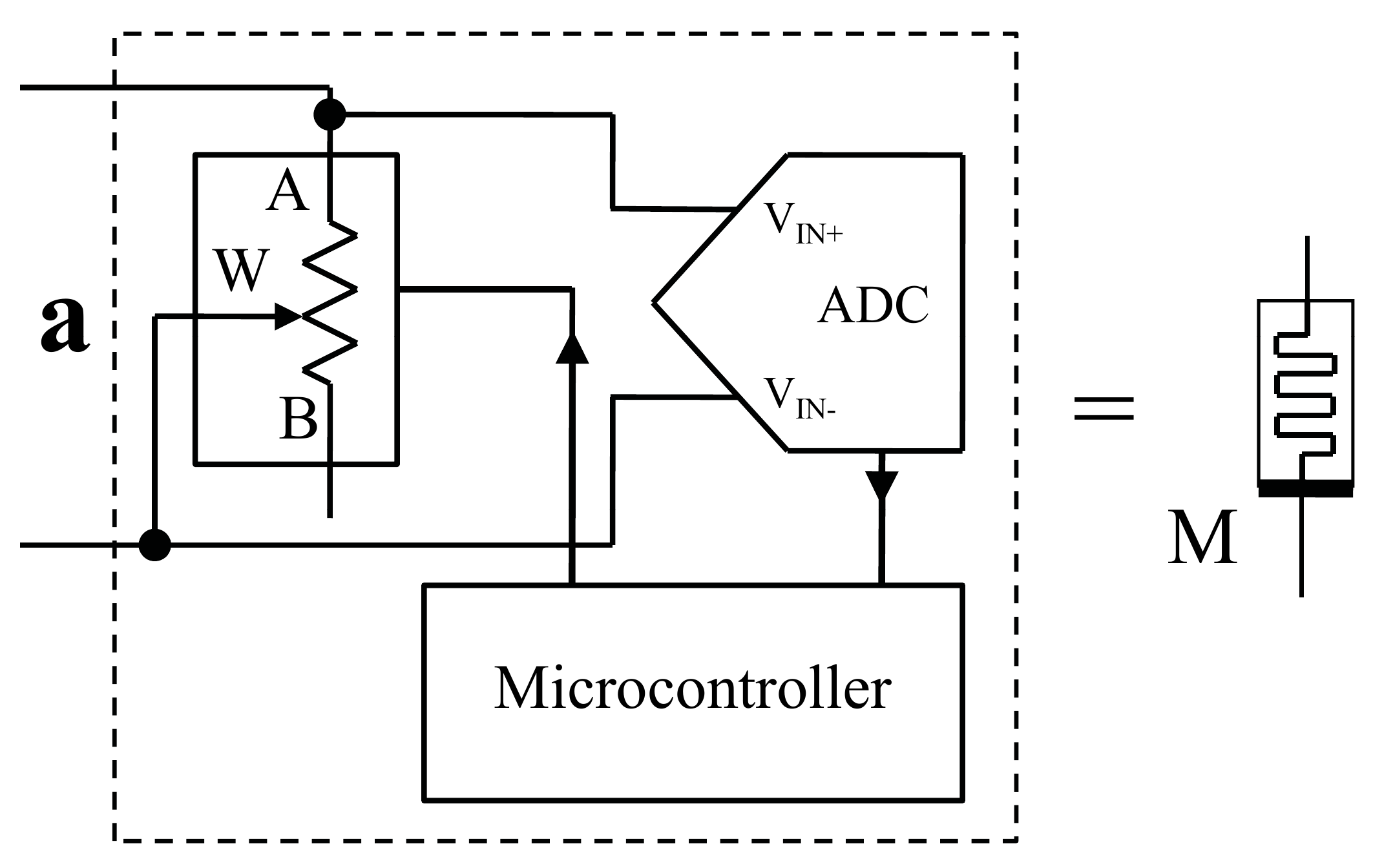}
\includegraphics[angle=0,width=7.0cm]{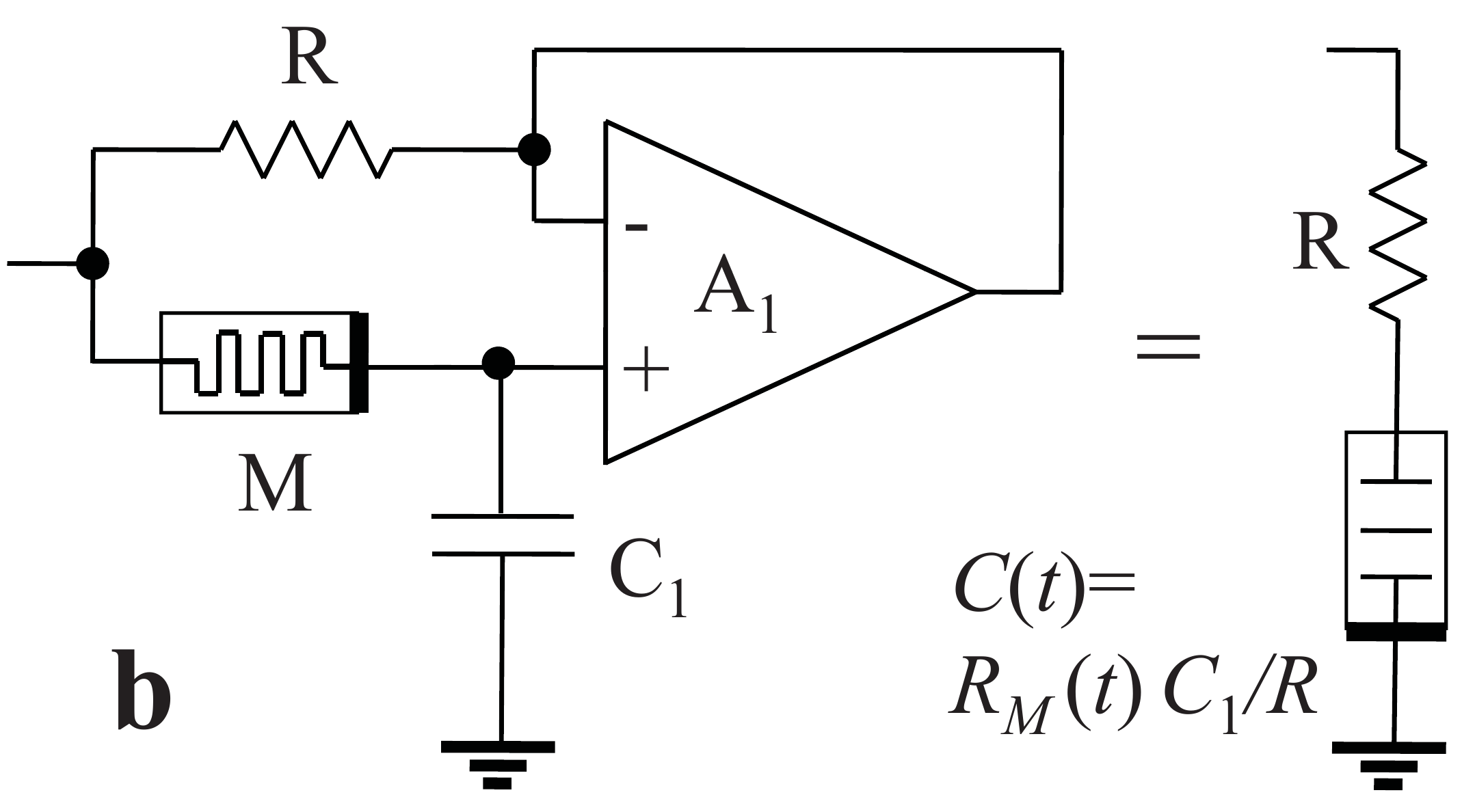}
\includegraphics[angle=0,width=7.0cm]{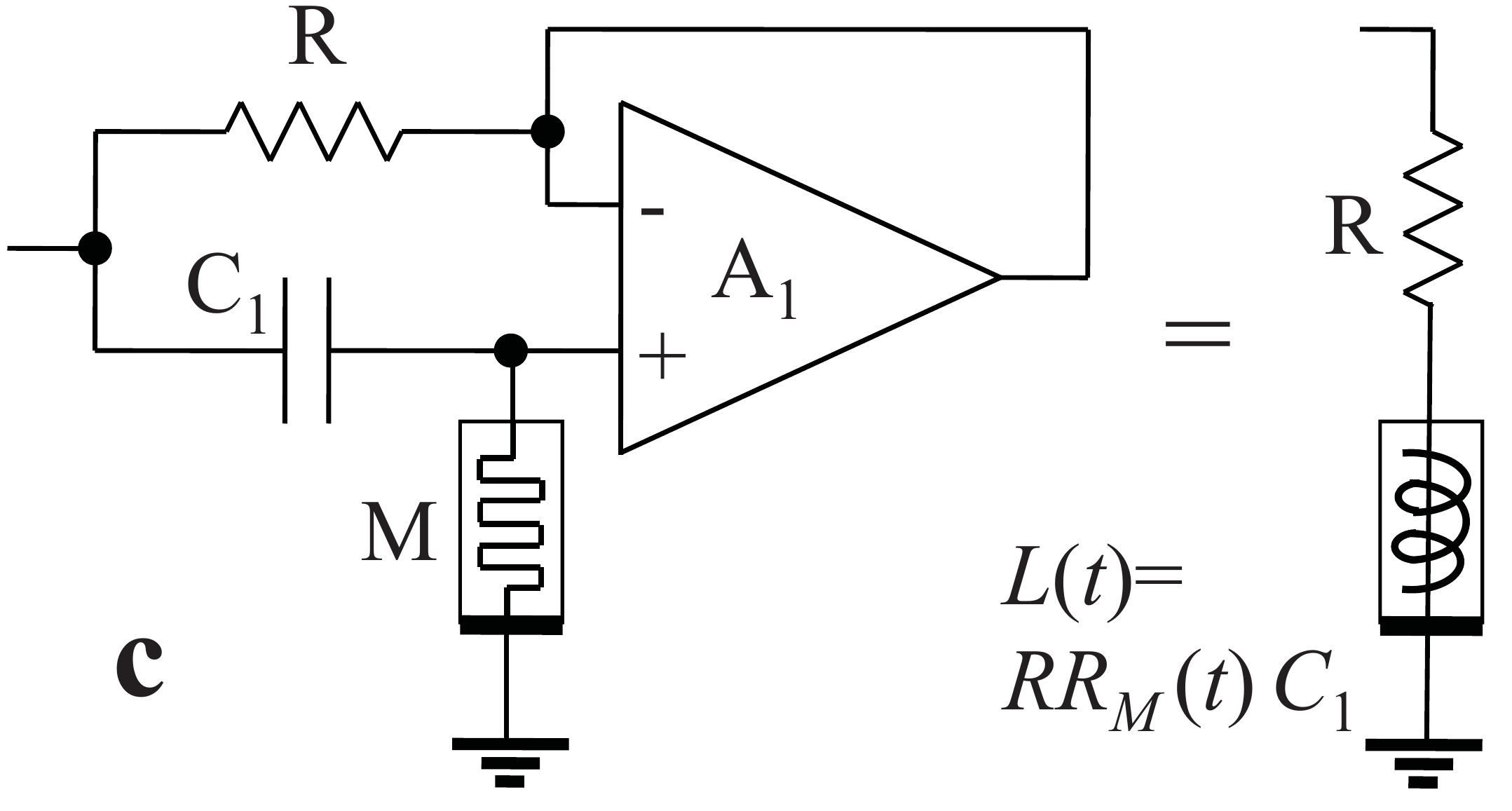}
\caption{Circuits simulating (a) memristive system, (b) memcapacitive system and
(c) meminductive system. Their approximate equivalent circuits are shown
on the right. Reprinted from \cite{pershin10b} with permission from IET.} \label{emul}
\end{center}
\end{figure}

All of the above mentioned programmable analog circuits were built
using a memristor emulator \cite{pershin09d} which is
schematically shown in figure \ref{emul}(a). It consists of a
microcontroller-controlled digital potentiometer whose resistance
value is calculated by the microcontroller using equations of a
current-controlled or voltage-controlled memristive system. For
these calculations, the value of the voltage applied to the
scheme is provided by an analog-to-digital converter. Since
virtually almost any set of memristive system equations can be
pre-programmed, the memristor emulator offers a unique opportunity
to simulate memristive behavior in electronic circuits.

Moreover, the present authors have suggested that specific circuits with memristive systems
can emulate an effective memcapacitive and meminductive behavior \cite{pershin10b}. The architecture of such circuits
(see figure \ref{emul} (b,c)) is based on the gyrator scheme.
In both memcapacitive system and memindictive system emulators, memory properties of the
memristor emulator described above are effectively transferred into an effective memcapacitive and meminductive
response.

For instance, let us consider the circuit of memcapacitive system emulator
depicted in figure \ref{emul} (b). In this scheme, since the
operational amplifier keeps nearly equal voltages at its positive
and negative inputs, the voltage on the capacitor $C_1$ is applied
to the right terminal of $R$. Therefore, we can think that an
effective capacitor with a time-dependent capacitance $C(t)$ is
connected to the right terminal of $R$, so that the relation
$RC(t)=R_M(t)C_1$ holds.

Let us assume that a voltage-controlled memristive system is used in the
memcapacitive system emulator. Then, equations describing the effective
memcapacitor are given by
\begin{eqnarray}
C(t) &=\frac{R_M(V-V_C,x,t)C_1}{R} \label{memcapemul1} \\
\frac{\textnormal{d}V_C}{\textnormal{d} t}
&=\frac{1}{C_1}\frac{V-V_C}{R_M(V-V_C,x,t)} \label{memcapemul2}, \\
\frac{\textnormal{d}x}{\textnormal{d} t} &=f(V-V_C,x,t),
\label{memcapemul3}
\end{eqnarray}
where $V$ is the applied voltage and $V_C$ is the voltage on the
capacitor $C_1$. In equations
(\ref{memcapemul1}-\ref{memcapemul3}), $V_C$ acts as an additional
state variable of the effective memcapacitive system. In fact, this is not
surprising since the capacitor $C_1$ (in figure \ref{emul}(b))
stores information in the form of charge. Consequently, the order
of the effective voltage-controlled memcapacitive system exceeds
the order of the memristive system by one.

The operation of the meminductive system emulator in figure \ref{emul}(c) can be
understood in a similar manner. These emulators also show an
interesting connection between the three memory elements. However, we need to stress that
since the memcapacitive system and meminductive system emulators involve an extra internal resistance, they do not
behave as single (perfect) memcapacitive or meminductive systems. Recently, Biolek and Biolkova \cite{biolek10b} have suggested
a different scheme (mutator) transforming memristive behavior into memcapacitive one. Their approach utilizes two
operational transimpedance amplifiers
operating as the current conveyors. The suggested scheme responds as an ideal memcapacitive system without an additional
resistance $R$ as that shown in the equivalent scheme in figure \ref{emul}(b).

\subsubsection{Other circuits}

There are a few other memristive system-based analog circuits discussed in
the literature \cite{Shima09a,Wey09a,Bray10a}. In particular, an
interesting method of using memristive systems as passive electromagnetic
switches was suggested and analyzed by Bray and Werner
\cite{Bray10a}. Way and Benderli \cite{Wey09a} have simulated and
analyzed an architecture of an amplitude-modulation circuit. A
circuit element combining diode and memory resistance properties, the {\it
switchable rectifier}, was proposed and built by Shima {\it et al.}
\cite{Shima09a}. Their device based on Pt/TiO$_x$/Pt structure
demonstrates a reproducible switching. Such a circuit element can
find useful applications in both analog and digital domains.

\begin{figure}
 \begin{center}
\includegraphics[angle=0,width=6.0cm]{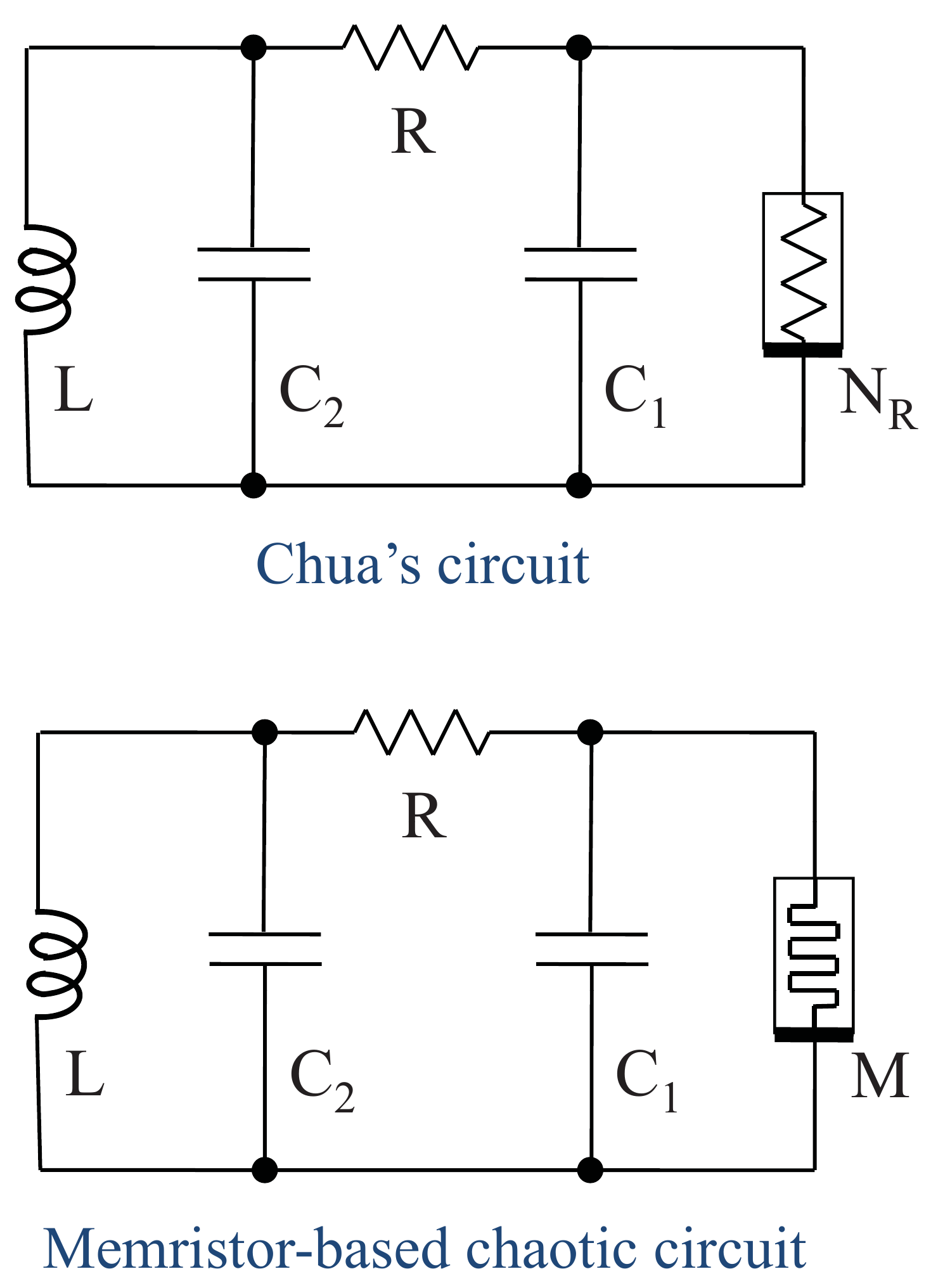}
\caption{Original Chua circuit and chaotic memristive circuit using an {\it active} memristor
\cite{Muthuswamy09a}.} \label{chaoscirc}
\end{center}
\end{figure}

Moreover, several authors have investigated chaotic circuits with
memristors
\cite{Itoh08a,Muthuswamy09a,Zhong10a,Bao10b,bao10a,messias10a}.
For the most part modified Chua's oscillators
\cite{Matsumoto84a,chuabook}
 were considered
\cite{Itoh08a,Muthuswamy09a,Bao10b,bao10a}. In such an approach,
the active nonlinear resistance element, Chua's diode
\cite{chuabook}, is replaced by an {\it active} memristor (as we show in
figure \ref{chaoscirc}) and novel features in chaotic behavior are
observed \cite{Itoh08a,Muthuswamy09a,Bao10b,bao10a}. An active memristor is one for which Eq.~(\ref{energM})
is not satisfied. In fact, one such circuit was implemented experimentally \cite{Muthuswamy10a}.
 We have, however, recently shown that
chaotic behavior does not require active memristors and even a single memristive system driven by an ac voltage
can manifest chaos \cite{Driscoll10b}. Moreover, chaos is observed when an ac-voltage is applied to the bistable membrane memcapacitive system considered in section \ref{sec:elasticmemc} \cite{pershin10a}. Apart from
interesting fundamental studies of non-linear systems, the field
of applications of such circuits includes secure communications
with chaos \cite{Muthuswamy09a}.

\section{Conclusions and outlook}
\label{sec:conclusions}

In this review we have shown that memory effects are ubiquitous in complex materials and nanoscale systems. This memory manifests itself under several
physical properties - related to charge, spin, and structural dynamics - which, however, can be generally categorized as memristive, memcapacitive and meminductive, or a combination of these. The advantage of these elements is
based on a combination of a history-dependent behavior with
properties of basic circuit elements - such as resistors,
capacitors or inductors.
The realization of this
link between the underlying physical properties that lead to memory and their description in terms of memory circuit elements opens up unprecedented opportunities
in both fundamental science and applications, since it guides the discovery of novel functionalities, inspires new concepts and connections between apparently diverse fields, and in certain
cases it also
allows a reconsideration of old concepts from a totally new perspective. \cite{Cagli09a}

We have provided a large set of examples of systems where these memory effects are observed or expected. Clearly, due to the necessary space limitations of this review we are far from
having exhausted all possible physical systems and devices that show these features. Indeed, as we have also emphasized in this review, we firmly think that due to the continued
miniaturization of devices, many more systems will be discovered that show memory. This confidence is supported by the physical fact that the change of state of electrons
and ions is not instantaneous, rather it generally depends on the past
dynamics.

We have also discussed many applications, ranging from information storage to learning/programmable circuits to biologically-inspired systems. In some cases, we also expect that
the cost of certain practical implementations of these elements should be relatively low in view of their simple structure. It is thus not too unlikely that some of the
systems we have discussed in this review may develop into commercially available products in the coming years, and find their own niche of applications. We therefore anticipate that novel exciting applications will be developed with beneficial impact not only for technology
but also for fundamental science.

\section*{Acknowledgments}

This work has been partially funded by the NSF grant
No. DMR-0802830.

\bibliographystyle{apsrmp}
\bibliography{memcapacitor}

\end{document}